%% file: scet_sv.tex
\documentclass[11pt,letter]{article}
\usepackage{jheppub,bm,booktabs,multirow}

\usepackage{axodraw4j,psfrag}
\usepackage{graphicx}
\usepackage{amssymb}
\usepackage{amsmath}
\usepackage{latexsym}
  
\makeatletter
\def\@fpheader{~}
\makeatother

\allowdisplaybreaks

\title{Introduction to Soft-Collinear Effective Theory}

\author[a]{T.~Becher}
\author[b]{A.~Broggio}
\author[c,d]{A.~Ferroglia}

\affiliation[a]{Albert Einstein Center for Fundamental Physics, Institut f\"ur Theoretische Physik, \\ Universit\"at Bern,
  Sidlerstrasse 5, CH-3012 Bern, Switzerland}

\affiliation[b]{Paul Scherrer Institut, CH-5232 Villigen PSI, Switzerland}

\affiliation[c]{Physics Department, New York City College of Technology,\\ The City University of New York,
 NY 11201 Brooklyn, USA}

\affiliation[d]{The Graduate School and University Center,\\
The City University of New York,
NY 10016 New York, USA}

\abstract{These lectures provide an introduction to Soft-Collinear
Effective Theory. After discussing the expansion of Feynman diagrams around the high-energy limit, the effective Lagrangian is constructed, first for a scalar theory, then for QCD. The underlying concepts are illustrated with the Sudakov form factor, i.e.\ the quark vector form factor at large momentum transfer. We then apply the formalism in two examples: We perform soft gluon resummation as well as transverse-momentum resummation for the Drell-Yan process using renormalization group evolution in SCET, and we derive the infrared structure of $n$-point gauge theory amplitudes by relating them to effective theory operators. We
conclude with an overview of the different applications of the effective theory.}

\emailAdd{becher@itp.unibe.ch}
\emailAdd{alessandro.broggio@psi.ch}
\emailAdd{aferroglia@citytech.cuny.edu}


\arxivnumber{1410.1892}

\begin{document}

\maketitle

\newcommand{\Dsl}{D\!\!\!\!\slash}
\newcommand{\MDsl}{{\mathfrak D}\!\!\!\!\slash}
\newcommand{\psl}{p\!\!\!\slash}
\newcommand{\nsl}{n\!\!\!\slash}
\newcommand{\ksl}{k\!\!\!\slash}
\newcommand{\nbsl}{\bar{n}\!\!\!\slash}
\newcommand{\scol}[1]{\Mahogany{#1}}
\newcommand{\lcol}[1]{\OliveGreen{#1}}
\newcommand{\pcol}[1]{\Blue{#1}}
%
\newcommand{\cc}{c}
\newcommand{\cb}{\bar{c}}
\newcommand{\be}{\begin{equation}}
\newcommand{\ee}{\end{equation}}
\newcommand{\bfm}[1]{\mbox{\boldmath$#1$}}
\newcommand{\bff}[1]{\mbox{\scriptsize\boldmath${#1}$}}
\newcommand{\al}{\alpha}
\newcommand{\bt}{\beta}
\newcommand{\lm}{\lambda}
\newcommand{\bare}{\mbox{{\tiny bare}}}
\newcommand{\cusp}{\mbox{{\tiny cusp}}}
\newcommand{\bea}{\begin{eqnarray}}
\newcommand{\eea}{\end{eqnarray}}
\newcommand{\bd}{\begin{displaymath}}
\newcommand{\ed}{\end{displaymath}}
\newcommand{\gm}{\gamma}
\newcommand{\Gm}{\Gamma}
\newcommand{\dl}{\delta}
\newcommand{\Dl}{\Delta}
\newcommand{\ep}{\varepsilon}
\newcommand{\vep}{\varepsilon}
\newcommand{\kp}{\kappa}
\newcommand{\Lm}{\Lambda}
\newcommand{\om}{\omega}
\newcommand{\pa}{\partial}
\newcommand{\nn}{\nonumber}
\newcommand{\dd}{\mbox{d}}
\newcommand{\grtsim}{\mbox{\raisebox{-3pt}{$\stackrel{>}{\sim}$}}}
\newcommand{\lessim}{\mbox{\raisebox{-3pt}{$\stackrel{<}{\sim}$}}}
\newcommand{\uk}{\underline{k}}
\newcommand{\gsim}{\;\rlap{\lower 3.5 pt \hbox{$\mathchar \sim$}} \raise 1pt \hbox {$>$}\;}
\newcommand{\lsim}{\;\rlap{\lower 3.5 pt \hbox{$\mathchar \sim$}} \raise 1pt \hbox {$<$}\;}
\newcommand{\Li}{\mbox{Li}}
\newcommand{\bc}{\begin{center}}
\newcommand{\ec}{\end{center}}
\newcommand{\ms}{{\overline{\mbox{{\rm MS}}}}}
\def\ff{f\hspace{-0.3cm}f}
\def\muf{\mu_f}
\def\muh{\mu_h}
\def\mui{\mu_s}

\def\lapprox{\lower .7ex\hbox{$\;\stackrel{\textstyle <}{\sim}\;$}}
\def\gapprox{\lower .7ex\hbox{$\;\stackrel{\textstyle >}{\sim}\;$}}

\makeatletter 
\renewcommand{\thefigure}{\@arabic\c@section.\@arabic\c@figure}
\makeatother

\newpage
\input{preface}

\newpage

\setcounter{page}{1}
\pagenumbering{arabic}

\input{1_introduction_sv}

\newpage

\setcounter{figure}{0}
\input{2_regions}

\newpage

\setcounter{figure}{0}
\input{3_scalarSCET}

\newpage

\setcounter{figure}{0}
\input{4_scetQCD}

\newpage

\setcounter{figure}{0}
\input{5_RGevolution}

\newpage

\setcounter{figure}{0}
\input{6_DY}
\newpage

\setcounter{figure}{0}
\input{7_pTresummation}

\newpage

\setcounter{figure}{0}
\input{8_IRDivergences}
\newpage

\setcounter{figure}{0}
\input{9_Outlook}

\newpage

%
\newpage
\appendix
\makeatletter 
\renewcommand{\thefigure}{\@Alph\c@section.\@arabic\c@figure}
\makeatother

\input{Appendix}

%
\newpage
\bibliography{bibscet}
\bibliographystyle{JHEP-2}

\end{document}

%% file: preface.tex
\section*{Preface}

Soft-Collinear Effective field Theory  (SCET) is the youngest member of the large family of low-energy effective field theories of the Standard Model and has been developed over the last fifteen years. By now, the effective theory has been applied to a large variety of processes, from $B$-meson decays to jet production at the LHC, but unfortunately there is still no introductory text available. The present work tries to fill this gap by providing an elementary and pedagogical introduction to the subject.

The original papers \cite{Bauer:2000ew,Bauer:2000yr,Bauer:2001ct,Bauer:2001yt,Beneke:2002ph,Beneke:2002ni,Hill:2002vw} are recommended reading, but not optimal as a first introduction for several reasons: They assume familiarity with effective field theory methods, in particular heavy-quark effective theory, and focus on $B$-physics applications. On top of this, different formalisms and notations are commonly used, and interested readers will need to work their way through several papers to become familiar with all of the ingredients necessary for an understanding of SCET. In in his recent book on the foundations of perturbative QCD, John Collins wrote that he found the SCET literature to be impenetrable, and we have heard similar complaints from other QCD experts. We hope that this self-contained introduction will alleviate these difficulties and make SCET accessible also to researchers outside the effective-field-theory community. We vividly remember our own difficulties in understanding SCET when we first studied it and have tried our best to make the subject accessible. Obviously, however, a course on quantum field theory and basic knowledge of perturbative QCD are prerequisites for these lectures.

The structure of the book is derived from the syllabus of a series of lectures given by one of us (T.B.) at the University of Z\"urich in 2010 and at Technische Universit\"at Dresden in 2011. When turning these lectures into a manuscript, we decided to include many detailed derivations and computations, to allow the reader to focus attention on the logic underlying SCET without being distracted and delayed by the need to reconstruct algebraic steps. On the other hand, in order not to overburden the text, we have relegated some of the computations to appendices, in particular explicit evaluations of loop integrals. The appendices also provide introductions to auxiliary topics such as Wilson lines and the color-space formalism, which will be familiar to expert readers, but are beyond most quantum field theory text books. Compared to the original lectures, we have expanded the scope to also include some more recent developments. On the other hand, to keep the text short and self contained we focus mostly on the basic formalism and have included only a small number of applications, aimed at illustrating how SCET is used in practice. To mitigate the impact of the choice of material we had to make, we conclude the book with a brief non-technical review of the many applications of SCET to $B$-physics and collider physics problems. This review can be found in Section~\ref{sec:OL} and should enable the reader to navigate the original literature.

Finally, we would like to thank a number of colleagues.
First and foremost, a special thank you goes to Matthias Neubert; while he was not involved in the writing of this introduction, he should nevertheless be considered a co-author: a lot of the understanding of SCET of the authors is due to collaborations with him and a lot of the material in this introduction is due to common work. 

In addition, T.B.\  would like to thank Martin Beneke and Dave Soper for discussions, Guido Bell, Ilya Feige, Xavier Garcia Tormo and Matt Schwartz for discussions and comments on the manuscript and Antonia Adler, Silvan Etter, Monika Hager, Stefanie Marti, Jan Piclum, Lorena Rothen and Arash Yunesi for pointing out typographical mistakes in earlier versions of the text.
A.B.\ would like to thank Robert Schabinger and Andrea Visconti for pointing out some misprints.
A.F.\ would like to thank Thomas L\"ubbert, Gil Paz, Ben Pecjak, and Lilin Yang for many useful discussions and clarifications which are reflected in part in the appendices of this work.  

T.B.\ acknowledges support by the Swiss National Science Foundation (SNF) under grant 200020-140978
and by the Munich Institute for Astro- and Particle Physics (MIAPP) of the DFG cluster of excellence ``Origin and Structure of the Universe". The research activity of A.F.\ is supported in part by the  National Science Foundation Grant No.~PHY-1068317 and No.~PHY-1417354.

\begin{flushright}
\vspace*{0.8cm}
\begin{minipage}{0.32\textwidth}
Thomas Becher\\
Alessandro Broggio\\
Andrea Ferroglia\\
\emph{Bern, Villigen, New York,\\
November 2014}
\end{minipage} 
\end{flushright}

\vspace*{0.8cm}

We would also like to thank Marcel Balsiger, Yaroslav Balytskyi, Monika Hager, Samuel Favrod, Ding Yu Shao, Kai Urban and Arash Yunesi for pointing out typos in the book which are now corrected in this new arXiv version.

%% file: 1_introduction_sv.tex
\section{Introduction}

Effective field theories (EFTs) are used whenever one encounters problems with two disparate scales, a high-energy scale  $\Lambda_h$ and a lower scale  $\Lambda_l$, in quantum field theory. EFTs allow one to expand physical quantities in the small ratio of the scales and to separate the low-energy contributions from the high-energy part. Performing the expansion usually greatly simplifies the problem and is often necessary in order to be able to attack a field-theory problem in the first place. In Quantum Chromodynamics (QCD), the low-energy part is usually non-perturbative, while the high-energy contribution can be computed perturbatively. Using an EFT one is able to separate the two pieces and compute them with appropriate techniques. For hadron-collider observables, the leading non-perturbative low-energy part is typically encoded in the parton distribution functions. However, even in cases where all scales in a given problem are in the perturbative domain, it is necessary to separate the contributions associated with different scales. If this is not done, higher-order corrections are enhanced by large logarithms of the scale ratios. In many physical problems, the leading logarithms at $n$-th order in perturbation theory are of the form $\alpha_s^n \ln^n(\Lambda_h/\Lambda_l)$, where $\alpha_s$ is the strong interaction coupling constant. The situation is different for processes described in Soft-Collinear Effective Theory (SCET) \cite{Bauer:2000ew,Bauer:2000yr,Bauer:2001ct,Bauer:2001yt,Beneke:2002ph,Beneke:2002ni,Hill:2002vw}, which involve energetic particles. In this case one encounters two logarithms for each power of the coupling constant, so that the leading logarithmic terms have the form $\alpha_s^n \ln^{2n}(\Lambda_h/\Lambda_l)$. These logarithms are also called Sudakov logarithms and were first observed in the electron form factor at large momentum transfer. 

While EFTs are commonly used in low-energy QCD, in particular in flavor physics, their application to high-energy processes is still fairly new. This is surprising, since processes at high-energy colliders are prime examples of multi-scale problems. A typical process at a hadron collider involves physics from large scales, such as the center-of-mass energy or the transverse momentum of a jet, down to very low scales such as the proton mass. Without disentangling the physics associated with these scales, it would be hopeless to try to obtain theoretical predictions for any such process. However, traditionally, this factorization is achieved with diagrammatic methods. From an analysis of the Feynman diagrams in the high-energy limit, one establishes that certain properties hold to all orders of perturbation theory. Based on such factorization theorems, also the resummation of Sudakov logarithms can be achieved. Reviews of the traditional diagrammatic techniques include \cite{Collins:1989gx} as well as the recent book \cite{Collins:2011zzd}. SCET provides an alternative formalism which allows one  to derive these factorization theorems and to perform the resummation of Sudakov logarithms. The use of  an effective Lagrangian makes it easier to derive the consequences of gauge invariance, which are not manifest on the level of the individual diagrams. The Lagrangian approach also provides a powerful tool for the resummation of Sudakov logarithms, which can be achieved using renormalization group (RG) evolution in the EFT. An effective Lagrangian provides a simple and systematic way of organizing computations. For more complex problems, such as power corrections, a purely diagrammatic approach seems prohibitively difficult. While we believe that the EFT approach has important advantages, we want to stress the close connection between the traditional approach and the diagrammatic techniques: the diagrams of SCET are in one-to-one correspondence with the expanded QCD diagrams. In fact, in our introduction we start by expanding diagrams around the high-energy limit and then build the Lagrangian such that the expansion of the diagrams is recovered. Let us also stress that an effective Lagrangian does not prevent one from making mistakes. SCET has not only been used to rederive results obtained earlier with traditional methods, but also to repeat previous mistakes and to come up with new ones: deriving all-order statements about perturbation theory is never a trivial task, irrespective of the formalism employed.

The following text aims to present the SCET basics in detail and then to illustrate them by means of a few sample applications. Starting from the expansion of Feynman diagrams describing the production of energetic particles, an effective Lagrangian is constructed which produces the different terms that contribute to the expanded diagrams. The technique we use for the expansion is called the strategy of regions and is based on dimensional regularization. There are two different low-energy regions contributing in processes with energetic particles. The particles can split into collinear particles and can emit soft particles. For this reason, and as its name suggests, SCET includes different low-energy fields, which describe the collinear and the soft emissions. The fact that the same QCD field is represented by different fields in the low-energy theory is a somewhat uncommon feature and makes SCET more complicated than other EFTs. In order  to simplify the construction of the relevant effective theory, we therefore first consider the case of a scalar theory before turning to QCD. We analyze the Sudakov problem in $\phi^3$ scalar theory, check that we reproduce the full theory result at one-loop order and then derive a factorization theorem for the $\phi^3$ form factor in $d=6$. After this, we extend the construction to QCD. The main complications compared to the scalar case are gauge invariance and the fact that different components of the gauge and quark fields scale with different powers of the expansion parameter. Gauge invariance leads to the appearance of Wilson lines. Also in the QCD case, we use the Sudakov form factor as an explicit example and show how the Sudakov logarithms can be resummed using RG techniques.

The main focus of this introduction is to explain the construction of the effective theory in detail. However, to see the method at work, we also include two example applications.  Since many of the applications of SCET  in the last few years were in collider physics, we choose our examples in this field. The first application concerns soft-gluon resummation for the inclusive Drell-Yan cross section $pp \to \gamma^*/Z + X \to \ell^+\ell^- + X$. This is one of the most basic processes at hadron-colliders, and one of the first for which resummation was performed in the traditional framework as well as in SCET \cite{Becher:2007ty}. 
During the last few years, important progress has been made to analyze also processes sensitive to small transverse momenta or small masses in the effective theory \cite{Becher:2010tm,Chiu:2011qc,Becher:2011dz,Chiu:2012ir}. In these cases, the SCET diagrams suffer from unregularized light-cone singularities in the individual sectors of the theory. These cancel when the different contributions are added, but need to be regularized at intermediate stages and lead to implicit dependence of the low-energy part on the high-energy scale. The structure of this collinear anomaly is understood to all orders and the corresponding formalism has been used to perform higher-log resummations. Therefore, we also discuss the application of this formalism to transverse-momentum resummation for the Drell-Yan process.

As a second application of SCET methods, we consider a process with energetic particles in many different directions. From the analysis of this process, one can derive the structure of infrared (IR) singularities in $n$-point gauge-theory amplitudes. Such singularities arise from regions where loop momenta become soft and collinear and can therefore be analyzed using SCET. Knowledge of these singularities provides a useful check on perturbative computations, and a necessary ingredient to perform Sudakov resummations for multi-jet processes.

We end our introduction with an overview of the different applications of the effective theory. These cover a wide range of topics, from heavy-quark physics, event shapes in $e^+e^-$ collisions, jet observables at hadron colliders, jet quenching in heavy-ion collisions, to decays of heavy dark-matter particles. We hope that this final chapter can serve as a guide to the SCET literature. Finally, the appendices of this work include a detailed discussion of Wilson lines, provide several detailed derivations of results needed in the main text, as well as a collection of perturbative results for the anomalous dimensions appearing in various renormalization group equations.

We remind the reader that our text aims to provide a first introduction to SCET and its applications to collider physics, rather than a comprehensive overview of the subject. While this choice limits the amount of material which can be presented, it allows us -- hopefully -- to write a self-contained and relatively brief introduction to the subject which should be accessible to graduate students with a background in quantum field theory.

%% file: 2_regions.tex
\section{The Strategy of Regions \label{sec:regions}}

The strategy of regions \cite{Beneke:1997zp} is a technique which allows one to 
carry out asymptotic expansions of loop integrals in dimensional regularization 
around various limits \cite{Smirnov:2002pj}. The expansion is obtained by splitting the integration in different regions and appropriately expanding the integrand in each case. In the effective theory, the different regions will be represented by different effective theory fields. The expanded integrals obtained by means of the strategy of regions technique are in one-to-one correspondence to the Feynman diagrams of  effective field theories regularized in dimensional regularization. 

If one is simply interested to expand some perturbative result in a small parameter, one can therefore work directly with the strategy of regions technique, without constructing an effective Lagrangian. However, the use an effective field theory offers some important advantages when one is interested in deriving all-order statements. In particular, one can use the effective Lagrangian 
\begin{itemize}
\item to derive factorization theorems and
\item to resum  logarithmically enhanced contributions at all orders in the coupling constant using Renormalization Group (RG) techniques.
\end{itemize}
In addition, in the effective field theory gauge invariance is manifest at the Lagrangian level, while this is not the case for individual diagrams. The effective Lagrangian also provides a systematic way to organize higher power corrections, by including subleading terms in the
effective Lagrangian. (In a collider physics context, higher-power contributions are also called higher twist corrections.)

\subsection{A Simple Example\label{sec:ASE}}

In order to illustrate the main idea of the strategy of regions we start by considering a simple integral, which we will expand using different methods, first using a cutoff to separate two different regions and then with dimensional regularization. The integral we will consider is 
\be \label{eq:ex}
I = \int_0^\infty dk \frac{k}{ (k^2+m^2) (k^2 +M^2)} = \frac{\ln{\frac{M}{m}}}{M^2 -m^2} \, .
\ee
This corresponds to a self-energy one-loop integral with two different particle masses at zero external momentum, evaluated in $d=2$. We will assume a large hierarchy between the masses, for example $m^2 \ll M^2$, and will discuss the expansion of the integral around the limit of small $m$. Since we know the full result, we can obtain the expansion simply by expanding the denominator on the r.h.s. of Eq.~(\ref{eq:ex})
\be \label{eq:exexp}
I = \frac{\ln{\frac{M}{m}}}{M^2} \left( 1 + \frac{m^2}{M^2} + \frac{m^4}{M^4} + 
\cdots \right) \, .
\ee
Note that the integral is not analytic in the expansion parameter $m/M$ because of the presence of the logarithm. Expansions of functions around points where they have essential singularities are also called asymptotic expansions. Our goal in the following is to obtain the expansion in Eq.~(\ref{eq:exexp}) by expanding the integrand in Eq.~(\ref{eq:ex}) before carrying out the integral. This is important in cases where the full result is not available. It will also tell us what kind of degrees of freedom the effective theory will contain. 

A naive expansion of the integrand leads to trouble, because it gives rise to IR divergent integrals. In fact
\be \label{eq:intexp}
\frac{k}{ (k^2+m^2) (k^2 +M^2)} = \frac{k}{ k^2 (k^2 +M^2)} \left(1 - \frac{m^2}{k^2} + \frac{m^4}{k^4} + \cdots  \right) \, 
\ee
cannot be used in the integrand of Eq.~(\ref{eq:exexp}):
\be \label{eq:naiveexp}
I \neq \int_0^\infty dk\, \frac{k}{ k^2 (k^2 +M^2)} \left(1 - \frac{m^2}{k^2} + \frac{m^4}{k^4} + \cdots  \right) \, .
\ee
This was to be expected: If it had been legitimate to simply Taylor expand the integrand in $m/M$ and integrate term by term, the result would necessarily be an analytic function of $m$ in the vicinity of $m=0$ because none of the integrals on the r.h.s. of Eq.~(\ref{eq:naiveexp}) depend on $m$ and so the integrals would simply give the Taylor coefficients of the expansion in $m$. But the result for $I$ is not analytic in $m/M$, as we stressed above. So just from the form of the result in Eq.~(\ref{eq:exexp}), it is clear that expansion and integration do not commute. The reason is simply that the series expansion in Eq.~(\ref{eq:intexp}) is valid only for $k \gg m^2$, while the integration domain in Eq.~(\ref{eq:ex}) includes a region in which $k^2 \sim m^2$, which contributes to the integral. To account for this fact, we should split the integration into two regions. We can do this by introducing a new scale $\Lambda$ such that $m \ll \Lambda \ll M$. We will call the scale $\Lambda$ a cutoff, even though the name is misleading, since we do not cut away any part of the integral. The role of $\Lambda$ is just to separate the two momentum regions. We then obtain
\be \label{eq:Iregions}
I  = \underbrace{\int_0^\Lambda dk \frac{k}{ (k^2+m^2) (k^2 +M^2)}}_{I_{(I)}} +
\underbrace{\int_\Lambda^\infty dk \frac{k}{ (k^2+m^2) (k^2 +M^2)}}_{I_{(II)}} \, .
\ee

We call the region $[0,\Lambda]$  the {\em low-energy} region.
 In this region $ k \sim m \ll M$, and therefore one can expand the integrand in the integral $I_{(I)}$ as follows
\be\label{eq:I_I}
I_{(I)} = \int_0^\Lambda dk \frac{k}{ (k^2+m^2) (k^2 +M^2)} = 
\int_0^\Lambda dk \frac{k}{ (k^2+m^2) M^2}\left(1 - \frac{k^2}{M^2} + \frac{k^4}{M^4} + \cdots \right) \, .
\ee
The scale $\Lambda$ acts as an ultraviolet cutoff for the integrals on the 
r.h.s.\ of the Eq.~(\ref{eq:I_I}).

The region $[\Lambda,\infty]$ is referred to as the {\em high-energy} region; in that region $m \ll k \sim M$, and one can expand the integrand according to 
\be\label{eq:I_II}
I_{(II)}  = \int_\Lambda^\infty dk \frac{k}{ (k^2+m^2) (k^2 +M^2)} = 
\int_\Lambda^\infty dk \frac{k}{ k^2 (k^2 +M^2)}\left(1 - \frac{m^2}{k^2}
+\frac{m^4}{k^4} +\cdots \right) \, .
\ee
In the equation above, $\Lambda$ acts as an infrared cutoff.

By integrating the first two terms on the r.h.s.\ of Eq.~(\ref{eq:I_I}) one finds
\be \label{eq:Ires}
I_{(I)} \approx \frac{M^2+m^2}{2 M^4} \ln{\left(1+\frac{\Lambda^2}{m^2} \right)} - \frac{\Lambda^2}{2 M^4} =
- \frac{1}{M^2}\ln{\left(\frac{m}{\Lambda}\right)} - \frac{\Lambda^2}{2 M^4} + {\mathcal O}\left(\frac{\Lambda^4}{M^6},\frac{m^2}{M^4} \log\left(\frac{\Lambda}{m}\right)\right) \, ,
\ee
since it was assumed above that $\Lambda \gg m$.
Similarly, by integrating the first term on the r.h.s.\ of Eq.~(\ref{eq:I_II})
one obtains
\be \label{eq:IIres}
I_{(II)} \approx \frac{1}{2 M^2} \ln{\left(1+ \frac{M^2}{\Lambda^2}\right)} =
-\frac{1}{M^2}\ln{\left(\frac{\Lambda}{M}\right)} + \frac{\Lambda^2}{2 M^4}
+ {\mathcal O}\left(\frac{\Lambda^4}{M^6}\log\left(\frac{M}{\Lambda}\right) \right)\, . 
\ee

Adding up the Eq.~(\ref{eq:Ires}) and (\ref{eq:IIres}) one finally obtains
\be \label{eq:cutoffres}
I  = I_{(I)} + I_{(II)} = -\frac{1}{M^2}\ln{\left(\frac{m}{M}\right)} +
{\mathcal O}\left(\frac{m^2}{M^4}\log\left(\frac{M}{m}\right) \right)\, , 
\ee
which is the expected result (see Eq.~(\ref{eq:exexp})). When summing the results for the low-energy and high-energy regions, the terms which depend on the cutoff $\Lambda$ cancel out; this has to happen, since the scale $\Lambda$ is not present in the original integral and was only introduced in order to split the original integral in a sum of two different terms. Since the final result cannot depend on $\Lambda$, there should be a way to obtain the expansion without introducing this additional scale. Our ultimate goal is to apply a similar technical expedient to the calculation of loop diagrams and it is well known that the use of hard cutoffs is impractical in such calculations. Fortunately it is possible to separate  the low- and high-energy regions using dimensional regularization. To see this, let us rewrite the original integral as follows
\be\label{eq:exDR}
I = \int_0^\infty dk\, k^{-\varepsilon} \frac{k}{ (k^2+m^2) (k^2 +M^2)} \, ,
\ee
where  we will eventually send $\varepsilon \to 0$ at the end of the calculation. (For simplicity, we did not introduce the $d$-dimensional angular integration so this is not exactly dimensional regularization.)

The integral in the low-energy region $k \sim m \ll M$ will be
\be \label{eq:rIDR}
I_{(I)} =
\int_0^\infty dk \,  k^{-\varepsilon}\frac{k}{ (k^2+m^2) M^2}\left(1 - \frac{k^2}{M^2} + \frac{k^4}{M^4} + \cdots \right) \, .
\ee
In Eq.~(\ref{eq:rIDR}) the integral is infrared safe in the region in which
$k \to 0$, the dimensional regulator $\varepsilon$ can be chosen positive, so
that the integrand is also ultraviolet finite. The integral in the high-energy
region will be 
\be\label{eq:rIIDR}
I_{(II)} =\int_0^\infty dk\,  k^{-\varepsilon} \frac{k}{ k^2 (k^2 +M^2)}\left(1 - \frac{m^2}{k^2}
+\frac{m^4}{k^4} +\cdots \right) \, .
\ee
The integral is ultraviolet safe, and we consider $\varepsilon < 0$, so that the integrand does not give rise to an infrared singularity in the region
where $k \to 0$.
By integrating the first term on the r.h.s.\ of Eq.~(\ref{eq:rIDR}) one finds, at leading power in the expansion around $m/M$,
\be \label{eq:IDRex}
I_{(I)} = \frac{m^{-\varepsilon}}{2 M^2}
\Gamma\left(1-\frac{\varepsilon}{2} \right) \Gamma\left(\frac{\varepsilon}{2}\right) = 
\frac{1}{M^2} \left(\frac{1}{\varepsilon} - \ln{m} +{\mathcal O}(\varepsilon) \right) \, .
\ee
The integral of the first term on the r.h.s.\ of Eq.~(\ref{eq:rIIDR}) is
\be \label{eq:IIDRex}
I_{(II)} = - \frac{ M^{-\varepsilon}}{2 M^2}
\Gamma\left(1-\frac{\varepsilon}{2} \right) \Gamma\left(\frac{\varepsilon}{2}\right) = 
\frac{1}{M^2} \left(-\frac{1}{\varepsilon} + \ln{M} +{\mathcal O}(\varepsilon) \right) \, .
\ee
The poles in $\varepsilon$ cancel in the sum of Eqs.~(\ref{eq:IDRex},\ref{eq:IIDRex}),
and the final result is again the one obtained by means of the cutoff method
in Eq.~(\ref{eq:cutoffres}).
The reader might be worried that we choose $\varepsilon > 0$ in the low-energy region and $\varepsilon < 0$ in the high-energy region and then combine the two. It is important to remember that the integrals in dimensional regularization are defined for arbitrary $\varepsilon$: we only choose $\varepsilon > 0$ to be able to evaluate $I_{(I)}$ as a standard integral, but by analytic continuation the resulting function on the right-hand side is uniquely defined for any complex-valued $\varepsilon$ and can be combined with $I_{(II)}$.

Also, the fact that in both Eq.~(\ref{eq:rIDR}) and Eq.~(\ref{eq:rIIDR}) the integration domain coincides with the full integration domain of the original integral might seem disturbing at first sight. Since we integrate the high-energy part over the low-energy region (and vice versa), one could fear that this leads to additional contributions which are already accounted for in the low-energy part. To see that this does not happen and that the two parts lead a life of their own, one should observe that the two integrals scale differently. The low-energy integral  $I_{(I)}$ factors out $m^{-\varepsilon}$, while the high-energy integral $I_{(II)}$ factors out
$M^{-\varepsilon}$. This statement remains true even if we consider the subleading terms. When keeping the complete dependence on $m$ and $M$
the result is
\be
I  = \frac{1}{2}\Gamma\left(1-\frac{\varepsilon}{2} \right) \Gamma\left(\frac{\varepsilon}{2}\right)
\,\frac{m^{-\varepsilon} - M^{-\varepsilon}}{M^2-m^2} \, .
\ee
The result clearly displays the low-energy and the high-energy part. Expanding in one region, one loses the other part and the full integral is recovered after adding the two contributions. Even though we integrate twice over the full integration domain, there is no double counting, since the two pieces scale differently: the low-energy integrals can never produce a term  $M^{-\varepsilon}$ since they depend analytically on the large scale, and vice-versa.

To demonstrate directly from the integral  that there is indeed no double counting, let us now see what happens if we insist in restricting the integration domain of the low- and high-energy region integrals when using dimensional regularization. The integral in the low-energy region would become in this case
\bea
I_{(I)}^\Lambda &=& 
\int_0^\Lambda dk \,  k^{-\varepsilon}\frac{k}{ (k^2+m^2) M^2}\left(1 - \frac{k^2}{M^2} + \frac{k^4}{M^4} + \cdots \right) \,  \nn \\
&=& \left[\int_0^\infty dk  - \int_\Lambda^\infty dk \right]k^{-\varepsilon}\frac{k}{ (k^2+m^2) M^2}\left(1 - \frac{k^2}{M^2} + \frac{k^4}{M^4} + \cdots \right) \,  \nn \\
& = & I_{(I)} - R_{(I)}\,.
\eea
The first integral in the second line of the equation above is the same as
the one in Eq.~(\ref{eq:rIDR}). In the integrand of $R_{(I)}$, which depends on the cutoff $\Lambda$, one can use the fact that $k \ge \Lambda \gg m^2$ to expand in the small $m$ limit:
\bea
R_{(I)} &=& \int_\Lambda^\infty dk k^{-\varepsilon}\frac{k}{ (k^2+m^2) M^2}\left(1 - \frac{k^2}{M^2}+  \cdots \right)\nn \\
& = & \int_\Lambda^\infty dk k^{-\varepsilon}\frac{k}{ k^2 M^2}\left(1 -\frac{m^2}{k^2} - \frac{k^2}{M^2}+  \cdots \right) \, .
\eea
For the remainder part $R_{(I)}$, we thus have performed two expansions. First the low-energy expansion, which is equivalent to expanding the integrand in the limit $M\to \infty$.  Then we have expanded the result around $m\to 0$, which is equivalent to the high-energy expansion.
At this point it is sufficient to observe that for dimensional reasons the integrals in the equation above must behave as follows
\be
\int_\Lambda^\infty dk\, k^{n-\varepsilon} \sim \Lambda^{n+1-\varepsilon} \, .
\ee
So the cutoff pieces scale as fractional powers of the cutoff. Since the $\Lambda$ dependent terms must cancel out completely in the calculation of $I$, one can as well drop the $\Lambda$ dependent integrals from the start. Therefore, when regulating divergences by means of dimensional regularization one can integrate over the complete integration domain, in this case  $k \in [0,\infty]$.

We can explicitly verify that the cutoff pieces vanish if we also consider the high-energy integral $I_{(II)}$ in Eq.~(\ref{eq:rIIDR}) with a lower cutoff $\Lambda$ on the integration. Proceeding in the same way as before, we can rewrite the high-energy integral as the expanded integral without a cutoff and a remainder which depends on the cutoff
\bea
R_{(II)} &=& \int_0^\Lambda dk k^{-\varepsilon}\frac{k}{k^2 (k^2+M^2) }\left(1 - \frac{m^2}{k^2}+  \cdots \right)\nn \\
& = & \int_0^\Lambda dk k^{-\varepsilon}\frac{k}{ k^2 M^2}\left(1 -\frac{m^2}{k^2} - \frac{k^2}{M^2}+  \cdots \right) \, .
\eea
In this remainder, we have again expanded the integrand in both the limit of small $m$ and also in the limit of large $M$, but in the opposite order as in $R_{(I)}$. However, the two expansions commute so that the integrands of $R_{(I)}$ and $R_{(II)}$ are identical. Adding up the two pieces, we find that
\begin{equation}\label{eq:R}
R = R_{(I)}+ R_{(II)} = \int_0^\infty dk k^{-\varepsilon}\frac{k}{ k^2 M^2}\left(1 -\frac{m^2}{k^2} - \frac{k^2}{M^2}+  \cdots \right) \,.
\end{equation}
This is manifestly independent of the cutoff. It is also manifestly zero, because it is given by a series of scaleless integrals. In the context of SCET, the overlap contribution $R$ is usually referred to as the ``zero-bin'' contribution \cite{Manohar:2006nz}, a name which will become clear when we discuss the label formalism in Section~\ref{sec:labelform}. There are two ways of obtaining the full overlap $R$. One can either expand the integrand of the high-energy integral $I_{(II)}$ around the low-energy limit, or the integrand of the low-energy integral $I_{(I)}$ around the high-energy limit. Since the overlap is obtained by expanding the single-scale integrals $I_{(I)}$ or $I_{(II)}$ it is given by scaleless integrals which vanish in dimensional regularization.

\subsection{The Sudakov Problem}

The example considered in the previous section had the purpose of illustrating some common features of the expansion of Feynman diagrams in the simplest possible setting. The general strategy to obtain the expansion of a given Feynman integral in a given kinematic limit is the following \cite{Smirnov:2002pj}:
\begin{itemize}
\item[i)] Identify all regions of the integrand which lead to singularities
in the limit under consideration,
\item[ii)] Expand the integrand in each region and integrate each expansion over the full phase space.
\item[iii)] Add the result of the integrations over the different regions to obtain the expansion of the original full integral.
\end{itemize}
In order for the procedure to work, it is  necessary to make sure that all of the expanded integrals are properly regularized. Sometimes dimensional regularization alone is not sufficient to regularize the integrals in every region, and one might need to employ additional analytic regulators or to perform subtractions. Below, we will discuss the massive Sudakov form factor, which is an example where this is necessary. It is also important to consider each region only once to avoid double counting. As stated above, one needs to identify all regions of integration which lead to singularities. Often, this is a simple task and the regions which one encounters at one loop are the same which are relevant at higher order. However, there are examples
in which new regions must be added to the list when increasing the number of loops present in the diagram \cite{Smirnov:1999bza}. We also stress that there is so far no general proof that the above procedure always produces the correct result. Recent work towards such a proof can be found in \cite{Jantzen:2011nz}.


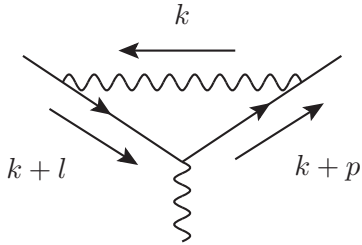
\begin{figure}[t]
\begin{center}
\vspace*{.5cm}
\[ 
\hspace*{1.5cm}
\vcenter{ \hbox{
  \begin{picture}(0,0)(0,0)
\SetScale{1}
\SetOffset(-22,0)
  \SetWidth{.8}

\ArrowLine(-60,40)(0,0)  
\ArrowLine(0,0)(60,40)  
\Photon(-45,30)(45,30){3}{9.5}
\Photon(0,0)(0,-30){3}{3.5}

\Text(-55,-7)[cb]{$k+l$}
\LongArrow(-50,20)(-20,1)
\Text(55,-7)[cb]{$k+p$}
\LongArrow(20,1)(50,20)
\LongArrow(20,42)(-20,42)
\Text(0,52)[cb]{$k$}
\end{picture}}
}
\]
\vspace*{.0cm}
\end{center} 
\caption{One-loop vertex corrections. The Feynman diagram is here shown in terms of fermions and photons, however, the spin structure is neglected in this section.} 
\label{fig:Sudakov}
\end{figure}


We want now to consider the simplest possible example relevant in the context of SCET, namely a one-loop vertex diagram. We neglect complications related to the spin of the particles, since the  momentum regions that one finds in the calculation of the tensor integrals are the same that one finds in the calculation of the scalar integral considered below. With reference to Figure~\ref{fig:Sudakov}, the vertex correction requires the evaluation of the 
following Feynman integral (all the internal propagators are considered  massless):
\be \label{eq:ISudakov}
I = i \pi^{-d/2} \mu^{4-d} \int d^d k \frac{1}{\left(k^2 +i0\right)
\left[(k+l)^2 +i 0 \right] \left[ (k+p)^2 +i 0 \right]} \, , 
\ee
where $d = 4 -2 \varepsilon$ is the dimensional regulator. The 't Hooft scale $\mu$ has been introduced to make the mass dimension of $I$ independent of the value of $d$. We introduce the following notation:
\be
L^2 \equiv -l^2 -i 0 \, , \qquad P^2 \equiv -p^2 -i 0 \,, \qquad Q^2 \equiv 
-(l-p)^2 -i 0 \, .
\ee
The goal is to calculate the integral in Eq.~(\ref{eq:ISudakov}) in the limit in which $L^2 \sim P^2 \ll Q^2$ that is, in the case in which the external legs carrying momenta $l$ and $p$ have large energies but small invariant masses.

Before going any further, we now need to introduce some basic notation used in SCET.
We choose two light-like reference vectors in the direction of the momenta $p$ and $l$ in the frame in which\footnote{In this lectures we employ the ``mostly minuses'' metric, and the components of a generic four-vector $x^\mu$ are $(t,x,y,z)$.} $\vec{Q} = 0$: 
\be
n_{\mu} = (1,0,0,1) \,  \quad \mbox{and} \quad \bar{n}_\mu = (1,0,0,-1) \, .
\ee
It is immediate to verify that
\be
n^2 = \bar{n}^2 = 0 \, , \quad \mbox{and} \quad n \cdot \bar{n} = 2 \, .
\ee
Any  vector can be then decomposed in a component proportional to $n$, a part proportional to $\bar{n}$, and a remainder perpendicular to both
\be
p^\mu = (n \cdot p) \frac{\bar{n}^\mu}{2} +  (\bar{n} \cdot p) \frac{n^\mu}{2} + p_\perp^\mu  \equiv p_+^\mu + p_-^\mu + p_\perp^\mu  \, .
\ee
Splitting the vectors into their light-cone components is useful to organize the expansion, since the different components scale differently. For the square of the vector $p$ one then finds
\be
p^2  = (n \cdot p)(\bar{n} \cdot p) + p_\perp^2 \, ,
\ee
while the scalar product between two vectors $p$ and $q$ becomes
\be
p \cdot q  = p_+ \cdot q_- +p_- \cdot q_+  + p_\perp \cdot q_\perp \, .
\ee

In the following we will often identify a vector by means of its components
in the $n$, $\bar{n}$, and $\perp$ basis, with the notation
\be
p^\mu = (\underbrace{n \cdot p}_{``+ \quad \mbox{comp.}"},
\underbrace{\bar{n} \cdot p}_{``- \quad \mbox{comp.}"},p_\perp^\mu) \, .
\ee
We warn the reader that in certain situations it is convenient to work with the scalar quantities
$p_+ \equiv n \cdot p$ and $p_- \equiv {\bar n} \cdot p$, which should not be mixed up with the 
related vector quantities $p_\pm^\mu$ introduced above. In the following we explicitly  indicate
what we mean by the symbols $p_\pm$ whenever the notation can give rise to ambiguities.

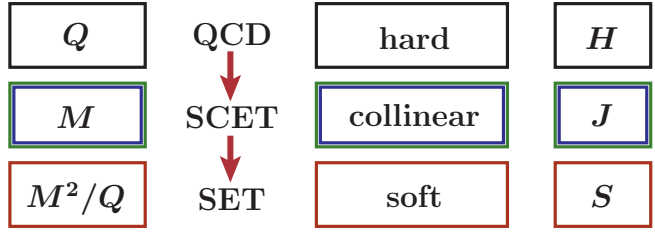
\begin{figure}[t]
\vspace*{.5cm}
\[ 
\hspace*{1.5cm}
\vcenter{ \hbox{
  \begin{picture}(0,0)(0,0)
\SetScale{0.6}
  \SetWidth{2.}

\CBoxc(-180,50)(85,40){Black}{White}  
\CBoxc(30,50)(120,40){Black}{White}  
\CBoxc(150,50)(60,40){Black}{White}  
  
\CBoxc(-180,0)(85,40){OliveGreen}{White} 
\CBoxc(30,0)(120,40){OliveGreen}{White} 
\CBoxc(150,0)(60,40){OliveGreen}{White} 
\CBoxc(-180,0)(80,35){Blue}{White} 
\CBoxc(30,0)(115,35){Blue}{White} 
\CBoxc(150,0)(55,35){Blue}{White}

\CBoxc(-180,-50)(85,40){Mahogany}{White}  
\CBoxc(30,-50)(120,40){Mahogany}{White}
\CBoxc(150,-50)(60,40){Mahogany}{White}
  

 \SetWidth{.8}

\Text(-109,24)[cb]{{\large \boldmath$Q$\unboldmath}}
\Text(19,25)[cb]{{\large {\bf hard}}}
\Text(-50,24)[cb]{{\large {\bf QCD}}}
\Text(90,25)[cb]{{\large \boldmath$H$\unboldmath}}

\Text(-109,-5)[cb]{{\large \boldmath$M$\unboldmath}}
\Text(19,-4)[cb]{{\large {\bf collinear}}}
\Text(-50,-5)[cb]{{\large {\bf SCET}}}
\Text(90,-4)[cb]{{\large \boldmath$J$\unboldmath}}

\Text(-109,-38)[cb]{{\large \boldmath$M^2/Q$\unboldmath}}
\Text(19,-35)[cb]{\large {\bf soft}}
\Text(-50,-36)[cb]{{\large {\bf SET}}}
\Text(90,-35)[cb]{{\large \boldmath$S$\unboldmath}}
  \SetWidth{3.8}
  \SetColor{Maroon}
\LongArrow(-84,40)(-84,17)
\LongArrow(-84,-13)(-84,-34)
\end{picture}}
}
\]
\vspace*{.8cm} 
\caption{Chart of regions and scales involved in the calculation. $Q$ indicates the hard scale, $M$ the scale characterizing collinear physics, and 
$M^2/Q$ the soft scale. SET stands for Soft Effective Theory.} 
\label{fig:scales1}
\end{figure}

We now introduce an  expansion parameter $\lambda$ which vanishes in the limit
in which we are interested in:
\be
\lambda^2 \sim \frac{P^2}{Q^2} \sim \frac{L^2}{Q^2} \, , \qquad 
\mbox{and} \qquad p^2 \sim l^2 \sim \lambda^2 Q^2 \, .
\ee
We choose the reference vectors in the directions of large momentum flow 
$p^\mu \approx Q n^\mu /2$ and $l^\mu \approx Q \bar{n}^\mu /2$. The components of $p$ and $l$ will then typically scale as follows
\be \label{eq:collScaling}
p^\mu \sim \left( \lambda^2, 1, \lambda \right) Q \, , 
\qquad \mbox{and} \qquad l^\mu \sim \left(1, \lambda^2, \lambda \right) Q \, ,
\ee
but the scaling is not completely unique. We could, for example, choose the reference vector $n^\mu$ such that the perpendicular components of $p^\mu$ are zero, which is compatible with Eq.~\eqref{eq:collScaling}, but also with $\left(1, \lambda^2, \lambda^n \right)\,Q$ for any $n>1$. However, when computing the loop diagram via the strategy of regions, one finds that only scalings $k^\mu \sim (\lambda^a,\lambda^b,\lambda^c) Q$, with $a+b=2c$ are important. For $c>0$,  these describe particles which go on shell as $\lambda\to 0$. In later sections, we will see that the corresponding propagators are associated with particles in the low-energy theory. Specifically, upon expanding the integrals, one finds that only the following four regions give non-vanishing contributions:
\begin{itemize}
\item {\bf Hard Region} (denoted by $h$ in the following) where the components of the integration momentum scale as $k^\mu \sim (1,1,1) \,Q$,
\item {\bf \boldmath \pcol{Region Collinear to $p$} \unboldmath} (denoted by $\cc
$)  where $k$ scales as $k^\mu \sim (\lambda^2,1,\lambda) \,Q$,
\item {\bf \boldmath \lcol{Region Collinear to $l$} \unboldmath} (denoted by $\cb
$)  where $k$ scales as $k^\mu \sim (1,\lambda^2,\lambda)\, Q$,
\item {\bf \boldmath \scol{Soft Region} \unboldmath} (denoted by $s$)  
where $k$ scales as $k^\mu \sim (\lambda^2,\lambda^2,\lambda^2)\, Q$.
\end{itemize}
All of the other possible scalings of the integration momentum, of the form 
$k^\mu \sim (\lambda^a,\lambda^b,\lambda^c)\, Q$ and with $a,b,c$ not matching one of the four cases listed  above, give rise upon expanding  to scaleless integrals only, and therefore they do not contribute to the final result. In SCET, each low-energy region listed above is represented by a different field; the situation is schematically illustrated in Figure~\ref{fig:scales1}.

In the following, we will compute the contribution of each of the non-vanishing regions in turn, but it is instructive to start by considering an example of a scaling which does not contribute for the case of the form factor, namely a soft scaling $k^\mu \sim (\lambda,\lambda,\lambda) Q$, which we will call semi-hard in order to distinguish it from the standard soft region, whose components scale as $\lambda^2$. The expansion of the propagator denominators takes the form
\be
(k+l)^2 = \overbrace{k^2}^{{\mathcal O}(\lambda^2)} + 2 (\overbrace{k_+\cdot l_-}^{{\mathcal O}(\lambda^3)} + \overbrace{k_-\cdot l_+}^{{\mathcal O}(\lambda)} + \overbrace{k_\perp\cdot l_\perp}^{{\mathcal O}(\lambda^2)}) +\overbrace{l^2}^{{\mathcal O}(\lambda^2)}  =  2 k_- \cdot l_+ + {\mathcal O}(\lambda^2)\,,
\ee
and analogously
\be
(k+p)^2  =  2 k_+ \cdot p_- + {\mathcal O}(\lambda^2)\,,
\ee
after which the hypothetical semi-hard contribution  becomes
\be \label{eq:shard}
I_{sh} = i \pi^{-d/2} \mu^{4-d} \int d^d k \frac{1}{\left(k^2 +i0\right)
\left(2 k_-\cdot l_+ +i 0 \right) \left(2 k_+\cdot p_- +i 0 \right)} \, .
\ee
This integrals vanishes: $I_{sh} =0$; the explicit calculation is performed in Appendix~\ref{ap:Is}.\footnote{The calculation proceeds through the same steps as the evaluation of the soft integral when the external legs are put on-shell, $I_{sh} =I_s(p^2=0,l^2=0)=0$, which also vanishes, as  discussed below.} 
As an exercise, we invite the reader to show that also the Glauber region $k^\mu \sim (\lambda^2,\lambda^2,\lambda) Q$ gives a vanishing contribution to the form factor integral.


It is interesting to observe that in the soft region the square of a four momentum is proportional to $\lambda^4$:
\be
p_s^2 \sim  \lambda^4 Q^2 \sim \frac{L^2 P^2 }{Q^2} \, .
\ee
The momenta scaling as $\lambda^4$ are often called {\em ultra soft} in the
literature to distinguish them from the semi-hard modes scaling as $p^2 \sim  \lambda^2$. Such modes contribute for example in exclusive $B$-decays and also in observables which are sensitive to small transverse momenta, such as transverse momentum spectra of electroweak bosons. The relevant theory in the presence of soft modes with $p^2 \sim  \lambda^2$ is usually called SCET$_{\rm II}$. The effective Lagrangian we construct here is also called SCET$_{\rm I}$. More important than the naming scheme is the basic fact that one always needs to check which momentum modes arise in a given problem, and then include all relevant ones in the effective Lagrangian. What is interesting about the presence of an ultra-soft contribution, is that it implies the loop diagrams involve a scale which is smaller than the invariants which can be formed by the external momenta. For example this implies that jet-production processes can involve non-perturbative physics, even when the invariant masses of the jets are perturbative.

In order to determine the integral that one needs to evaluate when 
the integration momentum is considered hard, we consider the way in which the terms in the propagators in Eq.~(\ref{eq:ISudakov}) scale. Clearly $k^2 \sim \lambda^0 Q^2$; for the other two propagators one finds
\be
(k+l)^2 = \overbrace{k^2}^{{\mathcal O}(1)} + 2 (\overbrace{k_+\cdot l_-}^{{\mathcal O}(\lambda^2)} + \overbrace{k_-\cdot l_+}^{{\mathcal O}(1)} + \overbrace{k_\perp\cdot l_\perp}^{{\mathcal O}(\lambda)}) +\overbrace{l^2}^{{\mathcal O}(\lambda^2)}  = k^2 + 2 k_- \cdot l_+ + {\mathcal O}(\lambda)\,,
\ee
and, similarly
\be
(k+p)^2 =k^2 + 2 k_+ \cdot p_- + {\mathcal O}(\lambda)\,.
\ee
The contribution of the {\bf hard region} to the integral $I$ is therefore given by
\be \label{eq:Ihard}
I_h = i \pi^{-d/2} \mu^{4-d} \int d^d k \frac{1}{\left(k^2 +i0\right)
\left(k^2 +2 k_-\cdot l_+ +i 0 \right) \left( k^2 + 2 k_+\cdot p_- +i 0 \right)} \, ; 
\ee
it coincides with the form factor integral with on shell external legs (i.e. calculated by setting  $p^2 = l^2 =0$ from the start).
The integral evaluates to
\bea\label{eq:Ihardres}
I_h &=& \frac{\Gamma(1+\varepsilon)}{2 l_+ \cdot p_-} \frac{\Gamma^2(-\varepsilon)}{\Gamma(1 -2 \varepsilon)} 
\left(\frac{\mu^2}{2  l_+ \cdot p_-} \right)^\varepsilon \,  \nn \\
&=& \frac{\Gamma(1+\varepsilon)}{Q^2} \left(\frac{1}{\varepsilon^2} + 
\frac{1}{\varepsilon} \ln\frac{\mu^2}{Q^2} + \frac{1}{2} \ln^2 \frac{\mu^2}{Q^2} - \frac{\pi^{2}}{6}\right) + {\mathcal O}\left(\varepsilon \right) \, .
\eea
The poles in $\varepsilon$ are of infrared origin. The detailed calculation of $I_h$ can be found in Appendix~\ref{ap:HardS}.

In the region collinear to $p$ the integration momentum scales as 
$k^\mu \sim (\lambda^2, 1, \lambda) Q$. In this region $k^2 \sim \lambda^2 Q^2$,
while
\be
(k+l)^2 = 2 k_-\cdot l_+ +{\mathcal O}(\lambda^2) \, , \qquad (k+p)^2 = {\mathcal O}(\lambda^2) \, .
\ee
The \pcol{{\bf collinear region}} integral is obtained by keeping only the leading term 
in each propagator
\bea \label{eq:Icp}
\pcol{I_{\cc}} &=&  \pcol{i \pi^{-d/2} \mu^{4-d} \int d^d k \frac{1}{\left(k^2 +i0\right)
\left(2 k_-\cdot l_+ +i 0 \right) \left[ (k+p)^2 +i 0 \right]}} \,  \nn \\
&=& \pcol{-\frac{\Gamma(1+\varepsilon)}{2 l_+ \cdot p_-} \frac{\Gamma^2(-\varepsilon)}{\Gamma(1- 2 \varepsilon)} \left( \frac{\mu^2}{P^2} \right)^\varepsilon} \,  \nn \\
&=& \pcol{\frac{\Gamma(1+\varepsilon)}{Q^2} \left( -\frac{1}{\varepsilon^2} -\frac{1}{\varepsilon} \ln{\frac{\mu^2}{P^2}}  - \frac{1}{2} \ln^2{\frac{\mu^2}{P^2}} + \frac{\pi^2}{6}\right) +{\mathcal O}(\varepsilon)} \, .
\eea
The calculations leading to the above result are collected in Appendix~\ref{ap:Icp}. 
We observe that the integral scales as $P^{-2 \varepsilon}$\,.
The calculation of the integral in the region collinear
to $l$ is identical to the calculation of the integral in the region collinear to $p$, Eq.~(\ref{eq:Icp}), except that
one needs to replace $P^2$ with $L^2$ in the final result.

In the soft region all of the components of the integration momentum are proportional to $\lambda^2$, therefore
\be
k^2 = {\mathcal O}(\lambda^4) \,, \quad (k+l)^2 = 2 k_-\cdot l_+ + l^2 + {\mathcal O}(\lambda^3)\, , \,\,\mbox{and} \quad(k+p)^2 = 2 k_+\cdot p_- + p^2 + {\mathcal O}(\lambda^3)\, ,
\ee
and therefore the integral in the \scol{{\bf soft region}} is
\bea \label{eq:Isres}
\scol{I_{s}} &=&  \scol{i \pi^{-d/2} \mu^{4-d} \int d^d k \frac{1}{\left(k^2 +i0\right)
\left(2 k_-\cdot l_+ +l^2 +i 0 \right) \left(2 k_+\cdot p_- +p^2  +i 0 \right)}} \,  \nn \\
&=& \scol{- \frac{\Gamma\left(1+\varepsilon\right)}{2 l_+ \cdot p_-} \Gamma(\varepsilon) \Gamma\left( - \varepsilon\right) \left( \frac{2  l_+\cdot p_- \mu^2}{L^2 P^2}\right)^\varepsilon}\,  \nn \\
&=& \scol{\frac{\Gamma(1+\varepsilon)}{Q^2} \left(\frac{1}{\varepsilon^2} + 
\frac{1}{\varepsilon} \ln\frac{\mu^2 Q^2}{L^2 P^2} + \frac{1}{2} \ln^2 \frac{\mu^2 Q^2}{L^2 P^2} + \frac{\pi^{2}}{6}\right) + {\mathcal O}\left(\varepsilon \right)} \, .
\eea
The poles in the last line of Eq.~(\ref{eq:Isres}) are of ultraviolet origin.
As expected, the result depends on the ``new'' soft scale $\Lambda_{\mbox{soft}}^2 \sim P^2 L^2/ Q^2$. The details of the calculation of $I_s$ can be found in  Appendix~\ref{ap:Is}. 

Following \cite{Manohar:2006nz}, many SCET papers worry about the overlap of the soft and collinear regions. To ensure that there is no double counting, they subtract from the collinear contribution $\pcol{I_{\cc}}$ its ``zero-bin contribution''. This zero-bin contribution is obtained by expanding the collinear integrand around the soft limit. This is completely analogous to the contribution $R$ in Eq.~(\ref{eq:R}), which was obtained by expanding the high-energy integrand around the low-energy limit. As in the case of $R$, this overlap contribution is given by scaleless integrals and vanishes in dimensional regularization. Since both the soft and collinear integrals only depend on a single scale ($P^2$ for the collinear integrals, $\Lambda_{\mbox{soft}}^2$ for the soft integrals), one is left with scaleless integrals if one performs any further expansions of the integrands. Therefore, if the integrands are systematically expanded in the different regions, one never needs to include zero-bin subtractions in dimensional regularization. If, on the other hand, higher-power terms are not systematically expanded away, one may end up with non-zero overlap contributions, which would then need to be subtracted to avoid double counting. The reader interested in a more detailed discussion of overlap contributions in loop integrals can consult \cite{Jantzen:2011nz}. Examples in which non-vanishing zero-bin contributions were encountered in SCET include computations which involve low-mass jets, defined with a jet-algorithm \cite{Cheung:2009sg,Hornig:2009kv,Hornig:2009vb}. In these cases, the soft and collinear phase-space integrals depend on jet algorithm parameters and contain several scales. This also complicates resummation: in the presence of several scales in the individual functions, one can end up with large logarithms which cannot be resummed by RG evolution. The presence of non-vanishing zero-bin contributions indicates that a full scale separation has not yet been achieved and one should then ask the question whether an effective theory can be constructed which achieves complete scale separation.
 
One can now sum the results obtained in the different regions to obtain what was the original goal of the calculation: an analytic expression for the integral in Eq.~(\ref{eq:ISudakov}) in the limit in which $L^2 \sim P^2 \ll Q^2$. One finds
\bea
I_{h} &=& \frac{\Gamma\left(1+\varepsilon\right)}{Q^2}
\left(\frac{1}{\varepsilon^2} +\frac{1}{\varepsilon} \ln\frac{\mu^2}{Q^2}
+\frac{1}{2} \ln^2 \frac{\mu^2}{Q^2} -\frac{\pi^2}{6} +{\mathcal O}(\lambda)\right) \nn \\
\pcol{I_{\cc
}} &=& \pcol{\frac{\Gamma\left(1+\varepsilon\right)}{Q^2}
\left(-\frac{1}{\varepsilon^2} -\frac{1}{\varepsilon} \ln\frac{\mu^2}{P^2}
-\frac{1}{2} \ln^2 \frac{\mu^2}{P^2} +\frac{\pi^2}{6} +{\mathcal O}(\lambda)\right)} \nn \\
\lcol{I_{\cb
}} &=& \lcol{\frac{\Gamma\left(1+\varepsilon\right)}{Q^2}
\left(-\frac{1}{\varepsilon^2} -\frac{1}{\varepsilon} \ln\frac{\mu^2}{L^2}
-\frac{1}{2} \ln^2 \frac{\mu^2}{L^2} +\frac{\pi^2}{6} +{\mathcal O}(\lambda)\right)} \nn \\
\scol{I_{s}} &=& \scol{\frac{\Gamma\left(1+\varepsilon\right)}{Q^2}
\left(\frac{1}{\varepsilon^2} +\frac{1}{\varepsilon} \ln\frac{\mu^2 Q^2}{L^2 P^2}
+\frac{1}{2} \ln^2 \frac{\mu^2 Q^2}{L^2 P^2} +\frac{\pi^2}{6} + {\mathcal O}(\lambda)\right)} \nn \\
& &\makebox[9.5cm]{\hrulefill}\nn \\
I \!\equiv\! I_h \!+\! \pcol{I_{\cc
}}\! +\! \lcol{I_{\cb
}} \!+\!\scol{I_s} &=& \frac{1}{Q^2}
\left(\ln{\frac{Q^2}{L^2}} \ln{\frac{Q^2}{P^2}}+ \frac{\pi^2}{3} + {\mathcal O}(\lambda)\right) \, .
\eea
The final result does not depend on the dimensional regulator $\varepsilon$ and the reader is invited to check that it coincides with the one that would be obtained 
by evaluating directly the integral in Eq.~(\ref{eq:ISudakov}) and then expanding the result in the $\lambda \to 0$ limit.
We stress the fact that the infrared divergences found in  the hard region cancel out against the ultraviolet divergences found in the sum of the soft and collinear contributions. This feature is general and requires a nontrivial interplay of the logarithms found in the various integrals:
\be
-\frac{1}{\varepsilon} \ln{\frac{\mu^2}{P^2}} -\frac{1}{\varepsilon} \ln{\frac{\mu^2}{L^2}} + \frac{1}{\varepsilon} \ln{\frac{\mu^2 Q^2}{L^2 P^2}} = -\frac{1}{\varepsilon} \ln{\frac{\mu^2}{Q^2}} \, .
\ee
The requirement that infrared divergences of the hard region should cancel against the ultraviolet divergences of the soft and collinear regions leads to constraints that must be satisfied by the infrared pole structure of a generic amplitude. This aspect will be further discussed in Section~\ref{sec:LV}.

\subsection{The Massive Sudakov Problem and the Collinear Anomaly \label{sec:MSPCA}}

For some observables the simple separation of the integral in hard, soft, and collinear regions breaks down because the different momentum regions are not well defined if one does not introduce additional regulators on top of dimensional regularization. This problem is referred to as the {\em Collinear Anomaly} in \cite{Becher:2010tm} and appears for example in processes with high momentum transfers and small but non negligible masses, such as in the resummation of electroweak Sudakov logarithms \cite{Chiu:2007yn,Chiu:2007dg}, and in observables sensitive only to transverse momenta such as the transverse momentum spectrum in Drell-Yan production \cite{Becher:2010tm} or in jet broadening \cite{Becher:2011pf,Chiu:2011qc}.

To illustrate this kind of situation we consider again the diagram in Fig.~\ref{fig:Sudakov} but this time we assume that the virtual particle which carries momentum $k$ has a mass $m$, such that $m^2 \sim \lambda^2 Q^2 \gg \lambda^4 Q^2$.
If the virtual momentum $k$ is soft ($k^2 \sim \lambda^4 Q^2$), the propagator carrying momentum $k$ has the following expansion
\be
\frac{1}{k^2-m^2} = -\frac{1}{m^2} \left( 1 - \frac{k^2}{m^2} + \cdots \right) \, .
\ee
The relevant integral for the soft region is then 
\bea \label{eq:IsresM}
I_{s} &=&  - i \pi^{-d/2} \mu^{4-d} \int d^d k \frac{1}{\left(m^2 -i0\right)
\left(2 k_-\cdot l_+ +l^2 +i 0 \right) \left(2 k_+\cdot p_- +p^2  +i 0 \right)} \, , 
\eea
and it can be proven that the above integral vanishes (see Appendix~\ref{ap:Is}).
One could then conclude that the complete integral is given by the sum of the hard region and the two collinear regions. Also, one could naively expect that the collinear integrals depend only on collinear scales such as the squared momenta $l^2$ and $p^2$ and the squared mass. However, this cannot be the case, since the hard region integral will have an infrared pole multiplied by a logarithm of the hard scale $Q$, and this infrared divergence must cancel in the final result. This apparent contradiction can be resolved after observing that the collinear integrals are not well defined unless one uses an additional regulator on top of dimensional regularization.

The simplest way to proceed consists in employing {\em Analytic Regulators} \cite{Smirnov:1997gx}; the complete integral can be written as
\be \label{eq:IcompL}
I  =  i \pi^{-d/2} \mu^{4-d} \int d^d k \frac{(-\nu^2)^\alpha}{(k^2-m^2+i0)
[(k+l)^2+i0] [(k+p)^2+i0]^{1+\alpha}} \, , 
\ee
where the power $\alpha$ is the analytic regulator which will be sent to zero at the end of the calculation, while $\nu$ is the 't Hooft scale associated to the analytic regulator. One could introduce a second analytic regulator for the collinear leg carrying momentum $l$; however, this is not necessary in our case. The complete integral $I$, as well as the hard region integral $I_h$ are well defined also if the analytic regulator is not present. On the other hand, the two collinear region integrals show poles for $\alpha \to 0$ which cancel in their sum. The two collinear integrals are 
\bea
\pcol{I_{\cc
}} &=&  \pcol{i \pi^{-d/2} \mu^{4-d} \int d^d k \frac{(-\nu^2)^\alpha}{\left(k^2-m^2 +i0\right)
\left(2 k_-\cdot l_+ +i 0 \right) \left[ (k+p)^2 +i 0 \right]^{1+\alpha}}} \, , \nn \\
\lcol{I_{\cb
}} &=&  \lcol{i \pi^{-d/2} \mu^{4-d} \int d^d k \frac{(-\nu^2)^\alpha}{\left(k^2-m^2 +i0\right)
\left[ (k+l)^2+i0\right] \left[2 k_+\cdot p_- +i 0 \right]^{1+\alpha}}} \, .
\label{eq:collregar}
\eea 
At this stage, we set  $p^2=l^2=0$ for simplicity. The calculation of the two collinear integrals gives (see Appendix~\ref{ap:IcAR} for the details of the calculation)
\bea
\pcol{I_{\cc
}} &=&  \pcol{\frac{\Gamma\left(1+\ep \right)}{Q^2}\left(\frac{\mu^2}{m^2} \right)^\ep\left(\frac{\nu^2}{m^2} \right)^\alpha 
\left( \frac{1}{\alpha \ep} -\frac{1}{\ep^2} +\frac{\pi^2}{3} +{\mathcal O}(\alpha,\ep)\right)\,} , \nn \\
\lcol{I_{\cb
}} &=& \lcol{\frac{\Gamma\left(1+\ep \right)}{Q^2}\left(\frac{\mu^2}{m^2} \right)^\ep\left(\frac{\nu^2}{Q^2} \right)^\alpha \left( -\frac{1}{\alpha \ep} +\frac{\pi^2}{6} +{\mathcal O}(\alpha,\ep) \right)}\, .
\label{eq:collregarRES}
\eea
In the sum of the two contributions the dependence on $\alpha$ and $\nu$ cancels:
\bea
\pcol{I_{\cc
}} + \lcol{I_{\cb
}} = \frac{\Gamma\left(1+\ep \right)}{Q^2} \Biggl[ -\frac{1}{\ep^2} -\frac{1}{\ep} \ln{\frac{\mu^2}{Q^2}} - \ln{\frac{\mu^2}{m^2}}\ln{\frac{\mu^2}{Q^2}} + \frac{1}{2} \ln^2{\frac{\mu^2}{m^2}} +\frac{\pi^2}{2} \Biggr] \, .
\eea
In order to obtain the complete result it is now sufficient to add the contribution of the hard region integral in Eq.~(\ref{eq:Ihard}):
\be
\pcol{I_{\cc
}} + \lcol{I_{\cb
}} +I_h =\frac{1}{Q^2} \Biggl[\frac{1}{2} \ln^2{\frac{m^2}{Q^2}} + \frac{\pi^2}{3} \Biggr] \, ,
\ee
which is indeed the correct result for the integral in Eq.~(\ref{eq:IcompL}).
It is important to remember that the original integral is only independent of the analytic regulator for non-zero 
$\ep$. In order  to have a result independent of  the analytic regulator one needs to send the analytic regulator to zero before expanding in $\ep$.

The collinear integrals before the introduction of the analytic regulator are analytic functions of $Q$, and therefore they cannot give rise to logarithmic dependence on $Q$.
The use of analytic regularization breaks this property, and in spite of the fact that the analytic regulator $\alpha$ can be sent to zero after the two collinear region integrals are summed, the property is not recovered. Consequently, the final result depends on a logarithm of $Q$.  This anomaly is not an anomaly of the full theory, but only an anomaly of the effective theory performing the region separation, which, as it will be shown in the following section, is SCET. This breakdown of naive factorization was observed earlier and called {\em Factorization Anomaly} by M.~Beneke
\cite{Beneke_lectures}.

The consequences  of the collinear anomaly for the factorization of hard and collinear contributions in scattering processes will be discussed in Section~\ref{sec:FCA}.

%% file: 3_scalarSCET.tex
\section{Scalar SCET \label{sec:scalarSCET}}

We now construct an effective field theory whose Feynman rules directly yield the hard, collinear, and soft
integrals for the Sudakov form factor that were considered in the previous
section. Initially we restrict our discussion to the case of scalar $\phi^3$ theory.
The procedure outlined in the following will be applied to  QCD in 
Section~\ref{sec:SCET_QCD}; however, since the different components of the quark and gluon fields scale differently, the effective Lagrangian derived from QCD will look more complicated than the one that we will derive in this section for a scalar theory.

\subsection{The Scalar SCET Lagrangian}
The starting point of our discussion is the Lagrangian
\be\label{eq:Lscalar}
{\mathcal L}\left(\phi\right) = \frac{1}{2} \partial_\mu \phi(x) \partial^\mu \phi(x) - \frac{g}{3!}
\phi^3(x) \, ,
\ee
where $\phi$ is the scalar field and $g$ the coupling constant of the theory. In
order to derive the SCET effective Lagrangian needed for the calculation of the
Sudakov form factor in this theory, one needs to split the scalar field in the
sum of a field collinear to the  momentum $p$, a field collinear to the momentum $l$, and a soft field:
\be \label{eq:phisplit}
\phi(x) \to \pcol{\phi_{\cc}(x)} + \lcol{\phi_{\cb}(x)} + \scol{\phi_s(x)} \, .
\ee 
It was not necessary to introduce in the sum above a field for the hard region, since these contributions are absorbed into the prefactors of the operators built from soft and collinear fields. These prefactors are called
{\em Wilson coefficients} and are the coupling constants of the effective theory. By writing down the most general set of operators and by adjusting their Wilson coefficients, one reproduces the full theory, as explained in detail below. When constructing the effective Lagrangian, we assume that the momenta of the different fields scale in the proper way. For the construction to make sense, it is important that the external momenta are chosen properly. For example, one must choose the external momentum flowing into a soft field to be soft.

\begin{figure}[t]
\vspace*{.5cm}
\[ 
\vcenter{ \hbox{
  \begin{picture}(0,0)(0,0)
\SetScale{1}
  \SetWidth{2}
\SetColor{Blue}
\Line(-45,0)(0,0)
\Text(-40,5)[cb]{{\Large \pcol{$\phi_{\cc
}$}}}
\Text(50,-40)[cb]{{\Large \pcol{$\phi_{\cc
}$}}}
\Line(0,0)(36,-36)
\SetColor{Mahogany}
\Text(50,27)[cb]{{\Large \scol{$\phi_{s}$}}}
\DashLine(0,0)(36,36){2}

\end{picture}}
}
\hspace{5cm}
\hspace*{1.5cm}
\vcenter{ \hbox{
  \begin{picture}(0,0)(0,0)
\SetScale{1}
  \SetWidth{2}
\SetColor{OliveGreen}
\DashLine(-45,0)(0,0){6}
\DashLine(0,0)(36,-36){6}
\Text(-40,5)[cb]{{\Large \lcol{$\phi_{\cb
}$}}}
\Text(50,-40)[cb]{{\Large \lcol{$\phi_{\cb
}$}}}
\SetColor{Mahogany}
\Text(50,27)[cb]{{\Large \scol{$\phi_{s}$}}}
\DashLine(0,0)(36,36){2}

\end{picture}}
}
\]
\vspace*{.8cm} 
\caption{Interaction vertices generated from the Lagrangian ${\mathcal L}_{c+s}$.} 
\label{fig:intvsc}
\end{figure}
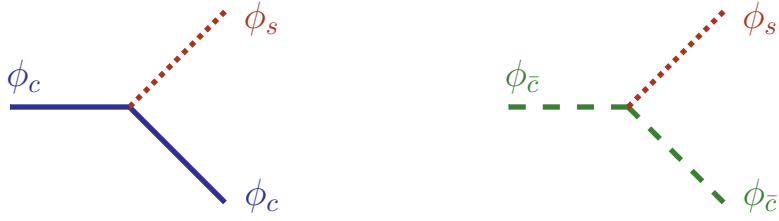

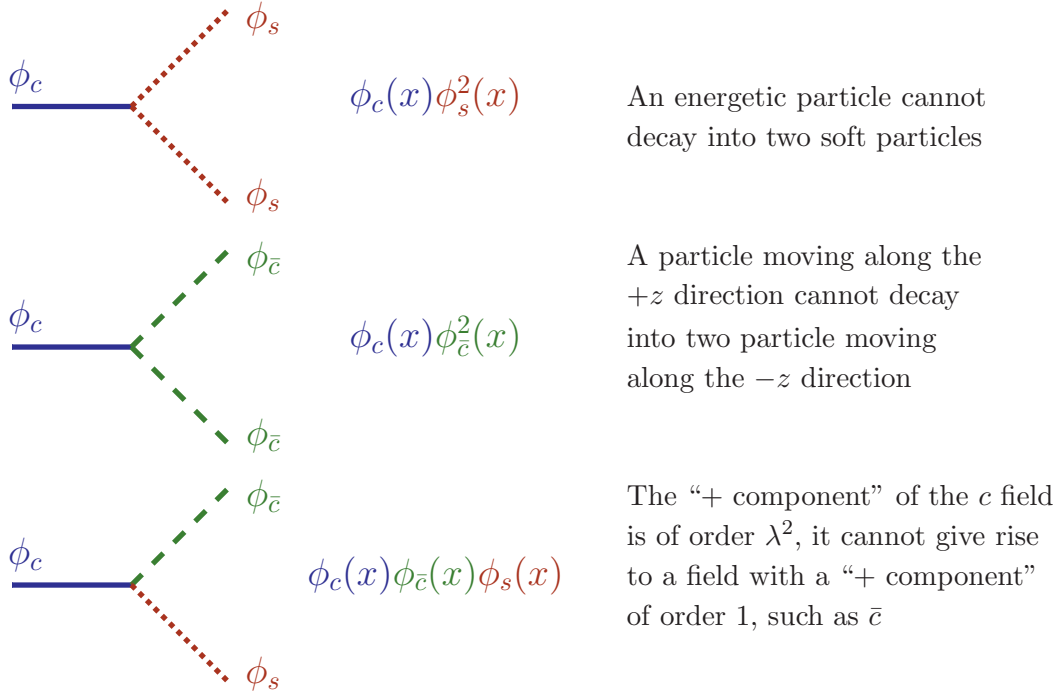
\begin{figure}[t]
\hspace*{3cm}
\vspace*{1cm}\\
\hspace*{2cm}
\begin{tabular}{ccccl}
  \begin{picture}(0,0)(0,0)
\SetScale{1}
  \SetWidth{2}
\SetColor{Blue}
\Line(-45,0)(0,0)
\Text(-40,5)[cb]{{\Large \pcol{$\phi_{\cc
}$}}}
\SetColor{Mahogany}
\DashLine(0,0)(36,-36){2}
\Text(50,27)[cb]{{\Large \scol{$\phi_{s}$}}}
\DashLine(0,0)(36,36){2}
\Text(50,-40)[cb]{{\Large \scol{$\phi_{s}$}}}

\end{picture} &\hspace*{1.5cm} &{\Large $ \pcol{\phi_{\cc
}(x)} \scol{\phi_s^2(x)}$}& & An energetic particle cannot\\
& & & &  decay into two soft particles \\
& & & & \\
& & & & \\
& & & &A particle moving along the  \\
& & & &$+z$ direction cannot decay  \\
\begin{picture}(0,0)(0,0)
\SetScale{1}
  \SetWidth{2}
\SetColor{Blue}
\Line(-45,0)(0,0)
\Text(-40,5)[cb]{{\Large \pcol{$\phi_{\cc
}$}}}
\SetColor{OliveGreen}
\DashLine(0,0)(36,-36){6}
\Text(50,27)[cb]{{\Large \lcol{$\phi_{\cb
}$}}}
\DashLine(0,0)(36,36){6}
\Text(50,-40)[cb]{{\Large \lcol{$\phi_{\cb
}$}}}

\end{picture}& &{\Large $\pcol{\phi_{\cc
}(x)} \lcol{\phi_{\cb
}^2(x)}$} & & 
into two particle moving \\
& & & & along the $-z$ direction\\
& & & & \\
& & & & \\
& & & & The ``$+$ component'' of the $\cc
$ field\\
& & & & is of order $\lambda^2$, it cannot give rise\\
\begin{picture}(0,0)(0,0)
\SetScale{1}
  \SetWidth{2}
\SetColor{Blue}
\Line(-45,0)(0,0)
\Text(-40,5)[cb]{{\Large \pcol{$\phi_{\cc
}$}}}
\SetColor{OliveGreen}
\DashLine(0,0)(36,36){6}
\Text(50,27)[cb]{{\Large \lcol{$\phi_{\cb
}$}}}
\SetColor{Mahogany}
\DashLine(0,0)(36,-36){2}
\Text(50,-40)[cb]{{\Large \scol{$\phi_{s}$}}}

\end{picture}& &{\Large $\pcol{\phi_{\cc
}(x)} \lcol{\phi_{\cb
}(x)} \scol{\phi_s(x)}$} & & to a field with a ``+ component" \\
& & & & of order $1$, such as $\cb
$ 
\end{tabular}
\vspace*{1.3cm}
\caption{Interaction forbidden by momentum conservation.} 
\label{fig:sl2-10}
\end{figure}

By splitting each one of the fields according to Eq.~(\ref{eq:phisplit}), the original Lagrangian can be written as the sum of four terms:%
\be \label{eq:scalSCET}
{\mathcal L}(\phi) = \underbrace{{\mathcal L}\left(\pcol{\phi_{\cc
}}\right)}_{
\equiv \pcol{{\mathcal L}_{\cc
}}} +  
\underbrace{{\mathcal L}\left(\lcol{\phi_{\cb
}}\right)}_{
\equiv \lcol{{\mathcal L}_{\cb
}}} + \underbrace{{\mathcal L}\left(\scol{\phi_{s}}\right)}_{\equiv \scol{{\mathcal L}_{s}}}
+ {\mathcal L}_{c+s}\left(\pcol{\phi_{\cc
}},\lcol{\phi_{\cb
}},\scol{\phi_{s}}\right)
\, .
\ee
The first three terms on the r.h.s.\ of the equation above are simply copies of the original Lagrangian, where all the fields are either collinear to $p$, collinear to $l$, or soft. The fourth term in Eq.~(\ref{eq:scalSCET}) describes
the interaction of collinear and soft fields
\be
{\mathcal L}_{c+s}\left(\pcol{\phi_{\cc
}},\lcol{\phi_{\cb
}},\scol{\phi_{s}}\right) = -\frac{g}{2} \pcol{\phi_{\cc
}}^2 \scol{\phi_{s}} - \frac{g}{2}\lcol{\phi_{\cb
}}^2 \scol{\phi_{s}} \, ,
\ee
which gives rise to the interaction vertices shown in Fig.~\ref{fig:intvsc}. At first sight, it looks like there should be many additional interaction terms, but the interactions between the fields which do not appear in Eq.~(\ref{eq:scalSCET}) are forbidden by momentum conservation, as it is shown in Fig.~\ref{fig:sl2-10}.

%

As a last step, one needs to expand each interaction term in the small momentum
components. This procedure is called derivative (or multipole) expansion
\cite{Beneke:2002ni}. Consider the Fourier transform of the fields in a given interaction term;
\be
\int\!\!d^d x \pcol{\phi^2_{\cc}(x)} \scol{\phi_{s}(x)} = 
\int\!\!d^d x \pcol{\!\!\int \!\!\frac{d^d p_1}{(2 \pi)^d} \int \!\! 
\frac{d^d p_2}{(2 \pi)^d}}\scol{ \int \!\! \frac{d^d p_s}{(2 \pi)^d}} \pcol{\tilde{\phi}_{\cc}(p_1)} \pcol{\tilde{\phi}_{\cc}(p_2)} \scol{\tilde{\phi}_s (p_s)}
 e^{-i(\pcol{p_1}+\pcol{p_2}+\scol{p_s})\cdot x} \, , 
\ee
where the tilde indicates the transformed fields. If, as we assumed, the momenta
\pcol{$p_1$} and \pcol{$p_2$} are collinear to \pcol{$p$}, while \scol{$p_s$} is soft, the sum of the three momenta scales as
\be
\pcol{p_1^\mu} + \pcol{p_2^\mu} + \scol{p_s^\mu} \sim \left( \lambda^2 ,1, \lambda \right) Q\, .
\ee
Consequently the components of $x$ must scale as
\be
x^\mu \sim \left(1, \frac{1}{\lambda^2}, \frac{1}{\lambda} \right) \frac{1}{Q} \, .
\ee
If one now considers the fact  that all of the components of the soft momentum scale as $\lambda^2$, one finds that
\be
p_s \cdot x = \underbrace{(p_s)_+\cdot x_-}_{{\mathcal O}(1)} +\underbrace{(p_s)_- \cdot x_+}_{{\mathcal O}(\lambda^2)} +
\underbrace{(p_s)_\perp\cdot x_\perp}_{{\mathcal O}(\lambda)} \, .
\ee
Since the derivatives of the soft field scale as the components of the soft momentum, the Taylor expansion of the soft field around the point $x^\mu_- = 
(x \cdot \bar{n}) n^\mu/2$ is
\be \label{eq:expPHIxm}
\scol{\phi_s(x)} = \scol{\phi_s(x_-)}+\underbrace{x_\perp \cdot \partial_\perp }_{{\mathcal O}(\lambda)} \scol{\phi_s(x_-)}  +\underbrace{x_+ \cdot \partial_- }_{{\mathcal O}(\lambda^2)} \scol{\phi_s(x_-)} + \frac{1}{2}\Bigl(\underbrace{x_{\mu \perp} x_{\nu \perp} \partial^\mu \partial^\nu }_{{\mathcal O}(\lambda^2)} \scol{\phi_s(x_-)} \Bigr) + \ldots \,
\ee
where we specify the $\lambda$ suppression relative to the leading term.
Consequently, up to first order in $\lambda$, the interaction term between the 
collinear and soft field can be rewritten as
\be \
\int d^d x \, \pcol{\phi^2_{\cc
}(x)} \scol{\phi_{s}(x)} = \int d^d x \,  \pcol{\phi^2_{\cc
}(x)} \scol{\phi_{s}(x_-)}\, 
(1 +{\mathcal O}(\lambda) )\, .  
\ee
Note that the expanded Lagrangian is only translation invariant up to terms 
of higher order in $\lambda$. 
An alternative to the position-space formalism used here and throughout this introduction is to treat the large momentum components as labels on the fields, see Section~\ref{sec:labelform}.

The leading power scalar SCET Lagrangian has then the following form 
\bea
{\mathcal L_{{\tiny \mbox{eff}}}} &=&
\pcol{\frac{1}{2} \partial_\mu \phi_{\cc} (x) \partial^\mu \phi_{\cc} (x) -
\frac{g}{3!}  \phi^3_{\cc} (x)}  +
\lcol{\frac{1}{2} \partial_\mu \phi_{\cb} (x) \partial^\mu \phi_{\cb} (x) -
\frac{g}{3!}  \phi^3_{\cb} (x) }\nn \\
&+& 
\scol{\frac{1}{2} \partial_\mu \phi_{s} (x) \partial^\mu \phi_{s} (x) -
\frac{g}{3!}  \phi^3_{s} (x)} -
\frac{g}{2} \pcol{\phi^2_{\cc} (x)} \scol{ \phi_{s} (x_-)} -
\frac{g}{2} \lcol{ \phi^2_{\cb} (x)} \scol{ \phi_{s} (x_+)} \, . \label{eq:Leff}
\eea

\subsection{Matching Procedure and Current Operator}

In an effective theory, the hard contributions lead to matching corrections.
The procedure which allows us to take the hard corrections into account is the following:
\begin{itemize}
\item[1)] Write down the most general form of the Lagrangian, including all the operators which are  compatible with the symmetry of the theory, each one of which will be multiplied by an arbitrary coefficient. These are called {\em Wilson coefficients}.
\item[ii)] Calculate a given interaction process both in the full theory and in the effective theory.
\item[iii)] Adjust the values of the Wilson coefficients in such a way that the results obtained in the full and in the effective theory coincide.
\end{itemize}
In general, such matching corrections modify the effective Lagrangian. However, for the case of SCET, it turns out that only the operators which involve collinear fields in different directions get matching corrections. For example, in order to describe the Sudakov form factor, we introduce an external current coupling to two scalar fields
\vspace*{2mm}\\
\be
J = \phi^2 =\hspace{0.5cm} \vcenter{ \hbox{
  \begin{picture}(0,0)(0,0)
\SetScale{.6}
  \SetWidth{2}
  \Line(-40,40)(0,0)
  \Line(0,0)(40,40)
  \Photon(0,0)(0,-15){3}{2}
\end{picture}}
}\hspace{0.5cm}\,,
\ee
and consider the current at large momentum transfer. In the following, we first explain why the matching corrections are absent for the Lagrangian derived in the last section and then compute them for the current operator, which will involve collinear fields in both directions.

To allow for the presence of matching corrections in the Lagrangian, we introduce Wilson coefficients which multiply the interaction terms in Eq.~(\ref{eq:Leff}); in particular the term involving three $\cc$ fields will become
\be
-
\frac{g}{3!}  \pcol{\phi^3_{\cc} (x)} \to -
\frac{g}{3!} C \pcol{\phi^3_{\cc}} (x) \equiv -
\frac{g}{3!} \left(1 + g^2 C^{(1)} +g^4 C^{(2)} +\cdots \right) \pcol{\phi^3_{\cc}} (x)
\, .
\ee
In order to fix the coefficient $C^{(1)}$ one requires that the corrections of order $g^2$ to the interaction of three scalar fields are the same in the full theory and in the effective theory. In the full theory these corrections coincide with
the one loop corrections to the $\phi^3$ vertex, while in the effective theory
one finds contributions originating from one loop graphs and contributions proportional to $C^{(1)}$. One obtains the following diagrammatic equation
\vspace*{.3cm}\\
\bea
\vcenter{ \hbox{
  \begin{picture}(0,0)(0,0)
\SetScale{.6}
  \SetWidth{2}
\Line(-45,0)(-15,0)
\CArc(0,0)(15,0,360)
\Text(-35,5)[cb]{{\Large \pcol{$\phi_{\cc}$}}}
\Text(40,-30)[cb]{{\Large \pcol{$\phi_{\cc}$}}}
\Line(11,-11)(36,-36)
\Text(40,17)[cb]{{\Large \pcol{$\phi_{\cc}$}}}
\Line(11,11)(36,36)
\end{picture}}
}
\hspace*{1.5cm} &=& \hspace*{1.cm}
\vcenter{ \hbox{
  \begin{picture}(0,0)(0,0)
\SetScale{.6}
  \SetWidth{2}
  \SetColor{Blue}
\Line(-45,0)(-15,0)
\CArc(0,0)(15,0,360)
\Line(11,-11)(36,-36)
\Line(11,11)(36,36)
\end{picture}}
}
\hspace*{.8cm}  + \hspace*{1cm} 
\vcenter{ \hbox{
  \begin{picture}(0,0)(0,0)
\SetScale{.6}
  \SetWidth{2}
  \SetColor{Blue}
\Line(-45,0)(-15,0)
\CArc(0,0)(15,-45,180)
  \SetColor{Mahogany}
\DashCArc(0,0)(15,180,315){2}
  \SetColor{Blue}
  
\Line(11,-11)(36,-36)
\Line(11,11)(36,36)
\end{picture}}
}\hspace*{1cm}+ \cdots + g^2 C^{(1)}\hspace*{1cm}
\vcenter{ \hbox{
  \begin{picture}(0,0)(0,0)
\SetScale{.6}
  \SetWidth{2}
  \SetColor{Blue}
\Line(-45,0)(0,0)
\Line(0,0)(36,-36)
\Line(0,0)(36,36)
\end{picture}}
}\hspace*{.9cm}   \, , \label{eq:MatCCC}
\eea
\vspace*{.3cm}\\
where all the external legs have momenta collinear to $p$. Blue lines indicate collinear fields in the effective theory and dotted red lines indicate the soft $\phi$ field. 
The dots in Eq.~(\ref{eq:MatCCC}) indicate two additional diagrams which can be obtained from the second diagram by moving the internal soft line in the other two possible positions. Note that the first diagram on the right-hand side of Eq.~(\ref{eq:MatCCC}) is identical to the original loop integral on the left-hand side of the equation. A non-zero one-loop matching coefficient would therefore only be needed to remove the contributions from the loops which involve a soft particle and have no analogue in the full theory. Fortunately, these additional one-loop diagrams all vanish, since the soft scale is only non-zero when both collinear and anti-collinear momenta are present. The matching coefficient therefore vanishes,
i.e. $C^{(1)} =0$. An even simpler way to see that no matching is necessary for purely collinear diagrams is to choose all external momenta $p_i^\mu$ in a single direction $p_i^\mu \propto p^\mu$, with an on-shell momentum $p^2=0$. In this case all loop integrals in Eq.~(\ref{eq:MatCCC}) are scaleless, since they do not have internal masses and all of the scalar products that can be generated with the external-leg momenta will be proportional to the square of the momentum $p$, which vanishes. Since in dimensional regularization scaleless integrals evaluate to zero, we can immediately conclude that the one loop matching condition is $C^{(1)} =0$. This argument also applies if one considers higher loops; we can conclude that $C = 1$ to all orders in perturbation theory. The same kind of reasoning can be applied to all of the interaction terms appearing in  Eq.~(\ref{eq:Leff}).

While the terms appearing in the Lagrangian in Eq.~(\ref{eq:Leff}) do not receive matching  corrections, the terms originating from the current operator $J$ do. The most general form that the current operator 
can have in the effective theory is 
\be \label{eq:auxJ}
J = J_2 + J_3 + \cdots = C_2 \pcol{\phi_{\cc}} \lcol{\phi_{\cb}} + \frac{C_3}{2!} \left( 
\pcol{\phi^2_{\cc}} \lcol{ \phi_{\cb}} +\pcol{ \phi_{\cc}} \lcol{\phi^2_{\cb}}\right) + \cdots \, ,
\ee
where the subscript in $J_i$ and $C_i$ indicates the number of fields involved in the corresponding operator.
In addition to operators with multiple fields, one should also consider operators involving derivatives on the fields. As it was shown above, the projection of the derivative of the collinear field in a given direction scales as the corresponding component of the momentum, therefore
\be
 n \cdot \partial \pcol{\phi_{\cc}(x)} \sim \lambda^2 \pcol{\phi_{\cc}(x)} \, , \quad
 \partial^\mu_{\perp} \pcol{\phi_{\cc}(x)} \sim \lambda \pcol{\phi_{\cc}(x)} \, ,
 \quad  \bar{n} \cdot \partial \phi_{\cc
}(x) \sim \lambda^0 \pcol{\phi_{\cc}(x)} \, ,
\ee 
and similarly
\be
 \bar{n} \cdot \partial \lcol{\phi_{\cb}(x)} \sim \lambda^2 \lcol{\phi_{\cb}(x)} \, , \quad
 \partial^\mu_{\perp} \lcol{\phi_{\cb}(x)} \sim \lambda \lcol{\phi_{\cb}(x)} \, ,
 \quad  n \cdot \partial \lcol{\phi_{\cb}(x)} \sim \lambda^0 \lcol{\phi_{\cb}(x)} \, .
\ee 
The derivatives $\bar{n} \cdot \partial \pcol{\phi_{\cc}}$ and $n \cdot \partial \lcol{\phi_{\cb}}$ are not power suppressed, because the collinear fields carry large energies in these directions. Even at leading power in $\lambda$, one needs to allow for the insertion of an arbitrary number of these derivatives in the current operators in the effective theory.
The expansion of a collinear field along the direction associated with the large momentum component can be written in terms of an infinite sum over the non-power suppressed derivatives
\be
\phi_c(x + t \bar{n}) = \sum_{i=0}^{\infty} \frac{t^i}{i!}  (\bar{n}\cdot \partial)^{i} \phi_c(x) \, .
\ee
Therefore, to include terms with arbitrarily high derivatives is equivalent to allowing non-locality of the collinear fields along the collinear directions. For example, the operator $J_2$ in Eq.~(\ref{eq:auxJ}) can be written as
\be \label{eq:J2}
J_2(x) = \int ds dt\,  C_2(s,t,\mu) \, \pcol{ \phi_{\cc}\left(x+ s \bar{n} \right)} \lcol{\phi_{\cb
} \left(x+ t n \right)} \, ,
\ee
the SCET operators are thus non-local along light-cone directions corresponding to large energies. The non-locality of the operators in position space is reflected in the dependence of the Wilson coefficients on the large energy scales present in the problem. In fact, the Fourier transform of the coefficient
$C_2(s,t)$ will be
\be
\tilde{C}_2\left(\bar{n} \cdot p \,, n \cdot l,\mu \right) = 
\int ds dt \, e^{i s \bar{n} \cdot p} e^{-i t n \cdot l} C_2 (s,t,\mu) \, .
\ee
To be precise, we have indicated that the Wilson coefficient $ C_2 (s,t,\mu)$ will depend on the renormalization scale. This dependence arises after renormalization in the effective theory and is governed by a renormalization group equation, which can be used to perform resummation. We will discuss this topic in detail in Section \ref{sec:LIV}.
The function $\tilde{C}_{2}$ must be expanded in powers of the coupling constant $g$ as follows
\be
\tilde{C}_2 =\tilde{C}^{(0)}_2  + g^2 \tilde{C}^{(1)}_2 + g^4 \tilde{C}^{(2)}_2
+ \cdots \,.
\ee
One can immediately see that the simple matching condition at order $g^0$ leads to the relation $\tilde{C}^{(0)}_2 =1$. 
Next, we write the matching equation which allows us to fix the value of $\tilde{C}_2$ at order $g^2$
\vspace*{8mm}
\be
\vcenter{ \hbox{
  \begin{picture}(0,0)(0,0)
\SetScale{.6}
  \SetWidth{2}
  \Line(-40,40)(0,0)
  \Line(0,0)(40,40)
  \Line(-30,30)(30,30)
  \Photon(0,0)(0,-15){3}{2}
\Text(-38,14)[cb]{{\Large $p$}}
\Text(38,15)[cb]{{\Large $l$}}
\LongArrow(-45,22)(-20,-3)
\LongArrow(20,-3)(45,22)
\end{picture}}
}
\hspace*{1.5cm} = \hspace*{1.cm}
g^2 \tilde{C}^{(1)}_2 (\underbrace{\bar{n} \cdot p \, n \cdot l}_{=Q^2} )
\hspace*{1.7cm}
\vcenter{ \hbox{
  \begin{picture}(0,0)(0,0)
\SetScale{.6}
  \SetWidth{2}
  \SetColor{Blue}
  \Line(-40,40)(0,0)
  \SetColor{OliveGreen}
  \DashLine(0,0)(40,40){6}
  \SetColor{Black}
  \Photon(0,0)(0,-15){3}{2}
\Text(-38,14)[cb]{{\Large \pcol{$\phi_{\cc
}$}}}
\Text(38,15)[cb]{{\Large \lcol{$\phi_{\cb
}$}}}
\end{picture}}
}\hspace*{1.cm}  \, .
\ee
The momenta $p$ and $l$ are both on-shell, and the diagram on the l.h.s.\ of the equation above coincides with the hard region integral introduced in Section~\ref{sec:regions}. On the r.h.s.\ of the matching equation, one should also include the contribution of the one-loop diagram with an internal soft leg multiplied by  $\tilde{C}_2^{(0)}$; however, that integral corresponds to the soft region integral  calculated in the previous section, but with on-shell external legs. The latter vanishes in dimensional regularization if
one sets $p^2 =l^2=0$ from the start, as it is shown in Appendix~\ref{ap:Is}. The same is true for all loop diagrams in the effective theory.

We now want to match the Feynman diagrams involving a current operator, two collinear fields of the type $\cb
$, and one collinear field of the type $\cc
$ to the effective theory at the lowest order in the coupling constant. The relevant diagrammatic equation is
\vspace*{8mm}
\bea
\vcenter{\hbox{
  \begin{picture}(0,0)(0,0)
\SetScale{.6}
  \SetWidth{2}
  \Line(-40,40)(0,0)
  \Line(0,0)(40,40)
  \Line(15,15)(-10,40)
  \Photon(0,0)(0,-15){3}{2}
\Text(-35,14)[cb]{$p$}
\Text(35,21)[cb]{ $l_1$}
\Text(10,21)[cb]{$l_2$}
\LongArrow(-45,22)(-20,-3)
\LongArrow(20,3)(45,28)
\LongArrow(2,15)(-23,40)
\end{picture}}
}
\hspace*{1.5cm} + \hspace*{1.cm} 
\vcenter{ \hbox{
  \begin{picture}(0,0)(0,0)
\SetScale{.6}
  \SetWidth{2}
  \Line(-40,40)(0,0)
  \Line(0,0)(40,40)
  \Line(-15,15)(10,40)
  \Photon(0,0)(0,-15){3}{2}
\Text(-35,14)[cb]{$p$}
\Text(35,21)[cb]{ $l_1$}
\Text(-5,21)[cb]{$l_2$}
\LongArrow(-45,22)(-20,-3)
\LongArrow(20,3)(45,28)
\LongArrow(-3,15)(22,40)
\end{picture}} 
}
\hspace*{1.cm} &=& \tilde{C}_2^{(0)} \hspace*{.8cm}
\vcenter{ \hbox{
  \begin{picture}(0,0)(0,0)
\SetScale{.6}
  \SetWidth{2}
  \SetColor{Blue}
  \Line(-40,40)(0,0)
  \SetColor{OliveGreen}
  \DashLine(0,0)(40,40){6}
  \DashLine(15,15)(-10,40){6}
  \SetColor{Black}
  \Photon(0,0)(0,-15){3}{2}
\end{picture}}
}\hspace*{1.cm} + \tilde{C}_3^{(0)} \hspace*{.8cm}
\vcenter{ \hbox{
  \begin{picture}(0,0)(0,0)
\SetScale{.6}
  \SetWidth{2}
  \SetColor{Blue}
  \Line(-40,40)(0,0)
  \SetColor{OliveGreen}
  \DashLine(0,0)(40,40){6}
  \DashLine(0,0)(0,40){6}
  \SetColor{Black}
  \Photon(0,0)(0,-15){3}{2}
\end{picture}}
}\hspace*{1.cm} \, . \label{eq:matchC3}
\eea
The diagrams on the l.h.s.\ of the Eq.~(\ref{eq:matchC3}) are easily evaluated, since they involve only single propagators carrying momenta
\be
(p-l_2)^2  = - 2 p \cdot l_2  + {\mathcal O}\left(\lambda^2 \right)= - \left( n\cdot l_2 \right)\left( \bar{n}\cdot p \right)  + {\mathcal O}\left(\lambda^2 \right)
\ee
and $(l_1 +l_2)^2$. Since $\tilde{C}_2^{(0)} =1$, the first diagram on the l.h.s.\ and the first diagram on the r.h.s.\ of Eq.~(\ref{eq:matchC3}) give identical contributions and drop out of the equation. The value of the coefficient $\tilde{C}_3^{(0)}$ is therefore determined by the second diagram on the l.h.s.\ of Eq.~(\ref{eq:matchC3}):
\be \label{eq:C30}
\tilde{C}_3^{(0)}\left( n \cdot l_1, n \cdot l_2 , \bar{n} \cdot p,\mu \right) = 
\frac{g}{ -  \left( n\cdot l_2 \right)\left( \bar{n}\cdot p \right) + i 0}\, .
\ee

What is the form of the operator giving rise to the Wilson coefficient
$\tilde{C}_3$ in the effective Lagrangian? Using the correspondence $p^\mu
\leftrightarrow i \partial^\mu$ (for an incoming momentum), we see that the Wilson coefficient originates from terms involving inverse derivatives on the effective theory fields:
\be
\frac{g}{ -  \left( n\cdot l_2 \right)\left( \bar{n}\cdot p \right) + i 0}
\longleftrightarrow - g \left(\frac{1}{\bar{n}\cdot \partial} \pcol{\phi_{\cc}} \right)
 \left(\frac{1}{n\cdot \partial} \lcol{\phi_{\cb}} \right) \lcol{\phi_{\cb}} \,.
 \label{eq:LagCorr}
\ee
The two derivatives scale both like $\lambda^0 Q$ and therefore the current operator $J_3$ is suppressed by a factor $1/Q^2$. The presence of an inverse derivative is at first sight disturbing, but it is again an effect of the non-locality mentioned above. Observe that 
the inverse derivative of a field can be written as an integral
\be
\frac{i}{i n \cdot \partial+i0^+} \, \phi(x) = \int_{-\infty}^0\! ds\, \phi(x + s n) \, ;
\label{eq:invder}
\ee
in fact the relation above can be checked by applying the derivative to the r.h.s.\
\be
n^\mu \int_{-\infty}^0\! ds\, \partial_\mu \phi(x + s n) =
n^\mu \int_{-\infty}^0\! ds\, \frac{1}{n^\mu} \frac{\partial}{ \partial s } \phi(x + s n) = \left. \phi(x + s n)\right|^0_{-\infty} = \phi(x) \, . 
\ee 
Note that Eq.~(\ref{eq:invder}) implies an infinitesimal imaginary part in the operator on the l.h.s.,   see Appendix~\ref{sec:definitioninvder} for details.

It is a characteristic feature of SCET that the operators are non-local along the directions of large light-cone momentum. In general, in order to write down the most general SCET operators, one smears the fields along the light cone.
Therefore the current operator in the full theory, which is quadratic in the fields, will be replaced as follows
\be
J = C \phi^2 (x)  = J_2(x) + J_3(x) +\cdots \, ,
\ee
where the operator $J_2$ has the form shown in Eq.~(\ref{eq:J2}), while 
\be
J_3(x) = \int_{-\infty}^{+\infty} ds \int_{-\infty}^{+\infty} dt_1 \int_{-\infty}^{+\infty} dt_2 \, 
 C_3(s,t_1 ,t_2,\mu) \pcol{\phi_{\cc}(x + s \bar{n})}  \lcol{\phi_{\cb}(x + t_1 n)  \phi_{\cb}(x + t_2 n)}  + \left( \cc \leftrightarrow \cb \right) \, .
\ee
The result for the position space Wilson coefficients is given by
\bea
C_2(s,t,\mu) &=& \delta(s) \delta(t) +{\mathcal O}(g^2) \, , \nn \\
C_3(s,t_1,t_2,\mu) &=& g \theta(-s)  \delta(t_1)\theta( t_2)+{\mathcal O}(g^3) \, ,
\eea
which can be verified by carrying out the Fourier integrals
\bea
\tilde{C}_2^{(0)} &=& \int_{-\infty}^{+\infty} ds \int_{-\infty}^{+\infty} dt e^{i s \bar{n} \cdot p} e^{-i t n \cdot l} \, \delta(s) \delta(t) =1 \, , \nn \\
\tilde{C}_3^{(0)} &=& g \int_{-\infty}^{+\infty} ds \int_{-\infty}^{+\infty} dt_1   \int_{-\infty}^{+\infty} dt_2e^{i s \bar{n} \cdot p} e^{-i t_1 n \cdot l_1}e^{-i t_2 n \cdot l_2}\theta(-s) \theta(t_2) \delta(t_1) \, \nn \\
&=& g \int_{-\infty}^{0} ds \int^{\infty}_0 dt_2 e^{i s \bar{n} \cdot p} e^{-i t_2 n \cdot l_2} = \frac{g}{-\left(\bar{n} \cdot p \right)\left(n \cdot l_2 \right) }\, .
\eea
The dependence of the functions $C_i$ on $s,t$ is equivalent to the dependence 
of the coefficients $\tilde{C}_i$ on the large energy scale in momentum space; the correspondence between the two notations is given by
\be
\delta(s) \leftrightarrow 1 \, , \qquad \theta(-s) \leftrightarrow \frac{1}{i Q} \, .
\ee

\subsection{Sudakov Form Factor in SCET}

At this point all of the elements needed for the calculation of the one-loop correction to the current operator in the $\phi^3$ theory in the limit in which 
$\lambda \to 0$ are available. By employing the Feynman rules derived from the SCET Lagrangian one finds
\vspace*{.8cm}
\bea
\vcenter{ \hbox{
  \begin{picture}(0,0)(0,0)
\SetScale{.6}
  \SetWidth{2}
  \Line(-40,40)(0,0)
  \Line(0,0)(40,40)
  \Line(-30,30)(30,30)
  \Photon(0,0)(0,-15){3}{2}
\Text(-38,14)[cb]{{\Large $p$}}
\Text(38,15)[cb]{{\Large $l$}}
\LongArrow(-45,22)(-20,-3)
\LongArrow(20,-3)(45,22)
\end{picture}}
}
\hspace*{1.5cm} &=& \hspace*{0.4cm}
g^2\,\tilde{C}^{(1)}_2 
\hspace*{1.4cm}
\vcenter{ \hbox{
  \begin{picture}(0,0)(0,0)
\SetScale{.6}
  \SetWidth{2}
  \SetColor{Blue}
  \Line(-40,40)(0,0)
  \SetColor{OliveGreen}
  \DashLine(0,0)(40,40){6}
  \SetColor{Black}
  \Photon(0,0)(0,-15){3}{2}
\Text(-38,14)[cb]{{\Large \pcol{$\phi_{\cc
}$}}}
\Text(38,15)[cb]{{\Large \lcol{$\phi_{\cb
}$}}}
\end{picture}}}
\hspace*{1.6cm} + 
\tilde{C}^{(0)}_3 
\hspace*{1.6cm}
\vcenter{ \hbox{
  \begin{picture}(0,0)(0,0)
\SetScale{.6}
  \SetWidth{2}
  \SetColor{Blue}
  \Line(-40,40)(0,0)
  \CArc(-30,0)(30,0,90)
  \SetColor{OliveGreen}
  \DashLine(0,0)(40,40){6}
  \SetColor{Black}
  \Photon(0,0)(0,-15){3}{2}
\Text(-38,14)[cb]{{\Large \pcol{$\phi_{\cc
}$}}}
\Text(38,15)[cb]{{\Large \lcol{$\phi_{\cb
}$}}}
\end{picture}}}
\hspace*{1.cm}  \,  \nn \\
& &  \nn \\
& &  \nn \\
&+& \tilde{C}^{(0)}_3 
\hspace*{1.6cm}
\vcenter{ \hbox{
  \begin{picture}(0,0)(0,0)
\SetScale{.6}
  \SetWidth{2}
  \SetColor{Blue}
  \Line(-40,40)(0,0)
  \SetColor{OliveGreen}
  \DashLine(0,0)(40,40){6}
  \DashCArc(30,0)(30,90,180){6}
  \SetColor{Black}
  \Photon(0,0)(0,-15){3}{2}
\Text(-38,14)[cb]{{\Large \pcol{$\phi_{\cc
}$}}}
\Text(38,15)[cb]{{\Large \lcol{$\phi_{\cb
}$}}}
\end{picture}}}
\hspace*{1.cm} + \hspace*{1.cm}
\tilde{C}^{(0)}_2 
\hspace*{1.4cm}
\vcenter{ \hbox{
  \begin{picture}(0,0)(0,0)
\SetScale{.6}
  \SetWidth{2}
  \SetColor{Blue}
  \Line(-40,40)(0,0)
  \SetColor{OliveGreen}
  \DashLine(0,0)(40,40){6}
  \SetColor{Mahogany}
  \DashLine(-30,30)(30,30){2}
  \SetColor{Black}
  \Photon(0,0)(0,-15){3}{2}
\Text(-38,14)[cb]{{\Large \pcol{$\phi_{\cc
}$}}}
\Text(38,15)[cb]{{\Large \lcol{$\phi_{\cb
}$}}}
\end{picture}}}
 \hspace*{1.cm}\, . \label{eq:SCETregions}
\eea
It is perhaps useful to repeat that in the relation above the squares of the external momenta $p$ and $l$ are small but not exactly equal to zero from the start as in the matching calculation. 
By employing the expressions of the Wilson coefficients provided in the previous sections it is possible to see that the four diagrams on the r.h.s.\ of Eq.~(\ref{eq:SCETregions}) are the hard-region integral, the two collinear region integrals, and the soft-region integral which were found with the strategy of regions.  For example, let us consider the third diagram on the r.h.s.\ of Eq.~(\ref{eq:SCETregions}); one finds that
\vspace*{6mm}\\
\be\tilde{C}^{(0)}_3 
\hspace*{1.6cm}
\vcenter{ \hbox{
  \begin{picture}(0,0)(0,0)
\SetScale{1}
  \SetWidth{2}
  \SetOffset(0,-10)
  \SetColor{Blue}
  \Line(-40,40)(0,0)
  \SetColor{OliveGreen}
  \DashLine(0,0)(40,40){6}
  \DashCArc(30,0)(30,90,180){6}
  \LongArrow(15,3)(35,23)
  \SetColor{Black}
  \Photon(0,0)(0,-15){3}{2}
\Text(-46,24)[cb]{ \pcol{$p$}}
\Text(46,25)[cb]{\lcol{$l$}}
\Text(0,25)[cb]{ \lcol{$k$}}
\Text(40,-5)[cb]{ \lcol{$k+l$}}
\end{picture}}}\hspace*{2.cm} \Longrightarrow 
i g \int d^d k \frac{1}{\left( k^2 +i 0\right) \left[ (k+l)^2 +i 0\right]}
\underbrace{\frac{g}{2 k_+\cdot p_- + i0}}_{= \tilde{C}^{(0)}_3 } \, .
\ee
Similarly, one can prove that the fourth integral on the r.h.s.\ of Eq.~(\ref{eq:SCETregions}) gives rise to the integral in Eq.~(\ref{eq:Isres}),
simply by observing that the momentum in the soft internal line scales like
$k^2 \sim \lambda^4$ and therefore one must neglect $k^2$ in the two collinear propagators.

For order-by-order calculations, the direct application of the strategy of regions is more efficient. However, SCET allows one to study all-order properties of scattering amplitudes, such as factorization theorems. We discuss a factorization theorem for the Sudakov form factor in the next section. Furthermore, the renormalization group equations in the effective field theory can be employed to resum to all orders large logarithms of the ratio $p^2/Q^2$ (where $p$ represents here one of the collinear momenta).  

\subsection[Factorization of the Sudakov Form Factor in $d=6$]{Factorization of the Sudakov Form Factor in \boldmath$d=6$\unboldmath}

In this section we want to employ the SCET Lagrangian derived from the $\phi^3$ theory to prove a factorization theorem for the Sudakov form factor.
In four dimensions, the analysis is complicated by the fact that the coupling constant $g$ is not dimensionless. To avoid this problem, we will consider the theory in six dimensions.  The action of the theory in $d$ dimensions is
\be \label{eq:action}
S = \int d^d x \left[\frac{1}{2} \partial_\mu  \phi  (x) \,
 \partial^\mu \phi (x) - g \phi^3(x) \right] \, ,
\ee
which is dimensionless when setting $\hbar =1$, as we do throughout this work. By looking at the kinetic term, one can see that the mass dimension of the field is
\begin{equation}
\left[ \phi \right] = \frac{d-2}{2} \, ; \qquad \left[ \phi \right] = 1 \, \quad \mbox{in} \quad d=4 \, , \qquad \left[ \phi \right] = 2 \, \quad \mbox{in} \quad d=6 \,.
\end{equation}
Similarly, by looking at the interaction term one can determine the mass dimension of the coupling
\be
\left[ g \right] = \frac{6-d}{2} \, ; \qquad \left[ g \right] = 1 \, \quad \mbox{in} \quad d=4 \, , \qquad \left[ g \right] = 0 \, \quad \mbox{in} \quad d=6 \,.
\ee

At this stage, we want to study how the various fields in the effective theory scale in terms of powers of $\lambda$ in $d=6$. This will allow us to assign a power of $\lambda$ to any operator in the effective theory to determine which parts of the effective Lagrangian contribute at a given order; this process is called {\em power counting}.
We consider now the two-point correlator for collinear fields\footnote{We remind the reader that in Eq.~(\ref{eq:croo}) $p$ is a collinear momentum in six dimension, where the component in the collinear direction scales as $\lambda^0$, the component anti-parallel to the collinear direction scales as $\lambda^2$ (as in the four dimensional case), and the four transverse directions  scale proportionally to $\lambda$.}
\be \label{eq:croo}
\langle 0 |\, T \!\left\{ \pcol{\phi_c (x) \phi_c(0)}\right\}  |0\rangle \sim \int \underbrace{d^6 p}_{\lambda^6} \,  \underbrace{e^{-i p\cdot x}}_{\lambda^0}\,
 \underbrace{\frac{i}{p^2}}_{\lambda^{-2}} \,
 \sim \lambda^4 \, ,
\ee
and conclude that the collinear fields scale as $\pcol{\phi_c} \sim \lambda^2$.
(In Eq.~(\ref{eq:croo}) and in what follows, $T$ indicates a time ordering.)
One can carry out the same analysis by considering the correlator of two soft fields (in this case all of the components of the soft momentum scale as $\lambda^2$)
\be
\langle 0 |\, T \!\left\{ \scol{\phi_s (x) \phi_s(0)}\right\} | 0 \rangle \sim \int \underbrace{d^6 p}_{\lambda^{12}} \,  \underbrace{e^{-i p\cdot x}}_{\lambda^0}\,
 \underbrace{\frac{i}{p^2}}_{\lambda^{-4}} \,
 \sim \lambda^8 \, ,
\ee
so that $\scol{\phi_s} \sim \lambda^4$.

Next, we determine the scaling of each of the terms which appear in the effective Lagrangian. Keeping in mind that the scaling of the integration measure is given by the components of $x$, the conjugate variable to $p$, one finds
\be
\begin{array}{lllll}
\int d^6 x \, \frac{1}{2}\pcol{\partial_\mu   \phi_c  (x) \,
 \partial^\mu \phi_c (x)} &\sim& \frac{1}{\lambda^6} \left(\lambda 
 \lambda^2 \right)^2 &=& \lambda^0 \, ,  \\
  & & & & \\
\int d^6 x \, \frac{1}{2}\scol{\partial_\mu   \phi_s  (x) \,
 \partial^\mu \phi_s (x)} &\sim& \frac{1}{\lambda^{12}} \left( \lambda^2 
 \lambda^4 \right)^2 &=& \lambda^0 \, ,  \\
  & & & & \\
 \int d^6 x \, g \,\pcol{ \phi^3_c  (x)} &\sim& \frac{1}{\lambda^6} \left( 
 \lambda^2 \right)^3 &=& \lambda^0 \, ,  \\
  & & & & \\
 \int d^6 x \, g\, \scol{ \phi^3_s (x)} &\sim& \frac{1}{\lambda^{12}} \left( 
 \lambda^4 \right)^3 &=& \lambda^0 \, ,  \\
  & & & & \\
 \int d^6 x \, g\, \pcol{\phi^2_c  (x)} \, \scol{\phi_s  (x)} &\sim& \frac{1}{\lambda^{6}} \left( 
 \lambda^2 \right)^2  \lambda^4&=& \lambda^2 \quad \Longrightarrow \quad \mbox{Suppressed}\, .
\end{array}
\ee
The terms originating from the current operator $J = \phi^2$  scale instead as follows
\begin{align}
\int d^6 x J_2(x) \propto \int d^6 x  \,\pcol{ \phi_{\cc} (x)} \,  \lcol{\phi_{\cb} (x)} & \sim 
\frac{1}{\lambda^4} \lambda^2 \lambda^2 = \lambda^0 \, ,  \label{eq:cursca2}
\end{align}
Combinations involving more fields are powers suppressed, as it can be seen below:
\begin{align}
\int  d^6 x  \, \pcol{\phi^2_{\cc} (x)}\, \lcol{\phi_{\cb} (x)} &\sim 
\frac{1}{\lambda^4} \left(\lambda^2\right)^2 \lambda^2 = \lambda^2\quad \Longrightarrow \quad \mbox{Suppressed} \, ,  \nonumber \\
  \int  d^6 x  \, \pcol{\phi_{\cc} (x)}\, \lcol{\phi_{\cb}(x)}\, \scol{\phi_{s} (x)} &\sim 
\frac{1}{\lambda^4} \left(\lambda^2\right)^2 \lambda^4 = \lambda^4\quad \Longrightarrow \quad \mbox{Suppressed} \, . \label{eq:cursca}
\end{align}
Observe that the integration measure in Eqs.~(\ref{eq:cursca2},\ref{eq:cursca}) scales as  $1/\lambda^4$ because  both the plus  and minus components of $x^\mu$ are of $\lambda^0$, since they are conjugate to a momentum which is a sum of $l$-collinear and $p$-collinear momenta. 
Therefore $d^6x \sim (p_\perp)^{-4} \sim \lambda^{-4}$.

In summary we conclude that 
\be
\int d^6 x \, {\mathcal L}_{{\mbox{\tiny SCET}}} = \int d^6 x \, 
\left[{\mathcal L}_{\cc
} + {\mathcal L}_{\cb
} + {\mathcal L}_{s} \right] +
{\mathcal O}\left( \lambda^2 \right) \, ,
\ee
while for the current operator one finds
\be
\int d^6 x \,J(x) \to  \int  d^6 x \, \int ds \, \int dt \, C(s,t,\mu) \, \pcol{\phi_{\cc}\left( x +s \bar{n} \right)} \,
 \lcol{\phi_{\cb}\left( x +t n \right)} \, +  {\mathcal O}\left( \lambda^2 \right) \, .
\ee
Since soft-collinear interactions are power suppressed, it is possible to obtain a factorization theorem.

Let us consider the following correlator

\bea \label{eq:Gpl}
G(p,l,\mu) &=& \int d^6 x_1 \, \int d^6 x_2 \, e^{-i p\cdot x_1 + i l \cdot x_2 }
\langle 0 | T \left\{ \pcol{\phi_{\cc} (x_1)} J (0) 
\lcol{\phi_{\cb} (x_2)} \right\} | 0 \rangle \, \nn \\
&=&\int d^6 x_1 \, \int d^6 x_2 \, e^{-i p\cdot x_1 + i l \cdot x_2 }
\int ds \, \int dt \, C(s,t,\mu) \nn \\
&\times&
\langle 0| T \left\{ \pcol{\phi_{\cc}(x_1)  \phi_{\cc}(s \bar{n}) } \right\}|0\rangle
\langle 0| T \left\{ \lcol{\phi_{\cb}(t n) \phi_{\cb}(x_2)} \right\}|0\rangle \, ,
\eea
Since the soft-collinear interactions are power suppressed, the fields $\pcol{\phi_{\cc}}$ and $\lcol{\phi_{\cb}}$ do not interact with each other. Up to power suppressed terms, we now deal with two separate theories and the matrix element in the first line reduces to a collinear matrix element of the $\pcol{\phi_{\cc}}$ fields times a matrix element of the $\lcol{\phi_{\cb}}$ fields.

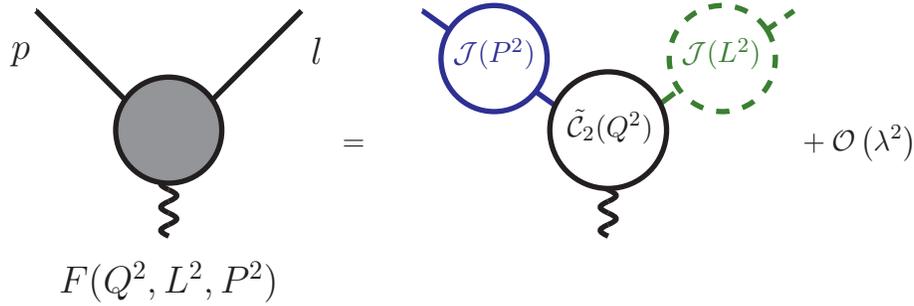
\begin{figure}[t]
\begin{center}
\[
\hspace*{18mm}
\vcenter{ \hbox{
  \begin{picture}(0,0)(0,0)
\SetScale{1}
  \SetWidth{2}
  \Line(-50,50)(0,0)
  \Line(0,0)(50,50)
  \Photon(0,0)(0,-35){3}{4}
  \CCirc(0,5){20}{Black}{Gray}
\Text(-56,29)[cb]{{\Large  $p$}}
\Text(56,30)[cb]{{\Large $l$}}
\Text(0,-60)[cb]{{\Large $F(Q^2,L^2,P^2)$}}
\end{picture}}} \hspace*{2.2cm} = \hspace*{3cm}
\vcenter{ \hbox{
  \begin{picture}(0,0)(0,0)
\SetScale{1}
  \SetWidth{2}
\SetColor{Blue}
  \Line(-70,50)(0,0)
  \CCirc(-43,32){20}{Blue}{White}
\SetColor{OliveGreen}
  \DashLine(0,0)(25,19){6}
    \DashLine(60,43)(70,50){6}
  \DashCArc(43,32)(20,33,393){5}
\SetColor{Black}
  \Photon(0,0)(0,-35){3}{4}
  \CCirc(0,5){22}{Black}{White}
\Text(-1,0)[cb]{ $\tilde{{\mathcal C}}_2(Q^2)$}
\Text(-45,28)[cb]{ \pcol{${\mathcal J}(P^2)$}}
\Text(41,28)[cb]{ \lcol{${\mathcal J}(L^2)$}}
\end{picture}}}\hspace*{2.5cm} + 
{\mathcal O}\left(\lambda^2 \right) 
\]
\vspace*{8mm}
\end{center}
\caption{Diagrammatic representation of the factorization theorem for the 
$\phi^3$ theory in $d=6$. \label{fig:factorization}}
\end{figure}

Translation invariance implies that
\be
\langle 0| T \left\{ \pcol{\phi_{\cc}(x_1)   \phi_{\cc}(s \bar{n})} \right\}|0\rangle =
\langle 0| T \left\{ \pcol{\phi_{\cc}(x_1 -s \bar{n})  \phi_{\cc}(0)} \right\}|0\rangle\, ,
\ee
and a similar relation for the other time ordered product. One can then  carry out the following changes of variables in Eq.~(\ref{eq:Gpl}):
\be
x_1 \rightarrow x_1 + s \bar{n} \, , \qquad \mbox{and} \qquad
x_2 \rightarrow x_2 + t n \, ,
\ee
to obtain
\be \label{eq:Gplfact}
G(p,l,\mu) = \int ds \, \int dt \, C\left(s,t,\mu \right) e^{- i s p\cdot \bar{n} + i t l \cdot n} \pcol{{\mathcal J}\!\left( p^2,\mu \right)} \lcol{{\mathcal J}\!\left( l^2 ,\mu\right)} \, ,
\ee
with 
\bea
\pcol{{\mathcal J}\!\left( p^2,\mu \right)} &\equiv& \int d^6 x_1 \, e^{-i p \cdot x_1}
\langle 0|T \left\{ \pcol{\phi_{\cc}(x_1) \phi_{\cc}(0)} \right\} |0\rangle  \, , \nn \\
\lcol{{\mathcal J}\!\left( l^2,\mu \right)} &\equiv& \int d^6 x_2 \,  e^{i l \cdot x_2}
\langle 0|T \left\{\lcol{ \phi_{\cb}(0) \phi_{\cb}(x_2)} \right\} |0\rangle \, . 
\eea
The functions ${\mathcal J}$ do not depend on $s$ and $t$, and therefore the integral in Eq.~(\ref{eq:Gplfact}) factors out. By introducing the notation
\be
\tilde{{\mathcal C}}_2\left( \bar{n} \cdot p , n \cdot l,\mu \right) \equiv \int ds \, \int dt \, C\left(s,t,\mu \right) e^{- i s p\cdot \bar{n} + i t l \cdot n} \, ,
\ee
one can rewrite the three-point correlator in Eq.~(\ref{eq:Gplfact}) as the product of three functions
\be\label{scalarsudakov}
G(p,l,\mu) =\tilde{{\mathcal C}}_2\left( \bar{n} \cdot p , n \cdot l,\mu \right) \pcol{{\mathcal J}\!\left( p^2,\mu \right)} \lcol{{\mathcal J}\!\left( l^2,\mu \right)} \, .
\ee
We have factorized the Green function $G$ into a product of a {\em hard function} $\tilde{{\mathcal C}}$ and two {\em jet functions} ${\mathcal J}$.
The jet function can be calculated within the full theory since
the collinear Lagrangian is identical to the complete $\phi^3$ Lagrangian. The content of the factorization theorem is summarized in diagrammatic form in Fig.~\ref{fig:factorization}. The nontrivial part of the factorization theorem is that the hard function can be calculated at $p^2 =l^2 =0$, so that we have managed to factor a function of three variables into a product of three functions of a single variable. The full Sudakov form factor is split into an high-energy contribution (the hard function), and two low-energy contributions (the jet functions).

It would be interesting to use the factorization theorem to resum Sudakov logarithms 
to all orders in the coupling constant; this can be done by employing renormalization group tools within the effective theory. We will return on this subject in Section~\ref{sec:LIV}. Let us note a particularity of $\phi^3$ theory in $d=6$. The Sudakov logarithms have the  form 
\begin{displaymath}
\left(g^2\right)^n \ln^{n} \left(\frac{p^2 l^2}{Q^4} \right) \, ,
\end{displaymath}
so there is only a single logarithm at each order in perturbation theory. This is due to the absence of a soft contribution to the Sudakov form factor (\ref{scalarsudakov}): the double logarithms arise in the interplay of soft and collinear contributions and will be present in the QCD case.

%% file: 4_scetQCD.tex
\section{Generalization to QCD \label{sec:SCET_QCD}}

The effective theory for QCD can be constructed following exactly the same procedure we employed in order to construct the SCET Lagrangian for the scalar $\phi^3$ theory in the previous section; in addition, many elements are the same as in the scalar case. In particular, the same momentum regions appear, since only the numerators of the diagrams differ between the $\phi^3$ theory case and the QCD case. However, in the QCD case three complications arise:
\begin{itemize}
\item[i)] Different components of the quark field $\psi(x)$ and of the gluon
field $A_\mu(x)$  scale differently with the expansion parameter $\lambda$;  
\item[ii)] the theory is invariant under gauge transformations, but it is necessary to make sure that they respect the scaling of the fields, and
\item[iii)] non-local operators involve Wilson lines to ensure gauge invariance. 
\end{itemize}

In order to keep the discussion as simple as possible, we start by considering only one
type of collinear fields, with a momentum which scales as
\be
p^\mu \sim \left(\lambda^2, 1, \lambda\right)Q \,.
\ee 
One then splits the gluon and quark fields in a collinear and a soft part
\be
A^\mu(x) \rightarrow \pcol{A_c^\mu(x)} +\scol{A_s^\mu(x)} \, ,
\qquad
\psi (x) \rightarrow \pcol{\psi_c(x)} + \scol{\psi_s(x)} \, .
\ee
We now consider the collinear part of the fermion field and we further split it into two components as follows
\be
\pcol{\psi_c(x)} \equiv \pcol{\xi(x)} + \pcol{\eta(x)} \, , 
\ee
where
\be
\pcol{\xi }= P_+ \, \pcol{\psi_c} \equiv \frac{\nsl \nbsl}{4} \pcol{\psi_c} \, , \qquad
\pcol{\eta} = P_- \, \pcol{\psi_c} \equiv \frac{\nbsl \nsl}{4} \pcol{\psi_c} \, .
\ee
As a consequence of the definition of the operators $P_\pm$ and of the fact that
$n^2 = \bar{n}^2 =0$ one finds that
\be
\nsl\, \pcol{ \xi(x)} = 0 \, ,  \quad \mbox{and} \quad \nbsl\, \pcol{\eta(x)} = 0 \, .
\ee 
It is easy to check that  $P_\pm$ are projection operators:
\bea
P_+ +P_- &=& \frac{\nsl \nbsl}{4}+ \frac{\nbsl \nsl}{4}=  \frac{2 \bar{n} \cdot n}{4} = 1 \, ,
\eea
and one can also immediately verify that $P_+^2=P_+$ and $P_-^2=P_-$.

%

\subsection{Power Counting}

As a first step, we want to determine the power of $\lambda$ with which the different field components of the SCET fields scale. As in the scalar case, this information can be obtained by looking at the two-point correlators. We start with the $\xi$ component\footnote{Observe that
$\overline{\left( \nsl \nbsl \psi \right)} = \psi_c^\dagger  \nbsl^\dagger \nsl^\dagger \gamma_0 = \bar{\psi}\nbsl \nsl$, which follows after inserting $\left(\gamma^0\right)^2=1$ between the Dirac matrices and using $\gamma^0 {\gamma^\mu}^\dagger \gamma^0 = \gamma^\mu$.}
\bea
\langle 0| T\left\{ \pcol{\xi(x) \bar{\xi}(0)} \right\} |0 \rangle &=& \frac{\nsl \nbsl}{4}\langle 0| T\left\{\pcol{ \psi_c (x) \bar{\psi}_c(0) }\right\} |0 \rangle\frac{\nbsl \nsl}{4} \,  \nn \\
&=& \int \frac{d^4 p}{(2 \pi)^4} \frac{i}{p^2 +i 0} e^{-i p\cdot x}\frac{\nsl \nbsl}{4} p\!\!\!\slash \frac{\nbsl \nsl}{4} \sim \lambda^4 \frac{1}{\lambda^2} =
\lambda^2 \, ,
\eea
where we employed the identity
\be
\frac{\nsl \nbsl}{4} p\!\!\!\slash \frac{\nbsl \nsl}{4} = 
\frac{\nsl \nbsl}{4}\left[ \bar{n} \cdot p \frac{\nsl}{2}+ n \cdot p \frac{\nbsl}{2}+{p\!\!\!\slash}_\perp\right]  \frac{\nbsl \nsl}{4}= \bar{n} \cdot p\frac{\nsl}{2} \sim \lambda^0 \, .
\ee
Therefore $\xi(x) \sim \lambda$. The correlator for the $\eta$ component is 
\bea
\langle 0| T\left\{ \pcol{ \eta(x) \bar{\eta}(0)} \right\} |0 \rangle &=& \frac{\nbsl \nsl}{4}\langle 0| T\left\{ \pcol{\psi_c (x) \bar{\psi}_c(0)} \right\} |0 \rangle\frac{\nsl \nbsl}{4} \,  \nn \\&=& \int \frac{d^4 p}{(2 \pi)^4} \frac{i}{p^2 +i 0} e^{-i p\cdot x}
\underbrace{\frac{\nbsl \nsl}{4} p\!\!\!\slash \frac{\nsl \nbsl}{4}}_{= n \cdot p\frac{\nbsl}{2}} \sim \lambda^4  \lambda^2 \frac{1}{\lambda^2} =
\lambda^4\, ;
\eea
the scaling of this component is $\pcol{\eta (x)} \sim \lambda^2$. The $\pcol{\eta}$ component is thus suppressed by one power of $\lambda$ with respect to the $\pcol{\xi}$ component. Finally for the soft field one finds
\bea
\langle 0| T\left\{\scol{ \psi_s (x) \bar{\psi}_s (0)} \right\} |0 \rangle &=& \int \frac{d^4 p}{(2 \pi)^4} \frac{ip\!\!\!\slash}{p^2 +i 0} e^{-i p\cdot x}  \sim (\lambda^2)^4 \lambda^2 \frac{1}{\lambda^4} =
\lambda^6 \, ,
\eea
so that $\psi_s \sim \lambda^3$.

The two-point correlator for the gluon field is 
\be
\langle 0| T\left\{ A^\mu (x) A^\nu (0) \right\} |0 \rangle =\int \frac{d^4 p}{(2 \pi)^4} \frac{i}{p^2 +i 0} e^{-i p\cdot x} \left[ - g^{\mu \nu} +\xi \frac{p^\mu p^\nu}{p^2}\right] \, .
\ee
A glance to the second term in the square bracket shows that the gluon field
scales like its momentum, therefore $\scol{A_s^\mu(x)} \sim p_s^\mu$ and 
$\pcol{A^\mu_c(x)} \sim p_c^\mu$, or equivalently
\be
\pcol{\bar{n} \cdot A_c} \sim \lambda^0 \, , \quad 
\pcol{n \cdot A_c} \sim \lambda^2 \, , \quad \pcol{A_\perp} \sim \lambda \, ;
\quad \scol{A_s^\mu} \sim \lambda^2 \, .
\ee
 The soft gluon field is power suppressed with respect to the collinear gluon field, except for what concerns the $n \cdot A_s$ component, which scales in the same way as the corresponding collinear gluon component. 
Only two of the four components of the gluon fields are physical and the reader might find it unnatural that the leading-power result for the collinear gluon field is given by the unphysical longitudinal polarization. This could be avoided by working in the light-cone gauge $\pcol{\bar{n} \cdot A_c}=0$ where this component vanishes instead of the $R_\xi$ gauges we consider. This gauge was used as a starting point in \cite{Beneke:2002ph,Beneke:2002ni} and gauge invariance was then recovered after relating the original fields to the light-cone-gauge fields using certain Wilson lines.

\subsection{Effective Lagrangian}

The collinear fermion Lagrangian has a special form since the $\pcol{\eta}$ components are of higher order in $\lambda$ with respect to the  $\pcol{\xi}$ components and can be integrated out. The covariant derivative is defined as usual as
\be
i D_\mu \equiv i \partial_\mu + g A_\mu = i \partial_\mu + g (\pcol{{A_c}^a_\mu} +\scol{{A_s}^a_\mu})\, t^a \, ,
\ee
where the matrices $t^a$ are the generators of $\text{SU}(3)$ in the fundamental representation. For the moment, we keep both the soft and collinear components of the gluon field even though ${A_s}_\perp$ and ${A_s}_-$ are power suppressed with respect to the collinear gluon field. We will come back to this point when discussing the soft-collinear interactions.
By using the relations $\nsl \pcol{\xi} =\pcol{ \bar{\xi}} \nsl = 0$ and $\nbsl \pcol{\eta} = \pcol{\bar{\eta}} \nbsl=0$, $\pcol{\bar{\xi}} \Dsl_\perp \pcol{\xi} = 0$ and
$\pcol{\bar{\eta}} \Dsl_\perp \pcol{\eta} =0$ one obtains
\footnote{
Note that $\pcol{\bar{\xi}}  \Dsl_\perp \pcol{\xi}  = \pcol{\bar{\xi}} P_-  \Dsl_\perp P_+ \pcol{\xi} = \pcol{\bar{\xi}}   \Dsl_\perp P_- P_+ \pcol{\xi} = 0$, where we have used $\{\nsl, \Dsl_\perp \} = \{\nbsl, \Dsl_\perp \} = 0$.
} 
\bea \label{eq:Lfc}
{\mathcal L}_c &=& \pcol{\bar{\psi}_c}\, i \Dsl\, \pcol{\psi_c} \,  \nn \\
&=& \left(\pcol{\bar{\xi}} + \pcol{\bar{\eta}}\right)
\left[\frac{\nsl}{2} i \bar{n} \cdot D  + \frac{\nbsl}{2} i n \cdot D +
i \Dsl_\perp \right] \left(\pcol{\xi} + \pcol{\eta}\right) \, \nn \\
&=& \pcol{\bar{\xi}} \frac{\nbsl}{2} i n \cdot D \pcol{\xi} + 
\pcol{\bar{\xi}} i \Dsl_\perp \pcol{\eta} + \pcol{\bar{\eta}}  i\Dsl_\perp \pcol{\xi} +
\pcol{\bar{\eta}} \frac{\nsl}{2} i \bar{n} \cdot D \pcol{\eta} \, .
\eea

Since the action is quadratic, one can integrate out $\pcol{\eta}$ exactly.
An easy way to obtain the Lagrangian after the field $\pcol{\eta}$ is integrated out consists in employing the equations of motion derived from the Lagrangian in Eq.~(\ref{eq:Lfc}).
The equations of motion for $\pcol{\bar{\xi}}$ are  
\be
\partial_\mu \frac{\partial {\mathcal L}_c}{ \partial (\partial_\mu \pcol{\bar{\xi}})} -
\frac{\partial {\mathcal L}_c}{\partial \pcol{\bar{\xi}} } = -
\frac{\partial {\mathcal L}_c}{\partial \pcol{\bar{\xi}} } = -\frac{\nbsl}{2} i n \cdot D \pcol{\xi} - i \Dsl_\perp \pcol{\eta }=0 \, ,
\ee 
or equivalently
\be
\frac{\nbsl}{2}  n \cdot D\pcol{ \xi} = -  \Dsl_\perp \pcol{\eta} \, .
\ee
Similarly for $\pcol{\bar{\eta}}$ one finds 
\be
\Dsl_\perp \pcol{\xi} = -\frac{\nsl}{2}  \bar{n} \cdot D \pcol{\eta} \, .
\ee
From the latter one obtains
\be
\frac{\nbsl}{2}\Dsl_\perp \pcol{\xi} = - \frac{\nbsl \nsl}{4}  \bar{n} \cdot D \pcol{\eta} =
 -  \bar{n} \cdot D \pcol{\eta} \,. 
\ee
Solving for $\pcol{\eta}$ one finds
\be \label{eq:gron}
\pcol{\eta}  = -\frac{\nbsl}{2 \bar{n} \cdot D} \Dsl_\perp \pcol{\xi} \, , 
\quad \mbox{and} \quad \pcol{\bar{\eta}} = - \pcol{\bar{\xi}} \, \overleftarrow{\Dsl_\perp} \frac{\nbsl}{2 \bar{n} \cdot \overleftarrow{D}} \, , 
\ee
where the arrow indicates that the covariant derivative is acting to the left.
At this stage one can insert Eqs.~(\ref{eq:gron}) in the collinear Lagrangian in order to eliminate $\eta$:
\bea
{\mathcal L}_c &=& \pcol{\bar{\xi}} \frac{\nbsl}{2} i n \cdot D \pcol{\xi}+
\pcol{\bar{\xi}} i \Dsl_\perp \frac{1}{i \bar{n} \cdot D} i \Dsl_\perp \frac{\nbsl}{2} \pcol{\xi} +\pcol{\bar{\xi}} i \overleftarrow{\Dsl}_\perp \frac{1}{i \bar{n} \cdot\overleftarrow{D} } i \Dsl_\perp \frac{\nbsl}{2} \pcol{\xi} \nn \\
&&+ \pcol{\bar{\xi}} i \overleftarrow{\Dsl}_\perp \frac{\nbsl}{2 i \bar{n} \cdot\overleftarrow{D} } \underbrace{\frac{\nsl}{2} i \bar{n} \cdot D \frac{\nbsl}{2 i \bar{n} \cdot D}}_{\frac{\nsl \nbsl}{4} = P_{+}} i \Dsl_\perp \pcol{\xi} \, \nn \\
&=& \pcol{\bar{\xi}} \frac{\nbsl}{2} i n \cdot D \pcol{ \xi} +
\pcol{\bar{\xi}} i \Dsl_\perp \frac{1}{i \bar{n} \cdot D} i \Dsl_\perp \frac{\nbsl}{2} \pcol{\xi} \, . \label{eq:colllag}
\eea
In deriving the equation above we repeatedly used the fact that $\{\nbsl, \Dsl_\perp \} = 0$ and in the last line we used that 
\bea
P_{+} \Dsl_\perp \pcol{\xi}  =  \Dsl_\perp P_{+} \pcol{ \xi} 
 & = & \Dsl_\perp \pcol{\xi} \, .
\eea

In the path integral, the integration over the fermionic fields $\pcol{\eta}$ and $\pcol{\bar{\eta}}$ gives (see for example \cite{Bohm:2001yx}, page 110)
\be
\int {\mathcal D} [\pcol{\eta}] {\mathcal D} [\pcol{\bar{\eta}}] 
\exp\left\{\int d^4 x \,\pcol{ \bar{\eta}} \frac{\nsl}{2} i \bar{n} \cdot D \pcol{\eta} \right\}
=\mbox{det}\left(\frac{\nsl}{2} i \bar{n} \cdot D \right) \, .
\ee
We now show that this overall determinant is irrelevant.
Observe that the determinant is gauge invariant; in fact, if we indicate with $V$ a $\text{SU}(N)$ matrix so that a quark field transforms according to $\psi \to V \psi$ under gauge transformations, the determinant's covariant derivative will transform as $D \to V D V^\dagger$. Therefore
\bea
\mbox{det}\left(\frac{\nsl}{2} i \bar{n} \cdot D \right) \rightarrow
\mbox{det}\left(\frac{\nsl}{2} V i \bar{n} \cdot D V^\dagger \right) &=& 
 \underbrace{\mbox{det}\left(V \right)}_{=1} 
\mbox{det}\left(\frac{\nsl}{2}\, i \bar{n} \cdot D  \right)
\underbrace{\mbox{det}\left(V^\dagger \right)}_{=1} \,  \nn \\
&=&\mbox{det}\left(\frac{\nsl}{2} i \bar{n} \cdot D \right)\, .
\eea
In the light cone gauge, where $\bar{n} \cdot D =\bar{n}\cdot \partial $, the determinant is trivially independent of the gluon field; since the determinant was just proven to be gauge independent, it does not depend on the gluon field in any gauge, and is therefore  an irrelevant factor multiplying the path integral. 
%
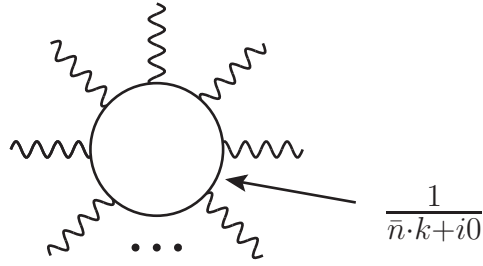
\begin{figure}[t]
\[
\vcenter{ \hbox{
  \begin{picture}(0,0)(0,0)
\SetScale{1}
  \SetWidth{1}
  \Photon(-40,40)(0,0){3}{7}
  \Photon(0,0)(40,40){3}{7}
  \Photon(40,-40)(0,0){3}{7}
  \Photon(0,0)(-40,-40){3}{7}
  \Photon(0,0)(0,55){3}{7}
  \Photon(-55,0)(0,0){3}{7}
  \Photon(-55,0)(0,0){3}{7}
  \Photon(0,0)(55,0){3}{7}
  \LongArrow(75,-20)(30,-12)
\CCirc(-8,-37){1}{Black}{Black}
\CCirc(0,-37){1}{Black}{Black}
\CCirc(8,-37){1}{Black}{Black}
  \CCirc(0,0){25}{Black}{White}
\Text(105,-35)[cb]{{\LARGE $\frac{1}{\bar{n}\cdot k + i 0}$}}
\end{picture}}} 
\]
\vspace*{8mm}
\caption{Diagrammatic representation of the diagrams corresponding to the 
determinant originating from the integration over $\eta$. \label{fig:det}}
\end{figure}
%
From the diagrammatic point of view the determinant corresponds to the graphs shown in Fig~\ref{fig:det}.
At a heuristic level, this can be understood by observing that since the $\pcol{\eta}$ field was integrated out, it cannot appear among the external legs and it can therefore contribute only through closed loops. In this aspect, the situation is the same as one encounters in the Euler-Heisenberg effective Lagrangian, where
the electron is considered a heavy field and is integrated out.
However,  in the case under study the diagrams in Fig.~\ref{fig:det} vanish, since $\mbox{Im}(\bar{n}\cdot k ) <0$ and all of the poles are on the same side of the $\bar{n} \cdot k$ axis.


While the collinear quark Lagrangian has a somewhat complicated structure, the collinear gluon Lagrangian is simply a copy of the QCD Lagrangian in which the gluon field $A^\mu$ is replaced by the collinear gluon field $A^\mu_c$. The same is true for the Lagrangian with the kinetic terms for the soft fields, which has the form 
\be \label{eq:softL}
\scol{{\mathcal L}_s} = \scol{\bar{\psi}_s i \Dsl_s \psi_s} -\frac{1}{4} \scol{(F^{a}_s)_{\mu \nu}
(F^{a}_s)^{\mu \nu}} \, ,
\ee 
where the covariant derivative and field strength are defined as
\bea
\scol{i D_s^\mu} &=& i \partial^\mu +g \scol{A_s^\mu } =i \partial^\mu +g \scol{(A^a_s)^\mu} t^a \, , \nn \\
 i g \scol{ (F^{a}_s)^{\mu \nu}} t^{a} &=& \left[ i \scol{D_s^\mu}, i \scol{D_s^\nu} \right]  =
 ig \left\{\partial^\mu \scol{A^\nu_s} - \partial^\nu \scol{A^\mu_s} - i g \left[\scol{A^\mu_s},\scol{A^\nu_s} \right]\right\} \, ,  \\
 &=& ig \left\{ \partial^\mu \scol{(A^a_s)^\nu} -\partial^\nu \scol{(A_s^a)^\mu} + g f^{abc}
 \scol{(A_s^b)^\mu} \scol{(A_s^c)^\nu}  \right\} t^{a}\nn  \, .
\eea
Therefore the kinetic terms of the SCET QCD Lagrangian are given in Eqs.~(\ref{eq:Lfc},\ref{eq:softL}) and by a standard kinetic term for the collinear gluons. Next, we consider the terms describing the interactions between soft and collinear fields.

\subsection{Soft-Collinear Interactions\label{sec:scint}}

The general construction of the soft-collinear interaction terms is somewhat involved and beyond the scope of these lectures; it can be found in \cite{Beneke:2002ni}. For collider physics applications it is usually sufficient to consider soft-collinear interactions at leading power. To obtain the leading power interactions, let us remind ourselves of the scaling of the different fields
\bea
\left( \pcol{n \cdot A_c} , \pcol{\bar{n} \cdot A_c}, \pcol{A_{c \perp}} \right) &\sim& 
\left(\lambda^2, 1,\lambda\right) \, , \nn \\
\left( \scol{ n \cdot A_s} , \scol{\bar{n} \cdot A_s}, \scol{A_{s \perp}} \right) &\sim& 
\left(\lambda^2, \lambda^2 ,\lambda^2\right) \, , \nn \\
\pcol{\xi} \sim \lambda \, ,&\qquad& \scol{\psi_s} \sim \lambda^3 \, .
\eea

In the case of the $\phi^3$ theory the soft-collinear interactions were obtained by replacing one of the fields in the interaction term with a soft field
\be
-\frac{g}{3!} \int d^4 x\, \phi^3 (x) \longrightarrow -\frac{g}{2!} \int d^4 x
\, \pcol{\phi^2_{\cc} (x) }\scol{\phi_{s}(x_-)}\, .
\ee
In the SCET Lagrangian for QCD, soft-collinear interactions involving soft quarks do not appear at leading order, since $\psi_s$ is power suppressed with respect to $\xi$. Furthermore, only the $n \cdot A_s$ component of the soft gluon field is not power suppressed with respect to the corresponding component of the collinear gluon field, so only this component enters the leading soft-collinear interactions. Therefore one can replace
\be
A^\mu(x) \longrightarrow \left( \pcol{n \cdot A_c (x)} +  \scol{n \cdot A_s (x_-)}\right)\frac{\bar{n}^\mu}{2} + \pcol{\bar{n} \cdot A_c (x) }\frac{n^\mu}{2} +\pcol{ A_{c \perp}^\mu (x)} \, . \label{eq:replA}
\ee
in the quark and gluon collinear Lagrangians discussed in the previous section. To summarize, the SCET Lagrangian for QCD can be written in a compact form as follows
\be \label{eq:LQCDSCET}
{\mathcal L}_{{\tiny \mbox{SCET}}} = \scol{\bar{\psi}_s i \Dsl_s \psi_s} + 
\pcol{\bar{\xi}} \frac{\nbsl}{2} \left[ i n \cdot D + i \pcol{\Dsl_{c \perp}} \frac{1}{i \pcol{\bar{n} \cdot D_c}} i \pcol{\Dsl_{c \perp}}\right] \pcol{\xi} -\frac{1}{4} \left( \scol{F_{\mu \nu}^{s, a}} \right)^2 -\frac{1}{4} \left( \pcol{F_{\mu \nu}^{c, a}} \right)^2 \, .
\ee
The various covariant derivatives which appear in Eq.~(\ref{eq:LQCDSCET}) are given by
\bea \label{eq:listcd}
i \scol{D^s_\mu} &=& i \partial_\mu + g \scol{A^{s }_\mu} = i \partial_\mu + g \scol{A_{\mu}^{s, a}}\, t^a \, , \nn \\
i \pcol{D^c_\mu} &=& i \partial_\mu + g \pcol{A^{c }_\mu} = i \partial_\mu + g \pcol{A_{\mu}^{c, a}}\, t^a \, ,\nn \\
i n\cdot D &=& i n \cdot \partial + g\, \pcol{n \cdot A_c (x)} + g\, \scol{n \cdot A_s (x_-)} \, .
\eea
The field strengths are
\bea \label{eq:fieldstr}
ig \scol{F^s_{\mu \nu}} &=& \left[ i \scol{D_\mu^s} ,  i \scol{D_\nu^s} \right] \, , \nn \\
ig \pcol{F^c_{\mu \nu}} &=& \left[ i D_\mu ,  i D_\nu \right] \, ,
\eea
where the covariant derivative appearing in the commutator in the last line of 
Eq.~(\ref{eq:fieldstr}) is
\be \label{eq:DforF}
D^\mu = n \cdot D \frac{\bar{n}^\mu}{2} +\pcol{\bar{n} \cdot D_c} \frac{n^\mu}{2} + \pcol{D^\mu_{c \perp}} \, .
\ee
As is evident from the last line of Eqs.~(\ref{eq:listcd}), $n \cdot D$ depends on the soft field $A_s$, so that one might wonder if the squared collinear field strength in the Lagrangian in Eq.~(\ref{eq:LQCDSCET}) gives rise to additional kinetic terms for the soft gluon field, in addition to the ones already included in the square of the soft field strength. This is not the case since the squared field strength is gauge invariant and one can choose to work in the gauge where $n \cdot A_s$ vanishes. In such a gauge the squared collinear field strength is clearly free from terms depending only on soft gluon fields.

The Lagrangian in Eq.~(\ref{eq:LQCDSCET}) includes only one collinear sector, but in practical applications one needs two or more. As it was done when discussing the scalar  $\phi^3$ theory, we will in the following consider two collinear momenta
$p \sim (\lambda^2,1, \lambda)$ and $l \sim (1, \lambda^2, \lambda)$.
The second collinear sector in the Lagrangian can be obtained by replacing $n^\mu \leftrightarrow \bar{n}^\mu$ (which implies $x_+ \leftrightarrow x_-$) in the first collinear sector.

\subsection{Gauge Transformations and Reparameterization Invariance}

We now discuss two symmetries of SCET. Both are not symmetries of nature but redundancies in our description. The first one is gauge symmetry which arises because we use four-component fields to describe the two physical polarizations of gauge bosons. The second one is called reparameterization invariance and arises because we have introduced two reference vectors, $n_\mu$ and $\bar{n}_\mu$, in the construction of the effective theory. The choice of these is not unique and physics is independent of their choice.

Let us start with reparameterization invariance, which was first explored in the context of Heavy Quark Effective Theory (HQET). This theory involves a reference vector in the direction of the heavy quark, whose direction can be changed by a small amount \cite{Luke:1992cs}.  In SCET, the set of transformations is richer.  Not only can one change the direction of both reference vectors by a small amount, but one can also rescale the light-like reference vectors. The most general infinitesimal transformation is a linear combination of the these three types of transformations \cite{Manohar:2002fd}
\begin{align}
{\rm (I)} \; &\left\{
\begin{aligned} n_\mu &\to n_\mu + \Delta_\mu^\perp \\ 
\bar{n}_\mu & \to \bar{n}_\mu \phantom{ + \Delta_\mu^\perp}
\end{aligned} \right. \,,&  
{\rm (II)} \; &\left\{
\begin{aligned} n_\mu & \to n_\mu \phantom{+ \epsilon_\mu^\perp} \\ 
\bar{n}_\mu & \to \bar{n}_\mu  + \epsilon_\mu^\perp
\end{aligned} \right. \,,&
{\rm (III)} \; &\left\{
\begin{aligned} n_\mu & \to (1+\alpha)\, n_\mu \\ 
\bar{n}_\mu & \to (1-\alpha)\, \bar{n}_\mu  
\end{aligned} \right. \,,
\end{align}
with $ \Delta^\perp \cdot n = \Delta^\perp \cdot \bar{n} =   \epsilon^\perp \cdot n = \epsilon^\perp \cdot \bar{n}=0$. In order for the transformations not to upset the power counting in the collinear sector one needs to ensure that the transformation parameter $ \Delta_\mu^\perp$ counts as $\mathcal{O}(\lambda)$ (or smaller). This can be seen, for example, by considering the transformation
\begin{equation}
n\cdot D_c \to n\cdot D_c + \Delta^\perp\cdot D^\perp_c \,.
\end{equation}
Requiring that the expression after the transformation remains of $\lambda^2$ implies 
$\Delta^\perp_\mu \sim \lambda$. Similarly, the power counting in the anti-collinear sector implies $\epsilon_\mu^\perp\sim \lambda$. The parameter $\alpha$ of the rescaling transformation (III), on the other hand, can be $\mathcal{O}(\lambda^0)$. The first two transformations connect operators at different orders in the power counting and become only relevant when power corrections are considered. The third transformation, on the other hand, is useful already when constructing operators at leading power. The simplest way to take the invariance under (III) into account is to construct all operators from building blocks which are invariant under this rescaling. The constraints from reparameterization invariance are linked to the Lorentz invariance of the underlying theory.  A powerful alternative approach to dealing with Lorentz invariance in effective field theories which involve reference vectors was recently developed in \cite{Heinonen:2012km}.

While reparameterization invariance is specific to the effective theory, gauge invariance was present already in the original QCD Lagrangian. However, in the same way in which we expanded the Lagrangian, it is necessary to expand the gauge transformations, and one must make sure that the gauge transformations respect the scaling of the fields. For example, we will see that transforming a soft field by means of a gauge function $\alpha(x)$ with collinear scaling would turn the soft field into a collinear field. 

We will consider two types of gauge transformations; the soft gauge transformation
\be
\scol{V_s(x)} = \scol{\exp\left[i \alpha_s^a(x) t^a \right]} \, ,
\ee
and the collinear gauge transformation
\be
\pcol{V_c(x)} = \pcol{\exp\left[i \alpha_c^a(x) t^a \right]} \, .
\ee
The function $\scol{\alpha_s}$ has soft scaling. i.e. $\partial \scol{\alpha_s} \sim \lambda^2 \scol{\alpha_s}$, while $\pcol{\alpha_c}$ has collinear scaling. We analyze the soft transformations first. Under a \scol{{\bf soft gauge transformation}} the soft fields transform in the standard way
\bea
\scol{\psi_s(x)} &\rightarrow& \scol{V_s(x)\psi_s(x)} \, , \nn \\
\scol{A_s^\mu(x)} &\rightarrow& \scol{ V_s(x) A_s^\mu(x) V^\dagger_s(x)} + \frac{i}{g} \scol{V_s(x)}
\left[\partial^\mu, \scol{V_s^\dagger (x)} \right] \, .
\eea

The collinear fields transform instead as follows
\bea \label{eq:scgaugetr}
\pcol{\xi(x)} &\rightarrow& \scol{V_s(x_-)} \pcol{\xi(x)} \, , \nn \\
\pcol{A_c^\mu(x)} &\rightarrow& \scol{V_s(x_-)}\pcol{ A_c^\mu(x)}\scol{ V^\dagger_s(x_-)}  \, .
\eea
The  gauge transformation matrices in Eq.~(\ref{eq:scgaugetr}) depend only on $x_-$ since, 
when transforming the collinear fields, one needs to expand the soft fields around $x_-$ in order to avoid inducing higher order corrections.
In fact the expansion of the full soft gauge transformation follows the same pattern already encountered in Eq.~(\ref{eq:expPHIxm})
\be
\scol{V_s(x)} = \scol{V_s(x_-)} + \underbrace{x_\perp \cdot \partial_\perp \scol{V_s(x_-)}}_{{\mathcal O}(\lambda)} + {\mathcal O}(\lambda^2) \, . 
\ee
A detailed discussion of the gauge transformation properties of the non-abelian gauge Lagrangian is provided in \cite{Beneke:2002ni}.

The transformation of the collinear gluon field differs from the standard one because it is missing the term $\scol{V_s} [\partial^\mu, \scol{V_s^\dagger}] \sim \lambda^2$ . This term is a higher power correction for the 
$\pcol{A_{c \perp}}$ and $\pcol{\bar{n} \cdot A_c}$ component of the collinear gluon field.
The component $\pcol{n \cdot A_c} \sim \lambda^2$ only appears in terms of the form
$n \cdot D$ (last line of Eq.~(\ref{eq:listcd})); the term $n \cdot D$
transforms as expected
\bea
\pcol{n \cdot A_c(x) }+ \scol{n \cdot A_s(x_-)} &\rightarrow& \scol{V_s(x_-)}\left[ \pcol{n \cdot A_c(x)} + \scol{n \cdot A_s(x_-)}\right] \scol{V^\dagger_s(x_-)} \nn \\
&  + & \frac{i}{g} \scol{V_s(x_{-})}\left[n \cdot \partial, \scol{V_s^\dagger (x_{-})} \right] \, , \\
i n \cdot D &\rightarrow&\scol{V_s(x_-)} \, i n \cdot D \, \scol{V^\dagger_s(x_-)} \, .
\eea

Since the \pcol{{\bf collinear gauge transformations}} involve a field with  large energy,
the soft fields cannot transform under them:
\be
\scol{\psi_s(x)} \rightarrow  \scol{\psi_s(x)} \, , \qquad
\scol{A_s^\mu(x)} \rightarrow \scol{A_s^\mu(x)}\, .
\ee 
The collinear fields instead transform as follows
\bea
\pcol{\xi(x)} &\rightarrow& \pcol{V_c(x) \xi(x)} \, ,  \nn \\
\pcol{A_c^\mu(x)} &\rightarrow& \pcol{V_c(x) A_c^\mu(x)  V^\dagger_c(x) } + \frac{1}{g} 
\pcol{V_c(x)} \left[i \partial^\mu + g \frac{\bar{n}^\mu}{2}  \scol{n \cdot A_s(x_-)} , \pcol{V^\dagger_c(x)} \right]\, , 
\eea
which implies 
\bea
\pcol{A^\mu_{c \perp}} &\rightarrow& \pcol{V_c A^\mu_{c \perp} V^\dagger_c} + \frac{i}{g}
\pcol{V_c} \left[ \partial^\mu_\perp,  \pcol{V^\dagger_c} \right] \, , \nn \\
\pcol{\bar{n} \cdot A_c} &\rightarrow& \pcol{V_c \bar{n} \cdot A_{c} V^\dagger_c} + \frac{i}{g}
\pcol{V_c} \left[ \bar{n} \cdot \partial,  \pcol{V^\dagger_c }\right] \, , \nn \\
\pcol{n \cdot A_c} &\rightarrow& \pcol{V_c n \cdot A_{c} V^\dagger_c} + \frac{i}{g}
\pcol{V_c} \left[ \scol{n \cdot D_s (x_-)},  \pcol{V^\dagger_c}\right] \, .  
\eea
The last transformation law in the equation above insures that
\be
i n \cdot D \rightarrow \pcol{V_c} \,  i n \cdot D \, \pcol{V^\dagger_c} \,.
\ee

It is easy to check that the Lagrangian in Eq.~(\ref{eq:LQCDSCET}) is separately invariant under soft and collinear gauge transformations. The various covariant derivatives all transform according to 
\begin{displaymath}
D_\mu \rightarrow V_i D_\mu V_i^\dagger \, ,
\end{displaymath}
where $i \in \{s,c\}$, and the fermions transform according to
\begin{displaymath}
\psi \rightarrow V_i \psi \, ,
\end{displaymath}
with the replacement $x \rightarrow x_-$ in the appropriate places. A complete discussion of the gauge transformations  and of the construction of the higher power terms can be found in \cite{Beneke:2002ni} (in the label formalism, which we discuss in Section~\ref{sec:labelform} below, the relevant reference is \cite{Bauer:2003mga}).

\subsection{Wilson Lines}

While discussing the scalar $\phi^3$ theory, we encountered non-local operators of the kind shown in Eq.~(\ref{eq:J2}). In a gauge theory, a product of fields at different space time points is only gauge invariant if the fields are connected by {\em Wilson lines}, defined as 
\be \label{eq:defWil}
\left[ x + s \bar{n} , x  \right] \equiv \mathbf{P} \exp\left[ ig \int_0^s ds' \, \bar{n} \cdot A (x + s' \bar{n}) \right] \, .
\ee
The operator $\mathbf{P}$ indicates the path ordering of the color matrices, such that 
\be
\mathbf{P}[A(x)A(x+s \bar{n})] = A(x+s \bar{n})A(x)\, , \qquad \mbox{for} \, \, \, s>0.
\ee
The conjugate Wilson line is defined with the opposite ordering prescription.
Under gauge transformations the Wilson lines transform as follows (see Appendix~\ref{app:Wil})
\be
\left[ x + s \bar{n} , x  \right] \longrightarrow V(x+s \bar{n}) 
 \left[ x + s \bar{n} , x  \right] V^\dagger(x) \, ,
\ee
therefore products of the form 
\bd
\bar{\psi}(x +s \bar{n}) \left[x +s \bar{n},x \right] \psi(x) \, 
\ed
are gauge invariant.

In SCET it is customary to work with Wilson lines which go to infinity:\footnote{To see that $W(x)$ corresponds to $[x,-\infty \bar{n}]$ let us start from the definition in Eq.~(\ref{eq:defWil}); by setting $x=x'-s \bar{n}$ one  obtains
\bd 
\left[ x'  , x'-s \bar{n}  \right] \equiv \mathbf{P} \exp\left[ ig \int_0^s ds' \, \bar{n} \cdot A (x' - s \bar{n} + s' \bar{n}) \right] \, .
\ed
One can then shift the integration variable according to $s' = t +s$ to obtain
\bd 
\left[ x'  , x'-s \bar{n}  \right] \equiv \mathbf{P} \exp\left[ ig \int_{-s}^0 dt \, \bar{n} \cdot A (x' + t \bar{n}) \right] \, .
\ed
Finally, one can send $s \to \infty$ and rename $x' \to x$ to obtain Eq.~(\ref{eq:Wx}).
}
\be \label{eq:Wx}
W(x) \equiv \left[x, -\infty \bar{n}\right] = 
\mathbf{P} \exp\left[ ig \int_{-\infty}^0 \!\!ds \bar{n} \cdot A(x + s \bar{n}) \right] \, .
\ee
To be precise, one should write $W(x) \equiv [x,x-\infty\, \bar{n}]$ in Eq.~(\ref{eq:Wx}); as long as the fields vanish at infinity, as they do in dimensional regularization, the difference is irrelevant.
The Wilson line along a finite segment can be written as a product of two Wilson lines extending to infinity:
\bea
\left[x +s \bar{n},x \right] &=& W\left( x + s \bar{n} \right) W^\dagger (x) \,  \nn \\
&=& \mathbf{P} \exp\left[ ig \int_{-\infty}^0\! \!\!dt\, \bar{n} \cdot A(x + s \bar{n} + t \bar{n}) \right]
\overline{\mathbf{P}} \exp\left[ -ig \int_{-\infty}^0 \!\!\!dt\, \bar{n} \cdot A(x  + t \bar{n}) \right] \,  \nn \\
&=& \mathbf{P} \exp\left[ ig \int_{0}^s \!\!\!dt\, \bar{n} \cdot A(x  + t \bar{n}) \right] \, ,
\eea
where the symbol $\overline{\mathbf{P}}$ indicates the  anti time ordering relevant for $W^\dagger (x)$.
The Wilson lines extending to infinity transform as follows under gauge transformations 
\be
W(x) \rightarrow V(x) W(x) V^\dagger(-\infty \bar{n}) \, .
\ee
If one considers gauge functions vanishing at infinity, such that $V(-\infty \bar{n}) =1$, the combinations
\be \label{eq:Chis}
\chi(x) \equiv W^\dagger (x) \psi(x) \, , \qquad \mbox{and} \quad
\bar{\chi}(x) \equiv \bar{\psi}(x)  W(x) \, ,
\ee
are gauge invariant and can be used as building blocks to construct non-local operators.

In Appendix~\ref{app:Wil} it is shown that the covariant derivative of the Wilson lines along the integration path in the exponent of the line vanishes;
in our particular case this implies that
\be
\bar{n} \cdot D W(x) = 0 \, .
\ee

Since in the SCET Lagrangian there are two kinds of gauge fields, the collinear and soft ones, it is necessary to consider two types of Wilson lines, which will be denoted as follows
\bea \label{eq:WLsc}
\pcol{W_c(x)} &=& \mathbf{P} \exp\left[ ig \int_{-\infty}^0 \!\!ds \, \pcol{\bar{n} \cdot A_c(x + s \bar{n})} \right] \qquad \mbox{(\pcol{collinear})}\, ,  \nn \\
\scol{S_n(x)} &=& \mathbf{P} \exp\left[ ig \int_{-\infty}^0 \!\!ds \, \scol{n \cdot A_s(x + s n)} \right] \qquad \mbox{(\scol{soft})}  \, .
\eea
The collinear Wilson lines are useful to construct operators, while the soft Wilson lines are useful because of the structure of the soft interaction.

\subsection{Decoupling Transformation\label{ssec:Dec}}

As seen above, the interaction between collinear quarks and soft gluons in the SCET Lagrangian takes the form
\be \label{eq:Lcs}
{\mathcal L}_{c+s} = \pcol{\bar{\xi}} \frac{\nbsl}{2} i n \cdot D \pcol{\xi}\, .
\ee
where the specific form of the covariant derivative in this case is given in Eq.~(\ref{eq:listcd}). We now redefine the fields $\pcol{\xi}$ and $\pcol{A^\mu_c(x)}$ employing the soft Wilson line defined in Eq.~(\ref{eq:WLsc}) 
\bea \label{eq:dectr}
\pcol{\xi(x)} &\rightarrow& \scol{S_n (x_-)} \pcol{ \xi^{(0)} (x)} \, , \nn \\
\pcol{A^\mu_c(x)} &\rightarrow&\scol{ S_n (x_-)}\pcol{ A^{(0) \mu}_c (x)}\scol{  S_n^\dagger (x_-)}\, .
\eea
As a consequence of the field transformations in Eq.~(\ref{eq:dectr}) one finds that
\bea
i n \cdot D \pcol{\xi(x)} &\rightarrow& i n \cdot D' \scol{S_n (x_-)}\pcol{\xi^{(0)}(x)} \,  \nn \\
&=&  \left(in \cdot \partial + g \pcol{n\cdot} \scol{S_n(x_-)}\pcol{A^{(0)}_c (x)}\scol{  S_n^\dagger (x_-)} + g \scol{n \cdot A_s(x_-)} \right) \scol{S_n (x_-)}\pcol{\xi^{(0)}(x)} \,  \nn \\
&=&  \Bigl(in \cdot \partial_- \scol{S_n(x_-) }+ \scol{S_n(x_-)} in \cdot \partial + 
\scol{S_n(x_-)}\,g \pcol{n\cdot A^{(0)}_c (x) }
\nn \\ &&
+ g \scol{n \cdot A_s(x_-)  S_n (x_-)}\Bigr)\pcol{\xi^{(0)}(x)}\,  \nn \\
&=&  \Bigl[ \underbrace{\left(i n \cdot D_- \scol{S_n(x_-)}\right)}_{=0}+ \scol{S_n(x_-)} in \cdot \partial  + 
\scol{S_n(x_-)} g \pcol{n\cdot A^{(0)}_c (x)}\Bigr] \pcol{\xi^{(0)}(x)}\,\,  \nn \\
&=& \scol{S_n(x_-)} \left( in \cdot \partial  + g \pcol{n\cdot A^{(0)}_c (x)}\right)\pcol{\xi^{(0)}(x)} \equiv  \scol{S_n(x_-)} i \pcol{n \cdot D^{(0)}_c \xi^{(0)}(x)} \, , \label{eq:decforD}
\eea
where we made use of the fact that the covariant derivative along the Wilson line is zero and that
\be
n^\alpha \frac{\partial}{\partial x^\alpha} \scol{S_n(x_-)} = 
n^\alpha \frac{\partial x_-^\beta}{\partial x^\alpha}
\frac{\partial}{\partial x_-^\beta} \scol{S_n(x_-)}  = \frac{n^\alpha \bar{n}_\alpha}{2}
n^\beta\frac{\partial}{\partial x_-^\beta} \scol{S_n(x_-)} \equiv n \cdot \partial_-
\scol{S_n(x_-)} \, .
\ee
(Remember that $x_-^\mu = \bar{n} \cdot x \, n^\mu/2$.)
In conclusion, under the field transformations in Eq.~(\ref{eq:dectr}), 
the Lagrangian in Eq.~(\ref{eq:Lcs}) changes as follows
\be \label{eq:decoupling}
{\mathcal L}_{c+s} \rightarrow \pcol{\bar{\xi}^{(0)}} \frac{\nbsl}{2} i \pcol{n \cdot D^{(0)}_c \xi^{(0)}(x)} \, ,
\ee
so that the soft gluon field no longer appears in the collinear Lagrangian (the subscript and superscript in the covariant derivative indicate that it depends on $\pcol{A_c^{(0)}}$ only). This kind of transformation is called {\em decoupling transformation}, since it decouples the soft gluon from the leading power collinear Lagrangian. However, it is important to stress that at subleading power soft collinear interactions are still present in the Lagrangian. It is possible to show that after a decoupling transformation also the interactions between soft and collinear gluons disappear from the leading power collinear Lagrangian (see Appendix~\ref{app:decFmunu}).

The decoupling transformation is an important element in proving factorization theorems, but does not imply that everything factorizes at leading power. For example, to analyze the Sudakov problem one needs to match the vector current operator; while the soft fields decouple from the Lagrangian, they are still present in the current operator. To deal with the Sudakov problem we need to deal with two collinear directions, as we did when considering the same problem in the $\phi^3$ theory. The QED current operator
\be
J^\mu(x) = \bar{\psi}(x) \gamma^\mu \psi(x) \, ,
\ee
corresponds to the SCET non-local operator
\be \label{eq:Source}
J^\mu(x) \rightarrow \int\! ds\! \int\! dt\, C_V(s,t) \pcol{\bar{\chi}_{\cc}\left(x+ s \bar{n} \right)} \gamma^\mu_\perp \lcol{\chi_{\cb}(x+t n)} \, ,
\ee
where the fields $\pcol{\chi_{\cc}}$ and $\lcol{\chi_{\cb}}$ are defined according to Eq.~(\ref{eq:Chis}):
\bea\label{eq:buildingblocks}
\pcol{\chi_{\cc}} = \pcol{W_{\cc}^\dagger \xi_{\cc}} \, , &\quad& \nsl \pcol{\chi_{\cc}} = 0 \, , \nn \\
\lcol{\chi_{\cb}} = \lcol{W_{\cb}^\dagger \xi_{\cb}} \, , &\quad& \nbsl \lcol{\chi_{\cb}} = 0 \, . 
\eea
Since 
\be
\gamma^\mu = \nsl \frac{\bar{n}^\mu}{2} + \nbsl \frac{n^\mu}{2} +\gamma^\mu_\perp \, ,
\ee
the only component surviving in Eq.~(\ref{eq:Source}) is $\gamma_\perp$. When applying the decoupling transformations
\bea
\pcol{\chi_{\cc}(x)} &\rightarrow& \scol{S_n\left(x_- \right)} \pcol{\chi^{(0)}_{\cc}(x)} \, , \nn \\
\lcol{\chi_{\cb}(x)} &\rightarrow& \scol{S_{\bar{n}}\left(x_+ \right)} \lcol{\chi^{(0)}_{\cb}(x)} \, , 
\eea
 the source term becomes
\begin{align} \label{eq:Jsep}
J^\mu(x) &= \int \! ds\! \int \! dt \,C_V(s,t) \pcol{\bar{\chi}^{(0)}_{\cc}\left(x+ s \bar{n} \right)}
\scol{S_n^\dagger\left(x_- \right)S_{\bar{n}}\left(x_+ \right)}
 \gamma^\mu_\perp \lcol{\chi^{(0)}_{\cb}(x+t n)} \, \nn \\
 & =  \int \! ds \!\int \! dt \,C_V(s,t) \pcol{\bar{\chi}^{(0)}_{\cc}\left(x_++ x_\perp+ s \bar{n} \right)}
\scol{S_n^\dagger\left(0 \right)S_{\bar{n}}\left(0 \right)}
 \gamma^\mu_\perp \lcol{\chi^{(0)}_{\cb}(x_- + x_\perp +t n)}\, +\dots \, .
\end{align}
In the second line, we have used the multipole expansion to drop power-suppressed dependence on $x^\mu \sim (1 , 1, 1/\lambda)$. The scaling follows because $x^\mu$ is conjugate to the sum of a collinear and an anti-collinear momentum.
%
\begin{figure}[t]
\begin{center}
\[
\hspace*{1.5cm}
\vcenter{ \hbox{
  \begin{picture}(0,0)(0,0)
\SetScale{1}
  \SetWidth{2}
  \Line(-50,50)(0,0)
  \Line(0,0)(50,50)
  \Photon(0,0)(0,-35){3}{4}
  \CCirc(0,5){20}{Black}{Gray}
\Text(-56,29)[cb]{{\Large  $p$}}
\Text(56,30)[cb]{{\Large $l$}}
\Text(0,-60)[cb]{{\Large $F(Q^2,L^2,P^2)$}}
\end{picture}}} \hspace*{2.2cm} = \hspace*{3cm}
\vcenter{ \hbox{
  \begin{picture}(0,0)(0,0)
\SetScale{1}
  \SetWidth{2}
  \SetOffset(0,-30)
\SetColor{Blue}
  \Line(-70,50)(0,0)
  \CCirc(-43,32){20}{Blue}{White}
\SetColor{OliveGreen}
    \DashLine(60,43)(70,50){6}
  \DashCArc(43,32)(20,33,393){5}
    \DashLine(0,0)(25,19){6}
\SetColor{Mahogany}
  \SetWidth{1}
  \Gluon(-43,71)(0,79){3}{6.5}
  \Gluon(-10,46)(0,77){3}{6}
  \Gluon(0,79)(33,63.5){3}{5.7}
  \SetWidth{1.5}
  \Line(-52,73.7)(0,33.7)
  \Line(-50,77)(0,38)
  \Line(0,33.7)(52,73.7)
  \Line(0,38)(50,77)
  \SetWidth{2}
  \CCirc(0,77){20}{Mahogany}{White}

\SetColor{Black}
  \Photon(0,0)(0,-35){3}{4}
  \CCirc(0,5){22}{Black}{White}
\Text(-1,0)[cb]{ $\tilde{C}_V(Q^2)$}
\Text(-45,28)[cb]{ \pcol{${\mathcal J}(P^2)$}}
\Text(41,28)[cb]{ \lcol{${\mathcal J}(L^2)$}}
\Text(-1,73)[cb]{ \scol{${\mathcal S}(\Lambda_s^2)$}}
\end{picture}}}\hspace*{3cm} + 
{\mathcal O}\left(\lambda^2 \right) 
\]
\vspace*{14mm}
\end{center}
\caption{Diagrammatic representation of the Sudakov form factor 
in QCD; the diagram illustrates  the separation of the different scales present in the problem. The soft scale  is \mbox{$\Lambda^2_s = L^2 P^2/ Q^2$}.
 \label{fig:scalesep}}
\end{figure}
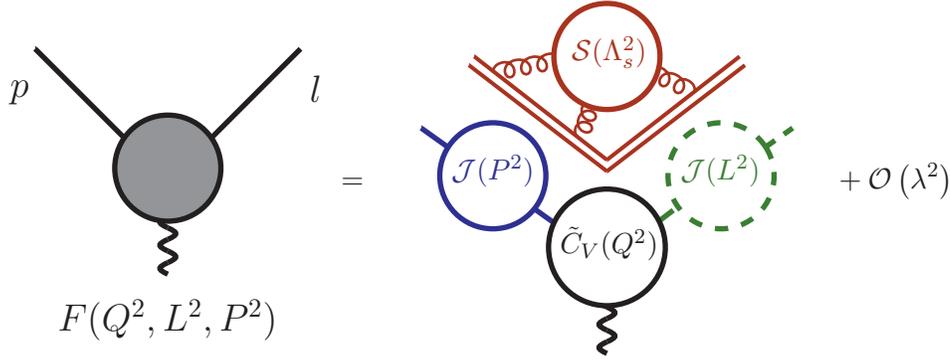
We see that the soft interactions do not cancel, and the Sudakov form factor receives low-energy  contributions which describe a long-range interaction between the fast moving ingoing and outgoing quarks.
The situation is summarized in diagrammatic form in Fig.~\ref{fig:scalesep}, where $p^\mu \sim i n^\mu$, $l^\mu \sim i \bar{n}^\mu$, and  the double lines represent the soft Wilson lines.

Do the soft corrections factorize? It depends on the precise meaning that one attributes to the word factorization. Unfortunately, there are two different definitions of the word factorization which are employed in this context:
\begin{itemize}
\item[i)] Factorization = {\em scale separation}. In the  source term in Eq.~(\ref{eq:Jsep}) the pieces associated to different scales are separated, so according to this definition the form factor is factorized.
\item[ii)] Factorization = {\em no low energy interactions}. The two collinear sectors in   Eq.~(\ref{eq:Jsep}) interact through soft interactions. The form factor is not factorized in this sense. 
\end{itemize}

\subsection{Factorization and Collinear Anomaly \label{sec:FCA}}

\begin{figure}[t]
\begin{center}
\[
\hspace*{1.5cm}
\vcenter{ \hbox{
  \begin{picture}(0,0)(0,0)
\SetScale{1}
  \SetWidth{2}
  \Line(-50,50)(0,0)
  \Line(0,0)(50,50)
  \Photon(0,0)(0,-35){3}{4}
  \CCirc(0,5){20}{Black}{Gray}
\Text(-56,29)[cb]{{\Large  $p$}}
\Text(56,30)[cb]{{\Large $l$}}
\Text(0,-60)[cb]{{\Large $F(Q^2,l^2,p^2,m)$}}
\end{picture}}} \hspace*{2.2cm} = \hspace*{3cm}
\vcenter{ \hbox{
  \begin{picture}(0,0)(0,0)
\SetScale{1}
  \SetWidth{2}
  \SetOffset(0,-30)
\SetColor{Blue}
  \Line(-70,50)(0,0)
  \COval(-43,32)(17,17)(0){Blue}{White}
\SetColor{OliveGreen}
\DashLine(60,43)(70,50){6}
\DashCArc(43,32)(17,33,393){5}
\DashLine(0,0)(25,19){6}
\SetColor{Mahogany}
  \SetWidth{1}
  \LongArrow(-25,70)(-30,50)
  \LongArrow(25,70)(30,50)
  \SetWidth{1.5}
  \SetWidth{2}
  \CBox(-38,65)(38,95){Mahogany}{White}

\SetColor{Black}
  \Photon(0,0)(0,-35){3}{4}
  \COval(0,5)(22,22)(0){Black}{White}
\Text(-1,0)[cb]{ $\tilde{C}_V(Q^2)$}
\Text(-45,27)[cb]{ \pcol{$J(p^2)$}}
\Text(42,27)[cb]{ \lcol{$J(l^2)$}}
\Text(-1,73)[cb]{ \scol{$F(m) \ln\frac{Q^2}{m^2}$}}
\end{picture}}}\hspace*{3cm} + 
{\mathcal O}\left(\lambda^2 \right) 
\]
\vspace*{14mm}
\end{center}
\caption{Diagrammatic representation of the Sudakov form factor 
in presence of a collinear anomaly; the diagram illustrates  the separation of the different scales present in the problem, after re-factorization. 
 \label{fig:scalesepM}}
\end{figure}
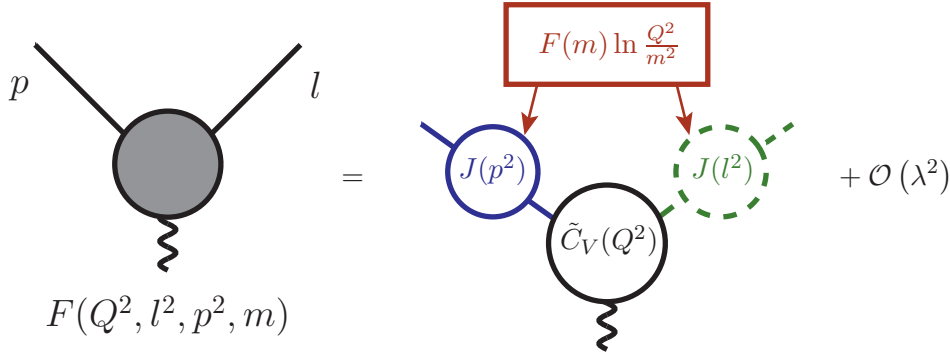

In the case analyzed in Section~\ref{sec:MSPCA}, in which the virtual propagator carrying momentum $k$ in the vertex correction has a small but non vanishing mass $m$, the integral over the soft region vanishes. One could naively think that this implies a factorization in $d = 4$ of the kind illustrated in Fig.~\ref{fig:factorization}. However, for $m^2 \sim \lambda^2$ the hard function is the same as in the massless case and is given by Eq.~(\ref{eq:Ihardres}). This function has an infrared divergence which depends on $Q$. Such a divergence cannot be canceled if the jet functions do not depend on $Q$ as well. In Section~\ref{sec:MSPCA} we have shown that this dependence is indeed present, and originates from the need to use an additional regulator to define in a proper way the collinear region integrals.
Here we want to study how the factorization is modified in this case. At all orders in perturbation theory, the product of the two jet functions must be independent of the analytic regulator, and therefore also independent  of the corresponding 't~Hooft scale $\nu$. Consequently, the quantity
\be
P = \pcol{{\mathcal J}_{\cc}\left(p^2,m^2,\ln{\frac{\nu^2}{m^2}},\mu  \right)}\lcol{{\mathcal J}_{\cb
}\left(l^2,m^2,\ln{\frac{\nu^2}{Q^2}},\mu \right)} \, ,
\ee
should satisfy the differential equation 
\be
\frac{d}{d \ln{\nu}}\ln P = \frac{d}{d \ln{\nu}}\left[\ln \pcol{{\mathcal J}_{\cc
} \left(p^2,m^2,\ln{\frac{\nu^2}{m^2}},\mu  \right)}
+ \ln\lcol{{\mathcal J}_{\cb
}\left(l^2,m^2,\ln{\frac{\nu^2}{Q^2}},\mu \right)}  \right] =0 \, .
\ee
This implies that the two terms in the square brackets in the equation above should be linear in $\ln(\nu^2/m^2)$ and $\ln(\nu^2/Q^2)$, respectively, and that the coefficients multiplying the logarithms should be independent from $p^2$ and $l^2$ \cite{Chiu:2007yn}.
One can then extract the terms depending on $\nu$ by defining  two new jet functions $J$ as follows:
\be
\ln P \equiv
\ln \pcol{J_{\cc
} \left(p^2,m^2,\mu  \right) }+ \ln \lcol{J_{\cb
} \left(l^2,m^2,\mu  \right)} - \scol{F(m^2,\mu) \ln{\frac{Q^2}{m^2}}} \, .
\ee
Thus one can re-factorize \cite{Chiu:2007yn,Becher:2010tm} the product of the two jet functions as follows
\be
P = e^{-\scol{F(m^2,\mu) \ln{\frac{Q^2}{m^2}}}} \pcol{J_{\cc
} \left(p^2,m^2,\mu  \right)} \lcol{J_{\cb
} \left(l^2,m^2,\mu  \right)} \, ,
\ee
which shows explicitly that the anomalous $Q$ dependence exponentiates. The factorization of hard and collinear physics in this case can be then schematically represented as shown in Fig.~\ref{fig:scalesepM}.

\subsection{Gauge Covariant Building Blocks \label{subsec:gaugeinvblocks}}

In the previous section in Eq.~(\ref{eq:Chis}), we introduced the notation
\be
\pcol{\chi(x)} \equiv \pcol{W^\dagger (x) \xi(x)} = \pcol{W^\dagger (x)} \frac{\nsl \nbsl}{4} \pcol{\psi(x)}
\, . 
\ee
It is convenient to work with the field $\pcol{\chi(x)}$ instead of the $ \pcol{\psi(x)}$ because $\pcol{\chi(x)}$ is invariant under collinear gauge transformations, which makes it easy to construct gauge invariant operators. Similarly, one introduces a gauge invariant building block $\pcol{{\mathcal A}^\mu}$  for the collinear gluon fields, which is defined as follows\footnote{Sometimes in the  literature a different definition, taking into account both the soft and collinear gluon fields, was adopted \cite{Becher:2005fg}:
\be
\pcol{{\mathcal A}^\mu} \equiv \pcol{ W^\dagger(x)} \Bigl(i \pcol{{D_c}^\mu W(x)} \Bigr) + \frac{\bar{n}^\mu}{2} \left( \pcol{W^\dagger(x)} g n\cdot \scol{A_s(x_-)}
 \pcol{W(x)} - g n\cdot \scol{A_s(x_-)} \right) \, .
\ee}
\be
\pcol{{\mathcal A}^\mu} \equiv \pcol{ W^\dagger(x)} \Bigl(i \pcol{{D_c}^\mu W(x)} \Bigr) \, .
\ee
From the definition above, it is possible to see that $\bar{n} \cdot \pcol{{\mathcal A}} =0$, since  $\bar{n} \pcol{\cdot D W} = 0$, as shown in  Appendix~\ref{app:Wil} .
 The component $n\cdot \pcol{{\mathcal A}} $ will
instead have the expression
\be
n \cdot \pcol{{\mathcal A}} = \pcol{ W^\dagger(x)} \Bigl(i n \cdot \pcol{{D_c} W(x)} \Bigr)\, .
\ee
Finally, the perpendicular component of the block $\pcol{{\mathcal A}}$ is
\be \label{eq:Acall}
\pcol{{\mathcal A}^\mu_{\perp}} = \pcol{ W^\dagger(x)} \Bigl(i \pcol{{D_c}^\mu_\perp W(x)} \Bigr) \, .
\ee
The notation above indicates that the covariant derivative acts only on the Wilson line. In the literature the fields $\pcol{{\mathcal A}_\perp}$ are sometimes also defined as
\be \label{eq:Acall2}
\pcol{{\mathcal A}^\mu_{\perp}} = \pcol{W^\dagger(x)} \Bigl[i \pcol{{D_c}^\mu_\perp}, \pcol{W(x)} \Bigr] \, .
\ee
The two definitions are equivalent, as it can be seen by multiplying the commutator by a test function $f$:
\be
\Bigl[D^\mu, W(x) \Bigr] f(x) =  D^\mu \Bigl(W(x) f(x)\Bigr)  - W(x) \Bigl( D^\mu f(x)\Bigr)  = \Bigl(D^\mu W(x) \Bigr) f (x)\, .
\ee

For leading-power operators, the perpendicular components of the field $\pcol{{\mathcal A}}$ are sufficient because $\bar{n} \pcol{\cdot {\mathcal A}}$ vanishes
and $n \cdot \pcol{{\mathcal A}}$ is power suppressed, since it involves the small component of the momentum and gluon field. 
The gauge independence of the fields $\pcol{\chi}$ and $\pcol{{\mathcal A}}$ follows immediately from the behavior of the fields $\pcol{\xi}$, the Wilson lines $\pcol{W}$, and the covariant derivatives under collinear gauge transformations.

It is  possible to rewrite the collinear Lagrangian ${\mathcal L}_c$ as a function of the gauge invariant fields \cite{Hill:2002vw}.
To do this, one needs to make use of the relation
\be
\pcol{W^\dagger} i \pcol{D_c^\mu W} = \pcol{W^\dagger} \Bigl(i \pcol{D_c^\mu W}\Bigr)  + 
\pcol{W^\dagger W} i \partial^\mu = \pcol{{\mathcal A}^\mu}+ i \partial^\mu \equiv i \pcol{{\mathfrak D}_\mu}\, ,
\ee
Moreover, the relation 
\be
\pcol{W^\dagger} i \bar{n} \cdot \pcol{D_c W} = \pcol{W^\dagger} \Bigl( \underbrace{i \bar{n} \cdot \pcol{D_c W}}_{ =0}\Bigr) + i \bar{n} \cdot \partial = i \bar{n} \cdot \partial  \, ,
\ee
leads to the identity
\be
\frac{1}{i \bar{n}\cdot \pcol{D_c}} = \pcol{W W^\dagger} \left(i \bar{n}\cdot \pcol{D_c}\right)^{-1} \pcol{W W^\dagger} = \pcol{W} \left(\pcol{W^\dagger} i \bar{n}\cdot \pcol{D_c W}\right)^{-1} \pcol{W ^\dagger} = \pcol{W} \frac{1}{i \bar{n}\cdot \partial} \pcol{W ^\dagger} \, .
\ee
By inserting repeatedly $W^\dagger W =\bm{1}$ between the fields, the collinear Lagrangian in Eq.~(\ref{eq:colllag}) can then  be rewritten as
\be
{\mathcal L}_c = \pcol{\bar{\chi}} \frac{\nbsl}{2} \left( i n \cdot \pcol{{\mathfrak D}}\right) \pcol{\chi} + \pcol{\bar{\chi}} i \pcol{\MDsl_\perp} \frac{1}{i \bar{n} \cdot \partial} i \pcol{\MDsl_\perp}\frac{\nbsl}{2}  \pcol{\chi} \, .  \label{eq:MK}
\ee
 In order to rewrite the collinear gluon Lagrangian in terms of the $\pcol{{\mathcal A}}$ fields we observe that
\be
\pcol{W^\dagger F_{\mu\nu} W} \equiv \pcol{W^\dagger F^a_{\mu\nu}} t^a \pcol{W} = \frac{1}{ig} \pcol{W^\dagger} \left[ i \pcol{D_{c, \mu}}, i \pcol{D_{c, \nu}}\right] \pcol{W} = \frac{1}{g} \left( \partial_\mu \pcol{{\mathcal A}_\nu} - \partial_\nu \pcol{{\mathcal A}_\nu} - i \left[
\pcol{{\mathcal A}_\mu}, \pcol{{\mathcal A}_\nu} \right] \right) \, .
\ee
Therefore, by defining 
\be
\pcol{{\mathcal F}_{\mu\nu}} \equiv  \partial_\mu \pcol{{\mathcal A}_\nu} - \partial_\nu \pcol{{\mathcal A}_\nu} - i \left[
\pcol{{\mathcal A}_\mu}, \pcol{{\mathcal A}_\nu} \right] \, ,
\ee
one finds that the kinetic term for the collinear gluons can be written as
\be
-\frac{1}{4} \pcol{F^a_{\mu \nu} F^{a , \mu \nu}}  = -\frac{1}{2} \mbox{Tr}\left[ \pcol{F_{\mu \nu} F^{ \mu \nu}} \right]  = -\frac{1}{2 } \mbox{Tr}\left[
\pcol{W^\dagger F_{\mu \nu} F^{ \mu \nu} W} \right] = -\frac{1}{2 g^2} \mbox{Tr}\left[ \pcol{{\mathcal F}_{\mu \nu}
{\mathcal F}^{\mu \nu}} \right] \, .
\ee

The leading soft-collinear interaction terms can be obtained by the replacement in Eq.~(\ref{eq:replA}). At the level of  invariant building blocks, this corresponds to the replacement
\be
\pcol{{\mathcal A}^\mu (x)} \rightarrow \pcol{{\mathcal A}^\mu(x)} + \frac{\bar{n}^\mu}{2} \pcol{W^\dagger (x)} g n \cdot \scol{A_s(x_-)} \pcol{W(x)} \, .
\label{eq:replAbeaut}
\ee

\subsection{Position Space Versus Label Formalism\label{sec:labelform}}

The Lagrangian we constructed was written directly in position space, and the expansion in small momentum components was translated into a derivative expansion of the Lagrangian, the so-called multipole expansion \cite{Beneke:2002ph,Beneke:2002ni}. The original papers on SCET 
\cite{Bauer:2000yr,Bauer:2001yt}, as well as a large fraction of the current literature instead use a hybrid position-momentum-space formalism known as label formalism. In order for the reader to be able to translate results between the two different formulations, we now briefly discuss the label formalism.

This formulation is motivated by HQET, where the momentum of a heavy quark inside a meson  is written as $p_Q^\mu = m_Q v^\mu + r^\mu$, where $v^\mu$ is the meson velocity. The small residual momentum arises from interactions of the heavy quark with the light constituents of the meson and is of order $\Lambda_{\rm QCD}\sim 1\,{\rm GeV}$, much smaller than the heavy-quark mass $m_Q$. To construct the effective Lagrangian, one then splits off the large part of the momentum from the field by redefining
\begin{equation}
Q(x) = e^{-i m_Q v\cdot x} h_v(x) \, .
\end{equation}
The field $h_v(x)$ is the heavy quark field in the effective theory and carries the label $v$, to make it clear that the large component $m_Q v$ has been removed from the momentum. The field $h_v(x)$ is slowly varying, since all derivatives on $h_v(x)$ are of order $\Lambda_{\rm QCD}$. This field is then used to construct the low-energy effective theory. 

In the same spirit, the collinear momenta in SCET are rewritten as 
\begin{equation}
p_c^\mu = q^\mu + r^\mu\,,\text{ with } q^\mu = q_-^\mu + q_\perp^\mu\,,
\end{equation}
where $q^\mu$ is the large label momentum, while the residual part $r^\mu$ scales like a soft momentum. The collinear gluon field introduced in Eq.~(\ref{eq:Acall}) is then written as
\begin{equation}
\pcol{{\cal A}^\mu(x)} = \sum_q e^{-i q\cdot x} \pcol{{\cal A}_{q}^\mu(x)}\,,
\label{eq:labA}
\end{equation}
so that all derivatives acting on the field ${\cal A}_{q}^\mu(x)$ scale as ${\mathcal O}(\lambda^2)$.
Similarly, for the fields $\chi$  introduced in Eq.~(\ref{eq:Chis}), one can define
\be
\pcol{\chi(x)} = \sum_q e^{-i q\cdot x}  \pcol{\chi_q(x)} \, .
\label{eq:labB}
\ee
The notation adopted in Eqs.~(\ref{eq:labA},\ref{eq:labB}) is symbolic, the label $q$ can assume a continuous set of values. 
An important difference to the HQET case is that one needs to sum over the different values of the label to obtain the full field. In contrast to the heavy-quark mass, the large momentum components of the collinear fields are not fixed, and change in interactions with other collinear fields. Only the sum of the label momenta is conserved in a given interaction. Since the collinear fields should carry large energy, it appears problematic to sum over all labels, since this sum can also run over regions where the label momentum is small. This happens for example in collinear loop integrals. In this region, which is called the {\em zero bin} in \cite{Manohar:2006nz}, the collinear fields become soft and one might worry about double counting the soft region. We have addressed the double-counting issue when we introduced the method of regions and have shown that in there is no double counting because the overlap region corresponds to scaleless integrals which vanish in dimensional regularization. This was demonstrated first for a simple example integral in Section~\ref{sec:ASE}, and then for the collinear integrals after Eq. (\ref{eq:Isres}). As stressed there, this is true as long as higher-power corrections are consistently expanded away, if not, one will need to subtract the overlap contributions. 

To extract the label of a given field, one introduces the label-operator \cite{Bauer:2001yt}
\begin{equation}
{\cal P}^\mu \pcol{{\cal A}_{q}^\mu(x)} =  q^\mu \, \pcol{{\cal A}_{q}^\mu(x)}  \,.
\end{equation}
In the label formalism, the collinear Lagrangian (\ref{eq:MK}) thus takes the form
\be\label{eq:Llabel}
{\mathcal L}_c = \pcol{\bar{\chi}_q}\, \frac{\nbsl}{2} \left(i n \cdot \partial  + n \cdot \pcol{{\cal A}_k}\right) \, \pcol{\chi_{q'}} + \pcol{\bar{\chi}_q}  \left (i {{{\cal P}\!\!\!\!/}^\perp} + \pcol{{{\cal A}\!\!\!/}^\perp_k} \right) \frac{1}{i \bar{n}\cdot {\cal P}} \left(i {{{\cal P}\!\!\!\!/}^\perp} + \pcol{{{\cal A}\!\!\!/}^\perp_{k'}} \right) \frac{\nbsl}{2}  \pcol{\chi_{q'}} \, , 
\ee
where it is implied that one sums over all the labels $q$, $q'$, $k$, $ k'$, while respecting conservation of the label momentum in interactions. We have also omitted the overall phase factors which ensure label conservation in each term. In the label formalism, the collinear Wilson line can be written as \cite{Bauer:2001yt}
\begin{equation}
\pcol{W} = \sum_{\rm perms.} \exp\left (-g \frac{1}{\bar{\cal P}} \bar{n}\cdot \pcol{A_{c,q}}\right)\,,
\end{equation}
where we have used the common abbreviation $\bar{\cal P}\equiv \bar{n}\cdot {\cal P}$, and $\pcol{A_{c,q}}$ is defined in analogy to Eq.~(\ref{eq:labA}).

It is instructive to rewrite the current operator in Eq.~(\ref{eq:Jsep}) in label notation. This operator contains the collinear fields $\pcol{{\chi}_{\cc}}$ and $\lcol{\chi_{\cb}}$ along the $p$ and $l$ directions. Since the labels do not involve the small momentum components, the operator $\bar{{\cal P}}$ only acts on the collinear fields, while the operator ${\cal P} \equiv n\cdot {\cal P}$ only acts on the anti-collinear fields in operator products.
\begin{align} \label{eq:JsepLabel}
J^\mu(0) &= \int \!ds \int\! dt\, C_V(s,t) 
\pcol{\bar{\chi}_{\cc}\left(s \bar{n} \right)}\, \gamma^\mu_\perp\, \lcol{\chi_{\cb}(t n)} \,  \nonumber \\
&= \sum_{q,k}\,\int \!ds \int\! dt\, C_V(s,t)  \,\pcol{\bar{\chi}_{\cc,q} \left(0 \right) }e^{i \bar{n}\cdot q s } \,\gamma^\mu_\perp \,  e^{-i n\cdot k t}  \lcol{\chi_{\cb,k}(0)} \,   \nonumber\\
  &= \sum_{q,k}\,\int \!ds \int\! dt\, C_V(s,t)  \,\pcol{\bar{\chi}_{\cc,q} \left(0 \right)} e^{i {\bar{\cal P}}^\dagger s }  
 \gamma^\mu_\perp e^{-i {\cal P} t} \lcol{\chi_{\cb,k}(0)} \,   \nonumber\\
 &=\sum_{q,k} \, \pcol{\bar{\chi}_{\cc,q}\left(0 \right)} \, {\tilde C}_V({\bar{\cal P}}^\dagger ,{\cal P} )\,
 \gamma^\mu_\perp \,\lcol{\chi_{\cb,k}(0)}\, ,
\end{align}
where we have explicitly written the sums over the labels $k$ and $q$ of the two fields. Instead of non-local operators in position space, one ends up with operators whose coefficients depend on the label momenta. The corresponding Wilson coefficients are just the Fourier transform of the position space coefficients. The results in Eq.~(\ref{eq:Llabel}) and Eq.~(\ref{eq:JsepLabel}) show that it is easy to map leading-power expressions in the two formalisms into each other. As a final remark, let us note that collinear gauge transformations act as convolutions on the label-formalism fields because these are essentially momentum-space quantities. We have side-stepped this issue by applying the label operator only to gauge invariant building blocks; the reader interested in gauge transformations and power corrections in the label formalism can consult \cite{Bauer:2003mga}.

%% file: 5_RGevolution.tex
\section{Resummation by RG Evolution \label{sec:LIV}}

The goal of this section is to discuss renormalization and 
 Renormalization Group (RG) evolution in the effective theory.
In the next section, we carry out the full computation of the resummed cross section in the case of Drell-Yan scattering. Here, in order to keep our discussion simple, we first consider the case
of the Sudakov form factor in QCD for which the relevant factorization
(in the sense of scale separation) theorem was obtained at the end of the previous section.
Obviously, this unphysical quantity is not of interest by itself, but this simple example illustrates the salient features which one also encounters in the analysis of physical processes. A characteristic property of
the RG equations in SCET is that the relevant anomalous dimensions are not just functions of the coupling constant, but involve a logarithmic dependence on the characteristic scale of the process. We now show how to solve such equations and then discuss why the anomalous dimensions only involve a single logarithm.

\begin{figure}[t]
\vspace{10 mm}
\[
\vcenter{ \hbox{
  \begin{picture}(0,0)(0,0)
\SetScale{1}
  \SetWidth{2}
  \ArrowLine(-50,50)(0,0)
  \ArrowLine(0,0)(50,50)
  \Photon(0,0)(0,-35){3}{4}
  \CCirc(0,5){20}{Black}{Gray}
\Text(-56,29)[cb]{{\Large  $p$}}
\Text(56,30)[cb]{{\Large $l$}}
\end{picture}}} \hspace*{2.2cm} = \hspace*{4.5cm}
\vcenter{ \hbox{
  \begin{picture}(0,0)(0,0)
\SetScale{1}
  \SetWidth{2}
  \SetOffset(0,0)
\SetColor{Blue}
  \Line(-70,50)(0,0)
\SetColor{OliveGreen}
  \DashLine(0,0)(70,50){6}

\SetColor{Black}
  \Photon(0,0)(0,-35){3}{4}
\Text(-80,-8)[cb]{{\LARGE $\tilde{C}^{\mbox{{\small bare}}}_V(Q^2) \times$}}
\end{picture}}}
\]
\vspace*{4mm}
\caption{Matching condition which allows to obtain
$\tilde{C}^{\mbox{{\tiny bare}}}_V (Q^2)$. In the calculation of the form factor one should  set from the start $p^2 = l^2 =0$.
 \label{fig:MatchingCV}}
\end{figure}
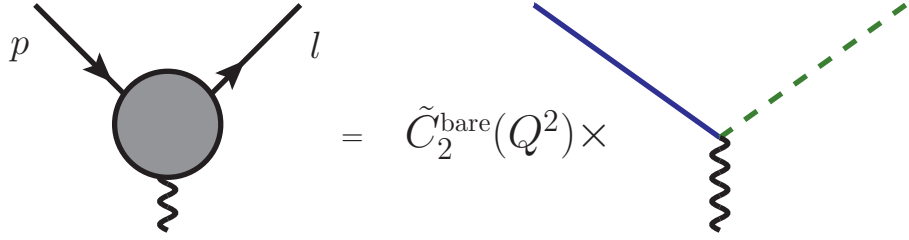

In the following, the Fourier transform of the matching coefficient of the current operator
$C(s,t)$ in Eq.~(\ref{eq:Jsep}) will be indicated  by $\tilde{C}^{\bare}_V (Q^2)$.
The value of this Wilson coefficient is determined in the same way as we discussed in the $\phi^3$-theory case, i.e.\ by matching it to the calculation of the on-shell form factor, as it is shown diagrammatically in 
Fig.~\ref{fig:MatchingCV}. The QCD on-shell form factors are known up to three-loop \cite{Baikov:2009bg,Gehrmann:2010tu}. The one-loop vector form factor translates into the following result for the bare matching coefficient
\be
\tilde{C}^{\mbox{{\tiny bare}}}_V (\ep,Q^2)
=  1+ \frac{\alpha^0_s}{4 \pi}
C_F \left(-\frac{2}{\varepsilon^2} - \frac{3}{\varepsilon} - 8 +\frac{\pi^2}{6} + {\mathcal O}(\varepsilon)\right) \left( \frac{e^{\gamma_E} Q^2}{4 \pi} \right)^{-\varepsilon}\, + {\mathcal O}\left(\alpha_s^2\right)\,,
\ee
where $\alpha^0_s=g^2_s/4\pi$ is the bare coupling constant and $\gamma_E\approx0.577$ is the Euler-Mascheroni constant. We first express the bare coupling $\alpha_s^0$ in terms of the $\ms$ renormalized coupling constant $\alpha_s(\mu)$ via the relation $Z_\alpha\, \alpha_s(\mu) \,\mu^{2\ep} = e^{-\ep \gamma_E}(4\pi)^\ep \alpha_s^0$, where $Z_\alpha  = 1+\mathcal{O}(\alpha_s)$ at the needed  accuracy. We obtain
\be
\tilde{C}^{\mbox{{\tiny bare}}}_V (\ep,Q^2)
=  1+ \frac{\alpha_s(\mu)}{4 \pi}
C_F \left(-\frac{2}{\varepsilon^2} - \frac{3}{\varepsilon} - 8 +\frac{\pi^2}{6} + {\mathcal O}(\varepsilon)\right) \left( \frac{Q^2}{\mu^2} \right)^{-\varepsilon}\, + {\mathcal O}\left(\alpha_s^2\right)\,,
\ee
At first order in $\alpha_s$, coupling constant renormalization does not change the divergences and is obviously not enough to arrive at a finite result. The remaining divergences are absorbed into a multiplicative $Z$ factor by defining a finite Wilson coefficient as follows
\be\label{eq:renormCV}
\tilde{C}_V (Q^2,\mu) = \lim_{\varepsilon \to 0} Z^{-1}\left(\ep,Q^2, \mu \right) \tilde{C}^{\mbox{{\tiny bare}}}_V (\ep,Q^2) \, ,
\ee
with
\be \label{eq:ZQ2m2}
Z\left(\ep,Q^2, \mu \right) = 1+ \frac{\alpha_s (\mu)}{4 \pi} C_F
\left(-\frac{2}{\varepsilon^2}+\frac{2}{\varepsilon}\ln\frac{Q^2}{\mu^2}
-\frac{3}{\varepsilon}  \right) + {\mathcal O}\left(\alpha_s^2\right) \, .
\ee
Consequently, the renormalized Wilson coefficient $\tilde{C}_V$ at order $\alpha_s$ is
\be \label{eq:CVren}
\tilde{C}_V (Q^2,\mu) = 1 + \frac{\alpha_s (\mu)}{4 \pi} C_F \left(
- \ln^2 \frac{Q^2}{\mu^2} +  3 \ln  \frac{Q^2}{\mu^2} + \frac{\pi^2}{6} -8\right)  + {\mathcal O}(\alpha_s^2)\, .
\ee
%

%
%

\subsection{Renormalization Group Equation}
\label{sec:RGEsud}

One immediately checks that at one-loop level the expression in Eq.~(\ref{eq:CVren}) satisfies
the differential equation
\be \label{eq:REGCV}
\frac{d}{d \ln \mu} \tilde{C}_V (Q^2,\mu) = \left[ C_F \gamma_{\cusp}(\alpha_s) 
\ln\frac{Q^2}{\mu^2}  + \gamma_V (\alpha_s)\right] \tilde{C}_V (Q^2,\mu) \, ,
\ee
where, at order $\alpha_s$, the functions $\gamma_{\cusp}$ and $\gamma_V$ are given by
\be
\gamma_{\cusp}(\alpha_s)  = 4 \frac{\alpha_s (\mu)}{4 \pi} \, , \qquad
\mbox{and} \qquad \gamma_V (\alpha_s) = - 6 C_F \frac{\alpha_s (\mu)}{4 \pi} \,.
\ee
(We remind the reader that $ d \alpha_s / d \ln \mu \propto \alpha^2_s$.) Eq.~(\ref{eq:REGCV}) is the {\em Renormalization Group Equation} satisfied by the Wilson coefficient $\tilde{C}_V$. The function $\gamma_{\cusp}$ is called the {\em Cusp Anomalous Dimension};  the origin of this name will be explained  below. The above RG equation is also valid beyond one-loop level; indeed it holds to all orders in perturbation theory as a consequence of factorization, as will be discussed in the next section. Since the on-shell form factor is now known up to three loops,
it is possible to extract the anomalous dimensions $\gamma_{\cusp}$ and $\gamma_V$ to order $\alpha_s^3$.
The RG equation in Eq.~(\ref{eq:REGCV}) contains an explicit logarithmic dependence on the scale $\mu$. This feature is characteristic of problems involving Sudakov double logarithms.

The solution of the RG equation in Eq.~(\ref{eq:REGCV}) sums the logarithmic terms to all orders in $\alpha_s$; in fact by separating  variables one obtains the solution 
\be\label{eq:RGsol}
\tilde{C}_V (Q^2,\mu) =  \exp\left\{ \int_{\mu_h}^\mu \left[C_F \gamma_{\cusp}(\alpha_s) 
\ln\frac{Q^2}{\mu^{\prime 2}}  + \gamma_V (\alpha_s) \right] d \ln \mu^\prime \right\} \, \tilde{C}_V (Q^2,\mu_h)\, ,
\ee
in which the logarithm appears in an exponential.
It is convenient to write the solution as the product of 
the Wilson coefficient calculated at a high scale $\mu_h$ and an evolution
matrix $U$ which ``runs down'' the scale from $\mu_h$ to $\mu$: 
\be \label{eq:Cexact}
\tilde{C}_V (Q^2,\mu) = U\left(\mu_h,\mu\right) \tilde{C}_V (Q^2,\mu_h) \, .
\ee 
To use the solution (\ref{eq:RGsol}) in practice, we rewrite the integration over the scale as an integration over the coupling by changing integration variables from $\mu$ to $\alpha_s(\mu)$ by using
\be
\frac{d \alpha_s(\mu)}{d \ln \mu} = \beta\left(\alpha_s(\mu) \right) \, ,
\ee
After rewriting also the logarithm in the exponent (\ref{eq:RGsol}) by employing the relation
\be\label{eq:logcoupl}
\ln\frac{\nu}{\mu} =  \int_{\alpha_s(\mu)}^{\alpha_s(\nu)} \frac{d\alpha}{\beta(\alpha)}\, ,
\ee
the evolution matrix can be written in the form 
\be \label{eq:evma}
U\left(\mu_h,\mu\right) = \exp\left[2 C_F S(\mu_h, \mu) - A_{\gamma_V}(\mu_h,\mu) \right] \left( \frac{Q^2}{\mu_h^2}\right)^{- C_F A_{\gamma_{\cusp}} (\mu_h,\mu)} \, ,
\ee
where the quantities $S$ and $A_\gamma$ are defined as
\bea \label{eq:SAdef}
S\left(\nu,\mu\right) &=& -\int_{\alpha_s(\nu)}^{\alpha_s(\mu)} d \alpha \frac{\gamma_{\cusp} (\alpha)}{\beta(\alpha)} \int_{\alpha_s(\nu)}^{\alpha} \frac{d \alpha'}{\beta(\alpha')} \, ,
\nn \\
A_{\gamma_i}(\nu,\mu)&=& -\int_{\alpha_s(\nu)}^{\alpha_s(\mu)} d \alpha \frac{\gamma_i (\alpha)}{\beta(\alpha)} \, ,
\eea
with $i \in \{V,\cusp\}$. It is straightforward to check that Eq.~(\ref{eq:Cexact}) with Eq.~(\ref{eq:evma}) indeed solves the RG equation 
Eq.~(\ref{eq:REGCV}) by observing that
\bea
\frac{d}{d \ln \mu} S\left(\nu,\mu\right) &=& - \gamma_{\cusp}\left(\alpha_s(\mu) \right) \int_{\alpha_s(\nu)}^{\alpha_s(\mu)}
\frac{d \alpha'}{\beta(\alpha')} \, , \nn \\
\frac{d}{d \ln \mu} A_{\gamma_i}\left(\nu,\mu\right) &=& - \gamma_i\left( \alpha_s(\mu)\right) \, .
\eea
Since $d \alpha_s/ \beta = d \ln \mu$, one can conclude from Eqs.~(\ref{eq:SAdef}) that the functions $A_i$ are responsible for the resummation of the single logarithms and the function $S$ for the resummation of the double logarithms. Explicit expressions for these functions can be obtained by inserting the perturbative expansion of the beta function and of the $\gamma$ functions in Eqs.~(\ref{eq:SAdef}). By parameterizing the expansions of the beta function and of the anomalous dimensions $\gamma_i$ as follows
\bea
\beta\left(\alpha_s \right) &=& -2 \alpha_s \left[ \beta_0 \left(\frac{\alpha_s}{4 \pi}\right)  + \beta_1 \left(\frac{\alpha_s}{4 \pi}\right)^2 + {\mathcal O}(\alpha^3_s) \right] \, , \nn \\
\gamma_{\cusp}(\alpha_s)  &=& \gamma_0^{\cusp} \left(\frac{\alpha_s}{4 \pi}\right) + \gamma_1^{\cusp} \left(\frac{\alpha_s}{4 \pi}\right)^2 + {\mathcal O}(\alpha^3_s) \, , \nn \\
\gamma_{V}(\alpha_s)  &=& \gamma_0^V \left(\frac{\alpha_s}{4 \pi}\right) + \gamma_1^V \left(\frac{\alpha_s}{4 \pi}\right)^2 + {\mathcal O}(\alpha^3_s) \, , 
\eea
and by inserting  these expansions in the integrands of Eqs.~(\ref{eq:SAdef}), one obtains
\bea \label{eq:SAA}
A_{\gamma_V}\left(\nu,\mu\right) &=& \frac{\gamma^V_0}{2 \beta_0} \ln{\frac{\alpha_s(\mu)}{\alpha_s(\nu)}} +{\mathcal O}(\alpha_s) \, , \nn \\
A_{\gamma_{\cusp}} \left(\nu,\mu\right) &=& \frac{\gamma^{\cusp}_0}{2 \beta_0} \ln{\frac{\alpha_s(\mu)}{\alpha_s(\nu)}} +{\mathcal O}(\alpha_s) \, ,\nn \\
S\left(\nu,\mu\right) &=& \frac{\gamma^{\cusp}_0}{4 \beta_0^2} \Biggl[
\frac{4 \pi}{\alpha_s(\nu)}\! \left(\! \frac{r-1}{r} - \ln{r} \!\right)\!+\! \left( \frac{\gamma^{\cusp}_1}{\gamma^{\cusp}_0} 
- \frac{\beta_1}{\beta_0}\right) \left( 1-r +\ln{r} \right)
\nn \\ && +\frac{\beta_1}{2 \beta_0} \ln^2 r
\Biggr] + {\mathcal O}(\alpha_s)\, ,
\eea
where $r = \alpha_s(\mu)/\alpha_s(\nu)$.
Note that $S\left(\nu,\mu\right)$ contains terms proportional to $1/\alpha_s$. By expanding $S\left(\nu,\mu\right)$ in terms of a single coupling $\alpha_s(\mu)$, one would find that this expansion produces terms of the form $\alpha_s^n(\mu) \ln^{2n}(\mu/\nu)$: $S\left(\nu,\mu\right)$ encodes the leading logarithmic terms. The way we organize the computation, which consists in eliminating large logarithms in favor of coupling constants at the different scales and then expanding in these couplings, is called {\em Renormalization Group Improved Perturbation Theory}. The large logarithm counts as $1/\alpha_s$, as it can be seen from Eq.~(\ref{eq:logcoupl}) remembering that $\beta(\alpha_s)\sim \alpha_s^2$.
 
We observe that the fixed order expression of the Wilson coefficient $\tilde{C}_V$ (Eq.~(\ref{eq:CVren})), becomes meaningless when $\mu \gg Q$ or $\mu \ll Q$, since in these cases the logarithms are large and the product
$\alpha_s \ln(Q^2/\mu^2) \sim 1$ cannot be used as an expansion parameter.
In contrast, if $\mu_h$ is taken approximately equal to the scale $Q$, the 
expression in Eq.~(\ref{eq:Cexact}) is valid for any value of $\mu$ for which 
$\alpha_s$ is perturbative.

\subsection{Resummation}

In the case of the Sudakov form factor, we integrated out the hard contribution and absorbed it into the Wilson coefficient $\tilde{C}_V\left(Q^2,\mu^2 \right)$,
and the decoupling also allows us to factorize soft and collinear interactions, as it is shown in Fig.~\ref{fig:scalesep}. The complete form factor can then be written as
\be \label{eq:SUDFK}
F\left(Q^2,L^2,P^2 \right) =  \tilde{C}_V\left(Q^2,\mu^2  \right) 
{\mathcal J}\left(L^2,\mu^2  \right)  {\mathcal J}\left(P^2,\mu^2  \right) 
{\mathcal S}\left(\Lambda_s^2,\mu^2  \right) \, , 
\ee
where the ${\mathcal J}$'s are the collinear functions and ${\mathcal S}$ is the soft function characterized by the scale $\Lambda^{2}_s = L^2 P^2 /Q^2$.
%
\begin{figure}[t]
\vspace*{3cm}
\[
\vcenter{ \hbox{
  \begin{picture}(0,0)(0,0)
\SetScale{1}
  \SetWidth{1}
 \SetOffset(20,0)
\CBox(-180,-50)(-62,111){Lavender}{Lavender}
\CBox(-62,-50)(63,111){Thistle}{Thistle}
\CBox(63,-50)(185,111){Orchid}{Orchid}
  \SetWidth{2}
\LongArrow(-180,-60)(-180,120)
  \Line(-200,-50)(185,-50)
  \Line(-183,-30)(-177,-30)
  \Line(-183,10)(-177,10)
  \Line(-183,50)(-177,50)
  \Line(-183,90)(-177,90)
  \DashLine(-180,-30)(185,-30){3}
\Text(-195,85)[cb]{{\Large $Q^2$}}
\Text(-123,85)[cb]{{\Large \boldmath$\tilde{C}_V(Q^2,\mu^2)$\unboldmath}}
\LongArrow(-123,82)(-123,-25)
\Text(-210,45)[cb]{{\Large $L^2 \sim P^2$}}
\Text(-1,55)[cb]{{\Large \boldmath${\mathcal J}(L^2,\mu^2)$\unboldmath}}
\Text(-1,35)[cb]{{\Large \boldmath${\mathcal J}(P^2,\mu^2)$\unboldmath}}
\LongArrow(-1,32)(-1,-25)
\Text(-195,5)[cb]{{\Large $\Lambda_s^2$}}
\Text(128,5)[cb]{{\Large \boldmath${\mathcal S}(\Lambda_s^2,\mu^2)$\unboldmath}}
\Text(-195,-35)[cb]{{\Large $\mu^2$}}
\LongArrow(128,2)(128,-25)
\end{picture}}} 
\]
\vspace*{14mm}
\caption{Schematic representation of the scale separation and of the calculational procedure in renormalization group improved perturbation theory.
 \label{fig:RG}}
\end{figure}
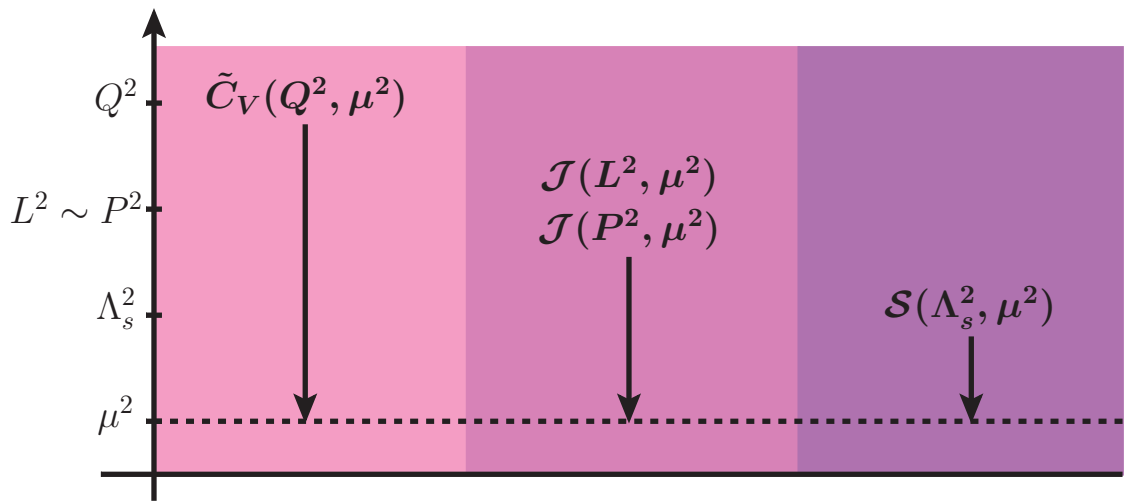

Above, we have resummed logarithms in the hard function by solving its RG equation. To achieve the resummation for the entire form factor, one solves the RG for each of the terms in the r.h.s.\ of Eq.~(\ref{eq:SUDFK}). All of them fulfill a RG equation of the same type as the one satisfied by the Wilson coefficient. Therefore, each factor in Eq.~(\ref{eq:SUDFK}) can be calculated perturbatively at its own characteristic scale, and then evolved to a common reference scale $\mu$. 
The procedure is summarized in Fig.~\ref{fig:RG}.
Since each factor is evaluated at its own natural scale, no large logarithms are present in the perturbative calculations; all of the large logarithms are resummed in the evolution factors originating from the solution of the RG equations.

The factorization formula puts constraints on the anomalous dimensions governing the RG equation of the various factors in Eq.~(\ref{eq:SUDFK}). The final result must be independent of the 't~Hooft scale, which is an artifact of the use of dimensional regularization:
\be
\frac{d}{d \ln \mu} \left[\tilde{C}_V\left(Q^2,\mu^2  \right) 
{\mathcal J}\left(L^2,\mu^2  \right)  {\mathcal J}\left(P^2,\mu^2  \right) 
{\mathcal S}\left(\Lambda_s^2,\mu^2  \right)  \right] =0 \, .
\ee
Consequently, one also finds
\bea \label{eq:const}
0 &=&\frac{d}{d \ln \mu} \ln \left[\tilde{C}_V\left(Q^2,\mu^2  \right) 
{\mathcal J}\left(L^2,\mu^2  \right)  {\mathcal J}\left(P^2,\mu^2  \right) 
{\mathcal S}\left(\Lambda_s^2,\mu^2  \right)  \right] \,  \nn \\
&=& \frac{1}{\tilde{C}_V\left(Q^2,\mu^2  \right)}\frac{d \tilde{C}_V\left(Q^2,\mu^2  \right)}{d \ln \mu} +
\frac{1}{{\mathcal J}\left(L^2,\mu^2  \right)}\frac{d{\mathcal J}\left(L^2,\mu^2  \right)}{d \ln \mu} 
\nn \\ && +
\frac{1}{{\mathcal J}\left(P^2,\mu^2  \right)}\frac{d{\mathcal J}\left(P^2,\mu^2  \right)}{d \ln \mu} + 
\frac{1}{{\mathcal S}\left(\Lambda_s^2,\mu^2  \right)}\frac{d{\mathcal S}\left(\Lambda_s^2,\mu^2  \right)}{d \ln \mu} \, .
\eea
The individual terms in this result are nothing but the anomalous dimensions of the different functions and the fact that the product Eq.~(\ref{eq:SUDFK}) is scale invariant thus implies that the sum of the anomalous dimensions vanishes.
The RG equation for the Wilson coefficient is Eq.~(\ref{eq:REGCV}), while the RG equation for the collinear and soft factors are
\bea
\frac{d}{d \ln \mu} {\mathcal J}\left(L^2,\mu^2  \right) &=&
-\left[C_F \gamma_{\cusp} \left(\alpha_s\right)  \ln{\frac{L^2}{\mu^2}} + \gamma_J\left(\alpha_s\right)\right]
{\mathcal J}\left(L^2,\mu^2  \right) \, , \nn \\
\frac{d}{d \ln \mu} {\mathcal S}\left(\Lambda_s^2,\mu^2  \right) &=&
\left[C_F \gamma_{\cusp} \left(\alpha_s\right) \ln{\frac{\Lambda_s^2}{\mu^2}}  + \gamma_S\left(\alpha_s\right)\right]
{\mathcal S}\left(\Lambda_s^2,\mu^2  \right) \, ;
\eea
therefore Eq.~(\ref{eq:const}) requires that
\bea
 C_F \gamma_{\cusp} \ln{\frac{Q^2}{\mu^2}} + \gamma_V -
C_F \gamma_{\cusp} \left(\! \ln{\frac{L^2}{\mu^2}} + \ln{\frac{P^2}{\mu^2}} 
\! \right) - 2 \gamma_J
 -  C_F \gamma_{\cusp} \ln{\frac{\mu^2}{\Lambda_s^2}}+ \gamma_S = 0 \, .
 \eea
For this cancellation to work, it is crucial that the scale dependence is logarithmic, with the same coefficient 
$\gamma_{\cusp}$ in all of the RG equations which enter in Eq.~(\ref{eq:const}). This explains why the anomalous dimensions are linear functions of the associated logarithms.

The label ``cusp'' in $\gamma_{\cusp}$ refers to the fact that the soft function is given by the matrix element of a Wilson line with a cusp. To see this, we note that the soft operator $S_n^\dagger(0) S_{\bar{n}}(0)$ in Eq.~(\ref{eq:Jsep}) can be viewed as a single Wilson line running first along $x^\mu(s) = s \bar{n}^\mu$, with $s=-\infty \dots 0$
and then back along $x^\mu(s) = s n^\mu$, with $s=0 \dots \infty$. This Wilson line has a cusp at the point $x^\mu(0)=0$, where the direction changes, see Figure \ref{fig:scalesep}.
Polyakov \cite{Polyakov:1980ca} and Brandt, Neri, and Sato \cite{Brandt:1981kf} proved that Wilson lines with cusps require renormalization and that the relevant anomalous dimension is proportional to the cusp angle. If the two lines forming the cusp are parallel to the vectors $n_1^\mu$ and $n_2^\mu$, the cusp angle $\beta_{12}$ is given by
\be\label{eq:cuspangle}
\cosh \beta_{12} = \frac{n_1 \cdot n_2}{\sqrt{n_1^2 n_2^2}} \, .
\ee
The angle above diverges for light-like Wilson lines. However, the anomalous dimension $\Gamma(\beta_{12})$ behaves as follows in the limit in which $n_i^2 \to 0$
\be
\Gamma(\beta_{12})\, \overbrace{=}^{n_i^2 \to 0} \, C_F \gamma^{\cusp}_i(\alpha_s) \ln{\frac{\mu^2}{\Lambda_s^2}} + \cdots \, ,
\ee
as was proven in \cite{Korchemskaya:1992je}, where the RG equation satisfied by light-like Wilson loops was derived.

%% file: 6_DY.tex
\section{Threshold Resummation in Drell-Yan Production \label{sec:DY}}

So far, we exclusively discussed the off-shell Sudakov form factor, which provides the simplest example to discuss Sudakov logarithms and their resummation. The off-shell form factor is, however, gauge dependent and therefore unphysical. With all the formalism in place, we are now ready to analyze a physical cross section. We will study the Drell-Yan process, which consists of the production of a lepton pair of momentum $q$, together with an arbitrary hadronic final state $X$ at a hadron collider. We consider a situation where we are close to the production threshold, and the energy $E_X$ of the radiation $X$ is much smaller than the momentum transfer, which is set by the invariant mass $M^2=q^2$ of the lepton pair. This leads to large Sudakov logarithms of the small ratio $E_X/M$, which we will resum using SCET.

Starting with the pioneering papers \cite{Sterman:1986aj,Catani:1989ne}, this type of threshold resummation has been performed for many hadron-collider cross sections, both with traditional methods and more recently using SCET. In particular, the effective theory framework has been used to resum large logarithmic terms  in Deep Inelastic Scattering (DIS) \cite{Becher:2006mr}, Drell-Yan\cite{Becher:2007ty,Becher:2010tm}, Higgs production \cite{Ahrens:2008qu, Ahrens:2008nc,Ahrens:2010rs,Becher:2012qa}, direct photon production \cite{Becher:2009th}, top-quark pair production at hadron colliders \cite{Ahrens:2009uz,Ahrens:2010zv,Ahrens:2011mw}, electroweak boson production \cite{Becher:2011xn,Becher:2012xr}, slepton pair production \cite{Broggio:2011bd} and top-squark pair production \cite{Broggio:2013uba, Broggio:2013cia}. The SCET resummation is based on RG evolution and is typically performed in momentum space \cite{Becher:2006nr}, while the traditional resummations are mostly done in moment space. However, one can show that the formalisms are equivalent and relate their ingredients order-by-order in perturbation theory \cite{Becher:2006mr,Becher:2007ty,Ahrens:2008nc}. Recently, there has been renewed interest in the comparison of the resummation methods \cite{Bonvini:2013td} and detailed studies were presented in \cite{Almeida:2014uva,Sterman:2013nya,Bonvini:2014qga}.
 
We will assume that $E_X\gg \Lambda_{\rm QCD}$ so that the coupling constant is still small enough to allow for a perturbative expansion. In cases, where the invariant mass $M$ is large, a conventional fixed-order expansion of the cross section will however be spoiled by the presence of large logarithms of the energy of the soft radiation $X$ over the invariant mass $M$. To address this problem, we first derive a factorization theorem, which separates the physics associated with the hard scale $M$ from soft physics, and then use RG evolution to resum the associated logarithms. The resummation of the associated logarithms was first achieved in the seminal papers \cite{Sterman:1986aj,Catani:1989ne}, the SCET analysis discussed below was performed in \cite{Becher:2007ty}. The relevant expansion parameter in the effective theory is $\lambda=E/M$. The soft fields are scaling as $(\lambda^2,\lambda^2,\lambda^2)$ and describe the radiation into the final state together with collinear modes in the directions of the incoming hadrons.\footnote{The expansion parameter was denoted by $\epsilon=\lambda^2$ in \cite{Becher:2007ty}.}

Our first task will be to write the cross section in a form which is suitable for the factorization analysis. We denote the two scattered hadrons by $N_1$ and $N_2$; the process of interest, $N_1(p)+ N_2(l) \to \ell^+(p_+)+ \ell^-(p_-) + X(p_X)$, is mediated by a virtual photon or a $Z$ boson. For simplicity, we consider the photon case, and compute the cross section as a function of the momentum of the lepton pair $q=p_+ + p_-$ and sum over the lepton spins. In the center-of-mass frame, the cross section is
\begin{multline} \label{eq:DYcross}
\frac{d\sigma}{d^4 q} = \frac{1}{2 s} \int \frac{d^3p_+}{(2\pi)^3 2E_+ } \frac{d^3p_-}{(2\pi)^3 2E_- } \delta^{(4)}(q-p_+-p_-) \\
\times \sum\hspace{-0.5cm}\int\limits_X\; \;\big | \langle \ell^+ \, \ell^-\, X   | N_1\, N_2 \rangle \big|^2 (2\pi)^{4} \delta^{(4)}(p+l-p_X-q)
\end{multline}
We are working to leading order in the electromagnetic interaction. The leptonic part thus factorizes from the hadronic part of the amplitude, and is given by
\begin{equation}
\langle \ell^+ \, \ell^-\, X  | N_1\, N_2 \rangle = \frac{e^2}{q^2} \,\bar{u}(p_-) \gamma_\mu v(p_+) \, \langle X | J_\mu(0) | N_1\, N_2 \rangle\, ,
\end{equation}
where $J^\mu = \sum_q e_q \bar\psi_q \,\gamma^\mu \, \psi_q$ is the electro-magnetic quark current. We now define the lepton tensor
\begin{align}
L_{\mu\nu} &=  \int \frac{d^3p_+}{(2\pi)^3 2E_+ } \frac{d^3p_-}{(2\pi)^3 2E_- } \delta^{(4)}(q-p_+-p_-) \sum_s  \bar{u}(p_-) \gamma_\nu v(p_+)  \bar{v}(p_+) \gamma_\mu u(p_-) \nonumber \\
& = \int \frac{d^3p_+}{(2\pi)^3 2E_+ } \frac{d^3p_-}{(2\pi)^3 2E_- } \delta^{(4)}(q-p_+-p_-) {\rm tr}\left[ {p\!\!\!/}_- \gamma_\mu \, {p\!\!\!/}_+ \gamma_\nu \right]\,   \nonumber \\
&= \frac{1}{(2\pi)^4} \frac{1}{6\pi}  \left( q_\mu q_\nu - g_{\mu\nu}\, q^2 \right)\,.
\end{align}
The tensor structure is fixed by current conservation, which implies that the tensor is transverse  $q^\mu L_{\mu\nu}=q^\nu L_{\mu\nu}=0$. To determine the overall prefactor, it is simplest to compute $g^{\mu\nu}  L_{\mu\nu}$. The cross section is then given by the product of the lepton tensor and a hadron tensor
\begin{equation}\label{eq:crossSection}
\frac{d\sigma}{d^4 q} = \frac{1}{2s} \frac{e^4}{(q^2)^2} L_{\mu\nu} W^{\mu\nu} = \frac{4\pi \alpha^2}{3 s q^2}  \frac{1}{(2\pi)^4} (-g_{\mu\nu}) W^{\mu\nu}\,,
\end{equation}
where we have used that also the hadron tensor is transverse. It is given by
\begin{align} \label{eq:hadTens}
W_{\mu\nu} &=\sum\hspace{-0.5cm}\int\limits_X\;\; \langle N_1 N_2 | J^\dagger_\mu(0)  |X \rangle \langle X   | J_\nu(0) | N_1\, N_2  \rangle (2\pi)^{4} \delta^{(4)}(p+l-p_X-q)  \nonumber \\
&= \int\! d^4 x\, e^{-iq x}\, \langle N_1 N_2 | J^\dagger_\mu(x) J_\nu(0) | N_1\, N_2  \rangle\, .
\end{align}
The second form is what we will use to derive the factorization theorem for the cross section. To show that the two forms are equivalent, one can insert a complete set of states between the two currents on the second line and then translate the current to zero using the momentum operator $J_\mu(x)=e^{i P x }J_\mu(0)e^{-i P x }$.

\subsection{Derivation of the Factorization Formula in SCET}

We are now ready to derive the factorization theorem for the hadronic tensor. In Section~\ref{ssec:Dec}, we analyzed the electromagnetic current operator of a quark in the effective theory.  The result reads
\be \label{eq:JsepAgain}
J^\mu(x) = \int\! dr \int\! dt \,C_V(r,t)\, \bar{\chi}_{\bar{c}}\left(x+ r n \right)
S_{\bar{n}}^\dagger\left(x \right)S_{n}\left(x \right)
 \gamma^\mu_\perp \chi_c(x+t \bar{n}) \, .
\ee 
This current describes an energetic quark in the direction of $N_1$ and an anti-quark in the direction of $N_2$. There is also a second contribution, 
shown in Eq.~(\ref{eq:Jsep}), in which the directions of the quark and anti-quark are interchanged. The above result for the current operator was obtained after the decoupling transformation and the collinear and soft fields do not interact, but for simplicity we drop the label on the fields and we write $\chi_{\bar{c}}$ instead of  $\chi_{\bar{c}}^{(0)}$. We have not yet multipole expanded the soft Wilson lines. The proper expansion will be performed at the end, after discussing the kinematics. In order to obtain the scaling of $x$, one must also consider the scaling of the photon field to which the current $J^\mu(x)$ couples. We will find below, that all spatial components of the soft fields can be expanded away so that the soft Wilson lines are evaluated at $\vec{x}=0$.

The result for the current can now be inserted into the expression in Eq.~(\ref{eq:hadTens}) for the hadronic tensor. Since the different fields do not interact, the hadronic tensor factorizes into a soft matrix element times collinear matrix elements. In order to obtain a simple form for the result, we first rearrange the collinear fields using the Fierz identity. The identity rearranges spinors as follows
\begin{equation}
\bar{u}_1  \Gamma_1 u_2\, \bar{u}_3  \Gamma_2 u_4 =  \sum C_{AB}\, \bar{u}_1  \Gamma_A u_4\, \bar{u}_3  \Gamma_B u_2 \, .
\end{equation}
Under Fierz transformation, the combination $ \Gamma_1 \otimes \Gamma_2= \gamma_\mu \otimes \gamma^\mu$ is mapped onto
\begin{equation}\label{eq:Fierz}
 \gamma_\mu \otimes \gamma^\mu \, \rightarrow \,-\frac{1}{2} \gamma_\mu \otimes \gamma^\mu -\frac{1}{2}  \gamma_\mu\gamma_5 \otimes \gamma^\mu \gamma_5+ 1\otimes 1 - \gamma_5 \otimes  \gamma_5\,.
\end{equation}
The SCET vector currents involve the matrix $\gamma^\mu_\perp$ instead of $\gamma^\mu$. The two are related by
\begin{equation}
\gamma^\mu= \gamma^\mu_\perp + n\!\!\!/ \frac{\bar{n}^\mu}{2} + \bar{n}\!\!\!/ \frac{n^\mu}{2} \,.
\end{equation}
However, since $\bar{n}\!\!\!/  \chi_{\bar{c}} = n\!\!\!/ \chi_c=0$, the additional terms do not contribute and we can use the Fierz relation (\ref{eq:Fierz}) for the full vector current. Using the same properties of the SCET spinors, we can then simplify the terms which appear on the right-hand side, which involve collinear spinors in the same direction,
\begin{align}
\bar{\chi}_c \gamma^\mu \chi_c & = n^\mu \bar{\chi}_c \frac{\bar{n}\!\!\!/ }{2} \chi_c  \, ,& \bar{\chi}_c   \chi_c & = \bar{\chi}_c \frac{n\!\!\!/ \bar{n}\!\!\!/}{4} \chi_c=0\, ,
\end{align}
and analogously for the spinor products involving $\gamma_5$. In the second relation, we have pulled the projection operator out of the collinear fermion field and then annihilated the anti-fermion with it. The final result for the Fierz identity for the two vector currents in SCET takes then  the simple form
\begin{equation}
\bar{\chi}_c \gamma_{\perp \mu} \chi_{\bar{c}} \, \bar{\chi}_{\bar{c}} \gamma^\mu_\perp \chi_c =  
\bar{\chi}_c \frac{\bar{n}\!\!\!/ }{2} \chi_c \, \bar{\chi}_{\bar{c}} \frac{n\!\!\!/ }{2} \chi_{\bar{c}} + \bar{\chi}_c \frac{\bar{n}\!\!\!/ }{2} \gamma_5 \chi_c \, \bar{\chi}_{\bar{c}} \frac{n\!\!\!/ }{2}  \gamma_5 \chi_{\bar{c}}\,.
\end{equation}
Note that this relation involves an extra minus sign compared to Eq.~(\ref{eq:Fierz}), which arises from anticommuting the fermion fields. The matrix element of the collinear fields will be the parton distribution function. Because of parity invariance of the strong interaction, the terms involving $\gamma_5$ have vanishing matrix elements and will be dropped in the following. 

Because the collinear and soft sectors no longer interact, each matrix element must be a color singlet. When taking a collinear matrix element, we can thus average over color
\begin{equation}
\bar{\chi}_{c,\alpha}\, \frac{\bar{n}\!\!\!/ }{2}\, \chi_{c,\beta} \to \frac{1}{N_c}\,  \delta_{\alpha\beta}\, \bar{\chi}_{c,\delta}\, \frac{\bar{n}\!\!\!/ }{2} \, \chi_{c,\delta}\,,
\end{equation}
where $\alpha, \beta, \delta$ are the color indices of the fields. 
After this averaging, the color indices of the soft Wilson lines are all contracted among themselves and the soft part of the matrix element takes the form
\begin{equation}\label{eq:softwilson}
 \hat{W}_{\rm DY} (x)  =\frac{1}{N_c}  {\rm tr}\, \langle 0 |  \bar{T}\left( S_n^\dagger\left(x \right)S_{\bar{n}}\left(x \right)  \right)
 T\left( S^\dagger_{\bar{n}}\left(0\right)S_{n}\left(0 \right)\right) | 0 \rangle \,.
\end{equation}
We have absorbed one of the factors of $N_c^{-1}$ into the definition of the matrix element so that $\hat{W}_{\rm DY} (x)= 1 + {\cal O}(\alpha_s)$.  We need to use anti-time ordering on the Wilson lines which arise from $J^\dagger_\mu(x)$. The reason is that we are computing an amplitude squared, see the first line of Eq.~(\ref{eq:hadTens}), so the propagators of the complex conjugate amplitude have the opposite $+i0^+$ prescription. A detailed discussion of this point is given in Appendix C of \cite{Becher:2007ty}. The soft matrix element is a vacuum matrix element since the initial state protons are composed of collinear fields and do not contain any soft partons. Soft partons cannot be part of the proton since the soft scale $E_X\gg \Lambda_{\rm QCD}$, while the proton constituents fulfill $p^2\sim l^2 \sim \Lambda_{\rm QCD}^2$.

Let us now put together the result after the simplifications. According to Eq.~(\ref{eq:crossSection}), the relevant quantity for the cross section is the hadronic tensor contracted with the metric. It takes the form
\begin{equation}\label{eq:g2formula}
\begin{aligned}
   & (-g_{\mu\nu})\, W^{\mu\nu}  = -  \frac{1}{N_c}\, \int\! d^4 x \, e^{-i q \cdot x}  \int \!dr \int \!dr' \int\!dt \int\! dt'\, C_V(r,t) \,C_V^*(r',t') \\[-2mm]
   &\hspace{7mm}\times \hat W_{\rm DY}(x)\,
\langle N_1(p)|\,\bar\chi_{c}(x+t' \bar{n})\,\frac{\rlap{/}{\bar n}}{2}\,\chi_{c}(t \bar{n})\,
|N_1(p)\rangle\,
\langle N_2(l)|\,\bar\chi_{\bar{c}}(r n)\,\frac{\rlap{/}{n}}{2}\,
\chi_{\bar{c}}(x +r' n)\,|N_2(l)\rangle \,.
\end{aligned}
\end{equation}
The final step in the derivation will be to perform the multipole expansion of these matrix elements. To perform the expansion, we need to know how the position-space variable $x^\mu$ scales. This variable is conjugate to the momentum  $q^\mu$ of the virtual photon, which is a sum of a collinear and an anti-collinear incoming momentum. We therefore infer that  $x^\mu$ generically scales as $(1,1,\lambda^{-1})$. At leading power, we can thus set $x_-$ to zero in the collinear fields and $x_+=0$ in the anti-collinear matrix element. From the generic scaling of $x^\mu$ one would also naively drop all $x^\mu$ dependence in the soft matrix element $\hat W_{\rm DY}(x)$. The generic scaling of $x^\mu$ is relevant for transverse momentum resummation, which will be discussed in Section~\ref{sec:pT}. Near the partonic threshold region the final state consists of soft gluons with definite energy. We will need to consider the energy and the three-momentum of the soft radiation instead of its light-cone components to perform the appropriate expansion. 

We will consider the expansion for the soft function below, and first turn to the collinear matrix elements which can be simplified further by noting that we do not have collinear  partons in the final state at the threshold (collinear radiation is present for transverse momentum resummation, see Section~\ref{sec:pT}). This implies that the collinear partons are part of the proton and as such their transverse momenta are of order $\Lambda_{\rm QCD}$, much smaller than the transverse momentum of the final state photon. Because of this, we can also expand away the transverse position dependence of the collinear matrix elements, after which they only depend on the position space variable conjugate to the large momentum and take the form 
\begin{equation}\label{eq:PDF}
\langle  N_1(p) | \bar \chi_c\left( x_+ +t' \bar{n}
\right) \frac{\bar{n}\!\!\!/}{2}   \chi_c(t \bar{n}) \,|   N_1(p) \rangle  =\bar{n}\cdot p\, \int_{-1}^1 dx_1\, f_{q/N_1}(x_1,\mu)\,e^{i\,x_1\, \left(x_+ +t' \bar{n} - t \bar{n} \right)\cdot p} \,.
\end{equation}
The variables $t$  and $t'$ appear in the convolutions with the Wilson coefficients in the currents. 
The non-perturbative quantities $f_{q/N_1}(x_1)$ are the usual parton distribution functions (PDFs) \cite{Soper:1996sn}. The variable $x_1$ is the fraction of the proton momentum carried by the quark field. Negative values correspond to the anti-quark distributions:  $f_{\bar{q}/N_1}(x_1)=f^*_{\bar{q}/N_1}(x_1) = - f_{q/N_1}(-x_1)$ \cite{Soper:1996sn}. The reason that these matrix elements are exactly the same as the  PDFs defined in QCD is that in the absence of soft interactions the collinear Lagrangian is completely equivalent to the standard QCD Lagrangian and the SCET collinear quark field is related to the standard quark field $\psi(x)$ simply by $\chi_c(x) = W^\dagger(x) \frac{ n \!\!\!/ \bar{n}\!\!\!/}{4} \psi(x)$. In terms of the QCD field the SCET matrix element (\ref{eq:PDF}) reads
\begin{displaymath}
\langle  N_1(p) | \bar \psi\left( t^{\prime \prime} \bar{n}
\right) \frac{\bar{n}\!\!\!/}{2}\, [t^{\prime \prime} \bar{n}, 0]  \psi(0) \,|   N_1(p) \rangle \,, \\
\end{displaymath}
where we set $x_+=0$ and $t^{\prime \prime}=t^\prime - t$. In this expression we combined $W(t^{\prime \prime} \bar{n}) W^\dagger(0) = [t^{\prime \prime} \bar{n}, 0]$ into a finite length Wilson line connecting the two quark fields.

We are now ready to combine all the ingredients to get the following form of the hadronic tensor
\begin{multline}\label{eq:DYfact}
(-g_{\mu\nu} W^{\mu\nu}) =\frac{1}{N_c} \int_{0}^{1} dx_1 \int_{0}^{1} dx_2\, s |\tilde{C}_V(-\hat{s} ,\mu)|^2\,\int d^4x\,\hat{W}_{\rm DY} (x,\mu)\,e^{i\,x \cdot ( x_1\,p_- + x_2\,l_+-q)}  \\
\times \left[ f_{q/N_1}(x_1,\mu) f_{\bar{q}/N_2}(x_2,\mu)  +\text{($q\leftrightarrow \bar{q}$)}\right]\,,
\end{multline}
where $s=\bar{n}\cdot p\,  n\cdot l$ and $\hat{s}= x_1 x_2 s$ are the hadronic and partonic squared center of mass energy, respectively. (The symbol $p_-$ in Eq.~(\ref{eq:DYfact}) and in the equations below refers to the minus light-cone component of the proton momentum and not to the anti-lepton momentum as in Eq.~\eqref{eq:DYcross}.) The second term, where $q$ and $\bar{q}$ are interchanged, arises from the contribution to the current matching Eq.~(\ref{eq:JsepAgain}), where the quark and anti-quark are interchanged which we did not explicitly write down. The hard function, soft function and PDFs in Eq.~\eqref{eq:DYfact} have been renormalized and depend on the scale $\mu$, see Eq.~\eqref{eq:renormCV}.

Note that the $x_1$ and $x_2$ integrations only run over positive values, while the integrations range from $-1 < x_{i} < 1$ in the PDF matrix elements Eq.~(\ref{eq:PDF}). The restriction to positive values arises from the fact that the final state in the hard scattering, which consists of the Drell-Yan pair and soft QCD radiation, has positive energy and invariant mass. It therefore has positive light-cone momenta both in the plus and minus direction. By momentum conservation these are equal to the incoming light-cone components $x_1\,p_-$ and $x_2\,l_+$ which enforces positive momentum fractions. To explicitly show this, one can use a Fourier representation
\begin{equation}
\hat{W}_{\rm DY} (x,\mu) = \int \frac{d^4 p_{X_s}}{(2\pi)^4} e^{-i x\, \cdot p_{X_s}}  \tilde{W}_{\rm DY} (p_{X_s},\mu)\,.
\end{equation}
Integration over $x$ then yields the four-momentum conservation $\delta$-functions and since $p_{X_s}$ is a sum of final state momenta $ \tilde{W}_{\rm DY} (p_{X_s},\mu)$ has only support for $p_{X_s}^{(0)} \geq 0$.

The result in Eq.~(\ref{eq:DYfact}) contains the Fourier transform of the hard matching coefficient 
\begin{equation}
\tilde{C}_V(-\hat{s} ,\mu) = \int\! dr \int\! dt \,C_V(r,t,\mu) e^{-i\,x_1\, t \bar{n}\cdot p} e^{-i\,x_2\, r n\cdot l}\, ,
\end{equation}
where the exponentials arise from the matrix element in Eq.~(\ref{eq:PDF}). We can further simplify Eq.~(\ref{eq:DYfact}) by replacing $\tilde{C}_V(-\hat{s} ,\mu)= \tilde{C}_V(-q^2 ,\mu)$ since we are close to the production threshold. For the cross section, we then obtain
\begin{multline}\label{eq:dsigmad4q}
d\sigma = \frac{d^4q}{(2\pi)^4} \frac{4\pi \alpha^2}{3 q^2 N_c} |\tilde{C}_V(-q^2 ,\mu)|^2  \int_{0}^{1} dx_1 \int_{0}^{1} dx_2\,
\sum_q e_q^2 \left[ f_{q/N_1}(x_1,\mu) f_{\bar{q}/N_2}(x_2,\mu)  +\text{($q\leftrightarrow \bar{q}$)}\right] \\
\times\int d^4x\, \hat{W}_{\rm DY} (x)\,e^{i\,x \cdot ( x_1\,p_- + x_2\,l_+-q)} \,.
\end{multline}  

Let us now be a bit more specific and compute the cross section differential in the boson mass $M^2=q^2$. To do so, we rewrite
\begin{equation}\label{eq:masssig}
\int\! d^4q = \int\! dM^2 \int\! \frac{d^3q}{2q_0} \, .
\end{equation} 
The electroweak boson (i.e. the virtual photon in the case considered here) near threshold is produced with small transverse momentum, since the transverse momentum has to be balanced by the soft radiation. We thus have $q^0 =\sqrt{\hat{s}} + {\mathcal O(\lambda^2)}$ and $|\vec{q}| \sim \lambda^2$. Since the denominator in Eq.~(\ref{eq:masssig}) does not depend on $\vec{q}$ to leading power, we can then perform the $\vec{q}$ integration. This yields $\delta^{(3)}(\vec{x})$, so that we need the soft function only for $\vec{x}=0$.
In addition, the following relation holds
\begin{equation}
 ( x_1\,p_- + x_2\,l_+ -q)^{(0)} =\frac{\sqrt{\hat{s}}}{2} (1-z) + {\mathcal O}(\lambda^4)\, \label{eq:knot} \, ,
\end{equation}
where we defined $z\equiv M^2/\hat{s}$. One finds that $1-z \sim {\mathcal O}(\lambda^2)$.
In order to prove Eq.~(\ref{eq:knot}) we start by observing that the l.h.s.~coincides with the energy of the additional final state partonic radiation, $p_x^{(0)}$. (We introduce the subscript $x$ in order to differentiate the final state radiation of the hard partonic process from the complete hadronic final state introduced above and indicated with the subscript $X$.) One can then take the square of the partonic momentum conservation to obtain
\be
\hat{s} = M^2 + 2 q\cdot p_x \, ,  \label{eq:irmknot}
\ee 
where we neglected the subleading term $p_x^2 \sim {\mathcal O}(\lambda^2)$. In the partonic center of mass frame, where $x_1 \vec{l} + x_2 \vec{p} =0$, one has that $\vec{q} = - \vec{p}_x$, so that 
$|\vec{q} | = |\vec{p}_x| = p_x^{(0)}$. Therefore Eq.~(\ref{eq:irmknot}) can be rewritten as
\be
\hat{s} = M^2 +2 p_x^{(0)} \sqrt{M^2 + \left(p_x^{(0)}\right)^2} + 2 \left(p_x^{(0)}\right)^2 \, .
\ee 
By solving the equation above with respect to $p_x^{(0)}$ one finds
\be
p_x^{(0)} = \frac{M (1-z)}{2 \sqrt{z}}  \equiv \frac{\omega}{2}\, , 
\ee
which coincides with Eq.~(\ref{eq:knot}) once the relation between $\hat{s}$, $M$, and $z$ is applied.

Our final result for the cross section then reads
\begin{multline} \label{eq:dsdM2}
\frac{d\sigma}{dM^2} =  H\left( M^2, \mu \right) \int_{0}^{1} dx_1 \int_{0}^{1} dx_2\, \\
\sum_q e_q^2 \left[ f_{q/N_1}(x_1,\mu) f_{\bar{q}/N_2}(x_2,\mu)  +\text{($q\leftrightarrow \bar{q}$)}\right] 
 \frac{1}{\sqrt{\hat{s}}} W_{\rm DY}\!\left( \sqrt{\hat{s}}(1-z),\mu\right)  \, , 
\end{multline} 
where the Fourier transformed soft function is defined as
\begin{equation}
W_{\rm DY}(\omega,\mu) =  \int \frac{dx^0}{4\pi} \, \hat{W}_{\rm DY} (x,\mu)\,e^{i\,x^0 \omega/2}\, ,
\label{eq:sffourier}
\end{equation}
and the hard function is given by
\be
H\left( M^2, \mu \right) \equiv  \frac{4\pi \alpha^2}{3 M^2 N_c} |\tilde{C}_V(-M^2 ,\mu)|^2  \, .
\ee
The result now shows that the perturbative expansion involves different scales. For the hard function, the natural scale choice for the renormalization scale is $\mu \sim M$, while the scale characterizing  the soft emissions is lower.

\subsection{Laplace Transformation and RG equation for the Soft Function}
\label{subsec:RGEsoft}

In this section we derive the RG equation satisfied by the soft function $W_{\rm DY}$ and solve it to resum  the large logarithms of $E_X/M$. To derive the equation, we will work in the strict threshold limit, where the collider energy is just barely enough to produce the lepton pair of mass $M$ and the kinematics is such that the entire final state is forced to be soft. In practice threshold resummation is used away from the machine threshold, but near the threshold $x_1 \sim x_2\sim 1$, full RG invariance of the resummed cross section is ensured. Near threshold $s \simeq \hat{s} \simeq M^2$ and the small energy of the hadronic final state is given by $E_X \simeq \sqrt{s} -M$. We now rewrite
\be\label{eq:Jaco}
\frac{d \sigma}{d M^2} \simeq  \frac{1}{2 \sqrt{s}} \frac{d \sigma}{ d E_X} \, .
\ee
The Jacobian from $M$ to $E_X$ has a minus sign since large invariant masses $M$ correspond to small energies $E_X$. In Eq.~(\ref{eq:Jaco}) we assume than one integrates from smaller to larger energies $E_X$ which cancels this sign. Near the machine threshold $M^2\approx s$, we can rewrite Eq.~(\ref{eq:dsdM2}) as follows
\bea \label{eq:dsdM22}
\frac{d\sigma}{dE_X} &=& H\!\left( s, \mu \right) \sum_{q} e_q^2\,\int _0^\infty d\bar{x}_1 \int_0^\infty d\bar{x}_2 \int_0^\infty d\omega\,  \delta\left(E_X -\frac{\sqrt{s}}{2} \bar{x}_1  -\frac{\sqrt{s}}{2} \bar{x}_2 -  \frac{\omega}{2} \right) \nn \\
&& \times
f_{q/N_1}(x_1,\mu)\, f_{\bar{q}/N_2}(x_2,\mu) \,W_{\rm DY}\!\left( \omega,\mu\right)  
+\text{($q\leftrightarrow \bar{q}$)}\,,
\eea
where we use the notation $\bar{x}_i = 1-x_i$. To arrive at this form, we have introduced an integration over the energy of the soft radiation and have rewritten the energy-conservation $\delta$-function in its original form by observing that the energy of the final state partonic radiation $p_x^{(0)}$ can be written as
\be
\frac{\omega}{2} \equiv p_x^{(0)} = \underbrace{\sqrt{s} -M}_{E_X}  - \left[ \bar{x}_1 +\bar{x}_2 \right] \frac{\sqrt{s}}{2} \, .
\ee
In words, this equation says that the small total energy $E_X$ in the final state is given by the energies of the two proton remnants and the energy of the soft radiation. We have furthermore extended the integration from $0 \leq \bar{x}_i \leq 1$ to $0 \leq \bar{x}_i \leq \infty$. This does not change the result since the energy conservation $\delta$-function restricts $\bar{x}_i \leq 2E_X/\sqrt{s} \ll 1 $ anyway.

In order to perform the resummation, we want to solve the RG equations of the soft function and the PDFs. Proceeding directly in momentum space is cumbersome because the RG equations involve convolutions of anomalous dimensions with the functions in momentum space. Furthermore, the anomalous dimensions and also the soft function itself are distribution valued. Both difficulties can be avoided by transforming into a space where the factorization theorem becomes a product instead of a convolution. This can be achieved by a Fourier, Laplace or Mellin transformation. Indeed, in the derivation of the factorization theorem we were working with a position-space matrix element, which factorized into a product of position-space matrix elements. To derive and solve the RG equations, it is most convenient to perform a Laplace transformation, which, together with its inverse, is given by
\begin{align}
\tilde f(s) &=\int_0^\infty\! d\omega\, e^{-\omega s} f(\omega)\,,  & &\text{ and } & f(\omega) &= \frac{1}{2\pi i } \int_{c-i \infty}^{c+i\infty} \!ds \,e^{\omega s} \tilde f (s)\,,
\end{align}
and is obviously closely related to a Fourier transform. In the inverse Laplace transform, the constant $c$ is chosen in such a way that it is larger than the real part of all singularities of $\tilde f (s)$.
It has the property that it turns convolutions of the form of Eq.~(\ref{eq:dsdM22}) into products. The Laplace transform of the convolution
\begin{equation}
h(\omega) =  \int_0^\infty\! d\omega_1 \int_0^\infty \!d\omega_2\, \delta(\omega-\omega_1-\omega_2) f(\omega_1) g(\omega_2)  
\end{equation}
is the product of the Laplace transforms:
\begin{equation}
\tilde h(s) = \int_0^\infty\! d\omega e^{-\omega s} h(\omega) = \int_0^\infty\! d\omega_1 \int_0^\infty \!d\omega_2\, e^{- \omega_1 s-\omega_2 s}\,f(\omega_1) g(\omega_2) =   \tilde f(s) \,\tilde g(s) \,.
\end{equation}
So, as promised, the Laplace transformation can be used to turn convolutions such as the one in the factorization theorem Eq.~(\ref{eq:dsdM22}) into products. Some useful properties of the Laplace transform are obtained by considering the Laplace transform
\begin{equation}\label{eq:LaplDist}
\int_0^\infty \!dx \,e^{-x s} \frac{1}{x^{1-\lambda}} =s^{-\lambda } \Gamma (\lambda )   = \frac{1}{\lambda} - \ln( s e^{\gamma_E} )+\frac{1}{2} 
\left( \frac{\pi^2}{6} +\ln^2( s e^{\gamma_E})\right) \lambda + \dots \, .
\end{equation}
When acting on a function with support from $x=0\dots 1$, the expansion of the original function yields a series of distributions
\begin{equation}\label{eq:distexp}
 \frac{1}{{x}^{1-\lambda}} =  \frac{1}{\lambda} \delta(x) + \left[ \frac{1}{x}\right]_+ + \lambda \left[ \frac{\ln x}{x}\right]_+ + \dots\,, 
 \end{equation}
which can be seen by using a test function $\varphi(x)$
\begin{equation}
\int_0^1dx\, \frac{1}{x^{1-\lambda}} \varphi(x) =  \int_0^1dx\, \frac{1}{x^{1-\lambda}} \left[\varphi(x) -\varphi(0) +\varphi(0)\right]  = \frac{\varphi(0)}{\lambda} + \int_0^1dx\, \frac{x^\lambda}{x} \left[\varphi(x) -\varphi(0)\right]\,.
 \end{equation}
 Comparing the individual orders in the expansion in $\lambda$, we can read off the Laplace transforms of $\delta$-functions and plus distributions.

Armed with these results, we now return to the cross section and take its Laplace transform. We follow \cite{Becher:2006nr} and set $s=1/(\kappa e^{\gamma_E})$. The variable $\kappa$ has the same mass dimension as the original variable and this choice eliminates the factors of $e^{\gamma_E}$ on the r.h.s\ of Eq.~(\ref{eq:LaplDist}). For the Laplace transform with respect to $E_X$ we obtain
\bea \label{eq:LapTot}
 \tilde{\sigma}(\kappa) & =& \int_0^\infty d E_X\, e^{- E_X/(\kappa e^{\gamma_E})} \,  \frac{d \sigma}{d E_X}  \,  \nn \\
&=& H (s,\mu) \sum_q e_q^2\, \tilde{f}_q \left( \frac{2 \kappa}{\sqrt{s}},\mu \right) \tilde{f}_{\bar{q}} \left( \frac{2 \kappa}{\sqrt{s}},\mu \right)
\tilde{s}_{\rm DY}(2\kappa,\mu)+\text{($q\leftrightarrow \bar{q}$)}  \, .
\eea
The Laplace transforms of the soft function and PDFs are given by
\be \label{eq:stilDY}
{\widetilde s_{\rm DY}}(\kappa,\mu) \equiv \int_0^\infty d \omega\, e^{-\omega/(\kappa e^{\gamma_E})} W_{\rm DY}(\omega,\mu) 
\ee
and
\be
\tilde{f}_{q/N}(\tau,\mu) =  \int_0^\infty d\bar{x} \exp\left( - \frac{\bar{x}}{\tau e ^{\gamma_E}}\right) f_{q/N}(x,\mu) \, .
\ee
The r.h.s.\ of Eq.~(\ref{eq:LapTot}) is simply the product of the hard, soft, and collinear functions, where the latter coincides in this case with the parton distribution functions. 

We can now obtain the RG-equation of the soft function ${\widetilde s_{\rm DY}}(\kappa,\mu)$ from the known equations for the hard function and the PDFs. For $x\to 1$, the PDFs satisfy a simplified Altarelli-Parisi equation 
\be \label{eq:RGEpdfx}
\frac{d f_{q/N}(z,\mu)}{d \ln \mu} = \int_z^1 \frac{dx}{x} P(x) f_{q/N}(z/x,\mu) \, ,
\ee
where the splitting function $P(x)$ is given by
\be \label{eq:splitting}
P(x) = 2 C_F \gamma_{\cusp}(\alpha_s) \left[ \frac{1}{\bar{x} } \right]_+ +2 \gamma_{f_q} (\alpha_s) \delta(\bar{x})  \, .
\ee
The splitting function $P(x)$ contains the part of the full Altarelli-Parisi kernel $P_{q\leftarrow q}(x)$ which becomes singular in the threshold limit $\bar{x} \equiv 1-x \to 0$. The remainder, as well as the splitting into other partons described by kernels such as $P_{g\leftarrow q}(x)$, is non-singular and can be neglected in the threshold limit. The anomalous dimension at first order in the strong coupling constant is $\gamma_{f_q} = 3 C_F \alpha_s/(4\pi)$, the two-loop result can be found in Appendix~\ref{app:AnDim}. Up to terms which are power suppressed in the limit $z\to 1$, we can rewrite Eq.~(\ref{eq:RGEpdfx}) in the form
\be \label{eq:RGEpdfx2}
\frac{d f_{q/N}(z,\mu)}{d \ln \mu} = \int_0^\infty \!d\bar{x} \int_0^\infty \!d\bar{y} \,P(x)\, f_{q/N}(y,\mu) \,\delta(\bar{z} - \bar{x} - \bar{y}) \, ,
\ee
which is precisely the type of convolution which the Laplace transform turns into a product. Using our result in Eq.~(\ref{eq:LaplDist}), the transformed equation reads
\be \label{eq:RGEpdf}
\frac{d \tilde{f}_{q/N}(\tau,\mu)}{d \ln \mu} = 2 \left[  C_F \gamma_{\cusp}(\alpha_s)  \ln \tau +  \gamma_{f_q} (\alpha_s) \right]  \tilde{f}_{q/N}(\tau,\mu) \, .
\ee

To derive the RG equation satisfied by the soft function, one observes that the differential cross section must be independent of the scale $\mu$, so that one finds
\be \label{eq:derRGEsoft}
\frac{d}{d \ln \mu} \tilde{\sigma}(\kappa)  = \left[\Gamma_H + 2 \Gamma_f+ \Gamma_s\right] \,\tilde{\sigma}(\kappa) = 0 \, ,
\ee
where the $\Gamma$'s indicate schematically the anomalous dimensions of the hard function, the parton distribution 
functions, and the soft function, respectively. The hard function is given by the absolute value squared of $C_V$. Its RG equation was discussed in detail in 
Section~\ref{sec:RGEsud}. For the Drell-Yan process, the function $C_V(Q^2,\mu^2)$  is evaluated at $Q^2=-M^2-i 0^+$ so that
\begin{equation}
\Gamma_H =\Gamma_{C_V}+\Gamma_{C_V}^* = 2 {\rm Re}\!\left[ \Gamma_{C_V} \right]= 2 \left[ C_F \gamma_{\cusp}(\alpha_s) \ln\frac{M^2}{\mu^2}  + \gamma_V (\alpha_s)\right]\,.
\end{equation}
Using the explicit form of the anomalous dimension of the PDF in Eqs.~(\ref{eq:RGEpdf}) 
and solving  Eq.~(\ref{eq:derRGEsoft}) with respect to $\Gamma_s$ one then finds
\bea
\Gamma_s &=& - 4 C_F \gamma_{\cusp}(\alpha_s) \ln\left( \frac{2 \kappa}{\sqrt{s}}\right) - 4 \gamma_{f_q} (\alpha_s) -2 
C_F \gamma_{\cusp} (\alpha_s) \ln\left( \frac{M^2}{\mu^2}\right) -2 \gamma_V \left( \alpha_s \right) \,  \nn \\
&\simeq& - 4 C_F  \gamma_{\cusp}(\alpha_s) \ln\left( \frac{2 \kappa}{\mu}\right) - 2\underbrace{\left(2 \gamma_{f_q} (\alpha_s)
+\gamma_V(\alpha_s) \right)}_{ \equiv  \gamma_W} \, .
\eea 
In the second line, we have used that $M^2 \simeq s$ in the threshold region to show that the dependence on the hard scale cancels out.

The above anomalous dimension is relevant for the cross section Eq.~(\ref{eq:LapTot}) which is proportional to $\tilde{s}_{\rm DY}(2\kappa,\mu)$. The RG equation satisfied by the Laplace transform of the soft function itself is thus 
\be
\frac{d\, {\widetilde s_{\rm DY}}(\kappa,\mu)}{d \ln \mu} = \left[ - 4 C_F \gamma_{\cusp} (\alpha_s) \ln{\left( \frac{\kappa}{\mu}\right)} - 2 \gamma_W(\alpha_s) \right] {\widetilde s_{\rm DY}}(\kappa,\mu) \, .
\ee
The RG equation above can be solved in the same way as the RG equation for the Wilson coefficient of the Sudakov form factor discussed in Section~\ref{sec:RGEsud}. One finds that the solution of the equation is
\be \label{eq:SRGEsol}
 {\widetilde s_{\rm DY}}(\kappa,\mu) = \exp{\left[ - 4 C_F S(\mu_s,\mu) + 2 A_{\gamma_W}(\mu_s,\mu)\right]}\, {\widetilde s_{\rm DY}}(\kappa,\mu_s)  \left(\frac{\kappa^2}{\mu_s^2} \right)^{\eta} \, , 
\ee 
where the functions $S$, $A_{\gamma_W}$, and $A_{\gamma_{\cusp}}$ are defined in Eqs.~(\ref{eq:SAdef}) and $\eta \equiv 2 C_F A_{\gamma_{\cusp}}(\mu_s,\mu)$. 

In order to compute the resummed cross section in momentum space, we would like to perform the inverse Laplace transform. To do so, we observe that the $\kappa$-dependence of the solution is very simple. To any order in perturbation theory, the function ${\widetilde s_{\rm DY}}(\kappa,\mu_s)$ is a polynomial in the logarithm
\be
L= \ln\frac{\kappa^2}{\mu_s^2} \, ,
\ee
which is multiplied by a factor $(\kappa^2/\mu_s^2)^\eta$ from the RG evolution. In fact, powers of logarithms can be obtained as derivatives with respect to $\eta$
\be
L^m \left(\frac{\kappa^2}{\mu_s^2}\right)^{\eta} = 
\partial_\eta^{(m)} \left(\frac{\kappa^2}{\mu_s^2}\right)^{\eta}  \, .
\ee
Because of this relation it is convenient to write the Laplace transformed function as a function of the logarithm $L$ and one can then replace ${\widetilde s_{\rm DY}}\left(L,\mu_s \right) \to {\widetilde s_{\rm DY}}\left(\partial_\eta,\mu_s \right)$ in Eq.~(\ref{eq:SRGEsol}). The computation of the inverse Laplace transform now boils down to obtaining the inverse of $\kappa^{2\eta}$. By dimensional analysis, the inverse must be given by a function of $\eta$ times $\omega^{2\eta-1}$. To determine the prefactor, let us compute the  Laplace transform of $\omega^{2\eta-1}$:
\be \label{eq:LapInt}
\int_0^\infty d \omega \, e^{-\omega/(\kappa e^{\gamma_E})} \omega^{2\eta-1} = \Gamma(2\eta) \,e^{2\eta \gamma_E}\, \kappa^{2\eta}\,.
\ee
From this result and our discussion above, we conclude that if one uses $L$ as the first argument 
in ${\widetilde s_{\rm DY}}$ the inverse transform can be written as \cite{Becher:2007ty}
\be \label{eq:WDYres}
W_{\rm DY}(\omega,\mu) =\exp{\left[ - 4 C_F S(\mu_s,\mu) + 2 A_{\gamma_W}(\mu_s,\mu)\right]}
{\widetilde s_{\rm DY}}\left(\partial_\eta,\mu_s \right)\frac{e^{-2 \gamma_E \eta}}{\Gamma(2 \eta)} \frac{1}{\omega}\left( \frac{\omega}{\mu_s}\right)^{2 \eta} \, .
\ee
This expression for $W_{\rm DY}(\omega,\mu) $ is well defined for $\eta > 0$, which is fulfilled for $\mu_s >\mu$. For the effective field theory, this ordering is natural: one would first compute the soft contributions at the relevant perturbative scale $\mu_s$ and then evolve down to a low scale $\mu$ where the PDFs are evaluated. However, in fixed-order computations the scale $\mu$ in the PDFs is typically chosen of order the hard scale and since the PDF fits were performed with this scale choice, it is preferable to adopt the same choice in the effective theory. Therefore, we need to be able to evaluate integrals with respect to $\omega$ for negative values of  $\eta$; this is done by analytic continuation. For instance, to obtain the result for $-1/2 < \eta <0$, it is necessary to employ the identity
\be
\int_0^{\Omega} d \omega \frac{f(\omega)}{\omega^{1-2 \eta}} = \int_0^\Omega d \omega 
\frac{f(\omega) - f(0)}{\omega^{1-2 \eta}} + \frac{f(0)}{2 \eta} \Omega^{2 \eta} \, , \label{eq:ancont}
\ee
where $f(\omega)$ is a smooth test function. If necessary,  it is possible to analytically continue the integral on the l.h.s.  of this Eq.~(\ref{eq:ancont}) to the region $\eta < -n/2$ for an arbitrary positive integer $n$. This can be done by
subtracting an increasing number of terms from the Taylor expansion of $f(\omega)$ at $\omega =0$ in the r.h.s. of Eq.~(\ref{eq:ancont}).

\subsection{Drell-Yan Soft Matrix Element\label{sec:soft}}

\begin{figure}[t]
\begin{center}
\vspace*{.5cm}
\[ 
\hspace*{1.5cm}
\vcenter{ \hbox{
  \begin{picture}(0,0)(0,0)
\SetOffset(-121,-45)
\SetScale{0.25}

    \SetWidth{1.5}
    \SetColor{Black}
    \Line[arrow,arrowpos=0.5,arrowlength=28.557,arrowwidth=11.423,arrowinset=0.2,flip,double,sep=4.5](192,271)(288,109)
    \Line[arrow,arrowpos=0.5,arrowlength=28.557,arrowwidth=11.423,arrowinset=0.2,flip,double,sep=4.5](288,111)(207,-48)
    \Line[arrow,arrowpos=0.5,arrowlength=28.557,arrowwidth=11.423,arrowinset=0.2,double,sep=4.5](480,111)(576,-51)
    \Line[arrow,arrowpos=0.5,arrowlength=28.557,arrowwidth=11.423,arrowinset=0.2,double,sep=4.5](560,271)(479,112)
     \Bezier(224,-17)(222.991,-13.388)(226.485,-8.406)(230.987,-7.035) \Bezier(230.987,-7.035)(239.993,-4.293)(244.987,-18.437)(238.315,-20.545) \Bezier(238.315,-20.545)(231.642,-22.652)(227.606,-8.205)(239.413,-4.049) \Bezier(239.413,-4.049)(251.221,0.107)(257.098,-13.693)(250.59,-16.225) \Bezier(250.59,-16.225)(244.081,-18.757)(239.087,-4.613)(250.64,0.292) \Bezier(250.64,0.292)(262.192,5.197)(268.87,-8.235)(262.543,-11.152) \Bezier(262.543,-11.152)(256.215,-14.069)(250.338,-0.268)(261.634,5.332) \Bezier(261.634,5.332)(272.93,10.931)(280.326,-2.119)(274.188,-5.38) \Bezier(274.188,-5.38)(268.05,-8.642)(261.373,4.789)(272.414,11.029) \Bezier(272.414,11.029)(283.455,17.269)(291.487,4.601)(285.542,1.034) \Bezier(285.542,1.034)(279.596,-2.532)(272.2,10.518)(282.993,17.343) \Bezier(282.993,17.343)(293.786,24.168)(302.376,11.872)(296.62,8.039) \Bezier(296.62,8.039)(290.863,4.205)(282.831,16.874)(293.385,24.229) \Bezier(293.385,24.229)(303.938,31.584)(313.015,19.642)(307.44,15.578) \Bezier(307.44,15.578)(301.865,11.515)(293.274,23.811)(303.602,31.643) \Bezier(303.602,31.643)(313.929,39.474)(323.426,27.864)(318.021,23.602) \Bezier(318.021,23.602)(312.617,19.34)(303.54,31.281)(313.656,39.538) \Bezier(313.656,39.538)(323.772,47.794)(333.628,36.487)(328.382,32.057) \Bezier(328.382,32.057)(323.136,27.626)(313.639,39.237)(323.561,47.868) \Bezier(323.561,47.868)(333.483,56.499)(343.643,45.464)(338.54,40.892) \Bezier(338.54,40.892)(333.437,36.32)(323.581,47.627)(333.328,56.584) \Bezier(333.328,56.584)(343.075,65.541)(353.49,54.746)(348.514,50.057) \Bezier(348.514,50.057)(343.539,45.368)(333.379,56.403)(342.971,65.639) \Bezier(342.971,65.639)(352.563,74.876)(363.187,64.287)(358.323,59.502) \Bezier(358.323,59.502)(353.459,54.717)(343.044,65.512)(352.503,74.984) \Bezier(352.503,74.984)(361.961,84.456)(372.752,74.038)(367.984,69.177) \Bezier(367.984,69.177)(363.215,64.315)(352.591,74.905)(361.938,84.569) \Bezier(361.938,84.569)(371.284,94.233)(382.205,83.951)(377.514,79.031) \Bezier(377.514,79.031)(372.824,74.112)(362.033,84.531)(371.29,94.345) \Bezier(371.29,94.345)(380.546,104.16)(391.56,93.977)(386.932,89.017) \Bezier(386.932,89.017)(382.304,84.056)(371.383,94.339)(380.573,104.263) \Bezier(380.573,104.263)(389.763,114.187)(400.837,104.069)(396.254,99.083) \Bezier(396.254,99.083)(391.672,94.097)(380.657,104.28)(389.804,114.274) \Bezier(389.804,114.274)(398.951,124.267)(410.05,114.177)(405.496,109.181) \Bezier(405.496,109.181)(400.943,104.185)(389.87,114.304)(398.998,124.327) \Bezier(398.998,124.327)(408.125,134.35)(419.216,124.251)(414.676,119.261) \Bezier(414.676,119.261)(410.135,114.27)(399.036,124.361)(408.169,134.374) \Bezier(408.169,134.374)(417.302,144.388)(428.352,134.243)(423.807,129.273) \Bezier(423.807,129.273)(419.263,124.303)(408.172,134.402)(417.336,144.366) \Bezier(417.336,144.366)(426.499,154.33)(437.472,144.102)(432.907,139.168) \Bezier(432.907,139.168)(428.343,134.234)(417.293,144.378)(426.513,154.252) \Bezier(426.513,154.252)(435.733,164.126)(446.592,153.778)(441.991,148.896) \Bezier(441.991,148.896)(437.39,144.014)(426.417,154.241)(435.719,163.984) \Bezier(435.719,163.984)(445.021,173.727)(455.725,163.219)(451.072,158.407) \Bezier(451.072,158.407)(446.418,153.595)(435.559,163.944)(444.97,173.512) \Bezier(444.97,173.512)(454.382,183.081)(464.888,172.375)(460.164,167.653) \Bezier(460.164,167.653)(455.441,162.93)(444.736,173.437)(454.285,182.788) \Bezier(454.285,182.788)(463.834,192.138)(474.092,181.194)(469.283,176.582) \Bezier(469.283,176.582)(464.474,171.969)(453.968,182.675)(463.682,191.76) \Bezier(463.682,191.76)(473.396,200.845)(483.351,189.625)(478.44,185.146) \Bezier(478.44,185.146)(473.529,180.667)(463.271,191.61)(473.179,200.379) \Bezier(473.179,200.379)(483.087,209.148)(492.677,197.614)(487.649,193.295) \Bezier(487.649,193.295)(482.62,188.976)(472.665,200.196)(482.796,208.595) \Bezier(482.796,208.595)(492.926,216.995)(502.08,205.111)(496.92,200.981) \Bezier(496.92,200.981)(491.76,196.852)(482.17,208.386)(492.551,216.357) \Bezier(492.551,216.357)(502.932,224.329)(511.569,212.065)(506.265,208.157) \Bezier(506.265,208.157)(500.961,204.25)(491.808,216.133)(502.465,223.613) \Bezier(502.465,223.613)(513.122,231.094)(521.154,218.425)(515.695,214.776) \Bezier(515.695,214.776)(510.237,211.127)(501.6,223.391)(512.574,230.307) \Bezier(512.574,230.307)(523.548,237.223)(530.829,224.109)(525.215,220.775) \Bezier(525.215,220.775)(519.601,217.441)(511.57,230.11)(523.518,236.698) \Bezier(523.518,236.698)(535.465,243.286)(541.824,229.701)(536.047,226.75) \Bezier(536.047,226.75)(530.27,223.799)(522.989,236.913)(531.905,241.353) \Bezier(531.905,241.353)(536.363,243.573)(542.41,242.396)(544,239)
    \SetWidth{2.7}
    \SetColor{Red}
    \Line(384,-65)(384,271)
    \Line(448,303)(384,271)
    \Line(320,-97)(384,-65)
    \SetColor{Black}
    \SetWidth{1.0}
    \GOval(288,111)(5,5)(0){0.882}
    \GOval(288,111)(6,6)(0){0.882}
    \GOval(480,111)(6,6)(0){0.882}
    \SetWidth{1.8}
        \Line[arrow,arrowpos=1,arrowlength=15.2,arrowwidth=6.08,arrowinset=0.2](352,111)(416,175)

      \Text(105,10)[cb]{$k$}
      
      \Text(45,-25)[cb]{$n$}
      \Text(148,-25)[cb]{$n$}
      \Text(45,70)[cb]{$\bar{n}$}
      \Text(148,70)[cb]{$\bar{n}$}

\end{picture}}
}
\]
\vspace*{1cm}
\end{center} 
\caption{One loop Feynman diagram contributing to the soft function. The vertical line (the ``cut'') indicates that the gluon crossing it is on-shell.
} 
\label{fig:SoftWilson2}
\end{figure}
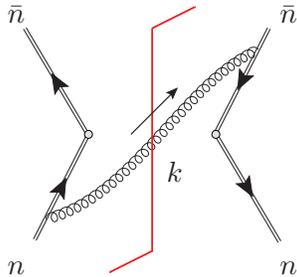


We now calculate the soft function $W_{\rm DY}$ at order $\alpha_s$. Together with the form factor $C_V$, we will then have all the one-loop ingredients of the factorization theorem. The  calculation outlined below is carried out in momentum space, however,  the soft function can also be calculated in position space, see Appendix \ref{ap:Sfps}. The  function $W_{\rm DY}$ at two loop order  can be found in  \cite{Belitsky:1998tc}. 
In this section, we compute the bare soft function. The  poles in $\ep$ will then be removed by renormalization in the modified minimal subtraction scheme.

To perform the calculation, we need Feynman rules for gluons emitted from the Wilson lines. 
The simplest form of these Feynman rules is obtained by treating a Wilson line in the $n$ direction as a particle flying along this direction. This particle has a Feynman rule $i g_s n^\mu t^a$ for the coupling to gluons and, after absorbing gluons of momentum $k$, an eikonal propagator $i/(n\cdot k)$. This kind of treatment is possible because the soft emissions encoded in the Wilson line indeed arise from a particle propagating along the $n$ direction. A formal derivation of these Feynman rules, obtained by expanding the Wilson line exponential in powers of the coupling constant and by subsequently performing a Fourier transform, is given in Appendix~\ref{ap:Wilson}.

In order to perform the momentum-space computation, it is best to first rewrite the Fourier transformed soft function Eq.~(\ref{eq:sffourier}) as an amplitude squared. To do so, we insert a complete set of states between the Wilson line products in Eq.~(\ref{eq:softwilson}) and translate the fields to $x=0$ using the momentum operator. The Fourier integral then takes the form
\begin{align}
W_{\rm DY}(\omega,\mu) &=  \int \!\frac{dx^0}{4\pi}\, e^{i\,x^0 \omega/2}    \nonumber \\
& \hspace{1.0cm}\times \frac{1}{N_c} \! \sum\hspace{-0.5cm}\int\limits_X {\rm tr} \langle 0 | e^{i P^0 x^0}  \bar{T}\!\left( S_n^\dagger\left(0 \right)S_{\bar{n}}\left(0 \right) \right) e^{-i P^0 x^0}  | X\rangle \langle X |
 T\!\left( S^\dagger_{\bar{n}}\left(0\right)S_{n}\left(0 \right)\right) | 0 \rangle \, \nonumber \\
 &=   \frac{1}{N_c}\sum\hspace{-0.5cm}\int\limits_X  \, {\rm tr} \langle 0 | \bar{T}\!\left(S_n^\dagger\left(0 \right)S_{\bar{n}}\left(0 \right)\right) | X\rangle \langle X | T\!\left(S^\dagger_{\bar{n}}\left(0\right)S_{n}\left(0 \right)\right) | 0 \rangle \delta(\omega - 2 E_X) .\label{eq:softfourier}
\end{align}
To get the second equation, we have acted with the energy operator on the states and have then performed the $x^0$-integral. Up to the fact that the color indices are contracted, the final result indeed has the form of an amplitude squared, describing emissions from two Wilson lines with energy $E_X$. The one-emission result for the soft function Eq.~(\ref{eq:softfourier}) is obtained from the diagram shown in Fig.~\ref{fig:SoftWilson2}, where a soft gluon is exchanged between the two quark Wilson lines, plus the corresponding contributions in which the gluon connects the other two lines, which gives the same result. We work in Feynman gauge, where the gluon propagator is proportional to $g_{\mu\nu}$. Because of this, diagrams where a gluon connects two Wilson lines in the same direction vanish, because they are proportional to $n^\mu\, n^\nu  g_{\mu\nu} = n^2= 0$. In addition to the real emission diagrams, one should also include loop corrections, but these lead to scaleless integrals and all vanish in dimensional regularization. 

The entire computation thus boils down to the evaluation of the following integral
\bea\label{eq:wilsonDY}
W_{\rm DY}^{\rm bare}\left( \omega \right) &=&  \delta(\omega)+ \frac{2}{N_c} \,{\rm tr} \int \frac{d^d k}{(2 \pi)^d} \underbrace{\left(  g_s \frac{n^\mu}{ n\cdot k} t^a\right)}_{\mbox{Wilson Line}} \times
\underbrace{\Bigl( - g_{\mu \nu}  2 \pi \delta(k^2) \theta(k_0) \Bigr)}_{\mbox{Cut Gluon Propagator}}
\nn \\
& & \nn \\
& & \times \left( - g_s \frac{\bar{n}^\nu}{ \bar{n}\cdot k} t^a\right)\, \delta\left(\omega - 2 k_0\right)\, , \label{eq:softth}
\eea
where $k$ is the momentum of the gluon crossing the cut in the diagram. The color trace is ${\rm tr}(t^a t^a) = C_F N_c$ and the integration measure can be rewritten as follows
\be
\int d^d k \, \theta\left( k_0\right) \delta(k^2) = \frac{1}{2} \int_0^\infty d k_+  \, \int_0^\infty d k_- \, \int d^{d-2} k_\perp\,\delta(k^2) \, , \label{eq:tagapp}
\ee
where here $k_+ = k_0 + k_z = n \cdot k$ and $k_- = k_0 - k_z = \bar{n} \cdot k$.
Consequently, the integral in Eq.~(\ref{eq:softth}) becomes
\bea\label{eq:intW}
W_{\rm DY}^{\rm bare} \left(\omega\right) &=& \delta(\omega) + \frac{2 g_s^2 C_F}{(2 \pi)^{d-1}} \int_0^\infty d k_+  \, \int_0^\infty d k_- \, \int d^{d-2} k_\perp \frac{1}{k_+  k_-}\nn\\
&& \times  \delta \left( k_+  k_- + k_\perp^2\right)
\delta \left( \omega - k_+ - k_-\right) \, .
\eea
The angular integrals are trivial and we rewrite
\bea
\int \! d^{d-2} k_\perp = \Omega_{d-2} \int_0^\infty \! dk_T \,k_T^{d-3}
\label{eq:ints1} \,, 
\eea
where $k_T$ is the magnitude of the transverse spatial momentum, $k_T^2 \equiv - k_\perp^2$ and the $d$-dimensional solid angle is $\Omega_d = 2 \pi^{d/2}/\Gamma(d/2)$. Next, we integrate over $k_T$ and $k_-$, which eliminates the two delta functions. After this, the integral in Eq.~(\ref{eq:intW}) can be rewritten as
\be
W_{\rm DY}^{\rm bare}\left( \omega \right) = 
\delta(\omega) +  \frac{\alpha_{s}^{0} (4 \pi )^{\ep} C_{F}}{\pi \Gamma(1-\ep)} \int^{\omega}_{0}d k_{+}\frac{1}{\left(k_{+} (\omega - k_{+})\right)^{1+\ep}}\, ,
\ee
where we have explicitly indicated that the coupling constant in the diagram is the bare one, $\alpha_{s}^{0} = g_s^2/(4\pi)$. The upper limit of integration is determined by the fact that the last delta function in Eq.~(\ref{eq:intW}) fixes $k_- = \omega -k_+$, where both $k_+$ and $\omega$ are positive. 
The remaining integral can easily be carried out, which yields the result
\be\label{eq:sfomega}
W_{\rm DY}^{\rm bare}\left( \omega\right) = \delta(\omega) + \frac{Z_\alpha \alpha_{s}}{ \pi} C_{F} \frac{1}{\omega} \left(\frac{\mu}{\omega} \right)^{2\ep}
\frac{e^{\ep\gamma_{E}} \Gamma(1-\ep)}{\ep^2 \Gamma\left(-2 \ep\right)}\,,
\ee
%
in which we have expressed the bare coupling through the renormalized coupling constant $\alpha_s\equiv \alpha_s(\mu)$  in the $\ms$ scheme via the relation $Z_\alpha\, \alpha_s \,\mu^{2\ep} = e^{-\ep \gamma_E}(4\pi)^\ep \alpha_s^0$. The renormalization factor  $Z_\alpha  = 1+\mathcal{O}(\alpha_s)$ can be set to one at the accuracy we are working. We can easily calculate the Laplace transformed soft function ${\widetilde s}_{\rm DY}$ defined in Eq.~(\ref{eq:stilDY}), using the result in Eq.~(\ref{eq:LapInt}). Expanding around $\ep \rightarrow 0$ we obtain
the bare function ${\widetilde s}_{\rm DY}$ at order $\alpha_s$, which reads
\be 
 {\widetilde s}^{\rm bare}_{\rm DY}(\kappa) = 1  +\frac{\alpha_s}{4 \pi} C_F  \left[ \frac{4}{\ep^2} -\frac{4 L}{\ep} +2 L^2 + \frac{\pi^2}{3}\right]  \, , \label{eq:bareSdy}
\ee
where $L=\ln \kappa^2/\mu^2$. One can also take the inverse Fourier transform with respect to $\omega$ of Eq.~(\ref{eq:sfomega}) to get the expression in position space for $\hat{W}^{\rm bare}_{\rm DY} (x_0)$:
\be
\hat{W}^{\rm bare}_{\rm DY} (x_0) =
\int_0^\infty d \omega e^{- i\frac{\omega}{2} x_0 }  W_{\rm DY}^{\rm bare}\left( \omega\right) 
= 1  + \frac{\alpha_{s}( -\mu^{2} x_0^2/4)^{\ep} C_{F}}{\pi } \frac{e^{\ep \gamma_E}\Gamma(1-\ep)}{\ep^2}\,,
\ee
which agrees with the result Eq.~(\ref{eq:sfpos}) in Appendix \ref{ap:Sfps} obtained by calculating the function directly in position space. Expanding in $\ep$ one finds
\be
\hat{W}^{\rm bare}_{\rm DY} (x_0)=  
1  +\frac{\alpha_s}{4 \pi} C_F \left[ \frac{4}{\ep^2} +\frac{4 L_{0}}{\ep} +2 L_{0}^2 + \frac{\pi^2}{3}\right] + {\mathcal O}(\alpha_s^2) \, , \label{eq:WDYhat}
\ee
where
\be
L_{0} = \ln\left(-\frac{1}{4}\mu^{2}x^{2}_{0} e^{2 \gamma_{E}}\right)\, .
\ee
It is easy to show, by applying the Laplace transform in Eq.~(\ref{eq:stilDY}) to Eq.~(\ref{eq:sffourier}), that the soft function in position space $\hat{W}_{\rm DY} (x_0,\mu)$ has the same functional form as the soft function in Laplace space ${ \widetilde s}_{\rm DY}(L,\mu)$. Indeed, one obtains one from the other by replacing the argument in the following way:
\be
 {\widetilde s}^{\rm bare}_{\rm DY}(\kappa)  = \hat{W}^{\rm bare}_{\rm DY} \left(x_0 = \frac{-2 i}{e^{\gamma_{E}} \kappa} \right)\, ,
\label{eq:widetsDY}
\ee
which is equivalent to replace $L_0$ with $-L$ in Eq.~(\ref{eq:WDYhat}).

In the following we will need the renormalized soft function, which is finite in the limit $\ep \to 0$. In momentum space the renormalization will involve a convolution with a $Z$-factor, since the soft function is distribution valued, but in Laplace (and position space), renormalization is multiplicative, as discussed in 
Section~\ref{subsec:RGEsoft}. The renormalized functions can be obtained by multiplying the bare functions by a renormalization factor $Z_s^{-1}(\alpha_s,L,\ep)$. At the one-loop level, the renormalized function is obtained by simply dropping the divergent parts of the bare function, and reads
\be 
 {\widetilde s}_{\rm DY}(\kappa,\mu) = 1  +\frac{\alpha_s}{4 \pi} C_F  \left[  2 L^2 + \frac{\pi^2}{3}\right]  \, . \label{eq:renSdy}
\ee
In contrast to the bare function, this function depends on $\mu$ and the reader can easily verify that it fulfills the RG equation derived in \ref{subsec:RGEsoft}.

\subsection{Resummation of Large Logarithms}

The partonic Drell-Yan cross section factors into the product of the squared Wilson coefficient and the soft function, as shown in Eq.~(\ref{eq:dsdM2}). The product of these two terms describes the hard partonic scattering; the physical (hadronic) cross section is obtained by integrating the product of the hard-scattering kernel and the parton distribution functions over the appropriate domain.
In Section~\ref{sec:LIV} we solved the RG equation satisfied by the Wilson coefficient $\tilde{C}_V$ (cf. 
Eqs.~(\ref{eq:Cexact},\ref{eq:evma})), while the solution of the RG equation satisfied by the soft function was presented
in Section~\ref{subsec:RGEsoft} above. By combining these two elements we obtain a resummed formula for the hard scattering kernel.

Equation~(\ref{eq:evma}) is valid for space-like momenta; 
the solution of the RG equation for the function $\tilde{C}_V$  needed in Drell-Yan scattering can be obtained from the one valid
for space-like momenta by analytic continuation. The sign of the imaginary part extracted from the logarithm in the RG equation can be determined by writing explicitly the infinitesimal imaginary part of $M^2$. The RG equation becomes \cite{Becher:2007ty}
\be
\frac{d}{d \ln \mu} \tilde{C}_V(-M^2 -i 0^+,\mu) = \left[ C_F \gamma_{\cusp}(\alpha_s) 
\left(\ln\frac{M^2}{\mu^2}  - i \pi \right)  + \gamma_V (\alpha_s)\right] \tilde{C}_V (-M^2-i 0^+,\mu) \, ,
\ee
and its solution is 
\bea \label{eq:tlRGEWilson}
\tilde{C}_V(-M^2 -i 0^+,\mu_f) &=& \exp\left[2 C_F S(\mu_h,\mu_f) -A_{\gamma_V}(\mu_h,\mu_f) + i \pi C_F A_{\gamma_{\cusp}}  (\mu_h,\mu_f) \right] \nn \\ 
& &\times \left( \frac{M^2}{\mu_h^2}\right)^{-C_F A_{\gamma_{\cusp}}(\mu_h,\mu_f)}\tilde{C}_V(-M^2,\mu_h) \, .
\eea
The functions $S$ and $A_{\gamma_i}$ are defined in Eq.~(\ref{eq:SAdef}).

Following the notation employed in \cite{Becher:2007ty}, one can define the hard-scattering kernel as 
\be \label{eq:HS6}
C(z,M,\mu_f) \equiv \left| \tilde{C}_V(-M^2,\mu_f)\right|^2 \sqrt{\hat{s}} \,W_{\rm DY}\left(\sqrt{\hat{s}} (1-z),\mu_f \right) \, . 
\ee
To get the resummed result, we simply insert the solutions of RG equations of the soft function, Eq.~(\ref{eq:WDYres}), and the hard function, Eq.~(\ref{eq:tlRGEWilson}), into Eq.~(\ref{eq:HS6}). The result can be simplified by making use of the relations
\bea
A_{\gamma_i}\left(\mu_h,\mu_f\right)  &=&  A_{\gamma_i}\left(\mu_h,\mu_s\right) + A_{\gamma_i}\left(\mu_s,\mu_f\right) \, , \nn \\
S\left(\mu_h,\mu_f\right) - S\left(\mu_s,\mu_f\right) &=& S\left(\mu_h,\mu_s\right) -  A_{\gamma_{\cusp}}\left(\mu_s,\mu_f \right) \ln{\frac{\mu_h}{\mu_s}} \, ,
\eea
as well as $\gamma_W=2\gamma_{f_q}+\gamma_V$. In this way one finds 
\be \label{eq:Caux1}
C(z,M,\mu_f)  = \left| \tilde{C}_V(-M^2,\mu_h)\right|^2 U(M,\mu_h,\mu_f,\mu_s) \frac{\sqrt{\hat{s}}}{\omega}\left(\frac{M}{\mu_s} \right)^{- 2 \eta}
{\widetilde s}_{\rm DY}\left(\partial_\eta,\mu_s \right)  
\left(\frac{\mu_s}{\omega} \right)^{- 2 \eta} \frac{e^{-2 \gamma_E \eta}}{\Gamma(2 \eta)} \, ,
\ee
where the evolution function $U$ is defined as 
\begin{align}
U(M,\mu_h,\mu_f,\mu_s) &= \exp\left[4 C_F S\left(\mu_h, \mu_s\right) +4 A_{\gamma_{f_q}}(\mu_s,\mu_f) - 2 A_{\gamma_V}(\mu_h,\mu_s)\right] 
\nonumber \\ & \times
\left(\frac{M^2}{\mu_h^2} \right)^{- 2 C_FA_{\gamma_{\rm cusp}}(\mu_h,\mu_s)} \, . \label{eq:Urun}
\end{align}
The factor $(\mu_s/\omega)^{-2 \eta}$ in Eq.~(\ref{eq:Caux1}) can be moved to the left of the soft function ${\widetilde s}_{\rm DY}$ to obtain
\be \label{eq:Caux2}
C(z,M,\mu_f)  = \left| \tilde{C}_V(-M^2,\mu_h)\right|^2 U(M,\mu_h,\mu_f,\mu_s) \frac{\sqrt{\hat{s}}}{\omega}\left(\frac{M}{\omega} \right)^{- 2 \eta} 
{\widetilde s}_{\rm DY}\left( \! \ln\frac{\omega^2}{\mu^2_s} + \partial_\eta,\mu_s \!\!\right)  
 \frac{e^{-2 \gamma_E \eta}}{\Gamma(2 \eta)} \, .
\ee
The explicit $z$ dependence of the hard-scattering kernel can be obtained by inserting the relation $\omega = M (1-z)/\sqrt{z}$. Finally, one obtains
\bea \label{eq:Caux3}
C(z,M,\mu_f)  &=& \left| \tilde{C}_V(-M^2,\mu_h)\right|^2 U(M,\mu_h,\mu_f,\mu_s) \, \frac{z^{-\eta}}{(1-z)^{1-2 \eta}}
\nn \\
&& \times 
{\widetilde s}_{\rm DY}\left(  \ln\frac{M^2 (1-z)^2}{\mu^2_s z} + \partial_\eta,\mu_s \right)  
 \frac{e^{-2 \gamma_E \eta}}{\Gamma(2 \eta)} \, .
\eea

As it was observed after Eq.~(\ref{eq:WDYres}), the formula above is well defined for $\eta > 0$, which corresponds to the case $\mu_s > \mu_f$. In the physically more relevant case in which $\mu_s < \mu_f$, one finds that $\eta <0$;
consequently, the integrals of 
$\ln^n(1-z)/(1-z)^{1-2 \eta}$ with test functions $f(z)$ must be defined using a subtraction at $z=1$ and analytic continuation in $\eta$. This procedure gives rise to plus distributions in the variable $1-z$. 

The resummed formula for the hard-scattering kernel, Eq.~(\ref{eq:Caux3}), is formally independent from the hard scale $\mu_h$ and the soft scale $\mu_s$. As long as $\mu_h \sim M$ and $\mu_s \sim \omega$, the Wilson coefficient $\tilde{C}_V$ and the soft function ${\widetilde s}_{\rm DY}$ in Eq.~(\ref{eq:Caux3}) are free of large logarithms and can be evaluated in perturbation theory. (We remind the reader that $\mu_s \gg \Lambda_{\rm QCD}$.) 
A residual dependence on $\mu_s$ and $\mu_h$ in the hard-scattering kernel arises precisely from the fact that  the matching coefficients and the anomalous dimensions are evaluated only up to a given order in perturbation theory. The residual higher-order scale dependence can be employed to asses the perturbative uncertainty, as we will discuss in detail in Section~\ref{sec:numDY}. The dependence on the factorization scale $\mu_f$ cancels formally in the convolution of the hard-scattering kernel with the parton distribution functions.

The fixed-order expression for the hard scattering kernel in perturbative QCD includes terms which are singular in the 
$z \to 1$ limit (plus distributions and Dirac delta functions). These singular terms can be obtained by setting $\mu_s = \mu_f = \mu_h$ in Eq.~(\ref{eq:Caux3}) and by expanding the formula in powers of $\alpha_s$. In particular this implies
that after taking the derivatives with respect to $\eta$, one should take the limit  $\eta = 0$. We further discuss the derivation of these approximate formulas in fixed order perturbation theory in Section~\ref{sec:approxNnLO}.

\begin{table}
\centerline{\parbox{14cm}{\caption{\label{tab:counting}
Different approximation schemes for the evaluation of the resummed 
cross-section formulae.}}}
\vspace{0.1cm}
\begin{center}
\begin{tabular}{ccrccc}
\hline
RG-impr.\ PT & Log.\ approx.\ & Accuracy $\sim\alpha_s^n L^k$
 & $\gamma_{\rm cusp}$
 & $\gamma_V$, $\gamma_\phi$ & $C_V$, $\widetilde s_{\rm DY}$ \\
\hline\\[-0.4cm]
--- & LL & $k= 2n$ & 1-loop
 & tree-level & tree-level \\
LO & NLL & $2n-1\le k\le 2n$& 2-loop
 & 1-loop & tree-level \\
NLO & NNLL & $2n-3\le k\le 2n$ & 3-loop & 2-loop 
 & 1-loop \\
NNLO & NNNLL & $2n-5\le k\le 2n$ & 4-loop & 3-loop
 & 2-loop \\[0.2cm]
\hline
\end{tabular}
\end{center}
\end{table}

The resummed expression for the hard-scattering kernel can be evaluated at any desired order in resummed perturbation theory. Different levels of accuracy require the evaluation of the matching coefficients and anomalous dimensions at different orders in perturbation theory; Table~\ref{tab:counting} summarizes the situation.
There are two different ways to label the level of accuracy at which a resummed formula is evaluated. In the counting scheme of RG-improved perturbation theory, the LO approximation includes all terms of ${\mathcal O}(1)$, the NLO approximation includes all of the terms of ${\mathcal O}(\alpha_s)$, and so on. In this framework, the large logarithms are eliminated in favor of coupling constants at the different scales in the problem by using the relation
\begin{equation}
\ln\left(\frac{\mu_h}{\mu_s}\right) = \int_{\alpha_s(\mu_s)}^{\alpha_s(\mu_h)} \frac{d\alpha}{\beta(\alpha)}
\end{equation}
and expanding in $\alpha_s(\mu_h)$ and $\alpha_s(\mu_s)$. This relation and the fact that $\beta(\alpha)\sim \alpha^2$ also makes it obvious that one has to count $\ln(\mu_h/\mu_s)$ as  $\sim 1/\alpha_s$. Traditionally, one instead expands in a single coupling constant, typically $\alpha_s(\mu_h)$, while treating $\alpha_s(\mu_h) \ln(\mu_h/\mu_s)$ as an ${\mathcal O}(1)$ quantity. One then counts how many towers of logarithms are resummed. N$^m$LO accuracy corresponds to N$^{m+1}$LL accuracy in the logarithmic counting. The logarithms are counted after their exponentiation. Because of this, a result at N$^{m}$LL accuracy actually predicts the first $2m$ logarithms in the cross section, see Table \ref{tab:counting}. It is necessary to organize the counting in the exponent because the LL and NLL terms count as $\mathcal{O}(1/\alpha_s)$ and $\mathcal{O}(1)$ in a region where $\ln(\mu_h/\mu_s)\sim 1/\alpha_s$ and cannot be expanded out. Since it is not immediately clear whether N$^n$LO accuracy refers to standard or RG-improved perturbation theory, it is by now common to also denote the SCET results by their logarithmic accuracy. The reader should however be aware that only if the  computation in RG improved perturbation theory is properly organized, all of the logarithms which can be predicted at this accuracy will be fully included in the result, see \cite{Almeida:2014uva} for a detailed discussion of this point. 

An analysis of the Drell-Yan resummed cross section at NNNLL (matched to NNLO fixed-order calculations) is carried out in \cite{Becher:2007ty} and we will now discuss some important aspects of this analysis which are also relevant for other processes analyzed by using SCET.
Before doing so, let us note in passing that it is not uncommon to include $C_V$, $\widetilde s_{\rm DY}$ one order higher than what is indicated in the Table \ref{tab:counting}. In the SCET literature, this is referred to as N$^{m}$LL$'$ accuracy and predicts all logarithms up to the power $2n-2m$ in the cross section at the $n$-order in perturbation theory; one logarithm more than in the standard counting. In traditional resummation literature N$^{m}$LL accuracy can indicate primed or unprimed counting, and one therefore needs to check on a case-by-case basis what is meant.

\subsection{Dynamical Threshold Enhancement\label{sec:DynamicalThresh}}

\begin{figure}
\begin{center}
\psfrag{la}[]{}
\psfrag{x}[]{$y$}
\psfrag{y}[b]{$\ff(y,\muf)$}
\includegraphics[width=0.5\textwidth]{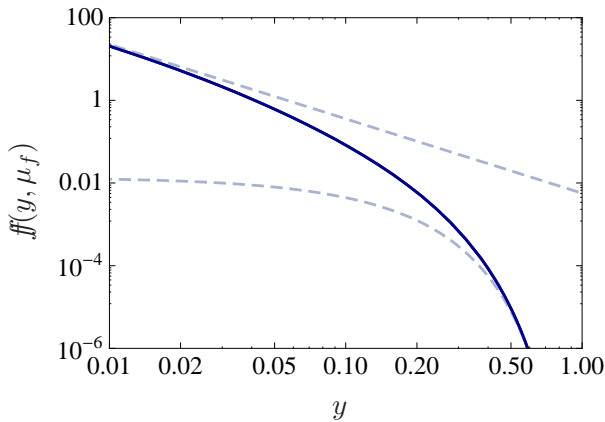} 
\end{center}
\vspace{-0.5cm}
\caption{
\label{fig:ffdrop}
Fall-off of the parton luminosity function $\ff(y,\muf)$ for $\muf=8$\,GeV. The dashed lines show the asymptotic behavior for small and large $y$. Figure taken from \cite{Becher:2007ty}.}
\end{figure}

We now discuss two closely related questions. The first one is whether the threshold approximation on which our computation is based yields a good approximation to the full cross section. The second one is how large the logarithms are which occur in the hadronic cross section. The answers to these questions are related because the relevant quantity is in both cases the typical energy of the soft radiation. The factorization formula Eq.~(\ref{eq:dsdM2}) is based on a leading-power expansion of the soft energy scale over the Drell-Yan mass $M$ and the corrections are thus suppressed by the ratio of these two scales. The same scale ratio also governs the size of the logarithms. Our RG-improved result Eq.~(\ref{eq:Caux3}) resums logarithms of $\mu_h/\mu_s$. To resum the logarithms in the cross section, these scales must be chosen appropriately in order to avoid large logarithms in the perturbative expansion of the hard and soft functions. For the hard function, it is obvious that this is achieved by setting $\mu_h\sim M$. We will now discuss in some detail what the appropriate choice for $\mu_s$ is. 

A naive way to avoid large logarithms in the soft function appearing on the r.h.s.\ of Eq.~(\ref{eq:Caux3}) would be to choose the soft scale so that $\mu_s\sim M (1-z)$ to avoid logarithms inside the $z$-integral. However, as $z\to1$ the scale becomes arbitrarily small which would lead to Landau singularities in the  integrand and would spoil the scale separation upon which the SCET approach is based. However, it is not necessary to avoid logarithms on the level of the integrand in order not to have logarithms in the result. To avoid logarithms in the cross section, one should choose $\mu_s$ as the average energy of the soft radiation. To see what the typical soft energy is, and whether the partonic threshold indeed yields the dominant contribution to the cross section, we need to analyze the convolution of the hard scattering kernel with the PDFs. To do so, we use Eqs.~(\ref{eq:dsdM2}) and Eq.~(\ref{eq:HS6}) and rewrite the threshold contribution to the  pair invariant mass distribution as
\begin{equation}\label{eq:dsigdM2}
   \frac{d\sigma^{\rm thresh}}{dM^2}
   = \frac{4\pi\alpha^2}{3N_c M^2 s} \sum_q e_q^2 
   \int\frac{dx_1}{x_1} \frac{dx_2}{x_2}\,C(z,M,\muf)
   \left[ f_{q/N_1}(x_1,\muf)\,f_{\bar q/N_2}(x_2,\muf)
    + (q\leftrightarrow\bar q) \right] \, , 
\end{equation}
where we introduced the quantity $\tau \equiv M^2/s$ so that  $z=\tau/(x_1 x_2)$, and the integration is restricted to the region where $x_1 x_2\ge\tau$. One can further rewrite this result by introducing the parton luminosity $\ff$, defined as the Mellin convolution of the PDFs:\footnote{We remind the reader that the Mellin convolution of two functions $f$ and $g$ is defined as 
\be
f \otimes g(y) \equiv \int_y^1 \frac{dx}{x} f(x) g\left(\frac{y}{x}\right) =   \int_y^1 \frac{dx}{x} g(x) f\left(\frac{y}{x}\right) \, . 
\ee
 }
\begin{equation}\label{eq:ffdef}
   \ff(y,\muf) = \sum_q e_q^2 \int_y^1\!\frac{dx}{x}
   \left[ f_{q/N_1}(x,\muf)\,f_{\bar q/N_2}(y/x,\muf)
    + (q\leftrightarrow\bar q) \right] \, .
\end{equation}
In this way, the lepton-pair invariant mass distribution can be brought into the simple form 
\begin{equation}\label{eq:dsigdM2alt}
   \frac{d\sigma^{\rm thresh}}{dM^2}
   = \frac{4\pi\alpha^2}{3N_c M^2 s} \int_\tau^1\!\frac{dz}{z}\,
   C(z,M,\muf)\,\ff(\tau/z,\muf) = \frac{4\pi\alpha^2}{3N_c M^2 s}  \left(C \otimes \ff\right) (\tau)\,,
\end{equation}
Eq.~(\ref{eq:dsigdM2alt}) shows explicitly that the calculation of physical observables such as the invariant mass distribution requires to evaluate an integral over $z$ in the range $z \in [\tau,1]$.  One might therefore wonder if calculations of the function $C$ in the $z \to 1$ limit, such as the one which we discussed in the previous section, are sufficient in order to obtain reliable predictions for physical quantities such as the pair invariant mass distribution.

There are two situations in which the contribution of the threshold region to the physical cross section is enhanced. The first is the strict threshold limit in which $\tau \approx 1$ so that the integration variable $z$ is necessarily in the threshold region. This situation is not relevant phenomenologically, because the partonic luminosity is extremely small in that region. A second, more interesting situation in which the $z \to 1$ region of the partonic cross section provides the numerically dominant contribution to the physical quantity arises when the partonic luminosity is a steeply falling function as $\bar{z} = 1-z$ increases. The behavior of the partonic luminosity as a function of its argument is found in Fig.~\ref{fig:ffdrop}. The figure refers to the case in which the factorization scale is set to $8$~GeV. The partonic luminosity is approximately equal to $y^{-a}$, with $a\approx 1.8$ for $y\to 0$, and approximately proportional to $(1-y)^b$ with $b \approx 11$ for $y \to 1$. The figure shows that these two simple functions of $y$ describe the partonic luminosity well in the regions $y< 0.05$ and $y > 0.3$, respectively.

Using the approximate form of the parton luminosity in the region of large $\tau > 0.3$, we can approximate
\begin{equation}
   \frac{d\sigma^{\rm thresh}}{dM^2}
   \approx \frac{4\pi\alpha^2}{3N_c M^2 s}\,\ff(\tau,\muf)
   \int_\tau^1\!\frac{dz}{z} 
   \left( \frac{1-\tau/z}{1-\tau} \right)^b C(z,M,\muf) \,. \label{eq:dinthenh}
\end{equation}
If in Eq.~(\ref{eq:dinthenh}) one expands the factor raised to the exponent $b$ in powers of $\bar{z}$ 
and treats $b$ as a large parameter, it is possible to see that the prefactor of the function $C$ is large (i.e. of order 1) if 
$\bar{z} < (1-\tau)/b$. Therefore, there is an enhancement of the partonic threshold region even if $\tau$ is not close to 1 and one should choose $\mu_s \sim M\,(1-\tau)/b$ to avoid large logarithms in the convolution of the hard-scattering kernel with the luminosity. This phenomenon goes under the name {\em Dynamical Threshold Enhancement}. In the region of small $\tau < 0.05$, the appropriate approximation is 
\begin{equation}\label{eq:lowtau}
   \frac{d\sigma^{\rm thresh}}{dM^2}
   \approx \frac{4\pi\alpha^2}{3N_c M^2 s}\,\ff(\tau,\muf)
   \int_\tau^1\!\frac{dz}{z}\,z^{a}\,C(z,M,\muf) \,,
\end{equation}
Extending the integration down to $\tau=0$, one obtains the $a$-th moment of the hard-scattering coefficient $C(z,M,\muf)$. The moment-space cross section is given in \cite{Becher:2007ty} and one finds that the appropriate soft scale in moment space is $M/(a e^{\gamma_E}) \approx M/3$, somewhat, but not much smaller than the Drell-Yan mass. In the intermediate regime $0.05< \tau < 0.3$ one can use a numerical procedure to determine the soft scale \cite{Becher:2007ty}. To do so, one analyzes the perturbative corrections to the soft function numerically as a function of the renormalization scale and then chooses the soft scale in the region where there are no large logarithms in this function, see \cite{Becher:2007ty} for details. An alternative method was put forward recently in \cite{Sterman:2013nya}. In this method the soft scale is determined from a moment expansion of the parton luminosity $\ff (y)$ around $y=\tau$. This corresponds exactly to the approximation Eq.~(\ref{eq:lowtau}) for small $\tau < 0.05$, but the authors of \cite{Sterman:2013nya}  perform a systematic expansion of the luminosity and show that this type of  treatment also works at higher $\tau$ if one allows for an exponent $a$ which depends on $\tau$. Numerically, the resulting value of $\mu_s$ is very similar to the one obtained in the numerical approach of \cite{Becher:2007ty}.

While the typical scale of the soft emissions is often not much lower than $M$ for moderate and small values of $\tau$, the numerical studies carried out in \cite{Becher:2007ty} show that the Drell-Yan cross section is nevertheless well approximated by keeping only the leading singular terms in the partonic hard scattering kernels, which are included in the function $C$ in Eq.~(\ref{eq:HS6}) and the same is true for many other processes. The reason is an inherent property of the hard-scattering kernels of hadronically inclusive  observables, which appear to receive the largest radiative corrections from the region of phase space corresponding to Born kinematics. In other words, the effects of hard real emissions appear to be suppressed compared with virtual corrections and soft emissions. For these reasons, threshold resummation is useful for a variety of processes of interest in collider physics. Among these are Higgs production \cite{Ahrens:2009uz}, top-quark pair production \cite{Ahrens:2010zv,Ahrens:2011mw}, slepton-pair production \cite{Broggio:2011bd}, stop-pair production   \cite{Broggio:2013cia}. In all these cases the resummation formalism discussed in this section can be employed in order to obtain physical predictions including soft gluon emission corrections at all orders in perturbation theory up to the desired logarithmic accuracy.

\subsection{Numerical Results\label{sec:numDY}}

Having discussed the value of the soft scale $\mu_s$, we now turn to numerical predictions. We will show some numerical results taken from the detailed phenomenological study of resummation to NNNLL in \cite{Becher:2007ty} and discuss how the theoretical uncertainties of the resummed results can be estimated. Table \ref{tab:counting} shows the ingredients which are needed at this accuracy. One of them, the four-loop cusp anomalous dimension $\gamma^{\rm cusp}_3$, is not yet available and is estimated using the simple Pad\'e estimate $\gamma^{\rm cusp}_3 =(\gamma^{\rm cusp}_2)^2/\gamma^{\rm cusp}_1$. Numerically, the effect of $\gamma^{\rm cusp}_3$ is tiny.

When performing a fixed-order computation of a cross section, one obtains a result which contains a single renormalization scale $\mu$. Formally the cross section is independent of $\mu$, but when computing to a fixed order, say NLO, a higher-order dependence on the scale $\mu$ remains. One can therefore use the $\mu$ dependence of the NLO result as an estimate of the size of the NNLO corrections. It is common practice to vary the scale $\mu$ by a factor two around a default value to estimate the perturbative uncertainty.\footnote{In fixed-order computations one often introduces a second scale by re-expanding $\alpha_s(\mu)$ in terms of the coupling $\alpha_s(\mu_r)$ at a different scale, the renormalization scale $\mu_r$. The original scale $\mu$ remains in the PDFs and is called the factorization scale $\mu_f$, and varied independently from $\mu_r$ by a factor of two.} This procedure is clearly somewhat arbitrary, but experience shows that it gives a reasonable uncertainty estimate in many (though not all) cases. When a given computation involves multiple scales, an immediate problem with this procedure is that it is not at all clear what one should adopt as the default scale. In such cases, a conservative approach in the context of a fixed-order computation, would be to vary the scale from the lowest to the highest scale in the problem.

\begin{figure}[t]
\begin{center}
\psfrag{la}[l]{\small ~$M<\muh<2M \phantom{\mu^{\rm I}}$}
\psfrag{x}[B]{\small $M^2/s$}
\psfrag{y}[]{\raisebox{0.5cm}{\small $K$}}
\begin{tabular}{cc}
\includegraphics[width=0.45\textwidth]{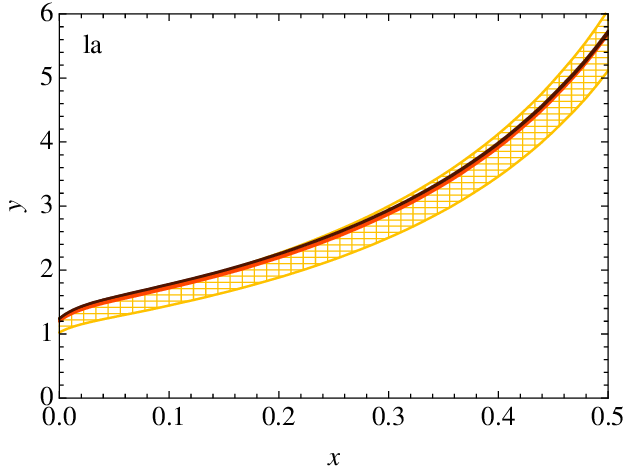} & 
\psfrag{la}[l]{\small ~$\mui^{\rm I}<\mui<\mui^{\rm II}$}
\includegraphics[width=0.45\textwidth]{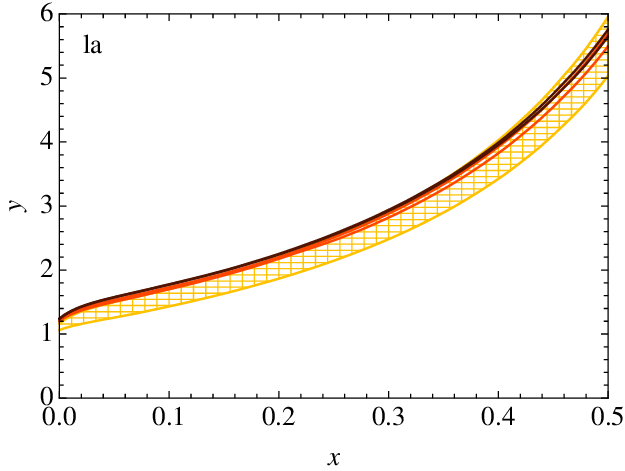}\\
\psfrag{la}[l]{\small ~$M/2<\muf<2M \phantom{\mu^{\rm I}}$}
\includegraphics[width=0.45\textwidth]{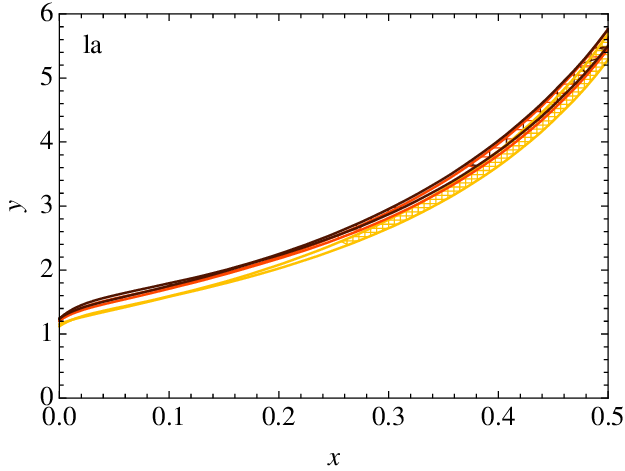} & 
\psfrag{la}[l]{\small ~$M/2<\muf<2M$ (fixed-order) 
 $\phantom{\mu^{\rm I}}$}
\includegraphics[width=0.45\textwidth]{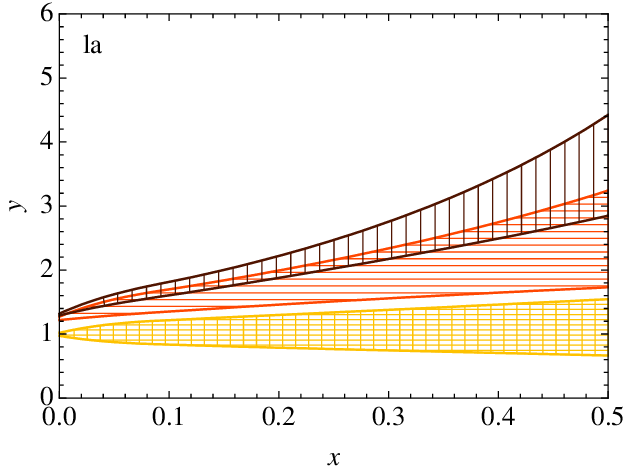} 
\end{tabular}
\end{center}
\vspace{-0.5cm}
\caption{\label{fig:scales}
Dependence of the resummed Drell-Yan cross section for $M=20$\,GeV on the scales $\muh$, $\mui$, and $\muf$. The bands show the $K$-factor obtained at NLL (light), NNLL (medium), and NNNLL (dark). The last plot shows for comparison the $\muf$ dependence in fixed-order perturbation theory at LO, NLO, and NNLO. 
}
\end{figure}

The uncertainty estimate in SCET is performed in exactly the same way, by scale variation. However, an important advantage of the effective theory framework is that we have separated the contributions associated with different scales and we can choose an appropriate value for the scale in each case. Instead of a single scale $\mu$, the resummed hard  scattering kernel Eq.~(\ref{eq:Caux3}) depends on the scale $\mu_h$, which governs the perturbative expansion of the hard function, on $\mu_s$, which is relevant for the soft corrections, and on the factorization scale $\mu_f$.  The stability of the resummed expression for the Drell-Yan cross section with respect to the variation of the hard, soft and factorization scale can be studied by looking at Fig.~\ref{fig:scales}, taken from \cite{Becher:2007ty}. The quantity plotted in the four panels is 
\begin{equation}
K\left(M^2, \tau\right) \equiv \left. \left( \frac{d \sigma}{d M^2} \right) \middle/ \left( \left.\frac{d \sigma}{d M^2}\right|_{\mbox{{\footnotesize LO}}} \right) \right. \, . \label{eq:Kfac}
\end{equation}
The pair invariant mass is set at $M = 20$~GeV and the plots were produced by employing MRST04NNLO PDFs. The numerator
of the fraction in Eq.~(\ref{eq:Kfac}) is obtained by evaluating Eq.~(\ref{eq:dsigdM2alt}); the denominator of that fraction is evaluated by keeping $\mu_f = M$ even when the factorization scale is varied in the numerator. The first-row panels show the excellent convergence of the perturbative expansion (after resummation) with respect to the variation of the hard and soft scales. The bands corresponding to the NLL, NNLL, and NNNLL resummation overlap, and the dependence on the matching scales $\muh$ and $\mui$ becomes negligible beyond NLL, indicating that the residual perturbative uncertainty is very small. The range for the soft scale variation $\mu_s^I < \mu_s < \mu_s^{II}$ was obtained using the numerical procedure to determine the soft scale, which we discussed above in Section~\ref{sec:DynamicalThresh}. The numerical values of $\mu_s^I$ and $\mu_s^{II}$can be found in \cite{Becher:2007ty}.

The lower left panel in Fig.~\ref{fig:scales} shows the dependence of the resummed Drell-Yan cross section at the various logarithmic accuracies with respect to variations of the factorization scale. One observes that both the convergence and the scale dependence are much improved by the resummation. The quality of the results can be further improved by performing a matching computation, i.e.\ by adding the non-threshold terms in fixed-order perturbation theory. This is also necessary to obtain results which are fully $\mu_f$-independent, because the threshold terms only cancel the end-point behavior of the Altarelli-Parisi equations, see Eq.~(\ref{eq:RGEpdfx}).

\begin{figure}[t]
\begin{center}
\begin{tabular}{c}
\psfrag{la}[lt]{\small $\phantom{ab}\sqrt{s}=14$\,TeV}
\psfrag{M}[B]{\small $M$\,[TeV]}
\psfrag{K}[B]{\small $K$}
\includegraphics[width=0.45\textwidth]{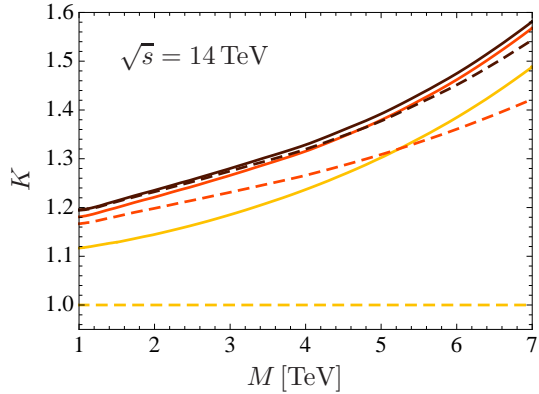} 
\end{tabular}
\end{center}
\vspace{-0.5cm}
\caption{\label{fig:plKfact}
Resummed (solid lines) versus fixed-order results (dashed lines) for the $K$-factor as a function of $M$. The light, medium, and dark lines correspond to LO, NLO, and NNLO, respectively. Default values are used for all scales.}
\end{figure}

Finally Fig.~\ref{fig:plKfact}, also taken from \cite{Becher:2007ty}, shows the comparison between resummed results at NLL, NNLL and NNNLL accuracy and the fixed order perturbation theory calculations at LO, NLO, and NNLO. The plot refers to the LHC running at a center of mass energy of $14$~TeV and shows calculations which include only contributions of diagrams mediated by a virtual photon.\footnote{Including the $Z^0$ channel would not alter our results for the $K$-factor significantly.} The differences between the two sets of curves show the effect of the resummation. The figure shows that resummation accelerates the convergence of the perturbative expansion. On the other hand, the plots also show that the most important logarithmic corrections are contained in the fixed-order NNLO results, at least for moderate lepton-pair masses. Similar conclusions apply to other processes where threshold resummation has been carried out. 

\subsection{Approximate Fixed-Order Formulas \label{sec:approxNnLO}}

In the previous sections we described the resummation up to NNNLL of the soft gluon emission corrections. For some processes in which the soft gluon corrections are numerically dominant but not large enough as to require all-order resummation, it can be convenient to employ Eq.~(\ref{eq:Caux3}) in order to extract approximate fixed-order formulas for the hard scattering kernel. In the recent past, this procedure was followed in several cases in which full NNLO calculations were not yet available, such as top-quark pair production and top-squark pair production.

Fixed-order formulas can be recovered from Eq.~(\ref{eq:Caux3}) by setting $\mu_f = \mu_h = \mu_s = \mu$. With this scale choice, the evolution factor $U$ is equal to unity and $\eta = 2 C_FA_{\gamma_{\cusp}}(\mu_s,\mu_f)$ goes to zero.
Therefore, in this case the hard scattering kernel $C$  can be written as 
\begin{equation}
C\left(z,M,\mu \right) = 
\tilde{c}\left(\partial_\eta ,M,\mu \right)
\left.
\left( \frac{M}{\mu}\right)^{2 \eta} \frac{e^{-2 \gamma_E \eta}}{\Gamma\left(2 \eta \right)}
\frac{z^{- \eta}}{(1-z)^{1-2 \eta}} \right|_{\eta \to 0} \, , \label{eq:approxDY}
\end{equation}
where
\be
\tilde{c}\left(L ,M,\mu \right) \equiv 
\left| \tilde{C}_V\left(-M^2,\mu \right) \right|^2 
\tilde{s}_{\rm  DY}\left(L, \mu \right) \, . \label{eq:tildec}
\ee
By inserting the analytic expressions of $C_V$ and $\tilde{s}_{\rm DY}$ up to order $\alpha_s^n$ in Eq.~(\ref{eq:approxDY}) it is possible to recover in full the fixed order expression of the hard scattering kernels up to order $\alpha_s^n$ in the soft emission approximation, i.e., in our case,  in the limit in which terms which are not singular when $z \to 1$ are ignored.

However, the benefit of this technique resides in the fact that, if one knows the Wilson coefficient $C_V$ and the soft function $\tilde{s}_{\rm DY}$ up to N$^{n}$LO, one can employ the RG equations satisfied by these two functions in order to calculate
all of the terms proportional to $\ln(\mu)$ in $C_V$ up to N$^{n+1}$LO as well as all of the terms proportional to $L$ in 
$\tilde{s}_{\rm DY}$ up to N$^{n+1}$LO. Equipped with this information, one can obtain an {\em approximate} N$^{n+1}$LO formula for the hard scattering kernel $C$ from Eq.~(\ref{eq:approxDY}). Such a formula will include all of the terms proportional to the plus distributions
\be
P_n(z) \equiv \left[ \frac{\ln^n(1-z)}{1-z}\right]_+ .
\ee
In other words, fixed order calculations of the hard and soft functions up to order $\alpha_s^n$ supplemented by the RG equations are sufficient in order to obtain information about a numerically large set of corrections of order $\alpha_s^{n+1}$ to $C$. However, this approach does not allow one to completely reconstruct the terms proportional to $\delta(1-z)$ at order N$^{n+1}$LO and these can have a non-negligible numerical impact.

Expanding the factor $(1-z)^{-1+2 \eta}$ in distributions by means of Eq.~(\ref{eq:distexp}), one can show that to take the derivatives with respect to $\eta$ and then to take the limit $\eta \to 0$
in Eq.~(\ref{eq:approxDY}) is equivalent to make the following set of replacements in $\tilde{c}$ in Eq.~(\ref{eq:tildec}):
\begin{align}
1 &\rightarrow \delta(1-z) \, , \nn \\
L &\rightarrow 2 P'_0(z) + \delta(1-z) L_M \, , \nn \\
L^2 &\rightarrow 4 P'_1(z) + \delta(1-z) L^2_M  \, , \nn \\
L^3 &\rightarrow 6 P'_2(z) - 4 \pi^2 P'_0(z) + \delta(1-z) \left( L_M^3 + 4 \zeta_3\right) \, , \nn \\
L^4 &\rightarrow 8 P'_3(z) -16 \pi^2 P'_1(z) +128 \zeta_3 P'_0(z) +\delta(1-z) \left( L_M^4 +16 \zeta_3 L_M \right) \, ,
\label{eq:repL}
\end{align}
where we introduced the notation $L_M = \ln(M^2/\mu^2)$ and made use of the distributions
\be
P'_n(z) \equiv \left[ \frac{1}{1-z} \ln^n\left(\frac{M^2 (1-z)^2}{\mu^2 z} \right)\right]_+ \, ,
\ee
which naturally arise from the r.h.s. of Eq.~(\ref{eq:approxDY}).
The replacements in Eq.~(\ref{eq:repL}) are sufficient in order to obtain NNLO formulas. Higher order calculations require one to derive replacement rules for higher powers of $L$.\footnote{The relation linking the distributions $P'_n$ to the more conventional distributions $P_n$ is 
\begin{align}
P'_n(z) &= \sum_{k=0}^n \Bigl(\begin{array}{c}
n \\
k 
\end{array}  
\Bigr) L_M^{n-k} \Biggl[2^k P_k(z)  \nn \\
& + \sum_{j=0}^{k-1} 
\Bigl(\begin{array}{c}
k \\
j 
\end{array}  
\Bigr) 2^j (-1)^{k-j} \Biggl(\frac{\ln^{k-j} z \ln^{j}(1-z)}{1-z}  - \delta(1-z) \int_0^1 dx \frac{\ln^{k-j} x \ln^{j}(1-x)}{1-x}\Biggr)\Biggr] \, .
\end{align}
}

As a sanity check, the reader can verify that by inserting into Eq.~(\ref{eq:tildec})  the NLO hard function $\tilde{C}_V$
(see Eq.~(\ref{eq:CVren}))
\be
\tilde{C}_V\left(-M^2,\mu \right) = 1 + C_F \frac{\alpha_s}{4 \pi} \left(-{\mathcal L}_M^2 +3 {\mathcal L}_M - 8 +\frac{\pi^2}{6} \right) +{\mathcal O}(\alpha_s^2) \, ,
\ee
(where ${\mathcal L}_M \equiv L_M - i \pi$) as well as the NLO renormalized soft function
in Laplace space 
\be
\tilde{s}_{\rm DY} (L,\mu) = 1+  C_F \frac{\alpha_s}{4 \pi} \left( 2 L^2 +\frac{\pi^2}{3} \right) + {\mathcal O}\left(\alpha_s^2\right) \, ,
\ee
and by applying the replacements in Eqs.~(\ref{eq:repL}), one indeed obtains the NLO hard scattering kernel in the soft limit:
\be
C(z,M,\mu) = \delta(1-z) + C_F \frac{\alpha_s}{4 \pi}  \left[8 P'_1(z) +  \delta(1-z) \left( 6 L_M + \frac{8}{3} \pi^2 -16 \right)\right] + {\mathcal O}\left(\alpha_s^2\right) \, .
\ee
This calculation can be easily extended to NNLO by employing the results collected in Appendix B of \cite{Becher:2007ty}.

%% file: 7_pTresummation.tex
\section{Transverse Momentum Resummation\label{sec:pT}}

In the previous section we employed  SCET methods in order to carry out the resummation of large logarithmic corrections in the total Drell-Yan cross section near threshold.
It is also interesting to measure and study the Drell-Yan production process in the case in which the produced lepton pair has a transverse momentum, $q_T$, with respect to the beam axis. In particular, the region in which  $q_T$ 
is small with respect to the pair-invariant mass, $q_T^2 \ll M^2$, is phenomenologically interesting. In that region the cross section is large and it is used in order to extract the $W$-boson mass and width. A closely related process is Higgs production via gluon fusion; in this case, the region in which the Higgs boson has a small transverse momentum is  important because one usually vetoes hard jets in order to enhance the signal over background ratio.

The leading logarithmic corrections in the Drell-Yan process in the region of small transverse momentum were resummed 
in \cite{DDT,Parisi:1979se,Curci:1979bg}, while  an all-order formula for the resummed cross section at small $q_T$ was obtained in  a seminal paper by Collins, Soper and Sterman (CSS) \cite{Collins:1984kg}. In spite of the fact that the vector boson transverse momentum spectrum is a classic example of a multi-scale process exhibiting logarithmic enhancements, the analysis of its factorization properties is rather subtle and particularly interesting, since it suffers from the collinear anomaly we encountered in the case of the massive Sudakov form factor in Section \ref{sec:MSPCA}. A factorization formula for the Drell-Yan cross section in the small $q_T$ region in SCET was derived in \cite{Becher:2010tm}. By using that formula, the resummation of large logarithmic corrections was studied in \cite{Becher:2011xn}. Here we want to present the salient features of that analysis.

The derivation of the factorization theorem follows the exact same steps as for the threshold resummation until the point where the multipole expansion is performed, so we can start with Eq.~\eqref{eq:g2formula}. In contrast to the threshold resummation case, the final state and the scaling of the momenta is now generic, we can therefore no longer neglect the transverse position dependence of the collinear matrix elements. Instead of the usual PDFs, one is thus left with generalized, $x_T$-dependent PDFs (with $x_T^2\equiv -x_\perp^2>0$) \cite{Collins:1981uk,Collins:1981uw}
\begin{equation}\label{Bdef}
   {\cal B}_{q/N}(z,x_T^2,\mu) 
   = \frac{1}{2\pi} \int_{-\infty}^{+\infty} dt\,e^{-izt\bar n\cdot p}\,\langle N(p)|\,
    \bar\chi(t\bar n+x_\perp)\,\frac{\rlap/\bar n}{2}\,\chi(0)\,|N(p)\rangle \,,     
\end{equation}
and similarly for the gluon and anti-quark cases. Their Fourier transforms with respect to $x_T$ are referred to as transverse-momentum dependent PDFs (TMPDFs). Since we are mostly concerned with the position-space functions, we will refer to both of these types of objects as transverse PDFs (TPDFs).  For soft fields which scale as $(\lambda^2,\lambda^2,\lambda^2)$, all $x$ dependence of the soft function can be dropped and one thus ends up with $\hat W_{\rm DY}(0)=1$ because the Wilson lines cancel by unitarity. The transverse momentum of these soft fields is too small to contribute to the observable. In order to contribute, the soft fields should scale as $(\lambda,\lambda,\lambda)$, a scaling we referred to as semi-hard earlier when we showed that such modes do not contribute to the off-shell Sudakov form factor. The operator definition of the soft function is the same for both types of soft scalings and we include for the moment a soft function with such semi-hard fields. After multipole expansion it only depends on $x_\perp$ and we denote it by ${\cal S}(x_\perp)$ to distinguish if from the usual ultra-soft function $\hat W_{\rm DY}$.

The derivation of the differential cross section follows similar steps as in Section~\ref{sec:DY}. Starting from Eq.~\eqref{eq:g2formula} we obtain a modified version of Eq.~\eqref{eq:dsigmad4q} where the usual PDFs are replaced by the generalized, $x_T$-dependent PDFs defined in Eq.~\eqref{Bdef}. The leading-power result for the cross section reads
\begin{multline}\label{eq:dsigmad4qbeam}
d\sigma = \frac{d^4q}{(2\pi)^4} \frac{4\pi \alpha^2}{3 q^2 N_c} |\tilde{C}_V(-q^2 ,\mu)|^2  \int_{0}^{1} d\xi_1 \int_{0}^{1} d\xi_2\, \int^{+\infty}_{-\infty}\frac{d x_+\,d x_-}{2}  \, e^{i \frac{x_+}{2} (\xi_1 p_- - q_-)}\, e^{i \frac{x_-}{2} (\xi_2 l_+ - q_+)}  
 \\ \quad \times \int d^2 x_\perp \,e^{-iq_\perp\cdot x_\perp} \sum_q\,e_q^2\,\bigg[ {\cal B}_{q/N_1}(\xi_1,x_T^2,\mu)\,
     {\bar {\cal B}}_{\bar q/N_2}(\xi_2,x_T^2,\mu) {\cal S}(x_\perp,\mu) + (q\leftrightarrow\bar q) \bigg]  \,,
\end{multline}
where we rewrote the $d^4 x$ integration in terms of the light-cone coordinates:
\be
\int d^4 x  = \frac{1}{2} \int_{-\infty}^{\infty} d x_+  
\int_{-\infty}^{\infty} d x_- \int d^2 x_\perp \, . 
\ee
By integrating over $x_+$ and $x_-$, one obtains two delta functions: 
\be 
 \int_{-\infty}^{\infty} d x_+  e^{i \frac{x_+}{2} (\xi_1 p_- - q_-)}
\int_{-\infty}^{\infty} d x_- e^{i \frac{x_-}{2} (\xi_2 l_+ - q_+)} =  \frac{4 (2\pi)^2}{s} \delta\left(\xi_1-\frac{q_-}{p_-}\right) \delta\left(\xi_2-\frac{q_+}{l_+}\right) \, . \label{eq:2dir}
\ee
The Dirac deltas in Eq.~(\ref{eq:2dir}) fix the momentum fractions $\xi_1$ and $\xi_2$. Finally, we use the fact that $d^4q\,\theta(q_0)\,\delta(q^2-M^2)=\frac12\,d^2q_\perp\,dy=\frac{\pi}{2}\,dq_T^2\,dy$, where the last identity holds after integration over the azimuthal angle. Naively, we thus end up with the factorization theorem
\begin{equation}\label{Bfactnaive}
\begin{aligned}
   \frac{d^3\sigma}{dM^2\,dq_T^2\,dy} 
   &= \frac{4\pi\alpha^2}{3N_c M^2 s} \left| \tilde{C}_V(-M^2,\mu)\right|^2 
    \frac{1}{4\pi} \int\!d^2x_\perp\,e^{-iq_\perp\cdot x_\perp} \\
   &\quad\times \sum_q\,e_q^2\,\bigg[ {\cal B}_{q/N_1}(\xi_1,x_T^2,\mu)\,
    {\bar {\cal B}}_{\bar q/N_2}(\xi_2,x_T^2,\mu) {\cal S}(x_\perp,\mu) + (q\leftrightarrow\bar q) \bigg] 
    + {\cal O}\bigg( \frac{q_T^2}{M^2} \bigg) \,,
\end{aligned}
\end{equation}
where
\begin{equation}\label{taudef}
   \xi_1 = \sqrt{\tau}\,e^{-y} \,, \qquad
   \xi_2 = \sqrt{\tau}\,e^{y} \,, \qquad
   \mbox{with} \quad
   \tau =  \frac{M^2+q^2_T}{s} \,.
\end{equation}
The above formula appears to achieve the desired factorization of the hard and hard-collinear scales, $M^2$ and $q_T^2\sim x_T^{-2}$.

Since the transverse separation is  $1/x_T \sim q_T $ and the transverse momentum is assumed to satisfy $q_T \gg \Lambda_{\rm QCD}$ so that it is in the perturbative domain, one can compute the differential cross section in perturbation theory. To do so, one considers an operator-product expansion of the form \cite{Collins:1984kg,Collins:1981uk,Collins:1981uw}
\begin{equation}\label{eq:OPE}
   {\cal B}_{i/N}(\xi,x_T^2, \mu) 
   = \sum_j \int_\xi^1\!\frac{dz}{z}\,{\cal I}_{i\leftarrow j}(z,x_T^2, \mu)\,
    f_{j/N}(\xi/z, \mu) + {\cal O}(\Lambda_{\rm QCD}^2\,x_T^2) \,.     
\end{equation}
The coefficient functions ${\cal I}_{i\leftarrow j}(z,x_T^2, \mu)$ contain the perturbatively calculable physics associated with $x_T$ and are convoluted with the standard PDFs. In the context of SCET, generalized PDFs defined in terms of hadron matrix elements in which collinear fields are separated by distances that are not light-like are referred to as beam functions. For such functions an analogous expansion was considered in \cite{Stewart:2009yx}.

\subsection{Phase-Space Regularization \label{sebsec:PSreg}}

\begin{figure}
\begin{center}
\begin{tabular}{cccc}
\includegraphics[height=0.142\textwidth]{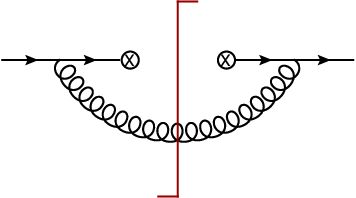} & 
\includegraphics[height=0.142\textwidth]{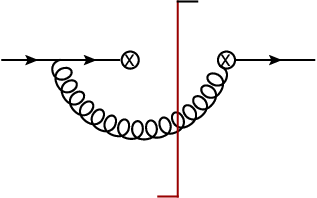} & 
\includegraphics[height=0.142\textwidth]{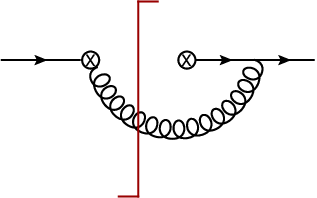} & 
\includegraphics[height=0.142\textwidth]{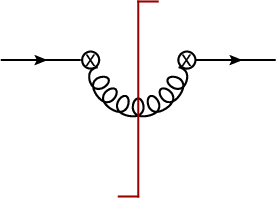} \\[0.1cm]
\end{tabular}
\includegraphics[height=0.142\textwidth]{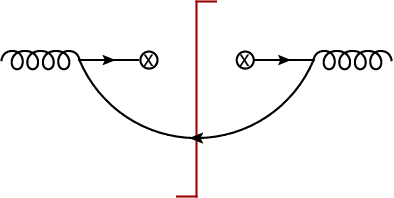}
\vspace{-0.5cm}
\end{center}
\caption{\label{fig:graphsIfun}
One-loop diagrams contributing to the matching coefficients ${\cal I}_{q\leftarrow q}$ (top row) and ${\cal I}_{q\leftarrow g}$ (bottom row). The vertical lines indicate cut propagators.}
\end{figure}

The coefficients ${\cal I}_{i\leftarrow j}(z,x_T^2, \mu)$ are obtained from a matching computation. The simplest possibility is to evaluate the collinear matrix elements with on-shell partonic states. The PDFs for such states are trivial $f_{i/k}(\xi) = \delta(1-\xi) \delta_{ik} $ and the computation gives directly the  coefficients ${\cal I}_{i\leftarrow k}(z,x_T^2, \mu)$. The relevant diagrams are shown in Figure \ref{fig:graphsIfun}.
Since the coefficients are independent of the states used in the matching, the same coefficients are relevant in the hadronic case. However, when trying to compute the $\mathcal{O}(\alpha_s)$ corrections to the functions ${\cal I}_{i\leftarrow j}(z,x_T^2, \mu)$ one encounters the same difficulty that was present for the massive Sudakov form factor, namely that some of the relevant integrals are not regularized in dimensional regularization. The unregulated singularities arise when integrating over the light-cone components $k_+$ and $k_-$ and only arise in the phase-space integrations  \cite{Becher:2011dz}. For loop integrals, the dimensional regularization parameter gets transmitted also to the integrals over the light-cone components after integrating over the transverse directions. This is no longer the case for the phase-space integrals which arise in transverse momentum resummation because the kinematic constraints fix the $(d-2)$-dimensional transverse momentum, which leaves the integration over the remaining light-cone components unregularized. A convenient way to regularize the ensuing singularities is to modify the phase-space integrals \cite{Becher:2011dz}
\begin{equation}\label{eq:regfinal}
\int\!d^dk  \,\delta(k^2) \theta(k^0) \to  \int\!d^dk \left(\frac{\nu}{k_+}\right)^\alpha \,  \delta(k^2) \theta(k^0) \,.
\end{equation}
The factor $(\nu/k_+)^\alpha$ regularizes the light-cone denominators which arise in SCET after expanding the QCD propagators. It suffices to regularize one light-cone component, since the product is fixed by the on-shell constraint  $k_+ k_- = k_T^2$. 

To understand how the problem arises and how the additional regularization solves it, let us compute a simple  toy-integral where the problem occurs. Let us consider the one-particle phase space with a cut on energy $k_0 < E$ and fixed transverse momentum $k_T=p_T \ll E$. The phase-space integral reduces to
\begin{equation}\label{origint}
I = \int_0^\infty\! dk_+  \int_0^\infty\! dk_- \,\delta(k_+ k_- - p_T^2)\,\theta(k^0)\, \theta( 2E - k_+ -k_- )\, \left(\frac{\nu}{k_+}\right)^\alpha = \ln\frac{4E^2} {p_T^2}+ \dots\, ,
\end{equation}
where the ellipsis denotes terms of ${\mathcal O}(\alpha)$ or ${\mathcal O}(p_T^2/E^2)$. Obviously, this integral is well defined, as are the phase-space integrals in ordinary QCD. The unregularized divergences only appear when the integral is expanded using the strategy of regions technique. Let us now expand on the level of the integrand. Since we restrict the phase space to small transverse momentum, the emissions must be collinear or soft. While the full integral is well defined for $\alpha=0$, this is no longer the case once the integrand is expanded in the different regions. The leading power of the expansion is obtained by dropping the small light-cone components in a given region from the $\theta$-function. In the anti-collinear region, for example, we expand $\theta( 2E - k_+ -k_- ) \to \theta( 2E - k_+ )$, which gives
\begin{equation}\label{eq:Ibarc}
I_{\bar{c}} 
= \int_0^{2E} \frac{dk_+}{k_+} \left(\frac{\nu}{k_+}\right)^\alpha = -\frac{1}{\alpha} \left(\frac{\nu}{2E}\right)^\alpha  \, . 
\end{equation}
In the collinear region, one obtains
\begin{equation} \label{eq:Ic}
I_{c}  
= \int_{\frac{p_T^2}{2E}}^\infty \frac{dk_+}{k_+} \left(\frac{\nu}{k_+}\right)^\alpha = +\frac{1}{\alpha} \left(\frac{2E\, \nu }{p_T^2}\right)^\alpha \, ,
\end{equation}
while in the soft region, one ends up with a scaleless integral, $I_s=0$. It is manifest that the individual integrals are only well defined with the additional regulator. However, the divergences in the regulator cancel once the two contributions are added and the large logarithm present in the original integral in Eq.~(\ref{origint}) is recovered:
\begin{equation}\label{eq:sumIcIbarc}
I_{\bar{c}}  + I_{c}   = -\frac{1}{\alpha} \left(\frac{\nu}{2E}\right)^\alpha +\frac{1}{\alpha} \ \left(\frac{2E\, \nu }{p_T^2}\right)^\alpha= \ln\frac{4E^2} {p_T^2}+ \dots\,.
\end{equation}
Precisely the same happens in the computation of the matching coefficients ${\cal I}_{q\leftarrow q}$. Both the left- and right-collinear function will suffer from divergences in the regulator $\alpha$, which cancel when the two pieces are put together. However, while the divergences cancel there is a remnant, namely the logarithm of the large energy $E$ in Eq.~(\ref{eq:sumIcIbarc}). It is surprising that the sum of the collinear contributions depends on the large scale $E$ in the problem. The example integral makes it clear where this logarithm is coming from. It arises from integrating over the large range in rapidity $y$, which is defined as
\begin{equation}
y=\frac{1}{2}\ln \frac{k_+}{k_-} \,,
\end{equation}
The anti-collinear fields have generically large $k_+$ and therefore large positive rapidities, while the collinear fields have large negative rapidities. Soft fields have small rapidities. The divergences arise in the overlap regions, when collinear rapidities become small, or soft ones large. These singularities cancel when the different contributions are added back together. The basic mechanism is the same as for the one-dimensional example integral discussed in detail in Section \ref{sec:ASE}, except that the singularities are associated with different rapidities instead of different energies.

Without the regulator, the soft and collinear integrals would be invariant under a rescaling of the hard scale $E \to \kappa\, E$; in fact, it would be sufficient to rescale the integration variable according to $k_+ \to k_+/\kappa$ in order to eliminate $\kappa$ from the integrand. The regulator breaks this invariance and the symmetry is not recovered after the regulator is switched off. This is trivial for the integrals in our example, but is also true for SCET in general. In the calculation of TPDFs, each collinear sector only knows about the momentum of one colliding hadron, but the hard scale is given by the product of the large components of both sectors.  This type of effect, that a quantum theory does not have the symmetries present at the classical level is called an anomaly. The present anomaly is called the collinear anomaly, or the factorization anomaly. It is not an anomaly of the full theory, but an anomaly of the low-energy effective theory. 

The calculation of the diagrams contributing to the transverse PDFs at NLO (first line in Fig.~\ref{fig:graphsIfun}) is discussed in some detail in Appendix~\ref{app:TPDFsNLO}. Here we report only the regularization independent, $\ms$ renormalized result for the product of coefficient functions at one-loop order
\begin{multline}
\left[{\mathcal I}_{q\leftarrow q}\left(z_1,x_T^2,\mu\right)  \bar{{\mathcal I}}_{\bar{q} \leftarrow\bar{q}} \left(z_2, x_T^2, \mu \right) {\mathcal S}(x_T^2) \right]_{q^2}
 = \delta(1-z_1) \delta(1-z_2)    \\  
 - \frac{C_F \alpha_s}{4 \pi}\Biggl\{ \delta(1-z_1) \delta(1-z_2) \Biggl(\! 4  L_\perp \ln\left( \frac{q^2}{\mu^2}\right) + 2 L_\perp^2  +\frac{\pi^2}{3}   \!\Biggr) \\
+ \Biggr[ 2 \delta(1-z_1) \Biggl( L_\perp \frac{1+z_2^2}{\left[ 1-z_2\right]_+} - (1-z_2) \Biggr)
+ \left(z_1 \leftrightarrow z_2 \right)   \Biggr] \Biggr\} \, . \label{eq:BBren}
\end{multline}
In Eq.~(\ref{eq:BBren}) we introduced the symbol $L_\perp$ which is defined as $L_\perp = \ln\left(x_T^2 \mu^2 e^{2 \gamma_E}/4 \right)$. The one-loop function for ${\mathcal I}_{q\leftarrow g}(z_1,x_T^2,\mu)$ can be computed without any additional regularization, it can be found in \cite{Becher:2010tm}. With the regulator in Eq.~(\ref{eq:regfinal}), the soft function ${\mathcal S}(x_T^2)=1$ to all orders in perturbation theory, because the relevant integrals are always scaleless, as in the toy example above. With other regulators, such as the one proposed in \cite{Chiu:2011qc,Chiu:2012ir} this would not be the case. The soft integrals would also be non-zero if we had chosen a left-right symmetric form 
\begin{equation}\label{eq:regsymm}
 \left(\frac{\nu}{k_+}\right)^\alpha \, \theta(k_+ - k_-) +  \left(\frac{\nu}{k_-}\right)^\alpha \, \theta(k_- - k_+)\,
\end{equation}
of the regulator. To extract the anomaly exponent defined below, the symmetric form can be advantageous since it reduces the calculation of the exponent to the evaluation of a soft instead of a collinear matrix element \cite{Becher:2013xia}.

\subsection{Refactorization and the Collinear Anomaly}

Because of the dependence of the TPDFs on the hard scale of the underlying process, Eq.~(\ref{Bfactnaive}) is not a useful factorization formula; part of the $q^2 = +M^2$ dependence is still hidden in the product of TPDFs. At one-loop order, the explicit dependence is shown in Eq.~\eqref{eq:BBren}. This implies that a complete separation of the hard and collinear scales $M^2$ and $q^2_T$ was not achieved. In order to complete the factorization and to carry out  the resummation of the large logarithms of the ratio $q_T^2/M^2$, it is necessary to control the dependence of the product of TPDFs at all orders in perturbation theory. In \cite{Becher:2010tm}, it was shown that in the $x_T$ space this product can be refactorized as follows: 
\begin{equation}\label{refact}
   \left[ {\cal B}_{q/N_1}(z_1,x_T^2,\mu)\,\bar{{\cal B}}_{\bar q/N_2}(z_2,x_T^2,\mu)
    \right]_{q^2}
   = \left( \frac{x_T^2 q^2}{b_0^2} \right)^{-F_{q\bar q}(x_T^2,\mu)}
  \!\! \!\! \!B_{q/N_1}(z_1,x_T^2,\mu)\,B_{\bar q/N_2}(z_2,x_T^2,\mu) \,,
\end{equation}
with $b_0 = 2 e^{-\gamma_E}$ and where the exponent $F_{q\bar{q}}$ depends only on the transverse coordinate $x_T$ and on the renormalization scale 
$\mu$. The functions $B_{i/N}$ of the r.h.s. of Eq.~(\ref{refact}) are independent from the hard momentum transfer. All the dependence on $q^2$ is explicit and has an extremely simple form: It is a pure power, with an exponent  $F_{q\bar q}$. which depends on the transverse separation $x_T$.
We observe that, if one chooses $\mu \sim x_T^{-1}$, the $q^2$ dependent prefactor resums all of the large logarithms of the hard scale, while $F_{q\bar q}(x_T^2, \mu)$ has a perturbative expansion in $\alpha_s(\mu)$ with coefficients of ${\mathcal O}(1)$.

Let us briefly review the derivation of the $q^2$ dependence. The argument relies on the fact that the divergences in the analytic regulator must cancel in the product of the beam functions. As a consequence, the product is independent of the scale $\nu$ associated with the regulator. Let us introduce the notation
\begin{align}
f\!\left(\ln\frac{\nu M x_T^2}{b_0^2}\right) &= \ln {\cal B}_{q/N_1}\!\left(\ln\frac{\nu M x_T^2}{b_0^2},z_1,x_T^2,\mu\right) \,, \nn\\
\bar{f}\!\left(\ln\frac{\nu }{M}\right) &= \ln \bar{{\cal B}}_{\bar{q}/N_2}\!\left(\ln\frac{\nu}{M},z_2,x_T^2,\mu\right)\, ,
\end{align}
where we added an extra argument to the functions ${\cal B}_{q/N_1}$ and $\bar{{\cal B}}_{\bar{q}/N_2}$ to make the dependence on $\nu$ explicit. The specific form of the $\nu$ dependence of the individual functions arises because of the dependence on the analytic regulator, which has the form $(\nu/k_+)^\alpha$, and the power counting of $k_+$ in the two collinear regions, see the example integral Eqs.~\eqref{eq:Ibarc} and \eqref{eq:Ic}. The factors of $b_0$ are inserted because they arise in the perturbative computation, see Appendix~\ref{app:TPDFsNLO}, but do not play any role for the argument we now want to make. Taking the logarithm of the product of the beam functions,
the independence from $\nu$ leads to the equations
\begin{align}
\frac{d^n}{d\ln \nu^n}\left[ f\!\left(\ln\frac{\nu M x_T^2}{b_0^2}\right) + \bar{f}\!\left(\ln\frac{\nu}{M}\right)   \right] = 0 \, , \nn \\
\frac{d}{d\ln M}\,\frac{d^n}{d\ln \nu^n}\left[ f\!\left(\ln\frac{\nu M x_T^2}{b_0^2}\right) + \bar{f}\!\left(\ln\frac{\nu}{M}\right) \right] = 0\, .
\label{eq:dnu}
\end{align}
These yield
\begin{align}
f^{(1)}\!\left(\ln\frac{\nu M x_T^2}{b_0^2}\right)  
+ \bar{f}^{(1)} \! \left(\ln\frac{\nu }{M}\right) &=0\,,  \nonumber\\
f^{(n)}(L) = \bar{f}^{(n)}(L) &=0 \;\text{ for }\; n>1\,,
\end{align}
Where $f^{(n)}(L)$ denotes the $n$-th derivative with respect to the argument of the function. 
This implies that the functions $f(L)$ and $\bar{f}(L)$ must be linear in their arguments, with a common coefficient, i.e.
\begin{align}
f(L) &= \ln B_q(z_1,x_T^2,\mu) - F_{q\bar q}(x_T^2,\mu)\, L \,, \nonumber\\
\bar{f}(L) &= \ln \bar{B}_q(z_2,x_T^2,\mu) + F_{q\bar q}(x_T^2,\mu)\, L\,.
\end{align}
The coefficient of the logarithm must be independent of $z_1$ and $z_2$ since it is identical in both functions. Plugging in the explicit form of the logarithms, we end up with
\begin{equation}
f\!\left(\ln\frac{\nu M x_T^2}{b_0^2}\right) + \bar{f}\!\left(\ln\frac{\nu }{M}\right)  = \ln\left( B_q(z_1,x_T^2,\mu) \bar{B}_q(z_2,x_T^2,\mu) \right) - F_{q\bar q}(x_T^2,\mu) \ln \frac{M^2 x_T^2}{b_0^2}\,,
\end{equation}
which is precisely Eq.~\eqref{refact}. We have thus proven that the anomaly logarithms exponentiate.

An alternative way to achieve this resummation is the rapidity RG \cite{Chiu:2011qc,Chiu:2012ir}. In this framework, one defines renormalized beam and soft functions by absorbing the $1/\alpha$ divergences into $Z$-factors. The renormalized functions are $\nu$-dependent and one then writes down RG-equations for these  functions, which describe their evolution under a change of $\nu$, in analogy to Eq.~(\ref{eq:dnu}) above. Solving these equations leads to the exponentiation of the anomaly logarithms which we just derived. With our regulator, the scale $\nu$ tracks the momentum component $k_+$, which is large in the anti-collinear sector and small in the collinear one. The RG evolution can thus be viewed as an evolution from large to small $k_+$, or equivalently from large to small rapidity. This provides a nice physical picture for the origin of these logarithms. The framework allows one to study the evolution in $\nu$ as well as $\mu$ and to have different values of $\mu$ in the beam and soft functions. On the other hand, certain aspects appear somewhat artificial, e.g. one introduces an additional coupling constant at intermediate stages which is set to one at end. Furthermore, the individual terms in the regularized factorization formula are highly scheme dependent (e.g.\ depending on the regulator one can have a soft function or not), so it is not clear whether the simultaneous $\mu$ and $\nu$ scale variations probe interesting physics.

Having derived the general form, we can now read off the anomaly exponent from Eq.~(\ref{eq:BBren}). 
Rewriting
\be
\ln\left(\frac{q^2}{\mu^2} \right) = \ln\left(\frac{M^2 x_T^2}{b_0^2}\right) - L_\perp \, 
\ee
in Eq.~(\ref{eq:BBren}) and looking at the prefactor of $\ln({M^2 x_T^2}/{b_0^2})$,
one finds
\be
F_{q\bar q}\left(x_T^2, \mu \right) = \frac{\alpha_s}{4 \pi} C_F \gamma^{\cusp}_0 L_\perp+ {\cal O}\left(\alpha_s^2 \right) \, . \label{eq:Fqqbas}
\ee
Furthermore, from the independence of the cross section  in Eq.~(\ref{Bfactnaive}) from the scale $\mu$ it follows that the ingredients in Eq.~(\ref{refact}) must satisfy the following RG equations:
\begin{equation}\label{Bevol}
\begin{aligned}
   \frac{dF_{q\bar q}(x_T^2,\mu)}{d\ln\mu} 
   &= 2 C_F\gamma_{\rm cusp}(\alpha_s) \,, \\
   \frac{d}{d\ln\mu}\,B_{q/N}(z,x_T^2,\mu)
   &= \left[ C_F \gamma_{\rm cusp} (\alpha_s)\,\ln\frac{x_T^2\mu^2}{b_0^2}
    - \gamma_V(\alpha_s) \right] B_{q/N}(z,x_T^2,\mu) \,.
\end{aligned}
\end{equation}

By inserting Eq.~(\ref{refact}) in Eq.~(\ref{Bfactnaive}), one finds a factorization formula in which the hard and collinear scales are completely separated, and all of the large logarithms can be resummed by setting $\mu \sim q_T$:
\begin{align} \label{eq:CSfact}
\frac{d^3 \sigma}{d M^2 d q_T^2 d y} & = \frac{4 \pi \alpha_s}{3 N_c M^2 s}\left|C_V\left(-M^2, \mu\right) \right|^2
\frac{1}{4 \pi} \int\! d^2 x_\perp e^{-i q_\perp \cdot x_\perp} \left(\frac{x_T^2 M^2}{b_0^2}\right)^{-F_{q\bar q}\left( x_T^2, \mu\right)}  \nonumber \\
& \times \sum_q e_q^2 \left[B_{q/N_1}\left(z_1,x_T^2, \mu \right)  B_{{\bar q}/N_2}\left(z_2,x_T^2, \mu \right) 
+ \left(q \leftrightarrow {\bar q} \right)\right] + {\cal O}\left(\frac{q_T^2}{M^2} \right) \, .
\end{align}

For a given transverse momentum $q_T$, the integral over $x_\perp$ receives numerically significant contributions 
from transverse separations $x_T \lessim 1/q_T$. For transverse momenta in the perturbative domain, i.e. for $q_T^2 \gg \Lambda^2_{\text{QCD}}$, the functions $B$ in Eq.~(\ref{eq:CSfact}) obey an operator-product expansion of the same 
form as the one obeyed by the ${\cal B}$ functions (see Eq.~(\ref{Bdef})). One finds 
\begin{equation}\label{eq:OPEB}
   B_{i/N}(\xi,x_T^2) 
   = \sum_j \int_\xi^1\!\frac{dz}{z}\, I_{i\leftarrow j}(z,x_T^2)\,
    f_{j/N_1}(\xi/z) + {\cal O}(\Lambda_{\rm QCD}^2\,x_T^2) \,.     
\end{equation}
The quantities $I_{i \leftarrow j}$ are related to the quantities ${\cal I}_{i \leftarrow j}$ by a refactorization formula analogous to Eq.~(\ref{refact}): 
\be
   \left[ {\cal I}_{q \leftarrow i}(z_1,x_T^2,\mu)\,\bar{{\cal I}}_{\bar q \leftarrow j}(z_2,x_T^2,\mu)
    \right]_{q^2}
   = \left( \frac{x_T^2 q^2}{b_0^2} \right)^{-F_{q\bar q}(x_T^2,\mu)}
  \!\! \!\! \!I_{q \leftarrow i}(z_1,x_T^2,\mu)\,I_{\bar q \leftarrow j}(z_2,x_T^2,\mu) \,. \label{eq:refacI}
\ee
By comparing Eq.~(\ref{eq:refacI}) with Eq.~(\ref{eq:BBren}), one finds the explicit expression for $I_{q \leftarrow q}$ at order $\alpha_s$, which is \cite{Becher:2010tm}
\be
I_{q \leftarrow q}(z,L_\perp,\alpha_s) = \delta(1-z) \left[ 1 + \frac{C_F \alpha_s}{4 \pi} \left( L_\perp^2 -\frac{\pi^2}{6}\right)\right] -
\frac{C_F \alpha_s}{2 \pi} \left[L_\perp \frac{1+z^2}{[1-z]_+} - (1-z)\right] + {\cal O} (\alpha_s^2) \,. \label{eq:Ias}
\ee

Neglecting power corrections of order $\Lambda_{\rm QCD}^2/q_T^2$, we can use the relation in Eq.~(\ref{eq:OPEB}) to express the differential cross section in Eq.~(\ref{eq:CSfact}) as a convolution of perturbative, factorized hard-scattering kernels 
\begin{equation}\label{Cdef}
\begin{aligned}
   C_{q\bar q\to ij}(z_1,z_2,q_T^2,M^2,\mu)
   &= \left| C_V(-M^2,\mu) \right|^2
    \frac{1}{4\pi} \int\!d^2x_\perp\,e^{-iq_\perp\cdot x_\perp}
    \left( \frac{x_T^2 M^2}{4e^{-2\gamma_E}} \right)^{-F_{q\bar q}(x_T^2,\mu)} \\
   &\quad\times I_{q\leftarrow i}(z_1,x_T^2,\mu)\,
    I_{\bar q\leftarrow j}(z_2,x_T^2,\mu)
\end{aligned}
\end{equation}
with ordinary PDFs. The result reads
\begin{eqnarray}\label{fact1}
   \frac{d^3\sigma}{dM^2\,dq_T^2\,dy} 
   &=& \frac{4\pi\alpha^2}{3N_c M^2 s}\,\sum_q\,e_q^2\,\sum_{i=q,g} \sum_{j=\bar q,g} 
    \int_{\xi_1}^1\!\frac{dz_1}{z_1} \int_{\xi_2}^1\!\frac{dz_2}{z_2} \\
   &&\mbox{}\times \bigg[ C_{q\bar q\to ij}(z_1,z_2,q_T^2,M^2,\mu)\,
    f_{i/N_1}(\xi_1/z_1,\mu)\, f_{j/N_2}(\xi_2/z_2,\mu) 
    + (q,i\leftrightarrow\bar q,j) \bigg] \,. \nonumber
\end{eqnarray}
This formula receives power corrections in the two small ratios $q_T^2/M^2$ and $\Lambda_{\rm QCD}^2/q_T^2$, which we do not  indicate explicitly. While the result looks different from the traditional CSS formula \cite{Collins:1984kg}, the two can nevertheless be shown to be equivalent. In \cite{Becher:2010tm} explicit relations between the ingredients in both approaches were derived. From these results, the three-loop coefficient $A^{(3)}$, the last missing ingredient for NNLL resummation in the CSS approach, was obtained.

\subsection{Transverse Momentum Spectra and the $q_T\to 0$ Limit}

\begin{figure}[t!]
\begin{center}
\begin{tabular}{rcr}
\includegraphics[width=0.455\textwidth]{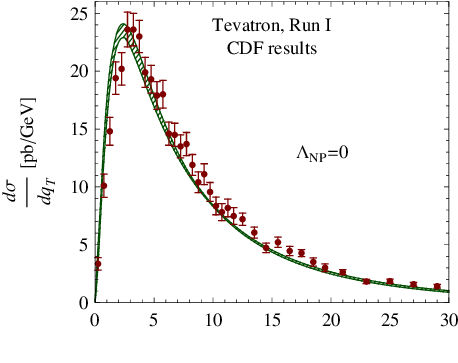} & &
\includegraphics[width=0.455\textwidth]{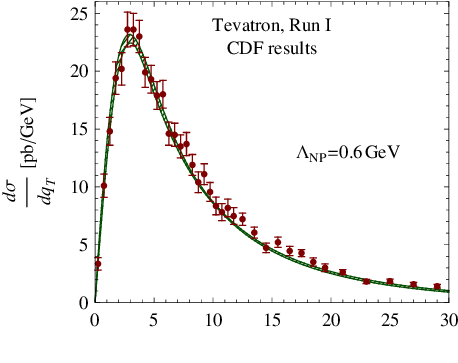} \\[-0.5cm]
\includegraphics[width=0.429\textwidth]{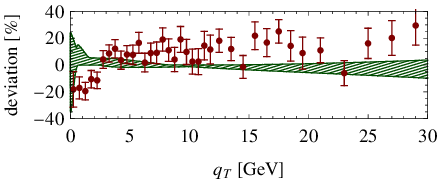} & &
\includegraphics[width=0.429\textwidth]{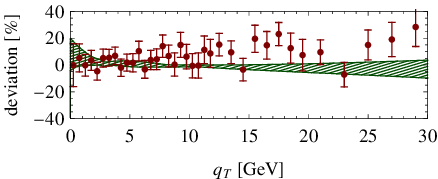} 
\end{tabular}
\end{center}
\vspace{-0.5cm}
\caption{\label{fig:qtSpectrum}
Comparison with Tevatron Run I data from {\sc CDF}, with and without long-distance corrections, taken from \cite{Becher:2011xn}. The lower panels show the deviation from the default theoretical prediction.}
\end{figure}

A detailed phenomenological analysis and comparison to data based on the factorization formula 
Eq.~(\ref{fact1}) was presented in \cite{Becher:2011xn}; here, we reproduce a plot from this reference in Figure \ref{fig:qtSpectrum}. With the factorized result and the known perturbative ingredients, it looks straightforward to obtain resummed predictions, but there is an interesting complication related to the Fourier integral at very small $q_T$. It can be understood by considering 
\begin{equation}\label{eq:Besselint}
K= \frac12 \int_0^\infty\!dx_T\,x_T\,J_0(x_T q_T)\,
    e^{-\eta L_\perp-\frac{a}{4} L_\perp^2} \,,
\end{equation}
which is the Fourier integral at NLL accuracy, where only the tree beam functions are needed, which are $\delta$-functions.\footnote{For reasons which are explained below, we also need to include the $\alpha_s L^2_\perp$ term of the $I_{q\leftarrow q}$ function in the limit $q_T\to 0$.} The Bessel function arises after rewriting $x_\perp \cdot q_\perp = - \cos \theta\, x_T \,q_T$ and integrating over the angle $\theta$. In order to obtain the exponential in the integrand of Eq.~(\ref{eq:Besselint}), one first rewrites  the anomaly in Eq.~(\ref{Cdef}) as
\be
\left( \frac{x_T^2 M^2}{b_0^2} \right)^{-F_{q\bar q}(x_T^2,\mu)}  = \exp \left[-F_{q\bar q}(x_T^2,\mu)  
 \ln \frac{M^2}{\mu^2}  -F_{q\bar q}(x_T^2,\mu) L_\perp\right] \, . \label{eq:anrew}
\ee
Then one replaces the factor $F_{q\bar q}$ multiplying $L_\perp$ in Eq.~(\ref{eq:anrew}) by its explicit expression which can be found in Eq.~(\ref{eq:Fqqbas}). In this way one obtains
\be \label{eq:anomexp}
\left( \frac{x_T^2 M^2}{b_0^2} \right)^{-F_{q\bar q}(x_T^2,\mu)}  = \exp \left[ -\eta L_\perp - \frac{1}{4} \frac{\alpha_s}{2 \pi} \left(2 \Gamma_0^F + \eta \beta_0\right) L_\perp^2 \right] \, ,
\ee
where $\Gamma_0^F \equiv C_F \gamma_0^{\cusp}$ and we introduced the quantity
\begin{equation}\label{etadef}
   \eta(M^2,\mu) = \frac{\alpha_s(\mu)}{4\pi}\,\Gamma_0^F \,\ln\frac{M^2}{\mu^2}\, .
\end{equation}
For low values of the scale $\mu$, the quantity $\eta$ counts as ${\mathcal O}(1)$ because the suppression by $\alpha_s$ is compensated by the large logarithm. Because of this, we have also included the two-loop term proportional to $\eta \beta_0$ in the $L_\perp^2$ terms in Eq.~(\ref{eq:anomexp}). To obtain this term, one can use the two-loop result for $F_{q\bar q}$ given in Eq.~(47) and Eq.~(50) in \cite{Becher:2010tm}. Alternatively, one can obtain this term by solving the one-loop RG equation for $F_{q\bar q}$ and expanding the result to two-loop order.
Finally, it is necessary  to include in the exponential in Eq.~(\ref{eq:Besselint}) also the terms proportional to $\alpha_s L_\perp^{2}$ coming from the functions $I$ in Eq.~(\ref{Cdef}), which can be read off Eq.~(\ref{eq:Ias}). After exponentiating these terms, one finally obtains the exponential in 
Eq.~(\ref{eq:Besselint}), where the terms proportional to $L_\perp^2$ are suppressed by the factor
\begin{equation}
a\equiv \frac{\alpha_s(\mu)}{2\pi}(\Gamma_0^F+\eta\beta_0) \,.
\end{equation}

For the scale choice $\mu \sim q_T \sim 1/x_T$, $L_\perp$ is small and it would thus appear that one can simply neglect the $L_\perp^2$ terms in Eq.~(\ref{eq:Besselint}). For not too small transverse momentum, this is indeed true: the integral over transverse separation is cut off by the oscillatory Fourier exponent, encoded in the Bessel function $J_0(x_T q_T)$ and the appropriate choice for $\mu$ is $\mu \sim q_T$. However, for $q_T\to 0$, the situation is different: in this case $J_0(x_T q_T) \to 1$ and the integration is cut off by the $L_\perp^2$ terms in the exponent. To see this, let us consider $q_T = 0$ and change integration variables from 
$x_\perp$ to $L_\perp$. The integral then takes the form
\begin{equation}\label{eq:Besselint2}
K= \frac{b_0^2}{4\mu^2} \int_{-\infty}^\infty\!dL_\perp
    e^{(1-\eta) L_\perp-\frac{a}{4} L_\perp^2} \,.
\end{equation}
This is a Gaussian integral with a peak at
\begin{equation}
L_\perp^{\rm peak} = \frac{2(1-\eta)}{a} \, ,
\end{equation}
and a width of order $1/\sqrt{a}$. The proper scale choice, which ensures that the logarithm $L_\perp$ is an order one quantity at the peak, is given by the condition $1-\eta = {\mathcal O}(\alpha_s)$. The solution of the equation $\eta =1$ defines a scale $\mu = q_*$. The equation is nontrivial, since $q_*$ occurs both as the argument of the logarithm and the coupling constant. Using the one-loop solution for the running coupling constant
\begin{equation}
\alpha_s(q_*) = \frac{\alpha_s(M)}{1+\frac{\alpha_s(M)}{4\pi} \beta_0 \ln(q_*^2/M^2)}\,,
\end{equation}
we obtain the approximate solution
\begin{equation}
q_* \approx M \exp\left( - \frac{2\pi}{\left(\Gamma_0^F+\beta_0\right)\alpha_s(M)}\right)\,.
\end{equation}
Numerically, for $M=M_Z$, one finds $q_* = 1.9 \,{\rm GeV}$. This value is quite low, but still in the short-distance domain. For Higgs production $\Gamma_0^F\to \Gamma_0^A = C_A \gamma_0^\text{\cusp}$, which leads to a higher value $q_* = 7.7 \,{\rm GeV}$. Since $q_*\sim M\,e^{-{\rm const}/\alpha_s(M)}$ the scale is non-perturbative, i.e.\ it cannot be obtained from a Taylor expansion at small coupling. The dynamic generation of a non-perturbative short-distance scale is quite remarkable. It arises due to the anomaly and shields the cross section from long-distance effects at very low $q_T$. This observation, that the spectrum can be computed at arbitrarily low $q_T$ with short-distance methods,  as long as the mass $M$ is large enough, was made already in \cite{Parisi:1979se}

The fact that the width of the Gaussian is of order $1/\sqrt{a}\sim 1/\sqrt{\alpha_s}$ implies that one has to count $L_\perp\sim 1/\sqrt{\alpha_s}$ at very low $q_T$ instead of order one. This explains why we included the $a L_\perp^2$ term in the exponent despite the fact that it only enters at NNLL accuracy for higher values of $q_T$. Since the counting changes at very low $q_T$, one has to reorganize the resummed result in this region. This reorganization was derived in \cite{Becher:2011xn} and a result was given which has NNLL accuracy both at small and very small $q_T$. The reader interested in a detailed discussion should consult this reference. 

The results plotted in Figure \ref{fig:qtSpectrum} include the terms relevant at very small $q_T$. In addition, the plot on the right-hand side includes a simple model for long-distance effects. The model used in \cite{Becher:2011xn} was to multiply each beam function by a Gaussian $e^{-\Lambda_{\rm NP}^2 x_T^2}$, which cuts off the integration over transverse position and to then adjust the parameter $\Lambda_{\rm NP}$ such that the data is reproduced. Figure \ref{fig:qtSpectrum} shows that agreement with the data is obtained with $\Lambda_{\rm NP}\approx 0.6 \,{\rm GeV}$ and that the long-distance effects only affect the spectrum for $q_T$ values below a few GeV. A systematic study of long-distance effects in anomalous observables was performed recently in \cite{Becher:2013iya}. The upshot of this analysis is that like perturbative corrections also non-perturbative effects to such observables are enhanced by anomaly logarithms. The leading non-perturbative effects are obtained by adding a non-perturbative correction $-{\Lambda'}_{\rm NP}^2 x_T^2$ to the anomaly exponent. The anomaly thus predicts the functional form of the correction and one can further show that it is independent of the flavor of the incoming quarks, because it can be obtained from a soft function \cite{Becher:2013iya}. It will be interesting to study these nonperturbative effects in detail with new precise LHC data for $Z$-production, which is already available in preliminary form \cite{CMS:2014sma,Aad:2014xaa}.

%% file: 8_IRDivergences.tex
\section{$\mathbf{n}$-Jet Processes and IR Divergences of Gauge Theory Amplitudes\label{sec:LV}}

So far, we have only considered processes which involve large energy flows in two directions, such as the Sudakov form factor or the inclusive Drell-Yan cross section. However, many processes involve multiple directions of large energy flow. These include collider processes with several jets of energetic particles in the final state. In this section, we will discuss the effective theory relevant for observables which involve $n$ directions of large energy flow. For simplicity, we will use the term $n$-jet processes when referring to such observables even in cases where the energetic particles are not clustered by a jet algorithm. It is easy to guess the structure of SCET for the $n$-jet case: one will need a collinear field for each direction of large momentum, and the different collinear sectors interact by exchanging soft particles.

The simplest $n$-jet quantities are Green's functions involving large momentum transfers, with external momenta close to the mass shell. In the $2$-jet case, the corresponding quantity is the Sudakov form factor which we studied in detail in the first few sections of our text. We have analyzed the factorization properties of this form factor in Section \ref{sec:SCET_QCD} and the result  is shown pictorially in Figure \ref{fig:scalesep}. If there are four or more relevant directions one encounters an interesting complication; namely that the hard and the soft functions have nontrivial color structure. RG invariance then imposes constraints on the anomalous dimensions of the hard, jet and soft functions.

In the case of the off-shell Sudakov form factor, the hard function was given by the on-shell form factor. We show below that the hard functions relevant for the case of off-shell $n$-particle Green's functions are simply the on-shell amplitudes of QCD. We will see that constraints on the anomalous dimension of the hard function then translate onto constraints on the structure of the IR divergences of gauge theory amplitudes.
Such constraints are valid at all orders in perturbation theory and they were analyzed in a series of papers in the last few years both using SCET \cite{Becher:2009cu,Becher:2009qa,Ahrens:2012he}, as well as with traditional diagrammatic methods \cite{Gardi:2009qi,Dixon:2009ur,DelDuca:2011ae,DelDuca:2012qg}.
This led to the formulation of a simple ansatz for the structure of the singularities in massless gauge theories, which is consistent with these constraints and in agreement with all available perturbative results. Furthermore, a formula which allows one to predict the structure of the IR singularities  in presence of massive particles up to two-loop order was also obtained by using effective field theory methods \cite{Becher:2009kw}.
   
In the following, we first discuss 
the application of SCET to $n$-jet processes and to the factorization of $n$-point 
off-shell Green functions in massless gauge theories. Subsequently, we show that the
IR-singularities of   gauge theory on-shell amplitudes can be absorbed into a
multiplicative $\bm{Z}$ factor, whose RG equation is governed by an anomalous
dimension $\bm{\Gamma}$. In the case of massless gauge theories, it can be shown that, up to two loops, $\bm{\Gamma}$ involves only two parton correlations. It was also conjectured that this statement could be valid at all orders.
 Several constraints on the structure of $\mathbf{\Gamma}$
can be obtained from considerations related to soft-collinear factorization, collinear limits, and non-abelian exponentiation.
With these tools it is possible to analyze the structure of the IR singularities in massless gauge theory amplitudes up to three loops \cite{Becher:2009cu,Becher:2009qa}. The results discussed in this section allow one to resum higher order logarithmic corrections in $n$-jet processes.

We  conclude the section by considering the case of gauge theories with massive particles; in this case it is possible to predict the structure of the IR divergences in a generic amplitude up to two loops.
Also the latter result allows one to implement the resummation of higher order logarithmic corrections in several collider processes with massive particles in the final state up to NNLL accuracy.

\subsection{Massless Amplitudes as Wilson Coefficients in SCET}

Consider a $n$-point off-shell Green's function in the limit in which the external momenta squared $p_i^2$ are small but the Mandelstam invariants 
\be \label{eq:Mandelstam}
s_{ij}  = 2 \sigma_{ij} \,  p_i\cdot p_j\, ,\quad (i\neq j)\, , 
\ee
are large; $p_k^2 \ll |s_{ij}|$. 
In Eq.~(\ref{eq:Mandelstam}), the sign factor $\sigma$ is given by $\sigma_{ij}=+1$ if both the momenta $p_i$ and $p_j$ are incoming or outgoing, while $\sigma_{ij}=-1$ applies to cases in which one of the momenta is incoming and the other one is outgoing. 
For each of the incoming and outgoing momenta, we introduce the usual reference vectors 
\be
n_{i \mu} = (1, \hat{\mathbf{n}}_i) \, , \quad \mbox{and} \quad \bar{n}_{i \mu}= (1, -\hat{\mathbf{n}}_i) \, ,\quad \mbox{with} \quad n_i^2 = \bar{n}_i^2 =0 \, , \quad  n_i \cdot \bar{n}_i = 2 \,.
\ee
The unit vector $\hat{\mathbf{n}}_i$ points in the same direction as the three momentum $\mathbf{p}_i$.
In the following we restrict our discussion to the case of a massless 
Yang-Mills theory, and we refer to quark and gluon fields, since we are mainly interested in the application of what follows to the QCD case. 
However, the methods and procedures outlined here can be applied to any unbroken gauge theory.

In order to study an $n$-jet process in SCET, it is necessary to consider quark and gluon collinear fields for each of the $n$ collinear directions (the fields are indicated by $\xi_i$ and $A_i^\mu$, respectively), as well as quark and gluon soft fields $\psi_s$ and  $A_s^\mu$, which interact with the various collinear sectors.
The complete SCET Lagrangian needed to describe an $n$-jet process  includes  several copies of the collinear Lagrangian in Eq.~(\ref{eq:colllag}) (supplemented by the kinematic term for the corresponding collinear gluon field); each copy corresponds to a different collinear
sector. The various copies differ only in the reference vectors $n_i$.
Therefore, the fields $\chi$ or ${\mathcal A}$ carry an additional index $i$ labeling their collinear direction.
Overall the SCET Lagrangian is given by
\bea \label{eq:lagnc}
{\mathcal L}_{{\tiny \mbox{SCET}}} &=& \bar{\psi}_s i \Dsl_s \psi_s  -  \frac{1}{4} \left( F^ {s,a}_{\mu \nu}\right)^2   \nn \\
& &+ 
\sum_{i = 1}^n \Biggl\{  \bar{\xi}_i \frac{\nbsl_i}{2} \left[ i n_i \cdot
D_i + i \Dsl_{c i \perp} \frac{1}{i \bar{n}_i \cdot D_{c i}}i \Dsl_{c i \perp}  \right] \xi_i - \frac{1}{4} \left( F^{c i,a}_{\mu \nu}\right)^2 \Biggr\} \, .
\eea
We remind the reader that the definitions of the covariant derivatives and collinear field strength appearing in the above Lagrangian in terms of the collinear gluon fields $A_i$ and soft gluon fields $A_s$ can be found in Eq.~(\ref{eq:listcd})
and in Eqs.~(\ref{eq:fieldstr},\ref{eq:DforF}), respectively.
 Soft fields can mediate low-energy interactions between various collinear fields. As discussed in Section \ref{sec:scint}, soft-collinear interactions only arise via the small component of the collinear covariant derivatives, which are given by
\be
i n_i\cdot D_i = i n_i\cdot \partial + n_i\cdot A_i(x) + n_i \cdot A_s(x_-) \, ,
\ee
see Eq.~(\ref{eq:listcd}). The decoupling of the soft and collinear field is achieved by the usual field redefinition by a soft Wilson line given in Eq.~(\ref{eq:dectr}). We now deal with $n$ different collinear sectors, and one has to redefine the fields in each sector with the appropriate Wilson line. For the construction of the operators, we will work with the building blocks $\chi_i(x)$ and ${\mathcal A}_{i \perp}^\mu (x)$ introduced in Section \ref{subsec:gaugeinvblocks}. For these fields, the decoupling transformation takes the form
\bea \label{eq:tagDB}
\chi_{i}(x) &=& S_i(x_-) \chi^{(0)}_i (x) \, , \nn \\
\bar{\chi}_{i}(x) &=&  \bar{\chi}^{(0)}_i (x) S^\dagger_i(x_-)\, , \nn \\
{\mathcal A}_{i \perp}^\mu (x) &=& S_i (x_-) {\mathcal A}^{(0) \mu}_{i \perp}(x) S^\dagger_i (x_-) \, ,
\eea
where the soft Wilson lines $S_i$ are defined in the same way as in Eq.(\ref{eq:WLsc}):
\be
S_i(x) = \mathbf{P} \exp\left[ ig \int_{-\infty}^0 \!\!ds \, n_i 
\cdot A^a_s(x + s n_i) t^a \right] \, . \label{eq:softW}
\ee
Let us stress that $x_-$ in Eq.~(\ref{eq:softW}) refers to the minus component along the relevant direction $x_-^\mu\equiv \bar{n}_i \cdot x\, n_i^\mu/2$. A proof of the decoupling was provided in Eqs.~(\ref{eq:dectr}-\ref{eq:decoupling}) above, where the fields $\xi$ and $A_c$ were employed. 
After the decoupling transformation, soft interactions manifest themselves as Wilson lines in operators built from collinear fields. We stress the fact that the different components of the quark and gluon fields are redefined in the same way: the soft gluons are insensitive to the spin of the collinear particles.

We observe that the last of Eq.~(\ref{eq:tagDB}), which is written in terms of 
two Wilson lines involving the $\text{SU}(3)$ generators in the fundamental representation, can be rewritten in terms of a single Wilson line involving
the generators in the adjoint representation $(\mathbf{T}^a_{A})_{bc} = - i f^{abc}$. In fact
\be \label{eq:AcallsingleS}
S(x_-) ({\mathcal A}^{(0) \mu}_{ \perp}(x))^a t^a S^\dagger (x_-) = 
t^a \left(S_A(x_-)\right)^{ab} ({\mathcal A}^{(0) \mu}_{ \perp}(x))^b \, , 
\ee
where 
\be
(S_A(x))^{ab} = \mathbf{P} \exp\left[ ig \int_{-\infty}^0 \!\!ds \, n 
\cdot A^c_s(x + s n) \left(-i f^{abc}\right) \right] \, .
\ee
In its infinitesimal form, the relation in Eq.~(\ref{eq:AcallsingleS}) can be proven by expanding the Wilson lines in powers of $g$ and by then applying the commutation relation for the group generators.

In order to have a unified treatment of quarks and gluons, we need to introduce some notation and use the color space formalism \cite{Bassetto:1984ik,Catani:1996jh}; the basics of this formalism are briefly reviewed in Appendix~\ref{app:colorspace}.
A generic collinear field will be represented by $\left(\phi_i \right)_{a_i}^{\alpha_i} (x)$, where $a_i$ is a color index and $\alpha_i$ is a Dirac or Lorentz index. The soft interaction can then be decoupled from this field by the redefinition
\be \label{eq:unifrep}
\left(\phi_i \right)_{a_i}^{\alpha_i} (x) = \left[ \mathbf{S}_i (x_-) \right]_{a_i b_i}
\left(\phi^{(0)}_i \right)_{b_i}^{\alpha_i} (x) \, .
\ee
The soft Wilson lines $\mathbf{S}_i$ are matrix-valued and defined as 
\be \label{eq:Smathbf}
\mathbf{S}_i(x) = \mathbf{P} \exp\left[ ig \int_{-\infty}^0 \!\!ds \, n_i 
\cdot A^a_s(x + s n_i) \mathbf{T}_i^a \right] \, ,
\ee
where $\mathbf{T}_i^a$ is the $a$-th color generator in the representation appropriate for the $i$th parton.
Also conjugate quark fields are treated according to this rule:  in this case $\mathbf{T}_i^a = -
(t^a)^T$, which translates in anti-path ordering in Eq.~(\ref{eq:Smathbf}).

Having discussed the Lagrangian, let us now turn to the construction of the operators. We want to describe $n$-jet processes, i.e.\ operators which involve energetic particles along $n$ different directions. As a consequence, the operator needs to contain at least one collinear field in each direction. Since additional soft fields in the SCET operators would lead to power
suppression, the leading order $n$-jet operators are built out of exactly $n$ collinear fields,
one for each direction of  large energy flow. Allowing for the most general operators, we can write the effective Hamiltonian  in the form
\be \label{eq:Heff}
{\mathcal H}_n^{\mathrm{eff}}= \int dt_1 \cdots dt_n\,{{\mathcal C}}_{a_1 \cdots a_n}^{\alpha_1 \cdots \alpha_n} \left(t_1, \cdots, t_n, \mu  \right)
\left(\phi_1 \right)^{\alpha_1}_{a_1} (x + t_1 \bar{n}_1)\cdots
\left(\phi_n \right)^{\alpha_n}_{a_n} (x + t_n \bar{n}_n) \, .
\ee
The Hamiltonian must be a color-neutral Lorentz scalar, so the coefficients must fulfill certain relations to ensure that this is the case. In fact, one  usually only writes down operators which are color-neutral Lorentz scalars. Writing the Hamiltonian  as in Eq.~(\ref{eq:Heff}) is highly redundant, but allows us to immediately rewrite the Hamiltonian in color-space notation as
\be
{\mathcal H}_n^{\mathrm{eff}}= \int dt_1 \cdots dt_n \langle\, O_n(\{\underline{t}\}) | {{\mathcal C}}_n(\{\underline{t}\},\mu) \rangle \, ,
\ee
where $\{\underline{t}\}$ is the set formed by the $n$ variables $t_i$.
The bra and ket states are related to the fields and Wilson coefficients through
the relations
\bea
\langle O_n(\{\underline{t}\}) | \{\underline{\alpha}\},\{\underline{a}\} \rangle &=&
\left(\phi_1 \right)^{\alpha_1}_{a_1} (x + t_1 \bar{n}_1)\cdots
\left(\phi_n \right)^{\alpha_n}_{a_n} (x + t_n \bar{n}_n) \, , \nn \\
\langle \{\underline{\alpha}\},\{\underline{a}\} | {{\mathcal C}}(\{\underline{t}\},\mu)\rangle &=& {{\mathcal C}}_{a_1 \cdots a_n}^{\alpha_1 \cdots \alpha_n} \left(t_1, \cdots, t_n, \mu  \right) \, ,
\eea
where the vectors $| \{\underline{\alpha}\},\{\underline{a}\} \rangle$ form a complete and orthonormal basis in the color and spin space.
When computing physical quantities in the effective theory, we will take matrix elements of the effective Hamiltonian, which we write symbolically as $\langle{\mathcal H}_n^{\mathrm{eff}} \rangle$. These are in turn obtained from the matrix elements of the effective  $n$-jet operators $\langle O_n(\mu) \rangle$, which are scale dependent because they need renormalization. In order for the physical quantities to be scale independent, their scale dependence must thus cancel against the scale dependence of the Wilson coefficients. This leads to the following equation
\begin{align}
\mu \frac{d}{d \mu} \Big\langle{\mathcal H}_n^{\mathrm{eff}} \Big\rangle = \mu \frac{d}{d \mu} \left(\int dt_1 \cdots dt_n \Big\langle\, \langle\, O_n(\{\underline{t}\},\mu) | {{\mathcal C}}_n(\{\underline{t}\},\mu)\rangle\,\Big\rangle\right) = 0\, .
\end{align}
This equation implies relations between the anomalous dimensions of the hard function and the anomalous dimensions of the SCET operators, which arise from soft and collinear interactions. To make the structure of the soft interactions explicit, let us apply the decoupling transformation to the operators.
The transformation produces a soft Wilson line for each collinear direction, after which the operator has the form
\be
{\mathcal H}_n^{\mathrm{eff}}= \int dt_1 \cdots dt_n \langle O^{(0)}_n(\{\underline{t}\}) |\mathbf{S}_1(0) \cdots \mathbf{S}_n(0) | {{\mathcal C}}(\{\underline{t}\},\mu) \rangle \, ,
\ee
where the operator $O^{(0)}_n$ has the same form as $O_n$, but is formed from the decoupled fields.
With this Hamiltonian, we can now compute off-shell Green's functions of collinear fields in the effective theory. After the decoupling transformation the different sectors no longer interact and when, as in our case, all external fields are collinear, the soft function corresponds to the vacuum matrix element
\be
{\mathcal S} (\{\underline{n}\},\mu) = \langle 0 |\mathbf{S}_1(0) \cdots \mathbf{S}_n(0) | 0 \rangle \, .
\ee
The collinear matrix elements yield a jet function for each direction, while the hard-scattering corrections are encoded in the Wilson coefficient $| {{\mathcal C}}(\{\underline{t}\},\mu) \rangle$. So we have factorized the Green's function into hard, jet and soft contributions. Using diagrammatic methods, the same result was obtained in \cite{Sen:1982bt,Kidonakis:1998nf}. 

Before exploring the consequences of the factorization, we now show that on-shell amplitudes are directly related to the Wilson coefficients $| {{\mathcal C}}(\{\underline{t}\},\mu) \rangle$. In order to determine the Wilson coefficients, we need to perform a matching computation, which amounts to a computation of the same quantity in the full and the effective theory. The simplest possibility is to calculate $n$-particle
on-shell amplitudes both in QCD and in SCET.
We use the color-space notation and denote the $n$-particle amplitudes by 
\begin{equation}
| {\mathcal M}_n (\varepsilon, \{\underline{p}\}) \rangle, \nonumber 
\end{equation}
where we have indicated explicitly the dependence on the regulator $d=4-2\varepsilon$.
Since the amplitudes are on-shell (i.e. $p_i^2 = 0$), all loop corrections in the
effective theory vanish; they consist of soft and collinear integrals, which become scaleless when $p_i^2$ is set to zero, see Section~\ref{sec:scalarSCET}. The on-shell matrix elements in the effective theory are thus given by their tree-level values. The latter are products of spinors and polarization vectors
which are in turn defined by the relations
\bea
\langle 0 | \left( \chi_j \right)^\alpha_a (t_j \bar{n}_j)|p_i; a_i, s_i  \rangle &=& \delta_{ij} \delta_{a_i a} e^{-i t_i \bar{n}_i \cdot p_i} u_\alpha(p_i, s_i) \, , \nn \\
\langle 0 | \left( A_{j \perp} \right)^a_\mu (t_j \bar{n}_j)|p_i; a_i, s_i  \rangle &=& \delta_{ij} \delta_{a_i a} e^{-i t_i \bar{n}_i \cdot p_i} \epsilon_\mu (p_i, s_i) \, . 
\eea
Because of the exponential factors, the integrals over $t_i$ produce the Fourier transform of the Wilson coefficient. The matching requirement, which states that the amplitudes in the full and effective theory must agree, thus yields the relation
\begin{equation}\label{eq:ampvswilson}
| {{\mathcal M}}_n (\varepsilon, \{\underline{p}\}) \rangle = | \tilde{{\mathcal C}}_n (\varepsilon,\{\underline{p}\}) \rangle \times (\text{"spinors and polarization vectors"})\,.
\end{equation}
The Fourier transformed bare Wilson coefficients $\tilde{{\mathcal C}}_n\left(\varepsilon,\{\underline{p} \} \right)$ depend on the large momentum components $\bar{n}_i \cdot p_i$, or equivalently, on the large momenta transfered $s_{ij}$ since
\be
s_{ij} = 2 \sigma_{ij} p_i \cdot p_j = \frac{1}{2} \sigma_{ij} n_i \cdot n_j
\bar{n}_i \cdot p_i \bar{n}_j \cdot p_j + {\mathcal O}(\lambda) \,,
\ee
where the last equality follows from the fact that the collinear momenta can be written as
\be
p_i^\mu = E_i n_i^\mu + {\mathcal O}(\lambda) = \bar{n}_i \cdot p_i \frac{n_i^\mu}{2} + {\mathcal O}(\lambda) \, .
\ee

The on-shell amplitudes on the left side of the Eq.~(\ref{eq:ampvswilson}) suffer from infrared singularities, which are regularized dimensionally. In contrast, the Wilson coefficients have ultra-violet divergences, which are also regularized by keeping $d\neq 4$. According to Eq.~(\ref{eq:ampvswilson}), these singularities must be equal: the
residual IR divergences in the on-shell amplitudes are identical to the ultraviolet (UV)
divergences in the Wilson coefficient. The equality comes about since the (vanishing) on-shell loop integrals in the effective theory suffer from both types of singularities. Schematically, the situation can be summarized by the
following relation:
\be
\underbrace{\frac{1}{\varepsilon_{\textrm{IR}}}}_{\textrm{on-shell amplitude}} = \underbrace{\frac{1}{\varepsilon_{\textrm{UV}}}}_{\textrm{Wilson coeff.}}  + 
\underbrace{\left( \frac{1}{\varepsilon_{\textrm{IR}}} - \frac{1}{\varepsilon_{\textrm{UV}}} \right)}_{\textrm{soft and coll. loop integrals}} \, .
\ee

\subsection{Renormalization}

From the discussion above we conclude that, up to a factor depending on spinors and polarization vectors, on-shell amplitudes in QCD coincide with bare Wilson coefficients of $n$-jets operators in SCET. The UV singularities in the Wilson coefficients can be subtracted by means of a multiplicative renormalization factor $\mathbf{Z}$, which is a matrix in color space \cite{Becher:2009cu,Becher:2009qa}.
The finite renormalized Wilson coefficient for the $n$-jet operator can be obtained through the relation
\be \label{eq:renCn}
| \tilde{{\mathcal C}}_n (\{\underline{p}\}, \mu) \rangle = \lim_{\varepsilon \to 0}
\mathbf{Z}^{-1} (\varepsilon, \{\underline{p}\}, \mu )| \tilde{{\mathcal C}}_n(\varepsilon, \{\underline{p}\}) \rangle \, ,
\ee
Because of the relation (\ref{eq:ampvswilson}), the same $\mathbf{Z}$ also makes the scattering amplitudes finite. We conclude that the IR singularities can be removed by a multiplicative factor and the structure of these singularities is governed by a renormalization group equation! The factor $\mathbf{Z}$ can be obtained starting from the RG equation satisfied by the Wilson coefficient. The RG equation can be written as
\be
\frac{d}{d \ln \mu} | \tilde{{\mathcal C}}_n (\{\underline{p}\}, \mu) \rangle = 
\mathbf{\Gamma}(\{\underline{p}\}, \mu) | \tilde{{\mathcal C}}_n (\{\underline{p}\}, \mu) \rangle \, ,
\ee
where $\mathbf{\Gamma}$ is the anomalous dimension, which is a matrix in color space. The anomalous dimension is related to the 
renormalization factor $\mathbf{Z}$ through
\be \label{eq:GammaZ}
\mathbf{\Gamma}(\{\underline{p}\}, \mu) = - \mathbf{Z}^{-1} (\varepsilon, \{\underline{p}\}, \mu ) \frac{d}{d \ln \mu}\mathbf{Z} (\varepsilon, \{\underline{p}\}, \mu ) \, .
\ee
The equation above simply follows from the fact that $| \tilde{{\mathcal C}}_n(\varepsilon, \{\underline{p}\})  \rangle$ in Eq.~(\ref{eq:renCn}) does not depend on the scale.
The Eq.~(\ref{eq:GammaZ}) can be formally inverted to obtain
\be \label{eq:invGZ}
\mathbf{Z} (\varepsilon, \{\underline{p}\}, \mu ) = \mathbf{P} \exp{\int_\mu^\infty \frac{d \mu'}{\mu'}\mathbf{\Gamma}(\{\underline{p}\}, \mu') } \, .
\ee
The validity of Eq.~(\ref{eq:invGZ}) can be proven by first expanding the exponential into a Taylor series and then following the same steps as in the Appendix \ref{app:Wil}, where we derive the differential equation for the Wilson lines, which are also a path-ordered exponential.

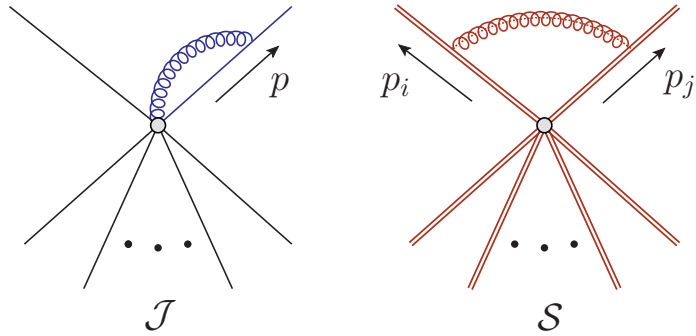
\begin{figure}[t]
\begin{center}
\vspace*{.5cm}
\[ 
\vcenter{ \hbox{
\begin{picture}(0,0)(0,0)
\SetScale{0.35}
\SetOffset(-140,0)
    \SetWidth{1.7}
    \SetColor{Blue}
    \GluonArc[clock](271.656,60.28)(80.597,-171.236,-285.356){7.5}{13}
    \SetWidth{1.7}
    \Line(192,47)(336,175)
        \SetColor{Black}
    \Line(192,47)(336,-81)
    \Line(192,47)(48,-81)
    \Line(192,47)(272,-129)
    \Line(192,47)(112,-129)
    \SetWidth{1.0}
    \Vertex(160,-81){4}
    \Vertex(160,-81){4}
    \Vertex(192,-85){4}
    \Vertex(224,-81){4}
    \SetWidth{1.9}
    \SetColor{Mahogany}
    \Line[double,sep=4.5](608,47)(528,-129)
    \Line[double,sep=4.5](608,47)(688,-129)
    \Line[double,sep=4.5](608,47)(752,175)
    \Line[double,sep=4.5](608,47)(448,175)
    \Line[double,sep=4.5](608,47)(752,-81)
    \Line[double,sep=4.5](608,47)(464,-81)
    \SetColor{Black}
    \SetWidth{1.0}
    \Vertex(576,-81){4}
    \Vertex(640,-81){4}
    \Vertex(608,-85){4}
    \SetWidth{1.7}
    \Line(192,47)(32,175)
    \GOval(192,47)(8,8)(0){0.882}
    \SetWidth{1.7}
    \SetColor{Mahogany}
    \GluonArc[clock](605.281,10.864)(151.145,128.592,52.64){7.5}{16}
        \Arc[dash,dashsize=2,clock](605,6.924)(156.089,127.954,52.046)
    \SetColor{Black}
    \SetWidth{1.7}
    \GOval(608,47)(8,8)(0){0.882}
    \SetWidth{1.7}
    \Line[arrow,arrowpos=1,arrowlength=12,arrowwidth=4.8,arrowinset=0.2](254,72)(317,128)
    \Text(110,27)[lb]{\Large{\Black{$p$}}}
    \Text(62,-61)[lb]{\Large{\Black{${\mathcal J}$}}}
    \Text(211,-61)[lb]{\Large{\Black{${\mathcal S}$}}}
    \Line[arrow,arrowpos=1,arrowlength=12,arrowwidth=4.8,arrowinset=0.2](671,71)(733,126)
    \Line[arrow,arrowpos=1,arrowlength=12,arrowwidth=4.8,arrowinset=0.2](530,73)(456,130)
    \Text(152,27)[lb]{\Large{\Black{$p_i$}}}
    \Text(259,27)[lb]{\Large{\Black{$p_j$}}}
\end{picture}}
}
\]
\vspace*{.8cm}
\end{center} 
\caption{SCET graphs contributing to the collinear and soft
functions ${\mathcal J}$ and ${\mathcal S}$. 
Solid lines denote collinear fields, dashed coily lines indicate soft fields, double lines indicate Wilson lines.
} 
\label{fig:JandSdiagrams}
\end{figure}

We can extract the anomalous dimension $\mathbf{\Gamma}$ by computing an infrared finite quantity in the effective theory. The simplest possibility is to consider off-shell Green's functions, for which the non-vanishing $p_i^2$'s screen the infrared singularities present in the on-shell case. The UV poles of the jet (for the quark and gluon case) and soft functions at order $\alpha_s$ are 
\bea \label{eq:JqJgS}
{\mathcal J}_q (p^2,\mu) &=& 1 +\frac{\alpha_s}{4 \pi} C_F \left( \frac{2}{\varepsilon^2} + \frac{2}{\varepsilon} \ln{\frac{\mu^2}{-p^2}} + \frac{3}{2 \varepsilon}\right) + {\mathcal O}(\varepsilon^0) \, , \nn \\
{\mathcal J}_g (p^2,\mu) &=&1 +\frac{\alpha_s}{4 \pi} 
\left[ C_A \left( \frac{2}{\varepsilon^2} + \frac{2}{\varepsilon}  \ln{\frac{\mu^2}{-p^2}} \right) + \frac{\beta_0}{2 \varepsilon}\right]
+ {\mathcal O}(\varepsilon^0) \, , \nn \\
\mathbf{{\mathcal S}}(\{\underline{p}\},\mu) &=&1 +\frac{\alpha_s}{4 \pi} \sum_{(i,j)}^n \frac{\mathbf{T}_i \cdot \mathbf{T}_j}{2}
\left(\frac{2}{\varepsilon^2} + \frac{2}{\varepsilon} \ln{\frac{- s_{ij} \mu^2}{(-p_i^2) (-p_j^2)}} \right) + {\mathcal O}(\varepsilon^0)\, .
\eea
The functions above are obtained by calculating the Feynman diagrams shown in 
Fig.~\ref{fig:JandSdiagrams}. Note that the field redefinitions in Eq.~(\ref{eq:tagDB}) change the off-shell behavior of the fields (while, of course,  they leave physical quantities, such as on-shell matrix elements, unchanged). Therefore, in order to compute the UV poles in Eq.~(\ref{eq:JqJgS}), one should employ the original non-decoupled SCET fields and Lagrangian.
The calculation of the relevant integrals is very similar to the calculation of the collinear- and soft-region integrals carried
out in Section~{\ref{sec:regions}} for the scalar theory.
%


%

The one-loop divergences of the complete effective theory $n$-particle matrix element can be obtained from Eqs.~(\ref{eq:JqJgS}) and are given by
\be \label{eq:SJJ}
\mathbf{{\mathcal S}}(\{\underline{p}\},\mu) \prod_{i=1}^n {\mathcal J}_i (p^2,\mu) = 1 -  \frac{\alpha_s}{4 \pi} \left[\sum_{\substack{i,j=1 \\ i\neq j}}^n \frac{\mathbf{T}_i \cdot \mathbf{T}_j}{2} \left(\frac{2}{\varepsilon^2} + \frac{2}{\varepsilon} \ln{\frac{\mu^2}{-s_{ij}}}\right) + \sum_i \frac{\gamma_0^i}{2 \varepsilon} + {\mathcal O}(\varepsilon^0)\right] \, ,
\ee
where $\gamma_0^q = -3 C_F$ and $\gamma_0^g = - \beta_0$. Observe that the off-shell momenta $p_i^2$ cancel from Eq.~(\ref{eq:SJJ}). This must be the case: 
One must be able to absorb the poles arising from the soft and jet functions in the unrenormalized Wilson coefficients, and the latter do not depend on the collinear momenta $p_i^2$. Consequently, the
renormalization factor $\bm{Z}$ and the associated anomalous dimension $\Gamma$ cannot depend on infrared scales. The one-loop anomalous dimension $\mathbf{\Gamma}_{0}$ can be directly extracted from the above result. It is given by minus twice the coefficient of the $1/\varepsilon$ terms in the above equation.

\subsection[A Conjecture for $\Gamma$]{A Conjecture for $\mathbf{\Gamma}$}

An all-order conjecture for the structure of $\mathbf{\Gamma}$ was proposed in 
\cite{Becher:2009cu}. The conjecture states that $\mathbf{\Gamma}$ has the following form
\be \label{eq:conj}
\mathbf{\Gamma} (\{\underline{p}\}, \mu)  = \sum_{(i,j)} \frac{\mathbf{T}_i \cdot \mathbf{T}_j}{2} \gamma_{\textrm{cusp}}(\alpha_s) \ln{\frac{\mu^2}{-s_{ij}}} + \sum_{i=1}^n \gamma_i (\alpha_s) \,.
\ee
The first sum on the r.h.s.\ of Eq.~(\ref{eq:conj}) runs over pairs $(i,j)$, ($i, j \in \{1,2,\cdots,n\}$ with $n$ the number of external legs) and excludes the case $i=j$. The factor $1/2$ in the cusp term takes care of the fact that we sum over both pairs $(1,2)$ and $(2,1)$, etc.
This dipole form is what we found in our one-loop calculation 
Eq.~(\ref{eq:SJJ}). The conjecture states that the same structure is valid also at higher order, and that the higher-order corrections only change the coefficients $\gamma_{\textrm{cusp}}(\alpha_s)$ and $\gamma_i (\alpha_s)$. For convenience, the explicit three-loop expressions of the anomalous dimensions appearing in Eq.~(\ref{eq:conj}) are collected in Appendix~\ref{app:AnDim}.

At one-loop level, it is trivial that only dipole terms can appear, 
since we obtain the one-loop corrections to the soft function by connecting two Wilson lines with a single gluon, as in Figure \ref{fig:JandSdiagrams}. At higher orders, we can connect several legs, and so one would expect that higher-order terms should appear, which would simultaneously involve the color charges of multiple legs. In a two-loop computation of the anomalous dimension of the soft function, it was observed that higher-order correlations do not appear at this order \cite{Aybat:2006wq,Aybat:2006mz}. What came as a surprise at the time is now understood as a consequence of the strong all-order constraints on the anomalous dimension, which we will discuss in detail below. 

The structure of the IR poles obtained by using Eq.~(\ref{eq:conj}) agrees with all perturbative results for scattering amplitudes to date. In particular, it agrees with the IR poles found in 
\begin{itemize}
\item the three-loop quark and gluon form factors \cite{Moch:2005tm}, which determine $\gamma_{\textrm{cusp}}(\alpha_s)$ as well as the functions $\gamma_i(\alpha_s)$ for quarks and gluons up to three-loop order in the expansion in $\alpha_s$ \cite{Becher:2009cu,Becher:2009qa},
\item the two-loop three-jet $V \to q q g$ amplitude \cite{Garland:2001tf,Garland:2002ak},
\item the two-loop four-jet amplitudes \cite{Anastasiou:2000kg, Anastasiou:2000ue,Anastasiou:2001sv, Bern:2002tk, Bern:2003ck},
\item the three-loop four-jet amplitudes in $N=4$ supersymmetric Yang-Mills theory in the planar limit \cite{Bern:2005iz}. 
\end{itemize}
While it is reassuring that the conjecture agrees with these results, it is also clear that they do not provide a strong test, since the two-loop form of the anomalous dimension follows from factorization constraints and the above list does not include any higher-order results which are sensitive to the presence of multi-leg color structures (these would not be visible in the planar limit, since they are color suppressed, see below). We discuss the constraints on $\mathbf{\Gamma}$ in detail below, and find that they are not strong enough to exclude terms beyond the dipole formula Eq.~(\ref{eq:conj}). Given that they are not excluded, one should expect additional terms to be present unless there are additional constraints which are not yet known. Indeed, a recent paper claims evidence for the presence of four-loop terms which violate the conjecture, based on a computation in the Regge limit \cite{Caron-Huot:2013fea}.

Before discussing the constraints on the anomalous dimension, we derive the $\mathbf{Z}$-factor which follows from Eq.~(\ref{eq:conj}). The expression for $\mathbf{Z}$ in terms of  $\mathbf{\Gamma}$ was given in Eq.~(\ref{eq:invGZ}). To compute the perturbative expansion of $\mathbf{Z}$, we will change variables from $\mu$ to $\alpha(\mu)$ by using 
\begin{equation}
   \frac{d\alpha_s}{d\ln\mu} = \beta(\alpha_s,\ep)
   = \beta(\alpha_s) - 2\ep\,\alpha_s \, ,
\end{equation} 
exactly as we did when solving the RG equation for the hard function of the Sudakov form factor in Section \ref{sec:RGEsud}. The only difference is that we need to work with the $d$-dimensional $\beta$-function $\beta(\alpha_s,\ep)$. As was the case for the Sudakov form factor, the anomalous dimension has a logarithmic term proportional to the cusp anomalous dimension and a remainder. For the coefficient of the logarithmic term, we introduce the notation
\begin{equation}\label{eq:Gampr}
   \Gamma'(\alpha_s) 
   \equiv \frac{\partial}{\partial\ln\mu}\,
   \bm{\Gamma}(\{\underline{p}\},\mu,\alpha_s) 
   = - \gamma_{\rm cusp}(\alpha_s)\,\sum_i\,C_i \,.
\end{equation}
The quantity defined above does not depend on the momenta, and in order to obtain the last identity in Eq.~(\ref{eq:Gampr}) from the ansatz in Eq.~(\ref{eq:conj}) one needs to employ the relation
\begin{equation}\label{colorrel}
   \sum_{(i,j)}\,\mathbf{T}_i\cdot\mathbf{T}_j
   = - \sum_i\,\mathbf{T}_i^2 = - \sum_i\,C_i \,,
\end{equation}
which follows from color conservation.
Rewritten as an integral over the coupling, the solution of Eq.~(\ref{eq:GammaZ}) reads
\begin{equation}\label{eq:reslnZ}
   \ln\bm{Z}(\ep,\{\underline{p}\},\mu)
   = \int\limits_0^{\alpha_s} \frac{d\alpha}{\alpha}\,
    \frac{1}{2\ep-\beta(\alpha)/\alpha} 
    \Bigg[ \bm{\Gamma}(\{\underline{p}\},\mu,\alpha) 
    + \int\limits_0^\alpha \frac{d\alpha'}{\alpha'}\,
    \frac{\Gamma'(\alpha')}{2\ep-\beta(\alpha')/\alpha'}
    \Bigg] \,,
\end{equation}
as it can be easily checked by inserting Eq.~(\ref{eq:reslnZ}) in Eq.~(\ref{eq:GammaZ}). The outermost integral in Eq.~(\ref{eq:reslnZ}) runs from $\alpha_s(\infty)=0$ to $\alpha_s(\mu)$.

In this form one can easily obtain the perturbative expansion of $\bf{Z}$ by inserting the expansions the anomalous dimensions and $\beta$-function
\begin{equation}\label{eq:Gbexp}
   \bm{\Gamma} = \sum_{n=0}^\infty\,\bm{\Gamma}_n 
    \left( \frac{\alpha_s}{4\pi} \right)^{n+1} , \quad
   \Gamma' = \sum_{n=0}^\infty\,\Gamma'_n 
    \left( \frac{\alpha_s}{4\pi} \right)^{n+1} , \quad
   \beta = -2\alpha_s\,\sum_{n=0}^\infty\,\beta_n 
    \left( \frac{\alpha_s}{4\pi} \right)^{n+1} .
\end{equation}
in the Eq.~(\ref{eq:reslnZ}). A comprehensive list of all of the factors appearing in Eq.~(\ref{eq:Gbexp})
can be found in Appendix~A in \cite{Becher:2009qa} and in Appendix~\ref{app:AnDim} of the present work. The perturbative expansion of $\bf{Z}$ in powers of $\alpha_s$ up to terms of order $\alpha_s^4$ is given by
\begin{align}\label{eq:lnZ}
   \ln\bm{Z} 
   &= \frac{\alpha_s}{4\pi} 
    \left( \frac{\Gamma_0'}{4\ep^2}
    + \frac{\bm{\Gamma}_0}{2\ep} \right) 
    + \left( \frac{\alpha_s}{4\pi} \right)^2 \! 
    \left[ - \frac{3\beta_0\Gamma_0'}{16\ep^3} 
    + \frac{\Gamma_1'-4\beta_0\bm{\Gamma}_0}{16\ep^2}
    + \frac{\bm{\Gamma}_1}{4\ep} \right]\nn \\
   &\!\!\!\mbox{}+ \left( \frac{\alpha_s}{4\pi} \right)^3\!
    \Bigg[ \frac{11\beta_0^2\,\Gamma_0'}{72\ep^4}
    - \frac{5\beta_0\Gamma_1' + 8\beta_1\Gamma_0' 
            - 12\beta_0^2\,\bm{\Gamma}_0}{72\ep^3}
+\frac{\Gamma_2' - 6\beta_0\bm{\Gamma}_1 
                - 6\beta_1\bm{\Gamma}_0}{36\ep^2}
 + \frac{\bm{\Gamma}_2}{6\ep} \Bigg] 
    + {\cal O}(\alpha_s^4) . 
\end{align}
Note that the leading singular term in $\ln\bm{Z}$ at the $n$-th order in $\alpha_s$ in perturbation theory diverges as $1/\varepsilon^{n+1}$. The leading singularities in $\bm{Z}$, on the other hand, are of order $1/\varepsilon^{n}$

\subsection[Constraints on $\Gamma$]{Constraints on $\mathbf{\Gamma}$}

Let us now discuss the  considerations leading to the ansatz in Eq.~(\ref{eq:conj}). The anomalous dimension must fulfill a set of all-order constraints. The most important one arises from soft-collinear factorization. Since physical observables must be scale independent, SCET operators matrix elements should evolve in the same way as the hard matching coefficients (which correspond to the on-shell scattering amplitudes). Therefore, the anomalous dimensions of the matching coefficients must be the sum of collinear  and soft contributions  $\Gamma_c$ and $\bm{\Gamma}_s$. Schematically
\be \label{eq:s26}
\bm{\Gamma}(s_{ij}) = \bm{\Gamma}_s(\Lambda_{ij}) + \sum_{i} \Gamma_c^i(p_i^2) \bm{1}.
\ee
The arguments of the functions in Eq.~(\ref{eq:s26}) indicate that, while the l.h.s. can depend only on the ``hard'' scalar products $s_{ij} = 2 \sigma_{ij} p_i \cdot p_j$, the soft contribution will depend on $\Lambda^2_{ij} = (-p_i^2) (-p_j^2)/ (-s_{ij})$ and the collinear 
contribution on the individual (slightly off- shell) squared momenta $p_i^2$. Moreover, the collinear term on the r.h.s.\ of Eq.~(\ref{eq:s26}) must be diagonal in color space, since  collinear interactions cannot lead to correlations between different partons.
Consequently, {\em $i)$} the dependence on $p_i^2$ should cancel in the sum of the soft and collinear terms, and {\em $ii)$} $\bm{\Gamma}$ and $\bm{\Gamma}_s$ should have the same color structure. Further constraints arise from non-abelian exponentiation, and from the factorization of amplitudes in the collinear and in the Regge limits. We will now discuss each of these constraints in turn.

\subsubsection{Non-Abelian Exponentiation}

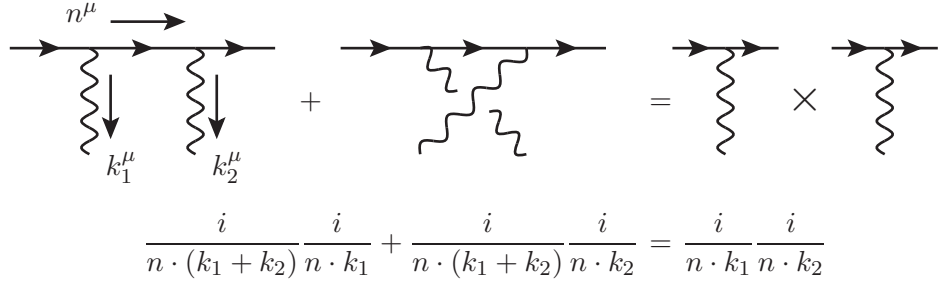
\begin{figure}[t!]
\vspace{10 mm}
\bea
\vcenter{ \hbox{
  \begin{picture}(0,0)(0,0)
\SetScale{1}
  \SetWidth{1}
  \SetOffset(0,-30)
  \ArrowLine(-50,50)(-20,50)
  \ArrowLine(-20,50)(20,50)
  \ArrowLine(20,50)(50,50)
  \LongArrow(28,40)(28,20)
  \LongArrow(-12,40)(-12,20)
  \Photon(-20,50)(-20,10){3}{4}
  \Photon(20,50)(20,10){3}{4}
\LongArrow(-12,60)(12,60)
\Text(-8,-1)[cb]{$k^\mu_1$}
\Text(32,-1)[cb]{$k^\mu_2$}
\Text(-23,60)[cb]{$n^\mu$}
\end{picture}}} 
\hspace*{2cm}
+
\hspace*{1.8cm}
\vcenter{ \hbox{
  \begin{picture}(0,0)(0,0)
\SetScale{1}
  \SetWidth{1}
  \SetOffset(0,-30)
  \ArrowLine(-50,50)(-20,50)
  \ArrowLine(-20,50)(20,50)
  \ArrowLine(20,50)(50,50)
  \Photon(-20,50)(-6,34){3}{2}
  \Photon(6,26)(20,10){3}{2}
  \Photon(20,50)(-20,10){3}{4}
\end{picture}}} 
\hspace*{2.2cm} &=& \hspace*{1.5cm}
\vcenter{ \hbox{
  \begin{picture}(0,0)(0,0)
\SetScale{1}
  \SetWidth{1}
  \SetOffset(0,-30)
  \ArrowLine(-50,50)(-30,50)
  \ArrowLine(-30,50)(-10,50)
  \ArrowLine(10,50)(30,50)
  \ArrowLine(30,50)(50,50)
  \Photon(-30,50)(-30,10){3}{4}
  \Photon(30,50)(30,10){3}{4}
\Text(0,25)[cb]{{\LARGE $\times$}}
\end{picture}}} \nn \\
& & \nn \\
& & \nn \\
\frac{i}{n \cdot \left(k_1 +k_2\right)}\frac{i}{n \cdot k_1} +
\frac{i}{n \cdot \left(k_1 +k_2\right)}\frac{i}{n \cdot k_2}  &=&
\frac{i}{n \cdot k_1} \frac{i}{n \cdot k_2} \nn
\eea
\caption{Diagrammatic relation between products of eikonal propagators in QED. $n^\mu$ identifies the direction along which the particle emitting soft photons is moving.
 \label{fig:eikid}}
\end{figure}

In QED, the identities satisfied by  eikonal propagators, such as
the one shown in Fig.~\ref{fig:eikid},
can be used to prove that the soft function exponentiates.\footnote{This simple exponentiation only holds at energies below the electron mass, i.e.\ after integrating out the massive fermions.}
%
%
Therefore, in QED the soft function, which is a matrix element of Wilson lines, can be written as 
\be \label{eq:Snop}
{\mathcal S}\left(\{\underline{n}\}, \mu \right) = 
\langle 0|\bm{S}_1(0)\cdots \bm{S}_n(0) |0\rangle = \exp{\left[\tilde{{\mathcal S}}\left(\{\underline{n}\}, \mu \right) \right]} \, .
\ee
The exponent $\tilde{{\mathcal S}}$ does not receive higher order corrections.
Therefore, the expression for the divergent part of $\tilde{{\mathcal S}}$ in QED will be of the form
\be
\tilde{{\mathcal S}}\left(\{\underline{n}\}, \mu \right) = \frac{\alpha_{\mbox{{\tiny QED}}}}{4 \pi} \sum_{(i,j)}^n
\frac{Q_i Q_j}{2} \left(\frac{2}{\ep^2} + \frac{2}{\ep} \ln \frac{-s_{ij} \mu^2}{(-p_i^2) 
(-p_j^2)}\right) \, ,
\ee
where $Q_i$ is the electric charge of the $i$-th external particle. 

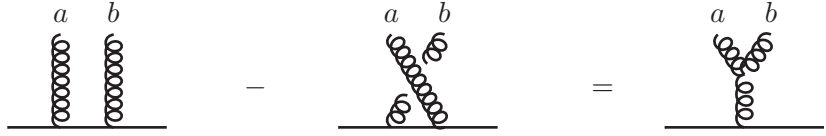
\begin{figure}[t!]
\vspace{10 mm}
\[
\vcenter{ \hbox{
  \begin{picture}(0,0)(0,0)
\SetScale{1}
  \SetWidth{1}
  \Gluon(-10,-15)(-10,20){3}{7}
  \Gluon(10,-15)(10,20){3}{7}
  \Line(-30,-15)(30,-15)
\Text(-10,25)[cb]{$a$}
\Text(10,25)[cb]{$b$}
\end{picture}}} 
\hspace*{2cm}
-
\hspace*{1.8cm}
\vcenter{ \hbox{
  \begin{picture}(0,0)(0,0)
\SetScale{1}
  \SetWidth{1}
  
  \Gluon(-10,-15)(-4,-3){3}{2.5}
  \Gluon(3.5,9)(10,20){3}{2.5}
  \Gluon(10,-15)(-10,20){3}{9}
  \Line(-30,-15)(30,-15)
\Text(-10,25)[cb]{$a$}
\Text(10,25)[cb]{$b$}
\end{picture}}} 
\hspace*{2.2cm} = \hspace*{1.5cm}
\vcenter{ \hbox{
  \begin{picture}(0,0)(0,0)
\SetScale{1}
  \SetWidth{1}
  \Gluon(0,-15)(0,5){3}{3}
  \Gluon(0,5)(-10,20){3}{3}
  \Gluon(0,5)(10,20){3}{3}
  \Line(-30,-15)(30,-15)
\Text(-10,25)[cb]{$a$}
\Text(10,25)[cb]{$b$}
\end{picture}}} 
\]
\caption{Diagrammatic form of the Lie commutator relation in the case in which the particle emitting soft radiation is a quark.
 \label{fig:comm}}
\end{figure}

\begin{figure}[t]
\vspace{10 mm}
\[
\vcenter{ \hbox{
  \begin{picture}(0,0)(0,0)
\SetScale{1}
  \SetWidth{1}
  \Gluon(-25,20)(-12,-2){3}{3.5}
  \Gluon(-5,-11)(0,-20){3}{1.5}
  \Gluon(0,20)(5,11){3}{1.5}
  \Gluon(12,2)(25,-20){3}{3.5}
  \Gluon(-25,-20)(25,20){3}{9.5}
  \Line(-40,20)(40,20)
    \Line(-40,22)(40,22)
  \Line(-40,-20)(40,-20)
    \Line(-40,-18)(40,-18)
\end{picture}}} 
\hspace*{2cm}
=
\hspace*{1.8cm}
\vcenter{ \hbox{
  \begin{picture}(0,0)(0,0)
\SetScale{1}
  \SetWidth{1}
  
  \Gluon(-25,20)(-25,-20){3}{6.5}
  \Gluon(0,20)(0,-20){3}{6.5}
  \Gluon(25,20)(25,-20){3}{6.5}
      \Line(-40,22)(40,22)
  \Line(-40,20)(40,20)
  \Line(-40,-20)(40,-20)
      \Line(-40,-18)(40,-18)
\end{picture}}} 
\hspace*{2.2cm} + \hspace*{2.2cm}
\vcenter{ \hbox{
  \begin{picture}(0,0)(0,0)
\SetScale{1}
  \SetWidth{1}
  \CBoxc(0,-10)(125,80){Lavender}{Lavender}
  \SetColor{Black}
   \Gluon(-25,-0)(-25,20){3}{3}
   \Gluon(-25,-20)(-25,0){3}{3}
  \Gluon(-25,0)(25,0){3}{9.5}
  \Gluon(25,20)(25,-0){3}{3}
  \Gluon(25,0)(25,-20){3}{3}
      \Line(-40,22)(40,22)
  \Line(-40,20)(40,20)
  \Line(-40,-20)(40,-20)
      \Line(-40,-18)(40,-18)
 
\Text(0,-35)[cb]{{\bf Single connected web}} 
\Text(0,-45)[cb]{ ``maximally non abelian''} 
\end{picture}}} 
\]
\vspace*{1.5cm}
\[
\hspace*{2.3cm}
-
\hspace*{1.8cm}
\vcenter{ \hbox{
  \begin{picture}(0,0)(0,0)
\SetScale{1}
  \SetWidth{1}
  
  \Gluon(-12.5,0)(-25,20){3}{3.5}
  \Gluon(0,20)(-12.5,0){3}{3.5}
  \Gluon(-12.5,-20)(-12.5,0){3}{3.5}
  \Gluon(25,20)(25,-20){3}{6.5}
      \Line(-40,22)(40,22)
  \Line(-40,20)(40,20)
  \Line(-40,-20)(40,-20)
      \Line(-40,-18)(40,-18)
\end{picture}}} 
\hspace*{2.2cm} - \hspace*{2.2cm}
\vcenter{ \hbox{
  \begin{picture}(0,0)(0,0)
\SetScale{1}
  \SetWidth{1}
  
  \Gluon(25,20)(12.5,0){3}{3.5}
  \Gluon(12.5,0)(0,20){3}{3.5}
  \Gluon(12.5,-20)(12.5,0){3}{3.5}
  \Gluon(-25,20)(-25,-20){3}{6.5}
      \Line(-40,22)(40,22)
  \Line(-40,20)(40,20)
  \Line(-40,-20)(40,-20)
      \Line(-40,-18)(40,-18)
\end{picture}}} 
\]

\vspace*{4mm}
\caption{Decomposition of a web in a sum of products of connected webs.\label{fig:connweb}}
\end{figure}
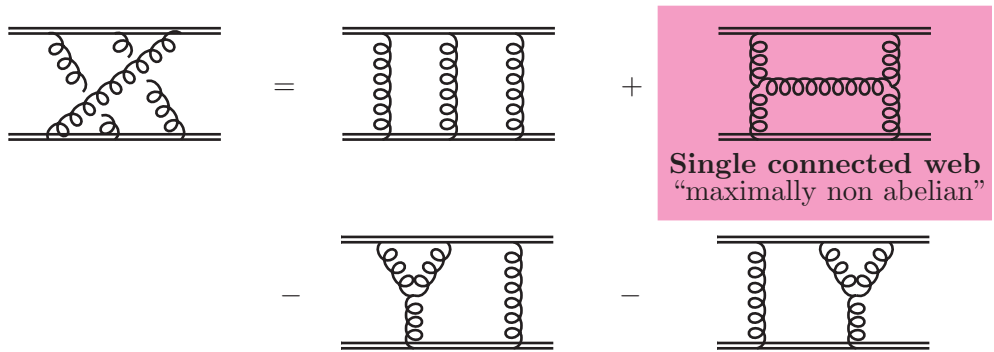

In QCD the situation is more complicated because the color matrices which appear in the quark-gluon vertices do not commute. However, it is possible to prove that, in the QCD case,  only Feynman diagrams with special color structure give corrections to the exponent $\tilde{{\mathcal S}}$. The corresponding color weights are called {\em color connected} or {\em maximally non abelian} \cite{Gatheral:1983cz, Frenkel:1984pz}. To define what they are, it is simplest to represent the color structure of a given diagram in diagrammatic form, in which the Lie commutator relation takes the form shown in Fig.~\ref{fig:comm}. The contributions to the exponent arise from diagrams in which a single connected web (i.e. a connected set of gluon lines, {\em not} counting crossed lines as being connected) is present. If one considers only the color factors in a Feynman diagram, it is possible to prove that by applying repeatedly the Lie commutator relation (see Fig.~\ref{fig:comm}) 
\be\label{eq:lie}
\mathbf{T}^a \mathbf{T}^b - \mathbf{T}^b \mathbf{T}^a = i f^{abc} \mathbf{T}^c \, ,
\ee
any ``web'' (i.e. a connected set of gluon lines, counting crossed lines as being connected) can be decomposed as a sum over products of connected webs.
 An example of this decomposition of a web is shown in Fig.~\ref{fig:connweb}. Only color structures corresponding to single connected webs contribute to the exponent $\tilde{{\mathcal S}}$ and therefore to the color structure of the soft anomalous dimension $\bm{\Gamma}_s$. If several Wilson lines are present in a soft function, the definitions in \cite{Gatheral:1983cz, Frenkel:1984pz} need to be slightly generalized. Instead of uncrossing lines, one will symmetrize the color generators arising on each leg in the form
\be
\mathbf{T}^{a_1} \mathbf{T}^{a_2} \ldots \mathbf{T}^{a_n} 
= \left(\mathbf{T}^{a_1}\,\mathbf{T}^{a_2} \ldots \mathbf{T}^{a_n}\right)_{\rm symm} + (\text{``commutator terms''})
\ee 
In the commutator terms one then uses the Lie commutator relation Eq.~(\ref{eq:lie}) to reduce the number of color generators and will then symmetrize again. After the symmetrization the distinction between crossed and uncrossed lines becomes irrelevant and one can directly obtain the diagrams as a sums over products of connected webs \cite{Becher:2009qa}. A detailed discussion of multiparton webs and their algebra can be found in the papers \cite{Mitov:2010rp,Gardi:2010rn,Gardi:2011wa,Gardi:2011yz}. 


\subsubsection{Soft-Collinear Factorization Constraints}

As stated above, the logarithms depending on the soft and collinear scales
should combine in order to give rise to logarithms of the hard scale. To regulate
collinear divergences we give a small off-shellness to the
external partons, so that the momentum of the external parton $i$ will satisfy
the relation $(-p_i^2)>0$. We then introduce the quantities
\be
\beta_{ij} = \ln \frac{-2 \sigma_{ij} p_i \cdot p_j \mu^2}{(-p_i^2) (-p_j^2)} \, ,
\qquad L_i = \ln \frac{\mu^2}{-p_i^2} \, .
\ee
The quantity $\beta_{ij}$ generalizes the definition of the cusp angle in Eq.~(\ref{eq:cuspangle}) to the case of light-like Wilson lines; in fact, for large values of the argument, $\text{arccosh}{x} \sim \ln (2 x)$. (We remind the reader that $\sigma_{ij} = 1$ if the partons $i,j$ are both incoming or outgoing, while $\sigma_{ij} = -1$ otherwise.) Then, the  collinear, soft, and hard logarithms satisfy the relation
\be
\underbrace{\beta_{ij}}_{\text{soft log}} = \underbrace{L_i + L_j}_{\text{collinear logs}} - \underbrace{\ln\frac{\mu^2}{-s_{ij}}}_{\text{hard log}} \, . \label{eq:bij2}
\ee 
It was already emphasized in Eq.~(\ref{eq:s26}) that the anomalous dimension matrix of $n$-jet SCET operators can be decomposed in soft and collinear pieces. Eq.~(\ref{eq:s26})  can be rewritten as
\be \label{eq:GsG}
\bm{\Gamma}\left(\{\underline{p}\},\mu \right) = \bm{\Gamma}_s \left(
\{\underline{\beta}\},\mu \right) + \sum_i \Gamma_c^i\left(L_i,\mu \right)
\bm{1}\, .
\ee
By employing Eq.~(\ref{eq:bij2}) it is possible to rewrite $\bm{\Gamma}_s$ as a function of $s_{ij}$ and $L_i$. By solving Eq.~(\ref{eq:GsG}) with respect to 
$\bm{\Gamma}_s$ one sees that $\bm{\Gamma}_s$ 
depends on $L_i$ only through $\Gamma_c$. Since the latter is known to be of the form
\be
\Gamma_c^i\left(L_i\right) = - \Gamma^i_{\text{cusp}}(\alpha_s) L_i + \gamma^i_c(\alpha_s) \, , 
\ee
$\bm{\Gamma}_s$ will satisfy the differential equation
\be \label{eq:diffeqGs}
\frac{\partial \bm{\Gamma}_s \left(\{\underline{s}\}, \{\underline{L}\}, \mu \right)}{ \partial L_i} = \Gamma^i_{\text{cusp}} (\alpha_s) \, . 
\ee
The differential equation satisfied by $\bm{\Gamma}_s$ suggests that the soft anomalous matrix should depend linearly on $\beta_{ij}$. The differential equation also restricts the kind of color structures which can appear in $\bm{\Gamma}_s$.

With four or more partons, it is possible to use exclusively soft logarithms to build conformal cross ratios, which do not depend on the small collinear squared momenta $p_i^2$ but only on the Mandelstam invariants $s_{ij}$ \cite{Gardi:2009qi}.
In fact, one has
\be \label{eq:concrossratio}
\beta_{ijkl} = \beta_{ij} +\beta_{kl} -\beta_{ik} -\beta_{jl} = \ln{\frac{(-s_{ij})(-s_{kl})}{(-s_{ik})(-s_{jl})}} \, .
\ee
The r.h.s.\ of the equation above shows explicitly that the argument of the logarithm is conformally invariant, i.e. it remains unchanged if all of the momenta $p_i$ are rescaled by the same factor.
A dependence of $\bm{\Gamma}_s$ on the ``conformal ratios'' $\beta_{ijkl}$ is not excluded by the constraint in Eq.~(\ref{eq:diffeqGs}).
However, any polynomial dependence on the conformal ratios can be excluded using other arguments, such as the consistency with collinear limits.

\subsubsection{Consistency with Collinear Limits \label{sec:concoll}}

When two partons become collinear, an $n$-point amplitude reduces to the product of an $(n-1)$-parton amplitude times a splitting amplitude \cite{Berends:1987me, Mangano:1990by, Bern:1995ix, Kosower:1999xi, Catani:2003vu}:
\be \label{eq:coll}
\left|{\mathcal M}_n\left(\{ p_1,p_2, \cdots, p_n \}\right) \right\rangle = 
 \text{{\bf Sp}}\left(\{p_1,p_2 \} \right)\left|{\mathcal M}_{n-1}\left(\{ P,p_3, \cdots, p_n \}\right) \right\rangle +\cdots \, ,
\ee
where $P= p_1+p_2$, with $p_1 \equiv z P$ and $p_2 \equiv (1-z) P$.
$\text{{\bf Sp}}$  is a matrix in color space, and it encodes the singular behavior of the $n$-point amplitude as $p_1$ and $p_2$ become collinear. The collinear factorization is valid in the limit $P^2 \to 0$, up to terms that are regular in the collinear limit, denoted by the ellipsis in Eq.~(\ref{eq:coll}).
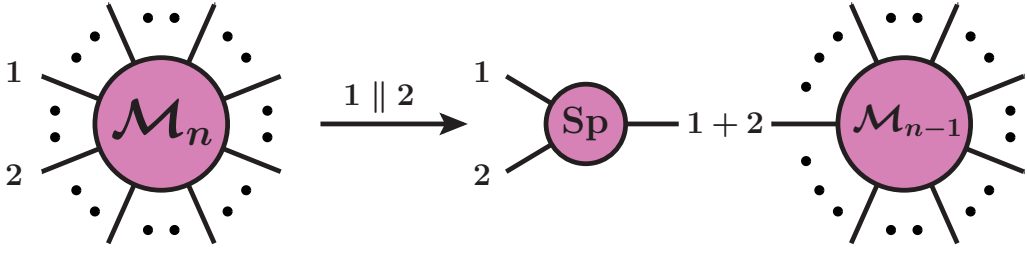
\begin{figure}
\vspace*{.5cm}
\begin{center}
\[ 
\vcenter{ \hbox{
  \begin{picture}(0,0)(0,0)
\SetScale{1}
  \SetWidth{1.8}
  \Line(-160,45)(-120,-45)
  \Line(-120,45)(-160,-45)
  \Line(-185,18)(-95,-18)
  \Line(-185,-18)(-95,18)
\GCirc(-180,5){1}{0}
\GCirc(-180,-5){1}{0}
\GCirc(-100,5){1}{0}
\GCirc(-100,-5){1}{0}
\GCirc(-135,40){1}{0}
\GCirc(-145,40){1}{0}
\GCirc(-135,-40){1}{0}
\GCirc(-145,-40){1}{0}
\GCirc(-165,33){1}{0}
\GCirc(-172,25){1}{0}
\GCirc(-165,-33){1}{0}
\GCirc(-172,-25){1}{0}
\GCirc(-115,-33){1}{0}
\GCirc(-108,-25){1}{0}
\GCirc(-115,33){1}{0}
\GCirc(-108,25){1}{0}
\CCirc(-140,0){25}{Black}{Thistle}
\Text(-142,-8)[cb]{ {\huge $\bm{{\mathcal M}_n}$}}
\Text(-198,16)[cb]{ {\large $\bm{1}$}}
\Text(-198,-23)[cb]{ {\large $\bm{2}$}}
\LongArrow(-80,0)(-30,0)
\Text(-60,4)[cb]{ {\large $\bm{1 \parallel 2}$}}

\Line(20,0)(-10,18)
\Line(20,0)(-10,-18)
\Line(20,0)(55,0)
\CCirc(20,0){15}{Black}{Thistle}
\Text(18,-6)[cb]{ {\Large $\bm{\mbox{Sp}}$}}
\Text(71,-5)[cb]{ {\large $\bm{1+2}$}}
\Text(-21,16)[cb]{ {\large $\bm{1}$}}
\Text(-21,-23)[cb]{ {\large $\bm{2}$}}

  \Line(160,45)(120,-45)
  \Line(120,45)(160,-45)
  \Line(185,18)(140,0)
  \Line(185,-18)(140,0)
  \Line(90,0)(140,0)
\GCirc(180,5){1}{0}
\GCirc(180,-5){1}{0}
\GCirc(103,14){1}{0}
\GCirc(103,-14){1}{0}
\GCirc(135,40){1}{0}
\GCirc(145,40){1}{0}
\GCirc(135,-40){1}{0}
\GCirc(145,-40){1}{0}
\GCirc(165,33){1}{0}
\GCirc(172,25){1}{0}
\GCirc(165,-33){1}{0}
\GCirc(172,-25){1}{0}
\GCirc(115,-33){1}{0}
\GCirc(108,-25){1}{0}
\GCirc(115,33){1}{0}
\GCirc(108,25){1}{0}
\CCirc(140,0){25}{Black}{Thistle}
\Text(140,-6)[cb]{ {\Large $\bm{{\mathcal M}_{n-1}}$}}

\end{picture}}
}
\]
\end{center}
\vspace*{10mm}
\caption{Schematic representation of the amplitude factorization in the collinear limit.
\label{fig:coll}}
\end{figure}
The relation in Eq.~(\ref{eq:coll}) is schematically shown in Fig.~\ref{fig:coll}, and it is valid both for regularized amplitudes
$|{\mathcal M}_n (\ep,\{p\} )\rangle$ and for minimally subtracted amplitudes
$|{\mathcal M}_n (\{p\},\mu )\rangle$. Since it is known that the divergences of a given amplitude can be renormalized by multiplying the bare amplitude by a $\bm{Z}$ factor, as it is shown in Eq.~(\ref{eq:renCn}), it is possible to derive the following constraint on the divergences of the splitting amplitude:
\be
\lim_{\ep \to 0} \bm{Z}^{-1}\left( \ep, \{p_1,\cdots,p_n \}\right)
\text{{\bf Sp}}\left(\ep,\{p_1,p_2\}\right) 
\bm{Z}\left( \ep, \{P,p_3,\cdots,p_n \}\right) = \text{{\bf Sp}}\left(\{p_1,p_2\},\mu\right) \, . 
\ee
The matrix $\text{{\bf Sp}}\left(\ep,\{p_1,p_2\}\right)$ does not depend on  the scale; therefore, as a consequence of Eq.~(\ref{eq:GammaZ}),  $\text{{\bf Sp}}\left(\{p_1,p_2\},\mu\right)$  satisfies the following RG equation \cite{Becher:2009qa}
\bea
\frac{d}{d \ln \mu} \text{{\bf Sp}} \left(\{p_1,p_2\}, \mu\right) &=& 
\bm{\Gamma}\left(\{p_1, \cdots, p_n \}, \mu \right) \text{{\bf Sp}} \left(\{p_1,p_2\}, \mu\right) \nn \\ & &
- \text{{\bf Sp}} \left(\{p_1,p_2\}, \mu\right) 
\bm{\Gamma}\left(\{P,p_3,\cdots,p_n\}, \mu \right) \, . \label{eq:diffSp}
\eea
Charge conservation implies that
\be \label{eq:commTp}
\left(\mathbf{T}_1 + \mathbf{T}_2 \right)
\text{{\bf Sp}} \left(\{p_1,p_2\}, \mu\right) 
= \text{{\bf Sp}} \left(\{p_1,p_2\}, \mu\right) \mathbf{T}_P \, ,
\ee 
where $\mathbf{T}_P$ is the color generator associated with the parent parton $P$. To see this, note that
\be
\left(\mathbf{T}_1 + \mathbf{T}_2 \right)\, \text{{\bf Sp}} \left(\{p_1,p_2\}, \mu\right) = -\sum_{i=3}^n \mathbf{T}_i \, \text{{\bf Sp}}\left(\{p_1,p_2\}, \mu\right) = \text{{\bf Sp}} \left(\{p_1,p_2\}, \mu\right) \mathbf{T}_P \,. \label{eq:T1T2TP}
\ee
For the second step in Eq.~(\ref{eq:T1T2TP}), one observes that the splitting amplitude is independent of the colors of the partons not involved in the splitting. One can thus commute the sum of the generators of the other partons to the right and then use color conservation in the $(n-1)$-parton space to replace it by $\mathbf{T}_P$.
Consequently, the differential equation (\ref{eq:diffSp}) can  be rewritten as
\be
\frac{d}{d \ln \mu} \text{{\bf Sp}} \left(\{p_1,p_2\}, \mu\right)  =
\bm{\Gamma}_{\text{Sp}} \left(\{p_1,p_2\},\mu \right) \text{{\bf Sp}} \left(\{p_1,p_2\}, \mu\right) \, ,
\ee 
with 
\be \label{eq:collcon}
\bm{\Gamma}_{\text{Sp}}\left(\{p_1,p_2 \} , \mu\right) = 
\bm{\Gamma}\left(\{p_1,\cdots,p_n \} , \mu\right) - 
\left.\bm{\Gamma}\left(\{P,p_3,\cdots,p_n \} , \mu\right)\right|_{\mathbf{T}_P \to \mathbf{T}_1+\bm{T_2}} \, . 
\ee
The matrix $\bm{\Gamma}_{\text{Sp}}$ must be independent from the momenta and colors of the partons  $p_3, \cdots, p_n$; the color dipole form 
suggested for $\bm{\Gamma}$ in Eq.~(\ref{eq:conj}) is consistent with Eq.~(\ref{eq:collcon}). This would not be the case if $\bm{\Gamma}$
would involve terms depending on higher powers of the color generators or momentum variables. 
The all order form of the anomalous dimension for the splitting amplitude is\cite{Becher:2009qa} 
\bea \label{eq:split}
\bm{\Gamma}_{\text{Sp}}\left(\{p_1,p_2 \} , \mu\right) &=& \gamma_{\cusp} \Bigl[ 
\mathbf{T}_1 \cdot \mathbf{T}_2 \ln{\frac{\mu^2}{-s_{12}}} +
\mathbf{T}_1 \cdot (\mathbf{T}_1 + \mathbf{T}_2) \ln{z} + 
\mathbf{T}_2 \cdot (\mathbf{T}_1 + \mathbf{T}_2) \ln{(1-z)}  
\Bigr] \nn \\ && + \gamma_{1} + \gamma_{2} - \gamma_P \, ,
\eea
where $\gamma_P$ is the anomalous dimension associated to the unresolved parton $P$. 
Equation~(\ref{eq:split}) can be obtained from the conjecture in Eq.~(\ref{eq:conj}) by using the commutativity of the color matrices acting on different partons to write
\bea
\bm{\Gamma}\left(\{p_1,\cdots,p_n \} , \mu\right) &=& \gamma_{\cusp} \Biggl[
 \mathbf{T}_1 \cdot \mathbf{T}_2 \ln{\left(\frac{\mu^2}{-s_{12}}\right)} +
 \sum^2_{i=1} \sum^n_{j=3} \mathbf{T}_i \cdot \mathbf{T}_j \ln{\left(\frac{\mu^2}{-s_{ij}}\right)} \nn \\
 & &+ \sum_{i=3}^{n-1}\sum_{j=i+1}^n \mathbf{T}_i \cdot \mathbf{T}_j \ln{\left(\frac{\mu^2}{-s_{ij}}\right)} 
 \Biggr] + \gamma_{1} + \gamma_{2} +\sum_{i=3}^n \gamma_{i}\, , \label{eq:Gspint1} \\
\bm{\Gamma}\left(\{P,p_3,\cdots,p_n \} , \mu\right)&=&\gamma_{\cusp} \Biggl[
\sum^n_{j=3} \mathbf{T}_P \cdot \mathbf{T}_j \ln{\left(\frac{\mu^2}{-s_{Pj}}\right)}+ \sum_{i=3}^{n-1}\sum_{j=i+1}^n \mathbf{T}_i \cdot \mathbf{T}_j \ln{\left(\frac{\mu^2}{-s_{ij}}\right)} \Biggr] \nn \\
& & + \gamma_P  +\sum_{i=3}^n \gamma_{i}\, . \label{eq:Gspint2} 
\eea
All of the terms not involving the partons labeled $1$, $2$, or the parent parton $P$ cancel trivially in the difference between Eqs.~(\ref{eq:Gspint1},
\ref{eq:Gspint2}). After replacing $\mathbf{T}_P = \mathbf{T}_1+\mathbf{T}_2$ one finds that the terms involving one of the collinear partons and one of the remaining partons combine according to
\bea
\mathbf{T}_1\cdot\mathbf{T}_j \left[\ln\left(\frac{\mu^2}{-s_{1j}}\right)  - 
\ln\left(\frac{\mu^2}{-s_{Pj}}\right)\right] &=& - \mathbf{T}_1\cdot\mathbf{T}_j \ln z \, , \\
\mathbf{T}_2\cdot\mathbf{T}_j \left[\ln\left(\frac{\mu^2}{-s_{2j}}\right)  - 
\ln\left(\frac{\mu^2}{-s_{Pj}}\right)\right] &=& - \mathbf{T}_2\cdot\mathbf{T}_j \ln (1-z) \, .
\eea
Finally, in order to recover Eq.~(\ref{eq:split}) it is sufficient to apply color conservation
\be
\mathbf{T}_i \cdot \sum_{j =3}^n \mathbf{T}_j = - \mathbf{T}_i \cdot \left( \mathbf{T}_1 + \mathbf{T}_2\right) \, . 
\ee

\subsubsection{Consistency with Regge Limits \label{sec:conRegge}}

Additional constraints on the structure of $\bm{\Gamma}$ can be obtained by considering the Regge limit of amplitudes \cite{Bret:2011xm,DelDuca:2011ae,DelDuca:2012qg}. We briefly summarize the features of Regge theory that are important for our discussion of infrared singularities in perturbative QCD. We start by considering a $2 \to 2$ partonic scattering process in massless QCD. In this context, the relevant Mandelstam invariants are $s = (p_1+p_2)^2$, $t = (p_1-p_3)^2$, and $u = (p_1-p_4)^2$, where $p_1, p_2$ are the incoming momenta and $p_3,p_4$ are the outgoing ones. These invariants are not independent but satisfy the relation $s+t+u = 0$. One then considers the forward scattering limit $s \gg -t$, $s \sim -u$. As expected, the presence of two different scales in this kinematic region induces large logarithms of the ratio $|t|/s$, which in turn spoil the convergence of the perturbative expansion. If certain conditions are satisfied, Regge theory implies that these large corrections can be resummed to all orders by replacing the tree level t-channel propagator of the particle responsible for the leading contribution to the high-energy limit according to the ``Reggeization'' prescription:
\be
\frac{1}{t} \rightarrow \frac{1}{t} \left(\frac{s}{-t} \right)^{\alpha(t)} \, ,
\label{eq:reggeans}
\ee 
where the exponent $\alpha(t)$ goes under the name of the \emph{Regge trajectory} of the particle involved in the $t$-channel process. In perturbative QCD one can prove that the ansatz in Eq.~(\ref{eq:reggeans}) does resum the large logarithms in forward scattering. For example the Reggeization of the gluon-gluon scattering amplitude was proven at LL accuracy in \cite{Balitsky:1979ap} and at NLL for the real part of the amplitude in \cite{Fadin:2006bj}. The Reggeization of quark gluon scattering at LL was considered in \cite{Bogdan:2006af}.

The Regge trajectory $\alpha$ is IR divergent since its calculation involves virtual corrections with soft gluons. Following the notation of \cite{DelDuca:2012qg}, its general structure can be written as
\be
\alpha(t,\ep) = \frac{\alpha_s(-t)}{4 \pi} \alpha^{(1)} +  \left(\frac{\alpha_s(-t)}{4 \pi}
\right)^2 \alpha^{(2)} + {\cal O}\left(\alpha_s^3 \right) \, , 
\ee
and 
\be
\alpha_s(-t) = \left(\frac{\mu^2}{-t}\right)^\ep \alpha_s(\mu^2) + {\cal O}(\alpha_s^3)\,.
\ee
For gluon $t$-channel exchanges, the coefficients $\alpha^{(1)}$ and $\alpha^{(2)}$ are
\begin{align}
\alpha^{(1)} = C_A \frac{\gamma^{\cusp}_0}{2 \ep} \, , \qquad
\alpha^{(2)} = C_A \left[-\frac{\beta_0}{\ep^2} +\frac{\gamma^{\cusp}_1}{4 \ep}
+ C_A \left(\frac{404}{27} - 2 \zeta_3 \right) -n_f \frac{56}{27}
\right] \, , \label{eq:gluonregget}
\end{align}
 see for example \cite{DelDuca:2001gu}. Not surprisingly if we consider what we already said about the structure of the IR poles in QCD amplitudes, the coefficients of the cusp anomalous dimension appear in the pole terms in Eq.~(\ref{eq:gluonregget}). Indeed, for the IR divergent part of the Regge trajectory, the IR structure of QCD amplitudes can be employed both to prove Reggeization at LL accuracy and to study the break down of Regge factorization at NNLL \cite{Bret:2011xm,DelDuca:2011ae}.
In this context, we are primarily interested in the constraints placed by Reggeization on possible additional terms depending on conformal ratios, which could contribute to Eq.~(\ref{eq:conj}) starting at three-loop order or higher. In fact some of the structures depending on the conformal ratio logarithms $\beta_{ijkl}$ which are not forbidden by other considerations have to be discarded because in the forward scattering they would give rise to super-leading logarithms $\alpha_s^n \ln^{n+1}(-s/t)$
staring from $n=3$. Those terms cannot be present because they would violate Regge factorization.

\subsection{Possible Violations of the Conjecture}

It is interesting to ask whether the above constraints are strong enough to prove the conjecture in Eq.~(\ref{eq:conj}). In the papers \cite{Becher:2009qa,Dixon:2009ur,Ahrens:2012he} a detailed order-by-order analysis of the possible terms in the anomalous dimension is performed. In particular, the most recent of these works, Ref.~\cite{Ahrens:2012he}, considers all possible terms up to four-loop order.
One finds that up to two-loop order non-abelian exponentiation and soft-collinear factorization exclude additional terms beyond the dipole formula Eq.~(\ref{eq:conj}). Beyond this order, an extra term $\bm{R}$ cannot be excluded. This remainder $\bm{R}$ is, however, strongly constrained:
%
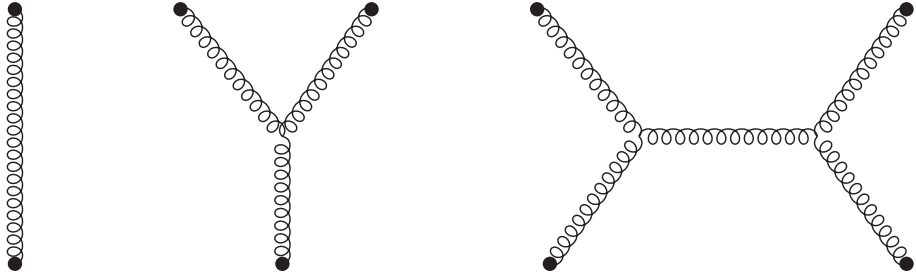
\begin{figure}
\begin{center}
\[ 
\vcenter{ \hbox{
  \begin{picture}(0,0)(0,0)
  \SetScale{0.3}
  \SetOffset(-216,0)
      \SetWidth{1.0}
      \SetColor{Black}
      \Vertex(176,190){9}
      \Vertex(176,-130){9}
      \Vertex(384,190){9}
      \Vertex(624,190){9}
      \Vertex(512,-130){9}
      \SetWidth{1.8}
      \Gluon(176,190)(176,-130){9.5}{20}
      \Gluon(384,190)(512,30){9.5}{11}
      \Gluon(1184,30)(1296,190){9.5}{11}
      \Gluon(512,30)(512,-130){9.5}{9}
      \Gluon(832,190)(960,30){9.5}{11}
      \Gluon(512,30)(624,190){9.5}{11}
      \Gluon(960,30)(848,-130){9.5}{11}
      \Gluon(1296,-130)(1184,30){9.5}{11}
      \Gluon(960,30)(1184,30){9.5}{12}
      \SetWidth{1.0}
      \Vertex(832,190){9}
      \Vertex(848,-130){9}
      \Vertex(1296,190){9}
      \Vertex(1296,-130){9}
\end{picture}}
}
\]
\end{center}
\vspace*{10mm}
\caption{The single connected web involving two, three, and  four particles. The dots represent color generators, which appear when the gluons are attached to Wilson lines. 
\label{fig:softwebs}}
\end{figure}
%
\begin{itemize}
\item It must fulfill the soft-collinear factorization constraint; that is the
logarithms appearing in the soft and collinear anomalous dimensions should
combine in such a way that the off-shell momenta $p_i^2$ cancel and combine into
functions depending only on the large invariants $s_{ij}$. This can be achieved
 by means of the conformal cross ratios introduced in Eq.~(\ref{eq:concrossratio}). If the reminder $\bm{R}$ exists, it must depend on conformal cross ratios and must therefore involve at least four particles.

\item Because of non-abelian exponentiation, the color structure of the soft anomalous dimension originates from single connected webs. The first single connected web involving four particles appears at three-loop; the corresponding web is shown in Figure~\ref{fig:softwebs}. The relevant contribution to the anomalous dimension must have the form \cite{Becher:2009qa}
\be
\bm{R} 
   =  \sum_{(i,j,k,l)}\!f^{ade} f^{bce} \mathbf{T}_i^a\,\mathbf{T}_j^b\,\mathbf{T}_k^c\,\mathbf{T}_l^d 
   F(\beta_{ijkl},\beta_{iklj}-\beta_{iljk})
\ee
where the conformal ratios $\beta$ are defined in Eq.~(\ref{eq:concrossratio}) and the sum extends over unordered four-tuples of indices.
The color structure of this term is subleading in the large $N_c$ limit and it can thus only arise from non-planar diagrams \cite{Becher:2009qa}. Consequently, in order to test the conjecture in Eq.~(\ref{eq:conj}) at three-loop level, it is necessary to compute the infrared structure of a three-loop four-point amplitude involving non-planar box diagrams.
It is interesting to observe that the full three-loop four-jet amplitudes in $N=4$ supersymmetric Yang-Mills theory were already reduced to linear combinations of a small number of scalar integrals  
\cite{Bern:2008pv}. Once these integrals are evaluated analytically or numerically, we will know if additional terms correlating four partons are present. 

\item The remainder term $\bm{R}$ should vanish in all the collinear limits, otherwise its existence would be excluded by the constraints dictated by the structure of the splitting function discussed in Section~\ref{sec:concoll}. Examples of functions $F(x,y)$ which fulfill this condition were constructed in \cite{Dixon:2009ur}. The three-loop function is the same in QCD and $N=4$ supersymmetric Yang-Mills, which implies that the remainder should have transcendentality five, the highest possible for a divergence at this order.
The simplest possibility is
\be\label{eq:collexample}
F(x,y) = x^3\,(x^2-y^2)
\ee

\item In addition, the remainder $\bm{R}$ must have the proper behavior in the Regge limit. The examples given in \cite{Dixon:2009ur}, in particular also the expression in Eq.~(\ref{eq:collexample}), violate the constraint from Regge factorization \cite{Bret:2011xm}. In the recent paper \cite{Ahrens:2012he} examples consistent also with the Regge constraint were constructed. The simplest is the function
\begin{equation}\label{testfun}
   F(x,y) = f_{1}(e^x,e^{-\frac12(x-y)},e^{-\frac12(x+y)}) -f_{2}(e^x,e^{-\frac12(x-y)},e^{-\frac12(x+y)})\,, \qquad
\end{equation}
where the functions $f_{n}$ are defined as
\begin{equation}\label{ansatz}
   f_{n}(\rho_{ijkl},\rho_{iklj},\rho_{iljk}) 
   = \frac{\ln\rho_{ijkl}}{2n^2}\,\big[ g(\rho_{iklj}^{n})\,g(\rho_{iljk})
    + g(\rho_{iklj})\,g(\rho_{iljk}^{n}) \big] \,,
\end{equation}
with $g(z) = \mbox{Li}_2(1-z) - \mbox{Li}_2(1-z^{-1})$.
\end{itemize}
Given that it is possible to construct explicit examples which are compatible with all known constraints, one would expect the corresponding terms to be present, unless the anomalous dimension $\bm{\Gamma}$ is subject to unknown additional constraints. As we mentioned earlier, the recent paper \cite{Caron-Huot:2013fea} claims to have found four-loop terms which violate the dipole form and thus contribute to the remainder, however, so far this computation has not yet been independently verified.

\subsection{Contributions from Higher Casimir Operators} 

\begin{figure}
\begin{center}
\[ 
\vcenter{ \hbox{
  \begin{picture}(0,0)(0,0)
  \SetScale{0.3}
  \SetOffset(-216,0)
      \SetWidth{1.0}
      \SetColor{Black}
\Vertex(60,190){9}
\Vertex(60,-130){9}
\Vertex(380,190){9}
\Vertex(380,-130){9}
\Vertex(560,190){9}
\Vertex(560,-130){9}
\Vertex(880,190){9}
\Vertex(880,-130){9}    
\Vertex(1060,190){9}
\Vertex(1060,-130){9}
\Vertex(1380,190){9}
\Vertex(1380,-130){9}    
     \SetWidth{1.8}
\ArrowArc(220,30)(85,-45,45)
\ArrowArc(220,30)(85,45,135)
\ArrowArc(220,30)(85,135,225)
\ArrowArc(220,30)(85,225,315)
\Gluon(60,190)(160,90){9.5}{8}
\Gluon(160,-30)(60,-130){9.5}{8}
\Gluon(380,-130)(280,-30){9.5}{8}
\Gluon(280,90)(380,190){9.5}{8}
\GlueArc(720,30)(85,-45,45){9.5}{8}
\GlueArc(720,30)(85,45,135){9.5}{8}
\GlueArc(720,30)(85,135,225){9.5}{8}
\GlueArc(720,30)(85,225,315){9.5}{8}
\Gluon(560,190)(660,90){9.5}{8}
\Gluon(660,-30)(560,-130){9.5}{8}
\Gluon(880,-130)(780,-30){9.5}{8}
\Gluon(780,90)(880,190){9.5}{8}
\DashArrowArc(1220,30)(85,-45,45){10}
\DashArrowArc(1220,30)(85,45,135){10}
\DashArrowArc(1220,30)(85,135,225){10}
\DashArrowArc(1220,30)(85,225,315){10}
\Gluon(1060,190)(1160,90){9.5}{8}
\Gluon(1160,-30)(1060,-130){9.5}{8}
\Gluon(1380,-130)(1280,-30){9.5}{8}
\Gluon(1280,90)(1380,190){9.5}{8}
\end{picture}}
}
\]
\end{center}
\vspace*{10mm}
\caption{Four-loop connected webs involving higher Casimir invariants; the three diagrams include a closed fermion, gluon and ghost loop, respectively. 
\label{fig:highercasimirs}}
\end{figure}
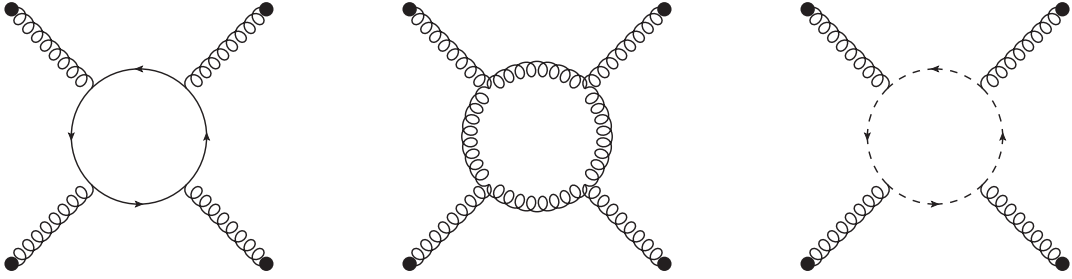

For the special case of two-jet operators, the simple form of Eq.~(\ref{eq:conj}) implies {\em Casimir scaling} of the cusp anomalous dimension, i.e., the cusp anomalous dimensions of quarks and gluons are related to each other by the ratio of the eigenvalues $C_i$ of the quadratic Casimir operators:
\begin{equation}
   \frac{\Gamma_{\rm cusp}^q(\alpha_s)}{C_F}
   = \frac{\Gamma_{\rm cusp}^g(\alpha_s)}{C_A}
   = \gamma_{\rm cusp}(\alpha_s) \,.
\end{equation}
The Casimir scaling holds up to three loops \cite{Moch:2004pa},
but it contradicts expectations from the AdS/CFT correspondence \cite{Armoni:2006ux,Alday:2007hr,Alday:2007mf}. 
Furthermore, the recent paper \cite{Grozin:2014hna} observes that up to three loops, the cusp anomalous dimension has an interesting iterative structure. If this structure persists at higher orders, it would imply a non-vanishing contribution involving the quartic Casimir invariant at four-loop order.
The invariant is constructed from symmetrized traces of four generators in a representation $R$
\be
d_R^{a_1 a_2 a_3 a_4} = \mbox{Tr} \left[ \left(\mathbf{T}_R^{a_1} \mathbf{T}_R^{a_2} \mathbf{T}_R^{a_3} \mathbf{T}_R^{a_4} \right)_+ \right] \, ,  \label{eq:4Casimir}
\ee
with
\be
\left(\mathbf{T}^{a_1} \mathbf{T}^{a_2} \mathbf{T}^{a_3} \mathbf{T}^{a_4} \right)_+ \equiv \frac{1}{4!}
\sum_{\text{permutations}} \mathbf{T}^{a_1} \mathbf{T}^{a_2} \mathbf{T}^{a_3} \mathbf{T}^{a_4} \,,
\ee
When this is contracted with four generators of an irreducible representation $R'$, one obtains the quartic Casimir invariant
\begin{equation}
d_R^{a_1 a_2 a_3 a_4}\, \mathbf{T}_{R'}^{a_1} \mathbf{T}_{R'}^{a_2} \mathbf{T}_{R'}^{a_3} \mathbf{T}_{R'}^{a_4}  = C_4(R,R') \, \bm{1}
\end{equation} 
To see that the object on the left-hand side defines an invariant, one shows that it commutes with all generators. Schur's lemma then implies that it must be proportional to $\bm{1}$ (see \cite{vanRitbergen:1998pn} for a detailed discussion of such invariants).

It is interesting to investigate whether contributions to the anomalous dimension $\bm{\Gamma}$ involving the structure $d_R^{a_1 a_2 a_3 a_4} $ exist, which are compatible with the factorization constraints discussed above. Non-abelian exponentiation implies that such contributions to $\bm{\Gamma}$ could first arise at four-loop order. Such contributions can arise from fermion, gauge-boson and ghost-loop diagrams of the kind shown in Figure~\ref{fig:highercasimirs}. An analysis of higher-Casimir contributions to $\bm{\Gamma}$ was first performed in \cite{Becher:2009qa}. This reference only considered terms linear in the cusp angle, the most general four-loop contribution was later analyzed in \cite{Ahrens:2012he}. These papers concluded that such terms are compatible with soft-collinear factorization but are ruled out by factorization in the collinear limit. A potential loop-hole arises from the fact that collinear limits can only be considered for amplitudes with $n\geq 4$. It would thus be conceivable (albeit strange) that such terms could be present for $n=2$ but absent for higher-point anomalous dimensions. Let us note that while collinear factorization was verified explicitly up to two loops \cite{Badger:2004uk}, an all-order proof was not available until recently. The paper \cite{Feige:2014wja} now presents such a proof.

\subsection{Massive Amplitudes}

If the external legs in a given amplitude are massive, it is possible to factor a cross section into the product of an hard function $H$ which depends on the large momentum transfers between jets, $s_{ij}$, and a soft function $S$ which depends on the maximum energy of unobserved soft emissions \cite{Yennie:1961ad, Weinberg:1965nx}.
Schematically, one can write the following factorization relation
\be
d \sigma = H \left(\{s_{ij}\},\{m_i\},\mu \right) \times S \left(\{v_i \cdot v_j\}, \mu \right) \, ,
\ee
where the four-velocities  of the massive particles are defined 
as
\be
v^\mu_i \equiv  \frac{p_i^\mu}{m_i} \, , \quad v_i^2 = 1 \, . 
\ee
The analysis of the IR divergences for massless QCD amplitudes was extended to the case in which also massive external legs are present in \cite{Becher:2009kw}.
The effective theory to be employed in this case is a combination of of SCET (for the massless partons) and  HQET for the massive partons \cite{Neubert:1993mb}. In such situations the soft function contains both massless and time-like Wilson lines
\be
{\mathcal S} \left(\{\underline{n}\},\{\underline{v}\},\mu \right) = 
\langle 0 | \bm{S}_{n_1} \cdots \bm{S}_{n_k} \bm{S}_{v_k+1}  \cdots \bm{S}_{v_n} | 0 \rangle \,, 
\ee
where $n_i$ are the light cone reference vectors of the $k$ massless external legs and $v_i$ are the velocities of the $n-k$ external massive legs. The massless Wilson lines are defined in Eq.~(\ref{eq:Smathbf}), while the massive Wilson lines are defined as 
\be \label{eq:Svel}
\mathbf{S}_i(x) = \mathbf{P} \exp\left[ - ig \int^{\infty}_0 \!\!dt \, v_i 
\cdot A^a_s(x + t v_i) \mathbf{T}_i^a \right] \, .
\ee

In the case in which massive partons are present, the constraints of the anomalous dimension which generalizes Eq.~(\ref{eq:conj}) are weaker
than in purely massless case. In particular, for massive legs  there are no constraints coming form soft-collinear factorization and from collinear limits.
For the purely massive case, all the color structures allowed by non-abelian exponentiation at a given order will be present.

When both massive and massless external legs are present, the anomalous dimension matrix governing the structure of the IR poles of QCD amplitudes has a part which depends on one- and two-parton correlations.
This part has the following form
\bea \label{eq:G2parmass}
\bm{\Gamma}\left( \{ \underline{p} \} , \{ \underline{m} \} ,\mu \right) \bigl|_{\text{2-parton}} &=& \sum_{(i,j)} \frac{\mathbf{T}_i \cdot\mathbf{T}_j}{2} \gamma_{\cusp} (\alpha_s) \ln{\frac{\mu^2}{-s_{ij}}} + \sum_i \gamma_i (\alpha_s) \nn \\
&& - \sum_{(I,J)}  \frac{\mathbf{T}_I \cdot\mathbf{T}_J}{2}\gamma_{\cusp} (\beta_{IJ},\alpha_s) + \sum_I \gamma_I (\alpha_s) \nn \\
&&+\sum_{(I,j)}  \frac{\mathbf{T}_I \cdot\mathbf{T}_j}{2}\gamma_{\cusp} (\alpha_s)\ln{\frac{m_I \mu}{-s_{Ij}}} \, .
\eea
In the equation above the last two lines contain the new terms required by the presence of massive partons.
The capital indices $I,J$ run over the massive legs, while the indices $i,j$ run over the massless ones. In all three cusp terms we sum over both orders of the indices as in Eq.~(\ref{eq:conj}). The hyperbolic angles formed by two time-like Wilson lines, $\beta_{IJ}$, are defined as
\be
\beta_{IJ} = \text{arccosh}{\left(-\sigma_{IJ} v_I \cdot v_J -i 0 \right)} \, ,
\ee
where $\sigma_{IJ} = 1$ if both partons $I$ and $J$ are incoming or outgoing and $\sigma_{IJ} = -1$ otherwise. The cusp anomalous dimension which appears in the second line of Eq.~(\ref{eq:G2parmass}) depends on $\beta_{IJ}$ and it is now known at up to three loops \cite{Grozin:2014hna}.  The anomalous dimensions for massive quarks
which appears in the last term on the r.h.s.\ in the second line of Eq.~(\ref{eq:G2parmass}) is known to two loops. The explicit two-loop expressions for both quantities are provided in Appendix~\ref{app:AnDim}. In the last line of Eq.~(\ref{eq:G2parmass}), we employ $s_{Ij} = 2 \sigma_{Ij}\,p_I \cdot p_j$ where, as usual, $\sigma_{Ij} = \pm 1$ depending on the parton directions.

However, in the massive case also three-parton correlations appear starting at two-loop order \cite{Mitov:2009sv}. The term of the anomalous dimension matrix arising from three parton correlations is 
\bea
\bm{\Gamma} \left( \{\underline{p}\}, \{\underline{m}\}, \mu \right)|_{\text{3-partons}} &=& i f^{abc} \sum_{(I,J,K)} \mathbf{T}_I^a\mathbf{T}_J^b\mathbf{T}_K^c F_1\left(\beta_{IJ}, \beta_{JK}, \beta_{KI}
 \right) \nn \\
 & &
+ i f^{abc} \sum_{(I,J,k)} \mathbf{T}_I^a\mathbf{T}_J^b\mathbf{T}_k^c f_2\left(\beta_{IJ}, \ln\left(\frac{-\sigma_{Ik} v_I \cdot p_k}{-\sigma_{Jk} v_J \cdot p_k}\right)
 \right)  . \label{eq:3part}
\eea
The first line in Eq.~(\ref{eq:3part}) describes  color correlations among three massive partons, while the second line arises from correlations among two massive partons and a massless one.
The functions $F_1$ and $f_2$ were calculated in \cite{Ferroglia:2009ep, Ferroglia:2009ii}. The function $F_1$ is completely antisymmetric in its arguments and has the following form
\bea
F_1 \left(\beta_{12},\beta_{23}, \beta_{31} \right) &=&
\left(\frac{\alpha_s}{4 \pi} \right)^2 \frac{4}{3} \sum_{(I,J,K)} \epsilon_{IJK} \, g\left(\beta_{IJ} \right) \beta_{KI} \coth \beta_{KI} \, ,
\eea
where 
\be
g \left(\beta \right) = \coth \beta \left[ \beta^2 +2 \beta \ln(1-e^{-2 \beta})
- \mbox{Li}_2(e^{-2 \beta}) + \frac{\pi^2}{6} \right] -\beta^2 -\frac{\pi^2}{6} \, .
\ee
The function $f_2$ is given by
\be
f_2 \left(\beta_{12},\ln\left(\frac{-\sigma_{23} v_2 \cdot p_3}{-\sigma_{13} v_1 \cdot p_3} \right) \right)  = - \left(\frac{\alpha_s}{4 \pi} \right)^2 4 g\left( \beta_{12} \right)\ln\left(\frac{-\sigma_{23} v_2 \cdot p_3}{-\sigma_{13} v_1 \cdot p_3} \right) \,.
\ee
Both functions are found
to be suppressed like ${\mathcal O} (m^4 /s^2 )$ in the limit in which the parton mass $m$ is much smaller than the hard scale(s) $s$, in accordance with
mass factorization theorems proposed in the literature \cite{Mitov:2006xs,Becher:2007cu}.

The anomalous dimension 
\be \label{eq:Gmastot}
\bm{\Gamma} \left( \{\underline{p}\}, \{\underline{m}\}, \mu \right) = 
\bm{\Gamma} \left( \{\underline{p}\}, \{\underline{m}\}, \mu \right)|_{\text{2-partons}}+\bm{\Gamma} \left( \{\underline{p}\}, \{\underline{m}\}, \mu \right)|_{\text{3-partons}} \, , 
\ee
is related to the renormalization factor $\bm{Z}$  (which  removes the IR poles from QCD amplitudes with massive and massless external legs) through the differential equation
\be
\bm{\Gamma} \left(\{\underline{p}\}, \{ \underline{m} \} ,\mu \right) = - \bm{Z}^{-1} \left(\ep,\{\underline{p}\}, \{ \underline{m} \} , \mu\right)  \frac{d}{d \ln \mu}\bm{Z} \left(\ep,\{\underline{p}\}, \{ \underline{m} \} , \mu\right)  \, ,
\ee
The equation above is formally identical to Eq.~(\ref{eq:GammaZ}) for the purely massless case. Once the anomalous dimension in Eq.~(\ref{eq:Gmastot}) became known, it was possible to calculate the expression of the  IR poles in the two-loop corrections to top-quark pair production \cite{Ferroglia:2009ii}, and to 
obtain NNLL resummation formulas for several observables related to the that process \cite{Ahrens:2009uz,Ahrens:2010zv,Ahrens:2011mw}.

%% file: 9_Outlook.tex
\section{Applications of SCET\label{sec:OL}}

SCET is the effective theory relevant for the description of energetic particles, so that, to first approximation, it should be possible to describe any process in high-energy physics within this framework. Not surprisingly then, there are a plethora of possible applications of the effective theory. So far, our focus was on the construction of the theory and we have included only a few basic applications which illustrate how it is used in practice. The main goal of these applications was to demonstrate how to use the framework to derive factorization theorems and to resum numerically large logarithmic corrections.  As an invitation and guide to further reading, we would like to close these lectures by briefly summarizing some of the many results obtained by means of SCET methods, as well as to point out some topics which are the subject of ongoing research. At the moment of this writing (summer of 2014), a search on {\tt Inspire} returns approximately 200 papers which include the word  SCET directly in the title; many more employ SCET methods in order to carry out calculations, or deal with particular technical aspects of SCET. It is neither possible nor useful to discuss here all of the publications on SCET which appeared since the theory was introduced at the beginning of the millennium.  We apologize in advance, if we overlooked your important paper in the short overview of the field which follows. Furthermore, for most of the applications discussed below, resummations are also available using traditional methods, but we restrict our discussions to 
work based on SCET.

Collider physics is an environment particularly suitable for the application of effective theory methods and SCET in particular: high-energy processes involve large scale hierarchies and are governed by soft and collinear emissions which can lead to Sudakov double logarithms. By now, the majority of SCET applications are in this area and we have aimed our presentation on collider physics applications of SCET. Nevertheless, before turning to these collider physics applications, we discuss in Section \ref{sec:Bdec} some of the work in heavy-quark physics, for which the effective field theory was originally developed, and where it has been used to analyze a variety of $B$-meson decays.

On the collider physics side, we by now have many high-precision calculations for simple inclusive final states. Hadronically inclusive cross sections are theoretically simpler and can in general be predicted with higher precision than more exclusive quantities. On the other hand, their physics content is limited and inclusive measurements do not exist. In order to make contact with experiments and to extract detailed information, one would therefore like to analyze also less inclusive observables in SCET and a lot of recent work is devoted to achieve this goal. We have therefore grouped the collider physics processes studied by means of SCET in three broad classes according to their exclusiveness. We'll first discuss hadronic inclusive cross sections in Section \ref{sec:inclusive}. A variety of such cross sections have been resummed with high logarithmic accuracy using the methods we discussed in Section \ref{sec:DY} and \ref{sec:pT}. Using SCET, many of the traditional event-shape variables were computed to higher accuracy and new event-shape variables were introduced that are useful in a hadron collider context; we review this topic in Section~\ref{sec:shapes}. There has also been a lot of work on jet observables in the past few year, which we discuss in Section~\ref{sec:jets}. We discuss a few recent developments in Section~\ref{sec:additional} and then conclude with a brief outlook in Section~\ref{sec:outlook}.

\subsection{Heavy-Quark Physics\label{sec:Bdec}}

For the reader not familiar with flavor physics, it might seem surprising that an effective theory for energetic particles can be relevant in the context of the low-energy processes such as $B$-decays. The effective theory becomes relevant when one considers $B$-decays in the heavy-quark limit $m_b \to \infty$, because the energy of the light decay products is of the order of the $b$-quark mass. For exclusive decays, or for inclusive decays where the decay of the heavy $B$-meson produces a hadronic ``jet'' of small invariant mass $m_X$, one can again perform the usual SCET expansion in the invariant mass over the energy of the jet. Doing so, one ends up with a standard hard times jet times soft function factorization theorem, as we encountered several times in these lectures. Such factorization theorems arise for exclusive two-body decays, such as $B\to \pi \pi$, where the $E_\pi \approx M_B/2 \gg M_\pi$, but also in inclusive decays such as $B\to X_s \gamma$ and $B\to X_u \ell \nu$ (here $X_f$ denotes any final state with the appropriate flavor quantum number). In the inclusive case, the low $m_X$ region is relevant because one needs hard photons and leptons to extract the inclusive decays from the background.  These cuts enforce that also the hadronic system has a large energy, close to the kinematic endpoint.

In order to analyze the factorization properties of $B$-decays, one uses the same formalism discussed here in conjunction with HQET, which is employed  to describe the $b$-quark (see \cite{Neubert:1993mb,Manohar:2000dt} for an introduction). While the basic method is the same as in collider applications, the numerical values of the scales involved  are much lower, which makes the practical applications quite different. The hard scale for $B$-decays is set by $m_b\approx 5\,{\rm GeV}$, which is still in the perturbative regime, so that the hard functions can be computed in perturbation theory. However, the soft and collinear functions are often non-perturbative. In exclusive decays, the virtuality of the collinear fields describing the final state mesons is of the order of the meson masses. Their matrix elements are called light-cone distribution amplitudes and are non-perturbative objects. In radiative and semileptonic inclusive decays near the kinematic end-point, the relevant soft matrix elements are called shape functions. Depending on the experimental cuts, these are either non-perturbative or at best barely perturbative. The two main difficulties in $B$-physics applications are the non-perturbative input, which needs to be taken from data (or computed with non-perturbative methods), and the presence of power corrections, which can make a leading-power treatment unreliable. Also in collider physics one needs to deal with non-perturbative input such as the PDFs, but the fact that the hard scale $Q$ is typically much higher and an external, adjustable parameter makes it easier to extract the necessary information from data. 

SCET was first proposed in the context of inclusive $B$-decays, as an alternative method to sum Sudakov logarithms in the end-point region \cite{Bauer:2000ew}. The physics of the end-point region and the appearance of a non-perturbative shape function was understood earlier \cite{Bigi:1993ex,Neubert:1993um} and a factorization theorem into a hard, a jet and a soft function was derived diagrammatically in \cite{Korchemsky:1994jb}, before the advent of SCET. However, the effective theory framework has allowed for detailed studies, not only of the leading-power factorization theorem, but also of the power corrections which affect the rate \cite{Bauer:2003pi,Beneke:2004in,Bosch:2004th,Lee:2004ja, Lee:2006wn, Benzke:2010js}. Furthermore, for the leading-power rate, the perturbative predictions were improved by computing the two-loop corrections to the hard \cite{Bonciani:2008wf,Asatrian:2008uk,Beneke:2008ei,Bell:2008ws,Greub:2009sv}, jet \cite{Becher:2006qw} and soft \cite{Becher:2005pd} functions. In addition, a dedicated framework for the transition region from a perturbative to a non-perturbative soft function was developed and applied to $B\to X_s \gamma$ \cite{Neubert:2004dd,Becher:2006pu}. The SCET results for $B\to X_u \ell \nu$ form the basis of the determination of $|V_{ub}|$ from inclusive decays \cite{Lange:2005yw}. 
For the case of $B\to X_s \gamma$ SCET made possible the analysis of the factorization properties of the full set of operators in the effective Hamiltonian, beyond the analysis of the factorization of the $b\to s \gamma$ dipole operator. This analysis has revealed nontrivial non-perturbative matrix elements, which are relevant not only in the endpoint region, but even for the total rate \cite{Lee:2006wn,Benzke:2010js}. The decay $B\to X_s \ell^+\ell^-$ is related to $B\to X_s \gamma$ but can probe additional new physics effects. Perturbative and non-perturbative effects in this decay in the low-$m_X$ region were analyzed in the papers \cite{Lee:2005pk,Lee:2005pwa,Lee:2008xc,Bell:2010mg}.

The factorization properties of exclusive two-body $B$-decays were first understood in \cite{Beneke:1999br}, following earlier results for the factorization of exclusive processes with light hadrons \cite{Efremov:1979qk,Lepage:1980fj}. The corresponding factorization formula was then used to analyze the large class of two-body decays in \cite{Beneke:2001ev,Beneke:2003zv}. This result again predates SCET and was originally justified at the two-loop level using diagrammatic methods \cite{Beneke:2000ry}. One of the first applications of SCET was an all-order derivation of factorization for the decay $B \to D \pi$ \cite{Bauer:2001cu}. Later, SCET was used to extend the factorization analysis to color-suppressed $B$ to $D$ decays \cite{Mantry:2003uz}. The more complicated decay into two light mesons was analyzed using SCET in \cite{Chay:2003zp,Chay:2003ju,Bauer:2004tj} and phenomenological results were presented in \cite{Bauer:2005kd,Williamson:2006hb}. There are ongoing efforts to compute the perturbative input in the factorization formula of charmless exclusive decays to two-loop accuracy and many of the necessary pieces are by now available \cite{Beneke:2005vv,Beneke:2006mk,Kivel:2006xc,Bell:2007tv,Pilipp:2007mg,Bell:2009nk,Beneke:2009ek}. Similar factorization theorems also hold for exclusive radiative decays \cite{Ali:2001ez,Beneke:2001at,Bosch:2001gv}. A detailed operator analysis of the relevant factorization theorem in SCET was given in \cite{Becher:2005fg,Ali:2006ew} and the most up-to-date computation of these decays can be found in \cite{Ali:2007sj}.

From the theoretical point of view, the  most interesting element of the factorization formula for the decay to light mesons is the heavy-to-light transition form factor. This was analyzed in \cite{Bauer:2002aj,Beneke:2003pa,Lange:2003pk}. An important question in the context of the form factor is whether it is possible to factorize it completely into a soft function for the $B$-meson and a collinear function for the light meson. Based on the formalism of soft-collinear messenger modes \cite{Becher:2003qh}, the analysis \cite{Lange:2003pk} concluded that this factorization is broken. This treatment has been criticized because the messenger modes which mediate the factorization breaking perturbatively have a very low virtuality, below the scale $\Lambda_{\rm QCD} \sim 1\, {\rm GeV}$ where QCD becomes nonperturbative. One can thus speculate whether non-perturbative effects would shield this factorization breaking effect.  Indeed, equipping all partons with masses of the order of  the typical QCD scale $\Lambda_{\rm QCD}$ eliminates these low-virtuality modes; however the factorization breaking would then arise from the additional regulators needed in the massive case \cite{Becher:2003qh,Beneke:2003pa}, i.e.\ via the collinear anomaly. Reference \cite{Manohar:2006nz} later conjectured a factorization of the form factor after zero-bin subtractions in the convolutions of the light-cone distribution amplitudes with the hard kernels, but the corresponding formalism was never fully fleshed out. Finally, \cite{Beneke:2008pi} analyzed $B \to \chi_{cJ} K$ as an example where the factorization breaking effects can be computed perturbatively using non-relativistic QCD and  generate a large rescattering phase, casting doubt on the statement in \cite{Manohar:2006nz,Arnesen:2006vb} that non-factorizable phases are absent in annihilation contributions to charmless $B$-decays. The result \cite{Beneke:2008pi} suggests that heavy-to-light form factors are indeed not factorizable into soft and collinear matrix elements. However, a full operator analysis of the heavy-to-light form factor is still missing at this point. 

\subsection{Inclusive Hadron-Collider Cross Sections\label{sec:inclusive}}

Among the first collider processes that were studied by means of SCET methods are resummations of hadronically inclusive cross section. In Sections \ref{sec:DY} and \ref{sec:pT} of this work we discussed in detail  the simplest example in this category, the Drell-Yan process $p p \to \ell^+ \ell^- \, X$, where $X$ is an arbitrary hadronic final state. In Section \ref{sec:DY}, we have resummed logarithms which arise when the invariant mass of the leptons is high, which enhances the partonic threshold region. Large logarithms also arise when the transverse momentum of the lepton pair is small, a situation we analyzed in Section \ref{sec:pT}.  While threshold resummation is based on a conventional factorization theorem, the factorization theorem relevant for transverse momentum suffers from a collinear anomaly. There are by now many examples of threshold and transverse momentum resummations performed in SCET. Here we briefly go over some examples and point out a few interesting features in each case.

Closely related to the Drell-Yan case is Higgs production, where logarithms which become large in the productions threshold limit $m_H^2/\hat{s} \to 1$ were resummed up to N$^3$LL \cite{Ahrens:2008nc,Ahrens:2008qu,Ahrens:2010rs}. Instead of the vector form factor of quarks, the relevant operator for Higgs production is the time-like scalar form factor of gluons. It turns out that this form factor receives very large higher-order  QCD corrections. In contrast, the space-like scalar form factor receives only small corrections. By choosing a negative value for the hard scale $\mu_h^2$ (i.e. by working with complex values of $\mu_h$) one can transition from the time-like to the space-like form factor and this choice was adopted in \cite{Ahrens:2008nc,Ahrens:2008qu,Ahrens:2010rs} to improve convergence. The formalism discussed in Section \ref{sec:DY} immediately applies to any colorless final state. Further examples include diboson production processes such as $WW$, $WZ$, $VH$ for which NNLL resummations have been performed in \cite{Dawson:2013lya,Wang:2014mqt,Li:2014ria}. Another application is slepton-pair production in supersymmetric extensions of the Standard Model for which the resummation was carried out up to N$^3$LL order  \cite{Broggio:2011bd}.

The computations discussed so far concern situations where there is only soft gluon radiation in the final state and the corresponding resummations are also called soft-gluon resummations. It is also interesting to consider  Drell-Yan type processes (i.e. $\gamma/W/Z/H$ production) at large transverse momentum. In this case, there has to be energetic QCD radiation in the final state to balance the large transverse momentum of the electroweak boson. In the partonic threshold region, this radiation consists of a single jet with low invariant mass $m_X$. The relevant hard function corresponds to a scattering amplitude instead of a form factor and the soft function now involves three Wilson lines: two in the directions of the incoming hadrons and one in the direction of the final-state jet. This situation was analyzed in \cite{Becher:2009th, Becher:2011fc, Becher:2012xr,Becher:2013vva, Becher:2014tsa} and the threshold resummation has now been carried out up to N$^3$LL accuracy. At this order, the result involves the full two-loop hard, jet \cite{Becher:2006qw,Becher:2010pd} and soft functions \cite{Becher:2012za}. The two-loop hard functions were extracted in \cite{Becher:2013vva} from the known results for the corresponding two-loop scattering amplitudes.

An important process at the LHC is top-quark pair production. In this case, the basic hard scattering process involves four colored partons. In contrast to the previously discussed examples, one therefore has to deal with nontrivial color structures
 in the hard and soft functions; as a consequence,  the RG equations satisfied by these functions become matrix valued. Three different singular limits of this process were studied within the SCET framework. These are the production 
threshold limit $\hat{s} \to 4 m_t^2$ \cite{Beneke:2009ye, Beneke:2010da, Beneke:2011mq}, which is employed in order to calculate the total top-quark pair-production cross section,  the soft  emission limit in Pair Invariant Mass kinematics (PIM), needed
for the calculation of the pair invariant-mass distribution 
 \cite{Ahrens:2009uz,Ahrens:2010zv}, and the soft limit in One Particle Inclusive (1PI) kinematics, which is employed in order to calculate 
 the top (or antitop) transverse momentum and rapidity distributions \cite{Ahrens:2011mw}. In all cases the resummation in momentum space was carried out up to NNLL order. The production threshold limit is interesting because the $t\bar{t}$ pair is non-relativistic in this region and the computation then involves an interesting interplay of SCET and non-relativistic QCD \cite{Beneke:2009ye, Beneke:2010da, Beneke:2011mq}. The hard scattering kernels obtained in PIM and 1PI kinematics can be combined with semi-leptonic decays of top quarks in the narrow-width approximation in a fully differential parton level Monte Carlo which allows for the study of any IR safe observable constructed from the momenta of the decay products of the top quark \cite{Broggio:2014yca}. The factorization of the cross section for highly boosted top-quark pairs was studied by means of SCET methods both in PIM \cite{Ferroglia:2012ku, Ferroglia:2012uy, Ferroglia:2013zwa} and 1PI \cite{Ferroglia:2013awa} kinematics. In terms of QCD effects, the production of heavy colored supersymmetric particles, such as gluinos and squarks is closely related to top production. Near the production threshold, these processes were studied  up to NLL and NNLL  accuracy in \cite{Falgari:2012sq, Beneke:2013opa}. The production of top-squark pairs was  evaluated within PIM and 1PI kinematics  up to approximate NNLO  \cite{Broggio:2013uba} and the corresponding soft gluon emission corrections were resummed up to NNLL accuracy \cite{Broggio:2013cia}. Several papers also considered threshold resummation for jet observables \cite{Bauer:2010vu,Bauer:2010jv,Liu:2012zg,Kelley:2010fn,Kelley:2010qs,Kelley:2011tj,Kelley:2011aa,Chien:2012ur,Broggio:2014hoa}. We will discuss jet variables and the associated challenges below.

In the last few years important progress was made in the analysis of  processes sensitive to small transverse momenta and processes whose factorization formulas in SCET involve a collinear anomaly. Such observables play an important role, in particular at hadron colliders. The simplest example is again the Drell-Yan process
at small  vector boson transverse momentum \cite{Becher:2010tm, Becher:2011xn, GarciaEchevarria:2011rb,Becher:2012fx}, which was discussed in Section~\ref{sec:pT}. As in the case of threshold resummation, the same formalism immediately applies to any colorless final state such as Higgs production \cite{Chiu:2012ir,Becher:2012yn}, where the resummation was carried out up to NNLL accuracy. The formalism was also applied to diboson production in \cite{Wang:2013qua}. A particularly interesting application is the production of a top-quark pair with low transverse momentum, whose cross section was resummed up to NNLL accuracy in  \cite{Zhu:2012ts, Li:2013mia}. This is an example of an observable, for which resummation was first performed using SCET. (In the meantime, the same resummation has also been achieved using traditional methods \cite{Catani:2014qha}.) An important ingredient for the matching to fixed-order computations and to increase the accuracy of the resummations are the two-loop beam functions relevant for this case, which were computed in \cite{Gehrmann:2012ze, Gehrmann:2014yya} using the analytic regulator introduced in \cite{Becher:2011dz}. The beam functions for transverse momentum resummation are nothing but TPDFs and this is the first two-loop computation for such objects. In addition to the beam functions for TPDFs, recently also the two-loop beam functions for virtuality dependent PDFs \cite{Gaunt:2014xga,Gaunt:2014cfa} and for fully unintegrated PDFs (which depend both on the virtuality and the transverse momentum) were computed \cite{Jain:2011iu,Gaunt:2014xxa}, so that the complete set of unintegrated PDFs is available.

Using SCET methods, we have also gained a better understanding of the form of the long-distance effects. The leading effects are enhanced by large logarithms and can be viewed as a non-perturbative correction to the anomaly \cite{Becher:2013iya}. Fits to extract non-perturbative effects in TPDFs have been performed in \cite{D'Alesio:2014vja}.

\subsection{Event Shapes\label{sec:shapes}}

The simplest observables that go beyond the hadronically inclusive cross sections we just discussed are event-shape variables. They classify events according to some simple geometric properties of the final-state hadron momenta and are designed in such a way that they probe properties which are insensitive to hadronization effects and can be computed in perturbation theory for sufficiently large center-of-mass energy. A necessary condition for this is that the variables are infrared safe: they must be defined in such a way that exactly collinear splittings and arbitrarily soft emissions do not change the value of the observable. 

The classic example of an event-shape variable at $e^+e^-$ colliders is the thrust $T$. To obtain the thrust of an event one first finds the axis where most of the momentum of the particles in the event flows. The thrust unit vector $\vec{n}_T$ points along this axis and the value of $T$ of an event is then given by the ratio of the momentum flowing along this axis over the total momentum  so that $T=1$ corresponds to an event where all particles fly exactly along the same direction. The precise definition reads
\begin{equation}\label{eq:thrust}
T = \frac{1}{P_{\rm tot}}\, \mathop{\rm max}_{\vec{n}_T}  \sum_i |\vec{n}_T\cdot \vec{p}_i| \,.
\end{equation}
The sum runs over all particles in the event and $P_{\rm tot} = \sum_i |\vec{p}_i|$. For massless particles $P_{\rm tot}$ is equal to the center of mass energy $Q$ of the collision. One immediately sees that the thrust remains the same if one splits a given momentum $\vec{p}_i$ into two collinear momenta, or emits an additional very soft particle: thrust is an infrared safe observable. 

\begin{figure}
\begin{center}
\begin{tabular}{lr}
 \includegraphics[height=0.29\textwidth]{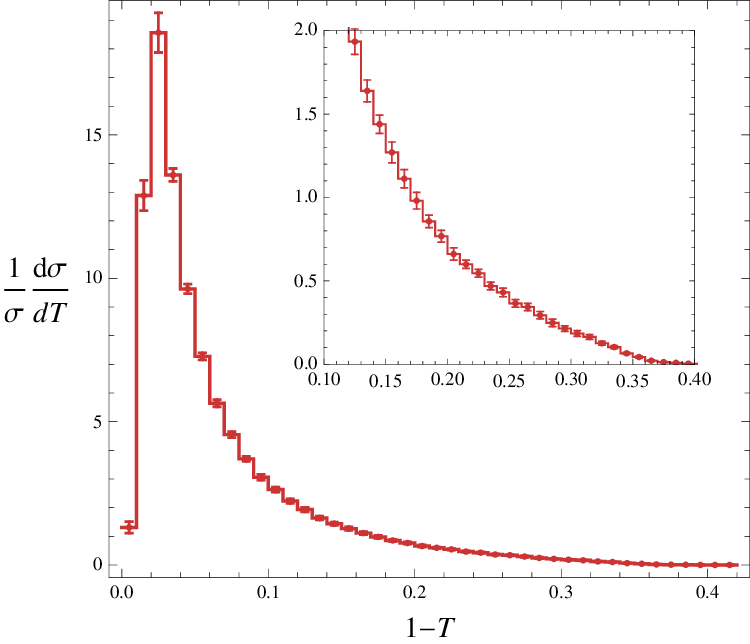}\hspace{0.5cm} & \raisebox{0.5cm}{\includegraphics[height=0.27\textwidth]{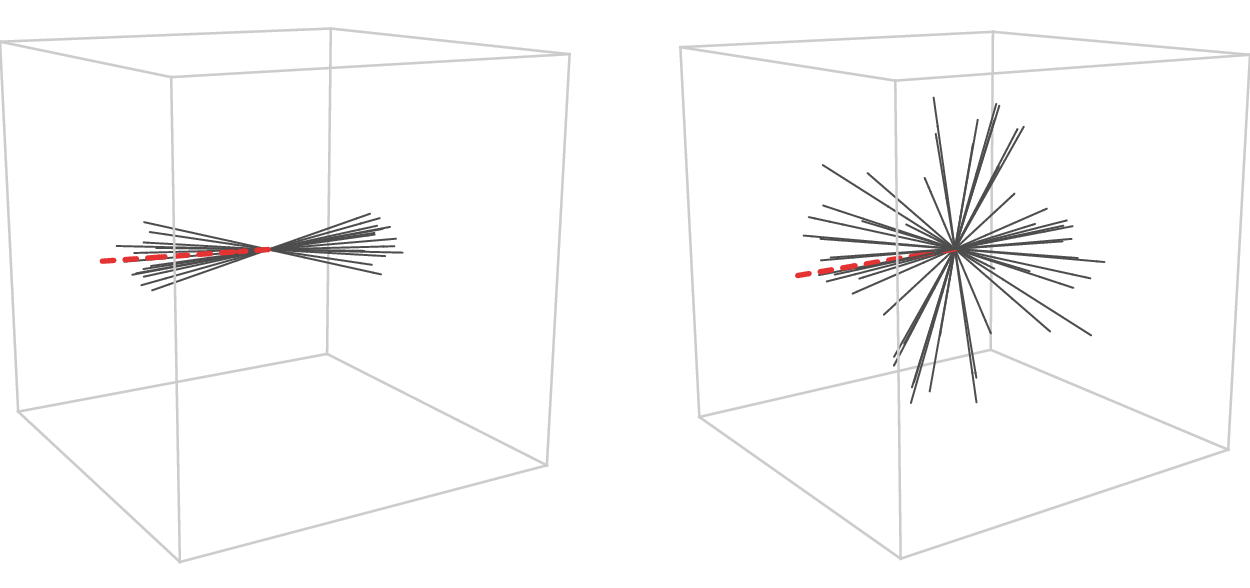} }
\end{tabular}
\end{center}
\vspace{-0.5cm}
\caption{
\label{fig:shape}
Left: The thrust distribution as measured by the ALEPH experiment at LEP I \cite{Heister:2003aj}. The inset shows the region relevant for the $\alpha_s$ determination. Right:  Sample collider events. The two-jet configuration on the left has a large thrust $T\approx  0.98$, while the multi-jet event on the right has $T \approx 0.65$ (note that a completely spherical event has $T=1/2$). The red dashed line indicated the thrust vector.}
\end{figure}

Thrust and a number of other event shapes have been measured with exquisite precision by the LEP experiments at CERN. As an example, Figure \ref{fig:shape} shows the thrust distribution as measured by the ALEPH experiment at LEP  \cite{Heister:2003aj}. One immediately observes that most events have large thrust. This is not surprising: the lowest order in perturbation theory consists of a back-to-back quark anti-quark pair and has $T=1$. Contributions which involve large-angle radiation are suppressed by the coupling constant $\alpha_s$. Most events therefore consist of two narrow jets formed by the $q\bar{q}$ pair and its accompanying soft and collinear radiation.  The typical mass of the jets at large thrust is $M_J^2 \sim Q^2(1-T)$ and perturbative corrections to the thrust distribution are enhanced by logarithms of $M_J^2/Q^2 \sim (1-T)$ which need to be resummed. One can analyze the two-jet region using SCET and can derive a factorization theorem for the cross section; this quantity  can be written in terms of a hard function, two jet functions and a soft function \cite{Fleming:2007qr,Schwartz:2007ib,Bauer:2008dt}. Using the RG methods we discussed in Section \ref{sec:DY} the thrust distribution was resummed  up to N$^3$LL accuracy in \cite{Becher:2008cf}, two orders in logarithmic accuracy higher than what had been achieved with traditional methods, and matched to NNLO fixed order results \cite{GehrmannDeRidder:2007hr,Weinzierl:2009ms}. Based on this result, Ref.~\cite{Abbate:2010xh} performed a precision determination of the strong coupling constant $\alpha_s$ from this variable. In this analysis, both the value of the coupling constant and non-perturbative effects are extracted from a fit to the available experimental data. The resulting value $\alpha_s(M_Z)=0.1135 \pm (0.0002)_{\rm expt} \pm (0.0005)_{\rm hadr} \pm (0.0009)_{\rm pert}$ has very small uncertainties and is significantly lower than the world average for $\alpha_s=0.1185\pm 0.0006$ \cite{Agashe:2014kda}, whose small error is due to the small uncertainty of the lattice QCD results. Let us note that hadronization effects play a significant role in the extraction of $\alpha_s$ from event shapes. Accounting for them lowers the extracted value of $\alpha_s(M_Z)$ by $8\%$ \cite{Abbate:2010xh}.
To obtain the above level of accuracy, hadronization as well as other small effects (such as hadron mass effects \cite{Salam:2001bd, Mateu:2012nk} and finite $b$-quark mass effects \cite{Gritschacher:2013pha,Pietrulewicz:2014qza}) need to be under good control. 
 
 It will be important to validate the above result for $\alpha_s$ by using other event-shape variables for which accurate predictions are available. One such example is the heavy jet mass, which was evaluated in \cite{Chien:2010kc} to N$^3$LL accuracy.  Another example are the total and wide jet broadenings for which a factorization theorem was obtained in \cite{Chiu:2011qc,Becher:2011pf} and for which NNLL resummation was performed in \cite{Becher:2012qc}. The definition of the total jet broadening is identical to the one for thrust in Eq.~(\ref{eq:thrust}), except that one measures the momentum transverse to the thrust axis instead of the longitudinal momentum. Since it involves small transverse momenta, the factorization theorem for broadening is characterized by the presence of a collinear anomaly, which generates an additional dependence on the large momentum transfer in the product of the jet and soft functions. The so-called angularity is an event shape which depends on a parameter $\alpha$ and reduces to thrust for $\alpha=0$ and broadening for $\alpha=1$. Lee and Sterman were able to show that the same parameter governs the leading non-perturbative corrections for all values $\alpha<1$ \cite{Lee:2006nr} and recently this result was extended to the broadening $\alpha=1$ \cite{Becher:2013iya}. Because of the anomaly, non-perturbative effects for the broadening are logarithmically enhanced. Since the same hadronic parameter generates the leading non-perturbative corrections to an entire class of event shapes, a simultaneous fit to multiple shapes should give a better handle on the hadronization effects.  Angularity distributions at NLL accuracy were studied in \cite{Hornig:2009vb} and a detailed comparison to the resummation for the same observables carried out by means of traditional (``direct'' QCD) methods was presented in \cite{Almeida:2014uva}. In the latter work it was found that the two resummation methods, direct QCD and SCET, are equivalent, and the origin of numerical differences in the implementation of the resummation formulas in the two frameworks was clarified.

A complication for the case of broadening is that this variable is sensitive to soft recoil \cite{Dokshitzer:1998kz}. While the transverse momentum of the soft radiation can be neglected for thrust, it has to be taken into account for broadening since it is of the same order of magnitude as the transverse momentum of the collinear radiation. These recoil effects complicate the computation of broadening and recently an alternative definition of the usual event shapes has been proposed which uses the broadening axis, the axis which minimizes the scalar sum of the transverse momenta, to define the shapes, and is insensitive to recoil effects \cite{Larkoski:2014uqa}. 

An event shape suitable for $e^+ e^-$, $e^- p$, as well as $p p$ collisions is the $N$-jettiness $\tau_N$. It provides a generalization of thrust and vanishes in the limit of $N$ infinitely narrow jets \cite{Stewart:2010tn}.
Instead of a single thrust vector, one introduces $N$ different reference vectors, one for each direction of a final state jet, and groups the momenta into $N$ groups, according to their largest component along the vectors. Finally, one minimizes over the directions of the vectors.  For $e^+ e^-$, two-jettiness $\tau_2 = 1-T$. For hadronic collisions, the reference vectors always include the two beam directions. The zero-jettiness (which considers only radiation along the beam directions) is also called beam thrust and has been computed to NNLL accuracy for the Drell-Yan and Higgs production processes in \cite{Stewart:2009yx,Berger:2010xi}. At the same accuracy, also 1-jettiness in Higgs production has been considered \cite{Jouttenus:2013hs}. The two-loop beam functions relevant for these cases are now available \cite{Gaunt:2014xga,Gaunt:2014cfa}. In addition, 1-jettiness for Deep-Inelastic Scattering (DIS), $e^- p \to e^- X$, has been considered and all the ingredients of the relevant factorization theorem were computed at one-loop accuracy \cite{Kang:2012zr,Kang:2013nha,Kang:2013lga,Kang:2014qba}.

\subsection{Jet Physics\label{sec:jets}}

Most of collider physics is discussed in terms of jet observables, which are much more common, but also more complicated than event-shape variables. The basic goal of jet definitions is to obtain observables which are closely related to the underlying hard-scattering process and provide detailed information about it, but are also inclusive enough to be perturbatively calculable. There are many different jet definitions, i.e.\ ways to group particles of a given event into jets, but a basic requirement is again infrared safety, as discussed in the context of event shapes in the previous subsection: the jet clustering should be insensitive to collinear splittings and soft emissions. Historically, not all jet algorithms used by experiments have fulfilled this requirement and only in recent years algorithms have been developed which are both practical at hadron colliders and theoretically sound \cite{Cacciari:2008gp,Salam:2007xv}. The jet algorithms currently in use fall in two categories: cone algorithms, which group particles moving inside a specified angular region into jets, and sequential recombination algorithms, which cluster particles according to a measure in momentum space, until a termination criterion is met at which point the combined particles define a jet. A review of different jet algorithms and their applications can be found in \cite{Salam:2009jx}. 

Given that jets are collinear sprays of particles surrounded by soft radiation, SCET is a natural framework to describe them, particularly in the limit where the invariant masses of the jets are much lower than the center-of-mass energy at which they are produced. In Section \ref{sec:LV}, we have discussed the effective theory for a process with $N$ jets: it involves a collinear field for each direction and a soft field which mediates interactions among the different directions. The relevant hard functions are given by the scattering amplitudes and we have shown in Section \ref{sec:LV} that factorization constraints completely fix their anomalous dimension to NNLL accuracy \cite{Becher:2009qa}. The jet and soft functions will depend on the jet clustering algorithm and need to be computed to one-loop accuracy if one aims at NNLL resummation. The basic factorization theorem for $N$-jet processes is discussed in more detail in \cite{Bauer:2008jx}.

While this broad-brush picture is correct, the standard hard-jet-soft factorization does not immediately translate into a resummation of all large logarithms. For complicated observables, it can happen that the soft and jet functions themselves suffer from large logarithms, for any choice of the renormalization scale. This happens, for example, if the soft radiation is not distributed uniformly. The simplest example of such a situation is a soft function where the radiation is split into two hemispheres and the radiation is forced to have energy $\omega_1$ in one hemisphere and $\omega_2$ in the other. If $\omega_1 \ll \omega_2$, this soft function has logarithms of the form $\alpha_s^n \ln^m \omega_1/\omega_2$ with $m \leq n$, starting at ${\mathcal O}(\alpha_s^2)$. These types of logarithms are called non-global logarithms because they appear in particular when an observable is insensitive to emissions into certain regions of phase space (the two-hemisphere soft function in our example is not very sensitive to emissions into the second hemisphere) \cite{Dasgupta:2001sh}. Except for the leading non-global logarithms in the large $N_c$ limit, which can be resummed using Monte Carlo methods \cite{Dasgupta:2001sh} or by solving an integro-differential equation \cite{Banfi:2002hw}, their resummation is not yet understood. The full two-loop hemisphere soft function was computed in \cite{Kelley:2011ng,Hornig:2011iu}, so that the full analytic dependence of this function on the ratio $ \omega_1/\omega_2$ is now known. Furthermore, the leading logarithms up to five-loop order were obtained in \cite{Schwartz:2014wha} analytically, by solving the equation derived in \cite{Banfi:2002hw} order by order. Hopefully these computations will provide a starting point for a better understanding  of these types of logarithms. Since we deal with a hierarchy of scales, it should be possible to construct an effective theory which factorizes the contributions from the two scales. However, constructing such an effective theory remains an interesting open problem.

All jet algorithms have a dimensionless parameter $R$, the jet radius, which determines the jet size. At small $R$, the jets are narrow and one encounters logarithms of $R$ in perturbative calculations. At one-loop order, the appearance of these logarithms in SCET for different $e^+e^-$ jet algorithms was studied in \cite{Cheung:2009sg}, the particular case of Sterman-Weinberg jets was discussed earlier in \cite{Trott:2006bk}. For recombination-style jet algorithms, such logarithms were studied in \cite{Kelley:2012kj,Kelley:2012zs}. These papers relate these logarithms to non-global logarithms and conclude that, due the complicated nature of the $n$-particle phase-space constraints, in recombination algorithms higher-logarithmic resummations appear difficult. 

Our discussion so far makes it clear that resummations for jet variables are challenging. On the other hand, there is a lot to gain from a better understanding of jet observables and jet substructure, in particular at the LHC. Because of the high center-of-mass energy, hadronic decay products of heavier particles are often very boosted and inside a single jet. The substructure of this jet can be used to identify the underlying particle. Many jet substructure techniques have been developed over the last years, but the validation and understanding of these methods is so far mostly based on parton shower Monte Carlo programs, which are only accurate at the leading-logarithmic level. In SCET, jet shapes for exclusive multi-jet events in $e^+e^-$ collisions were considered in \cite{Ellis:2009wj, Ellis:2010rwa}; the factorization of these observables was studied by means of SCET method and the resummation of large logarithms was carried out up to NLL accuracy. These observables suffer, however, from non-global logarithms which enter at the same accuracy \cite{Banfi:2010pa}. The resummation of the jet mass in $e^+ e^- \to 2$ jets with a jet veto was analyzed by means of SCET in \cite{Kelley:2011tj, Kelley:2011aa}. In these works the dominant dependence on $R$ and the leading non-global logarithms were obtained. The full two-loop result for this quantity was recently obtained in \cite{vonManteuffel:2013vja}. One promising way to study the structure of jets is to use events shapes defined for the particles inside the jet. For this purpose ``N-subjettiness'' was introduced in \cite{Thaler:2010tr}. 

In many cases, measurements and searches at the LHC are performed using jet bins, because the background composition can be quite different if the final state contains jets. An example is Higgs production with subsequent decay $H\to W^+ W^-$, which receives a large background from $t\bar{t} \to W^+ W^- b \bar{b}$. Since it comes with two $b$-jets, the top-quark background can be significantly reduced by imposing a veto on jets. (An alternative method is to use an event shape, such as beam thrust, to suppress additional jets \cite{Berger:2010xi}.) Since the recombination jet algorithms used by the experiments cluster all particles into jets, one cannot completely avoid jets, but one can veto jets with transverse momentum larger than a threshold $p_T^{\rm jet} > p_T^{\rm veto}$, where $p_T^{\rm veto}$ is chosen to be of order $20 - 30 {\rm GeV}$. This veto induces logarithms of the ratio $m_H/p_T^{\text{veto}}$. With traditional methods, the resummation of these logarithms was achieved to NLL accuracy in \cite{Banfi:2012yh} and extended to NNLL in \cite{Banfi:2012jm}. In between these two papers, an all-order resummation formula for the cross section was obtained using SCET in \cite{Becher:2012qa}. This factorization theorem again suffers from a collinear anomaly. Based on this theorem, numerical results at N$^3$LL$_{\rm partial}$ were given in \cite{Becher:2013xia}, where ``partial'' refers to the fact, that the three-loop anomaly constant, which is needed at this accuracy, is not yet available and its effect was estimated numerically. This constant is a function of the jet radius and the leading logarithmic piece at three-loop accuracy was obtained in \cite{Alioli:2013hba} and turns out to be small. The factorization theorem obtained in 
\cite{Becher:2012qa} is based on the fact that soft and collinear radiation separately clusters into jets at low $p_T^{\rm veto}$. This was called into question in \cite{Tackmann:2012bt} where it was claimed that the independent clustering of the radiation is only guaranteed at small jet radius $R \sim p_T^{\text{veto}}/m_H$ and that NNLL resummation at finite $R$ is therefore not possible. The numerical results  \cite{Banfi:2012jm} and the analytic studies of \cite{Becher:2013xia} show that resummation at NNLL is possible, but the authors of \cite{Tackmann:2012bt} maintain that the factorization at finite $R$ beyond this accuracy is an open question. They have released an independent numerical analysis at NNLL$'$  level, where the prime indicates that the analysis includes all the two-loop ingredients in the factorization formula  \cite{Stewart:2013faa}. The three results \cite{Banfi:2012jm,Becher:2013xia,Stewart:2013faa} differ in formalism and scheme choices, but are equivalent at NNLL accuracy and are matched to the NNLO fixed-order result. In addition, also resummation for the cross section with a single hard jet was considered \cite{Liu:2013hba,Liu:2012sz}.  This suffers from the non-global logarithms discussed above, but it was argued that their effects are small. A perhaps more important limitation is that the resummation is only valid for large $p_T$ of the jet, but most jets in the one-jet bin have transverse momentum just above the veto scale. The consistent combination of results in different jet bins was addressed in \cite{Boughezal:2013oha}. Jet vetoes play a role also in other processes. The resummation of the associated logarithms was also studied for off-shell Higgs bosons decaying in $W$ boson pairs \cite{Moult:2014pja}, associated production of a Higgs and a vector boson \cite{Shao:2013uba} and $W^+W^-$ production \cite{Jaiswal:2014yba}. 

All resummations described so far were performed analytically, on a case-by-case basis. It would be desirable to have a flexible numerical framework, similar to a Monte-Carlo event generator, which resums not only leading, but also subleading logarithms. Within traditional resummation, such an automated resummation was achieved at NLL accuracy with a computer code called CAESAR \cite{Banfi:2004yd}. This code is restricted to observables which are global, so most jet observables cannot be resummed. So far, no such code based on SCET has been constructed, but there are ongoing efforts to improve parton showers using SCET. The relation between SCET and parton showers was first investigated in \cite{Bauer:2006mk} where the parton shower was derived from a sequence of effective theories. This paper did, however, not address the role of soft gluons and its analysis is therefore incomplete. A full analysis for the case of a hierarchical three-jets configuration, where two jets are close to each other, was later given in \cite{Bauer:2011uc}. A SCET improved parton shower is GENEVA  \cite{Bauer:2008qh,Bauer:2008qj,Alioli:2012fc,Alioli:2013hqa}. (At the time of this writing, a public code is not yet available.) Based on a standard parton shower, it is accurate at LL accuracy, but implements matching to fixed order at different jet multiplicities. To distinguish the different jet multiplicities, it uses the $N$-jettiness event shape, which is implemented at NNLL accuracy. The fact of having the resolution parameter resummed has the advantage that one does not suffer from large corrections due to logarithms of the resolution parameter. One subtle issue in this approach is that one wants to add showering to get a good description of other observables, but needs to avoid that the shower destroys the logarithmic accuracy of $N$-jettiness. In other words, one needs to consistently match the parton shower and the resummed result.

SCET has also been used to analyze jet-physics observables which are not infrared safe and therefore need non-perturbative input; the papers \cite{Bauer:2002ie,Bauer:2003di} provide an early example of this type of analysis. More recent work includes cross sections with identified hadrons inside a jet, which require non-perturbative fragmentation functions \cite{Neubert:2007je,Procura:2009vm,Jain:2011xz,Procura:2011aq,Jain:2012uq,Bauer:2013bza}, jet charge distributions \cite{Waalewijn:2012sv} and track-based observables \cite{Chang:2013rca,Chang:2013iba}.

\subsection{Electroweak Sudakov Logarithms, Glauber Gluons, and Gravity\label{sec:additional}}

We close our overview by going over a few additional interesting applications, which do not fit the classification of the previous subsections. These include some newer developments, where first steps towards an effective-theory analysis of a given problem were achieved, but open questions remain.

For high-energy collisions at energies which exceed the masses of electroweak bosons, large electroweak Sudakov logarithms arise. We have discussed the massive scalar form factor integral in Section \ref{sec:MSPCA}, which provides the simplest example where such a logarithm is present. Electroweak logarithms can be numerically large at the LHC and their resummation by means of SCET was studied in \cite{Chiu:2007yn,Chiu:2007dg, Chiu:2008vv, Chiu:2009mg, Chiu:2009ft}.  The example in Section \ref{sec:MSPCA} shows that such processes suffer from collinear anomalies and the resummation of the anomalous logarithms was first understood in this context \cite{Chiu:2007dg}. The papers \cite{Chiu:2007yn,Chiu:2007dg, Chiu:2008vv, Chiu:2009mg, Chiu:2009ft} focused on logarithms due to virtual diagrams and computed them for a number of different processes. An example where this formalism has been applied to a physical cross section is  \cite{Becher:2013zua}: In that work the resummation of electroweak logarithms in single $Z,W,\gamma$  production at large transverse momentum was carried out. In \cite{Becher:2013zua}, the real emissions were included using threshold resummation. SCET has also been used to analyze electroweak effects in Higgs production via vector-boson fusion \cite{Fuhrer:2010vi, Siringo:2012mi} and to the $t\bar{t}$ asymmetry \cite{Manohar:2012rs}. An interesting property of electroweak Sudakov logarithms is that they persist even in inclusive cross sections \cite{Ciafaloni:2000df,Bell:2010gi,Manohar:2014vxa}. This arises because the initial states of the collisions are color-neutral but charged under weak SU(2).

In the factorization proofs for the Drell-Yan process, an important part of the analysis was to show that the so-called Glauber momentum region does not contribute \cite{Collins:1981tt,Bodwin:1984hc,Collins:1985ue}. The Glauber gluons have transverse momenta $p_T$ much larger than their light-cone components $p_+$ and $p_-$ and can induce Coulomb-like interactions among soft and collinear particles. Since their $p_+$ and $p_-$ momentum components are negligible compared to the transverse momentum, these Glauber modes are off the mass shell. Similar to Coulomb gluons, they should be described by a potential, not by a dynamical field. Glauber interactions can only arise in forward scattering and naively one might think that they cannot play a role for processes such as inclusive Drell-Yan production. However, any hadron collider cross section includes a forward-scattering part, because the proton remnant (i.e. the proton without the parton which participates in the hard-scattering) moves in the forward direction. A strategy of region analysis of Drell-Yan diagrams which include the spectator quarks was performed in \cite{Bauer:2010cc} which showed that a contribution from such a region indeed arises in individual diagrams. At the same time, this analysis also revealed that this contribution is not unambiguously defined without additional regulators, as is characteristic for processes which suffer from a collinear anomaly. An important feature of forward scattering amplitudes is Regge behavior, the statement that the forward scattering amplitude develops a power-like dependence on the momentum transfer ${\cal M} \sim s^{\alpha(t)}$, where $s$ and $t$ are the usual Mandelstam invariants and $\alpha(t)$ is called the Regge exponent. Recently, there has been progress in analyzing this behavior using SCET and understanding the role Glauber modes play in it. In particular, two papers have shown how to obtain Regge behavior from SCET. The paper \cite{Donoghue:2014mpa} performed a region analysis of forward scattering diagrams, while the paper \cite{Fleming:2014rea} has derived the Balitsky-Fadin-Kuraev-Lipatov (BFKL) equation \cite{Fadin:1975cb,Balitsky:1978ic} using the rapidity RG \cite{Chiu:2012ir}, an alternative framework to resum anomaly logarithms. In both papers, the Glauber region plays a prominent role. Nevertheless, a complete treatment of Glauber contributions in the effective theory is not yet available. The inclusion of Glauber gluons would open the door to analyze what role these modes play in different observables. The paper \cite{Gaunt:2014ska} showed that the standard arguments used to show their absence fail for less inclusive variables and argues that Glauber contributions might be responsible for multi-parton interactions. These are modeled in parton-shower Monte Carlos to describe the data but lack a clear field theoretic interpretation. Glauber gluons also play a role when one describes the propagation of an energetic particle through a medium. This process is relevant to understand jet quenching in heavy-ion collisions, which was analyzed in the context of SCET in \cite{Idilbi:2008vm,D'Eramo:2010ak,Ovanesyan:2011xy,Benzke:2012sz}.

In addition to the soft multi-parton interactions, which we mentioned in the context of Glauber contributions, one can encounter double hard scattering at colliders, a situation which was studied in SCET in \cite{Manohar:2012jr,Manohar:2012pe,Chang:2012nw}. The corresponding factorization is again anomalous.

Finally, SCET has also been used to analyze the physics of soft and collinear gravitons \cite{Beneke:2012xa}. This connects back to a paper of Weinberg we mentioned earlier and in which he showed that the infrared structure of gravity is very similar to QED \cite{Weinberg:1965nx}. The structure is simpler than the one encountered in non-abelian gauge theories because the graviton couples proportional to energy and the spin structure prohibits singular collinear splittings. The proof of the absence of collinear singularities in gravity is nevertheless nontrivial, because one component of the collinear field scales with a negative power $1/\lambda$ (for a diagrammatic proof of the absence of collinear singularities, see \cite{Akhoury:2011kq}).

\subsection{Outlook \label{sec:outlook}}

Our overview shows that SCET has been applied to a broad range of processes. Often, also a treatment with traditional methods is available, but by now there are many examples where higher-log resummation was first achieved within the effective theory framework and where the accuracy obtained using SCET is higher than what was obtained using diagrammatic methods. At this year's (2014) Loopfest conference, the summary speaker observed that most of the presented work on resummation was done using effective field theory methods, so SCET is becoming a standard tool in this field. That the method has reached maturity can also be deduced from the fact that one of the chapters of the Review of Particle Physics \cite{Agashe:2014kda} is devoted to SCET and that the effective theory now has found its way into a quantum field theory textbook \cite{Schwartz:2013pla}.

On the other hand, in many respects, we are still at the beginning. There are still new applications of SCET which appear in the literature, a very recent example is the computation of the annihilation rates of heavy dark matter particles \cite{Baumgart:2014vma,Bauer:2014ula,Ovanesyan:2014fwa}. As we discussed in detail above, there are also many questions which remain open. In particular, there are large classes of observables, such as those which involve non-global logarithms, for which we do not yet know how to resum to higher accuracy, and even for basic hadron collider observables, there are still open questions about their factorization. Even in cases where we know how to perform resummations in the effective field theory, they are currently still done analytically on a case-by-case basis. We eventually would like to have automated tools to perform such calculations. On an even more basic level, there is also ongoing work on an alternative, more physical formulation of the effective theory for energetic particles \cite{Georgi:2013lza,Georgi:2014sxa}.

We hope our introduction will make SCET accessible to a wider audience and invite our readers to contribute to answering some of the open questions!

%% file: Appendix.tex
\setcounter{figure}{0}
\section{Summary of Notations and Conventions}

For the reader's convenience, we collect here some basic notation and conventions used throughout these lectures and introduced at various points in the main text. 

\subsubsection*{Kinematics}

We make use of the traditional mostly minus metric with signature $\left(+,-,-,-\right)$, so that for an on-shell massive particle of four-momentum $p^\mu \equiv(E, \vec{p}) = (p_0,p_x,p_y,p_z)$ one finds 
\be
p^2 \equiv p \cdot p = E^2 -|\vec{p}|^2 = m^2 >0 \, .
\ee

The light vectors $n$ and $\bar{n}$ are defined as
\be
n_\mu = (1,0,0,1) \, , \quad \text{and} \quad \bar{n}_{\mu} = (1,0,0,-1) \, ,
\ee
so that 
\be
n^2 = n \cdot n = 0\, , \qquad \bar{n}^2 = \bar{n} \cdot \bar{n} = 0\, , 
\qquad n \cdot \bar{n} = 2 \,.
\ee

For each four-vector $p^\mu$ we can define the \emph{four-vectors} $p_+^\mu$ and $p_-^\mu$  by using the light-like vectors $n$ and $\bar{n}$:
\be
p_+^\mu \equiv \left(n \cdot p \right)\frac{\bar{n}^\mu}{2} \, , \qquad
p_-^\mu \equiv \left(\bar{n} \cdot p \right)\frac{n^\mu}{2} \, , \qquad\label{eq:appnot2}
\ee
so that the vector $p^\mu$ can be written in terms of these vectors or as a list of components in the light-cone basis defined by $n$ and $\bar{n}$:
\be
p^\mu = p_+^\mu + p_-^\mu  + p_\perp^\mu \equiv \left(n \cdot p\, , \bar{n} \cdot p\, , p_\perp^\mu \right) \, .
\ee
The four-vector $p_\perp^\mu$ is related to \emph{two-dimensional} vector  $\vec{p}_T$ by $p_\perp^2 = - |\vec{p}_T|^2 \equiv - p_T^2$.

The scalar product of two vectors $p^\mu$ and $q^\mu$ can then be written as
\begin{align}
p \cdot q &= p_+ \cdot q_- +  p_- \cdot q_+ + p_\perp \cdot q_\perp 
= \frac{1}{2} \left( n \cdot p \right)\left(\bar{n} \cdot q \right) + \frac{1}{2} \left( n \cdot q \right)\left(\bar{n} \cdot p \right)- \vec{p}_T \cdot \vec{q}_T \, ,
\end{align}
Consequently, the square of $p$ is
\be
p^2 = 2 p_+ \cdot p_- +p_\perp^2 = \left( n \cdot p \right)\left(\bar{n} \cdot p \right) -p_T^2 \,. \label{eq:appnot1}
\ee

In order to match the literature and to keep the notation as supple as possible, on a few occasions, explicitly indicated in the text, we employ the superscripts $\pm$ to indicate the light-cone \emph{components} rather than the four-vectors. In those cases $p_+ \equiv n \cdot p = p_0 + p_z$ and $p_- \equiv \bar{n} \cdot p= p_0 - p_z$. We make use of this notation for example in Section~\ref{sec:soft} after Eq.~(\ref{eq:tagapp}). If $p_+$ and $p_-$ indicate the light-cone  components, then $p^2 = p_+ p_- -p_T^2$. This last identity should be compared with Eq.~(\ref{eq:appnot1}) which is written in terms of the light cone vectors defined in Eq.~(\ref{eq:appnot2}). Using light-cone components, the integration over momentum space can be written as
\begin{align}
\int d^d k &= \frac{1}{2} \int_{-\infty}^{+\infty} dk_+ \int_{-\infty}^{+\infty} dk_- \int d^{d-2} k_\perp\,, \nn \\
\int d^d k\, \delta(k^2) \theta(k^0) &= \frac{1}{2} \int_{0}^{\infty} dk_+ \int_{0}^{\infty} dk_- \int d^{d-2} k_\perp \,\delta(k_+ k_- - k_T^2)\, .
\end{align}
We also use the common notation $\delta^+(k^2) = \delta(k^2) \theta(k^0)$.

\subsubsection*{Scaling}

We summarize the scaling of the various types of momenta in terms of a generic hard scale $Q$ and a small dimensionless expansion parameter $\lambda$:
\begin{align}
&\text{{\bf Hard momentum}} & \hspace*{.1cm}& p \sim \left(1,1,1 \right) Q \, , \nonumber \\
&\pcol{\text{{\bf Collinear to   }} {\bm{n}}} & \hspace*{.1cm}& p \sim \left(\lambda^2,1,\lambda \right) Q \, , \nonumber \\
&\lcol{\text{{\bf Collinear to   }} {\bm{\bar{n}}}} & \hspace*{.1cm}& p \sim \left(1,\lambda^2,\lambda \right) Q \, , \nonumber \\
&\text{{\bf Semi-hard (a.k.a. soft)}}  & \hspace*{.1cm}& p \sim \left(\lambda,\lambda,\lambda \right) Q \, , \nonumber \\
&\scol{\text{{\bf Soft (a.k.a. ultra-soft)}}}  & \hspace*{.1cm}& p \sim \left(\lambda^2,\lambda^2,\lambda^2 \right) Q \, , \nonumber \\
&\text{{\bf Glauber modes}}  & \hspace*{.1cm}& p \sim \left(\lambda^2,\lambda^2,\lambda \right) Q \, .\nonumber
\end{align}

The collinear and soft fields in QCD scale as follows
\begin{align}
&\pcol{\text{{\bf Collinear quark component   }} \bm{\xi}} & \hspace*{.1cm}&
\pcol{\xi} = P_+ \psi_c = \frac{\nsl \nbsl}{4} \psi_c \sim \lambda \, , \nonumber \\
&\pcol{\text{{\bf Collinear quark component   }} \bm{\eta}} & \hspace*{.1cm}&
\pcol{\eta} = P_- \psi_c = \frac{\nbsl \nsl}{4} \psi_c \sim \lambda^2 \, , \nonumber \\
&\pcol{\text{{\bf Collinear gluon fields  }} \bm{A_c}} & \hspace*{.1cm}&
\pcol{A_c} \sim p \sim (\lambda^2, 1, \lambda) \, , \nonumber \\
&\scol{\text{{\bf Soft quark field  }} \bm{\psi_s}} & \hspace*{.1cm}&
\scol{\psi_s} \sim \lambda^3 \, , \nonumber \\
&\scol{\text{{\bf Soft gluon fields  }} \bm{A_s}} & \hspace*{.1cm}&
\scol{A_s} \sim p \sim (\lambda^2, \lambda^2, \lambda^2) \, . \nonumber
\end{align}

\subsubsection*{Regularization and Ultraviolet Renormalization}

We employ dimensional regularization in order to regulate both UV and IR divergences. The dimensional regulator $\ep$ is defined through the equation  $d = 4 - 2 \ep$, where $d$ is the number of dimensions.

At several points throughout the lectures we carry out UV renormalization in massless QCD by absorbing the UV poles in the bare coupling constant $g_s$.
The relevant relation to achieve this goal is 
\be
g_s^2 = 4 \pi \alpha_s^0 = 4 \pi \left(\frac{e^{\gamma_E}}{4 \pi} \mu^2\right)^\ep Z_\alpha \alpha_s \left(\mu \right) \, ,
\ee
with
\be
Z_\alpha = 1- \frac{\alpha_s(\mu)}{4 \pi} \frac{\beta_0}{\ep} 
+ \left(\frac{\alpha_s(\mu)}{4 \pi} \right)^2 \left(\frac{\beta_0^2}{\ep^2} - \frac{\beta_1}{2 \ep} \right)+{\mathcal O} \left(\alpha_s^3 \right) \, .
\ee
With these definitions, the bare (squared) coupling $\alpha_s^0$ (which is of course scale independent) has mass dimension $2 \ep$, while the renormalized coupling $\alpha_s$ is dimensionless. The explicit expression of the coefficients $\beta_0,\beta_1$ can be found in Appendix~\ref{app:AnDim} together with the explicit expressions of the other anomalous dimensions employed in this work.

\setcounter{figure}{0}
\section{One-Loop Integrals}

In this Appendix we collect some details concerning the explicit calculation of the loop integrals discussed in Sections~\ref{sec:regions} and \ref{sec:DY}. The computation of these loop integrals can be done using standard methods, such as Feynman parameterization. The only difference between QCD and SCET integrals are that the latter can also involve propagator denominators which are linear in the loop momentum, while QCD only involves quadratic denominators in covariant gauges. The linear propagators arise from expanding away small momentum components in the soft and collinear regions. To combine linear and quadratic propagators, it is convenient to use the representation
\begin{equation}
\frac{1}{a b} = \int_0^\infty \!dy\, \frac{1}{(a+b y)^2}\,, \label{eq:parlin}
\end{equation}
in cases where $a$ is a standard propagator, and $b$ linear in the loop momentum. Note that the $y$ integral runs up to infinity. By performing a variable change from $y$ to $x$, with $y= x/(1-x)$, one recovers the standard Feynman parameterization. 
Eq.~(\ref{eq:parlin}) can be generalized to the case of $n$- propagators as follows
\begin{equation}
\frac{1}{a_1 a_2 a_3 \cdots a_n} = \int_0^\infty \! dy_1 \int_0^\infty \! dy_2 \cdots  \int_0^\infty \! dy_{n-1}
\frac{(n-1)!}{(a_1 +a_2 y_1 +a_3 y_2+ \cdots a_n y_{n-1})^n} \, . \label{eq:parlinN}
\end{equation}
For higher power of propagators, one uses
\begin{equation}
\frac{1}{a^n b^m} = \frac{\Gamma (m+n)}{\Gamma (m) \Gamma (n)} \int_0^\infty \!dy\, \frac{ y^{m-1}}{(a+y b)^{n+m}}\,.
\end{equation}
Using this relation together with standard Feynman parameterization, one can easily bring all SCET loop integrals in a form where the momentum integration can be carried out.

\subsection{Integral \boldmath$I_h$\unboldmath \label{ap:HardS}}
 In order to obtain the result in Eq.~(\ref{eq:Ihardres}) for the integral $I_h$, we start by applying the Feynman parameterization
 \be
 \frac{1}{a b c} = 2 \int_0^1 dx \int_0^x dy \frac{1}{\left[a y + b (x-y) + c (1-x)\right]^3} \, , 
 \ee
  to the integral in Eq.~(\ref{eq:Ihard}). We then obtain
\bea \label{eq:Ihardcalc}
I_h &=& i \pi^{-d/2} \mu^{4-d} \int d^d k \frac{1}{k^2 
\left(k^2 +2 k_-\cdot l_+ \right) \left( k^2 + 2 k_+\cdot p_- \right)} \,  \nn \\
&=&  i \pi^{-d/2} \mu^{4-d} \int_0^1 dx \int_0^x dy \int d^d k 
\frac{2}{\chi^3 (x,y,k)} \ , 
\eea
where
\bea
\chi (x,y,k) &=& (k^2 + 2 k_+ \cdot p_-) y + k^2 (x-y) + 
(k^2 + 2 k_-\cdot l_+) (1-x) \,  \nn \\
&=& k^2 +2 k \cdot \left[p y +l (1-x)\right] +{\mathcal O}(\lambda) \, .
\eea

The integral over the virtual momentum can be evaluated by employing the formula
\be
\int d^d k \frac{1}{(k^2 +2 k \cdot  Q - M^2)^\alpha} = (-1)^\alpha \frac{i \pi^{\frac{d}{2}}}{\left(M^2 +Q^2 \right)^{\alpha - \frac{d}{2}}} \frac{\Gamma\left( \alpha - \frac{d}{2}\right)}{\Gamma\left(\alpha \right)} \, . \label{eq:MF}
\ee
In this way one finds
\be
\int d^d k \frac{1}{\chi^3 (x,y,k)} = -\frac{i \pi^{\frac{d}{2}}}{2}
\Gamma\left( 3 - \frac{d}{2}\right) V^{\frac{d}{2} -3} \left(x,y \right) \, ,
\ee
with
\be
V(x,y) = p^2 y^2 + l^2 (1-x)^2 +2 p\cdot l y (1-x) 
= 2 l_+ \cdot p_- y (1-x) + {\mathcal O}(\lambda^2) \, .
\ee
Therefore, the integral $I_h$ becomes
\be \label{eq:Ihpar}
I_h =  \frac{\Gamma(1+\varepsilon)}{2 l_+ \cdot p_-} 
\left( \frac{\mu^2}{2 l_+ \cdot p_-} \right)^\varepsilon
\int_0^1 dx \int_0^x dy\frac{1}{\left[ y (1-x) \right]^{1+\varepsilon}} \, .
\ee
The integral over the Feynman parameters $x$ and $y$ gives
\bea
\int_0^1 dx \int_0^x dy \frac{1}{\left[ y (1-x) \right]^{1+\varepsilon}} &=& 
-\frac{1}{\varepsilon} \int_0^1 dx x^{-\varepsilon} (1-x)^{-1-\varepsilon} \, 
\nn \\
 &=& -\frac{1}{\varepsilon} \frac{\Gamma(1-\varepsilon) \Gamma(-\varepsilon)}{\Gamma(1- 2 \varepsilon)} = \frac{\Gamma^2(-\varepsilon)}{\Gamma(1- 2 \varepsilon)} \, .
\eea
By inserting the equation above in Eq.(\ref{eq:Ihpar}) one obtains Eq.~(\ref{eq:Ihardres})\,.

\subsection{Integral \boldmath$I_{\cc}$\unboldmath \label{ap:Icp}}

In this appendix we evaluate the collinear region integral in Eq.~(\ref{eq:Icp}). This integral now involves a linear propagator and we choose to employ the parametrization of the integrand in Eq.~(\ref{eq:parlinN}) for the case $n=3$
\be \label{eq:C1}
\frac{1}{a b c} = \int_0^\infty d x_1 \, \int_0^\infty dx_2 \, \frac{2}{(a +b x_1 +c x_2)^3}\,,
\ee
where we identify the denominators as follows: $a =k^2$, $c = 2 l_+\cdot k$, and $b=(k+p)^2$. In this way one finds that
\be
I_{\cc} = i \pi^{-\frac{d}{2}} \mu^{4-d} \int_0^\infty dx_1 \int_0^\infty dx_2 \int d^d k\, \frac{2}{\left[(1+x_1) \left( k^2 + 2k\cdot V -M^2\right)\right]^3}\,,
\ee
with 
\be
V^\mu = \frac{x_1 p^\mu +x_2 l_+^\mu}{(1+x_1)} \, , \qquad M^2 = \frac{-x_1 p^2}{(1+x_1)} \,. 
\ee
At this stage it is possible to evaluate the integral over the virtual momentum 
by employing the master formula Eq.~(\ref{eq:MF}); in this way one finds
\bea
I_{\cc} &=& \mu^{2 \varepsilon} \Gamma\left(1+\varepsilon \right)\int_0^\infty dx_1  \frac{1}{(1+x_1)^3}\int_0^\infty dx_2 \left( 
\frac{ P^2 x_1 + 2 l_{+} \cdot p_- x_1 x_2 }{(1+x_1)^2}\right)^{-1-\varepsilon}
\nn \\
&=& \mu^{2 \varepsilon} \Gamma\left(1+\varepsilon \right)\int_0^\infty dx_1  \frac{x_1^{-1-\varepsilon}}{(1+x_1)^{1- 2 \varepsilon}}\int_0^\infty dx_2 \left( 
P^2  + 2 l_{+} \cdot p_- x_2 \right)^{-1-\varepsilon} \, ,
\eea
where $p^2 = -P^2$.
The integrals over $x_1$ and $x_2$ factor and one finds
\be
I_{\cc} = \mu^{2 \varepsilon} \Gamma\left(1+\varepsilon \right) P^{-2-2 \varepsilon} \frac{\Gamma(1-\varepsilon)\Gamma(-\varepsilon)}{\Gamma(1-2 \varepsilon)}\int_0^\infty dx_2 \left( 
1  + r x_2 \right)^{-1-\varepsilon} \, ,
\ee
with  $r = 2 l_{+} \cdot p_-/P^2$. By replacing $x_2 \to x'_2/r$ one arrives at
\bea
I_{\cc} &=& \left(\frac{\mu^{2 }}{P^2}\right)^{\varepsilon} \frac{\Gamma\left(1+\varepsilon \right)}{2 l_{+} \cdot p_-} \frac{\Gamma(1-\varepsilon)\Gamma(-\varepsilon)}{\Gamma(1-2 \varepsilon)}\int_0^\infty dx'_2 \left( 
1  + x'_2 \right)^{-1-\varepsilon} \nn \\
&=&\left(\frac{\mu^{2 }}{P^2}\right)^{\varepsilon} \frac{\Gamma\left(1+\varepsilon \right)}{2 l_{+} \cdot p_-} \frac{\Gamma(1-\varepsilon)\Gamma(-\varepsilon)}{\varepsilon\Gamma(1-2 \varepsilon)}\nn \\
&=&- \frac{\Gamma\left(1+\varepsilon \right)}{2 l_{+} \cdot p_-} \frac{\Gamma^2(-\varepsilon)}{\Gamma(1-2 \varepsilon)}\left(\frac{\mu^{2 }}{P^2}\right)^{\varepsilon}\, ,
\eea
which is the result found in Eq.~(\ref{eq:Icp}).

\subsection{Integral \boldmath$I_s$\unboldmath \label{ap:Is}}

Finally, let us also evaluate the soft region integral in Eq.~(\ref{eq:Isres}).
As a first step we again apply the Feynman parametrization in Eq.~(\ref{eq:C1}).
By choosing $a =k^2$, $b = 2 p_-\cdot k +p^2$, and $c=2 l_+\cdot k +l^2$, the denominator of the integrand of Eq.~(\ref{eq:C1}) becomes
\be \label{eq:pols}
a +b x_1 +c x_2 = k^2 + 2 k \cdot \left(p_- x_1 + l_+ x_2 \right) + p^2 x_1 + l^2 x_2 \,.
\ee
It is now possible to integrate over the virtual momentum by employing the master formula in Eq.~(\ref{eq:MF})
%
%
so that one finds
\be \label{eq:Is}
I_s = \mu^{2 \ep}\Gamma\left( 1+ \ep\right) \int_0^\infty dx_1  \int_0^\infty dx_2 \left( x_1 P^2 +x_2 L^2 +2 l_+ \cdot p_- x_1 x_2\right)^{-1-\ep} \, ,
\ee
where $P^2 = -p^2$ and $L^2=-l^2$.
To complete the evaluation of the integral we need to resort to a series of changes of variables. One starts by replacing $x_1 \to x'_1/P^2$ and $x_2 \to
x'_2/L^2$; in this way the integral becomes (neglecting the prime superscript)
\be
I_s = \mu^{2 \varepsilon} \frac{\Gamma\left(1+\varepsilon\right)}{P^2 L^2}  \int_0^\infty dx_1  \int_0^\infty dx_2 \left( x_1  +x_2  +a x_1 x_2\right)^{-1-\varepsilon} \, ,
\ee
with $a = 2 l_+ \cdot p_-/(P^2 L^2)$. It is now convenient to separate the integration variables; we send $x_2 \to x_1 x'_2$ to obtain 
\be
I_s = \mu^{2 \varepsilon} \frac{\Gamma\left( 1+\varepsilon\right)}{P^2 L^2}  \int_0^\infty dx_1  \int_0^\infty dx_2 x_1^{-\varepsilon} \left( 1  +x_2  +a x_1 x_2\right)^{-1-\varepsilon} \, .
\ee
Then, we replace $x_2 \to x'_2/(1+a x_1)$; in this way one finds
\be
I_s = \mu^{2 \varepsilon} \frac{\Gamma\left( 1+\varepsilon\right)}{P^2 L^2}  \int_0^\infty dx_1  \frac{x_1^{-\varepsilon}}{1+a x_1} \int_0^\infty dx_2 \left( 1  +x_2 \right)^{-1-\varepsilon} \, ;
\ee
the two integrals are now factored. To complete the calculation we replace
$x_1 \to x'_1/a$ to obtain
\be
I_s = \mu^{2 \varepsilon} \frac{\Gamma\left( 1+\varepsilon\right)}{P^2 L^2} a^{-1+\varepsilon} \underbrace{\int_0^\infty dx_1  \frac{x_1^{-\varepsilon}}{1+x_1}}_{=\Gamma\left(1-\varepsilon \right)
\Gamma\left(\varepsilon \right)
} \, \underbrace{\int_0^\infty dx_2 \left( 1  +x_2 \right)^{-1-\varepsilon}}_{=\frac{1}{\varepsilon}} \, .
\ee
Finally, one finds
\bea
I_s &=&  \frac{\Gamma\left( 1+\varepsilon\right)}{P^2 L^2} \left( \frac{P^2 L^2}{2 l_+ \cdot p_-} \right) \left( \frac{2 l_+ \cdot p_- \mu^2}{P^2 L^2} \right)^\varepsilon \frac{(-\varepsilon) \Gamma(\varepsilon) \Gamma(-\varepsilon)}{\varepsilon}\, , \nonumber \\
&=& 
-\frac{\Gamma\left( 1+\varepsilon\right)}{2 l_+ \cdot p_- }\Gamma(\varepsilon) \Gamma(-\varepsilon)\left( \frac{2 l_+ \cdot p_- \mu^2}{P^2 L^2} \right)^\varepsilon \, ,
 \eea
which is the result in Eq.~(\ref{eq:Isres}).

If the external legs are set on-shell at the beginning of the calculation ($p^2 = l^2 = 0$) the integral vanishes, even if $p \cdot l \neq  0$. This can be readily proven by setting $p^2 = l^2 = 0$ in Eq.~(\ref{eq:Is}). By doing this one obtains
\be \label{eq:Intsl}
I_s\left(p^2=0,l^2=0\right) = \mu^{2\varepsilon} \Gamma\left( 1 + \varepsilon \right) \int_0^\infty dx_1  \int_0^\infty dx_2 \left( 2 l_+ \cdot p_-\right)^{-1-\varepsilon} \left( x_1 x_2\right)^{-1-\varepsilon} \, ,
\ee
where the two integrals in $x_1$ and $x_2$ factorize. It is now sufficient to prove that one of the two integrals vanishes. Let us consider the $x_1$ integration:
\be
\int_0^\infty dx \frac{1}{x^{1+\varepsilon}} \, ,
\ee
it develops an ultraviolet divergence for $\varepsilon < 0$ and an infrared divergence for $\varepsilon > 0$. In order to give a mathematical meaning to this integral we split the integration region into two parts using a regulator $\Lambda$: the infrared region for $x < \Lambda$ and the ultraviolet region for $x > \Lambda$:
\be
\int_0^\infty dx \frac{1}{x^{1+\varepsilon}} = \int_0^\Lambda dx \frac{1}{x^{1+\varepsilon}} + \int_\Lambda^\infty dx \frac{1}{x^{1+\varepsilon}} \, .
\ee
On the r.h.s. the first integral is convergent for $\varepsilon < 0$, while the second one is convergent for $\varepsilon > 0$. To distinguish the nature of the two divergences we can use two different regulators in the two different regions, by working out the integration for $\varepsilon_{\mbox{{\tiny IR}}} < 0$ and for $\varepsilon_{\mbox{{\tiny UV}}}  > 0$ we find 
\be \label{eq:Intnull}
\int_0^\infty dx \frac{1}{x^{1+\varepsilon}} = -\frac{\Lambda^{-\varepsilon_{\mbox{{\tiny IR}}}}}{\varepsilon_{\mbox{{\tiny IR}}}} + \frac{\Lambda^{-\varepsilon_{\mbox{{\tiny UV}}}}}{\varepsilon_{\mbox{{\tiny UV}}}} \, ,
\ee
where both integrals develop poles for $\varepsilon_{\mbox{{\tiny IR}}}  = \varepsilon_{\mbox{{\tiny UV}}}  = 0$. The r.h.s. can be analytically continued for arbitrary values of $\varepsilon_{\mbox{{\tiny IR}}}$ and $\varepsilon_{\mbox{{\tiny UV}}} $ without any constraint, therefore we are free to identify 
$\varepsilon_{\mbox{{\tiny IR}}} $ and $\varepsilon_{\mbox{{\tiny UV}}} $. As a consequence of this, the integral in Eq.~(\ref{eq:Intnull}) vanishes.
Another interesting way of proving that
\be \label{eq:stat}
\int d^d k \frac{1}{k^2 (k \cdot p_1) (k \cdot p_2)} = 0 \, ,
\ee
for any $p_1, p_2$ involves integration by parts identities. One starts from the fact that in dimensional regularization
\be
\int d^d k \,  \frac{\partial}{\partial k^\mu} \frac{ v^\mu }{k^2 (k \cdot p_1) (k \cdot p_2)} = 0 \, ,
\ee
for any $v^\mu$. In a standard integral, one could have boundary terms at infinite momentum, but in dimensional regularization these are absent because one can always choose the dimension in such a way that the integrand goes to zero sufficiently fast for $k^\mu \to \infty$. By choosing $v^\mu = k^\mu$ applying the derivative to the integrand one obtains
\bea
0 &=& \int d^d k \,  \Biggl[ \frac{d-4}{(k^2) (k \cdot p_1) (k \cdot p_2)} \Biggr]\, . 
\eea
Since in dimensional regularization one works in $d \neq 4$ (and then one takes the limit $\ep \to 0$) the relation above implies Eq.~(\ref{eq:stat}).

Finally, we want to consider the soft integral in Eq.~(\ref{eq:IsresM}) and show that it vanishes. For this purpose, we define the integral
\be
I(\alpha,\beta) = \int d^dk \frac{1}{(2 k \cdot l +l^2)^\alpha (2 k \cdot p +p^2)^\beta} \, 
\ee
for $\alpha, \beta \ge 1$. By considering the integration by parts identities

\be \label{eq:IBPs_sp}
\int d^dk \frac{\partial}{\partial k^\mu}\frac{v^\mu}{(2 k \cdot l +l^2)^\alpha (2 k \cdot p +p^2)^\beta} = 0\, ,
\ee
with $v \in \{l,p\}$ one finds the relations 
\bea
\alpha l^2 I(\alpha+1,\beta) + \beta \, (l\cdot p) \, I(\alpha,\beta+1) &=& 0 \, , \nn \\
\alpha \, (l \cdot p) \,  I(\alpha+1,\beta) + \beta p^2 I(\alpha,\beta+1) &=& 0 \,. 
\eea
The equations above imply that
\be \label{eq:relI}
I(\alpha+1,\beta) = I(\alpha,\beta+1) = 0 \quad \mbox{if} \quad \left(l \cdot p \right)^2 - l^2 p^2 \neq 0 \,.
\ee
If in Eq.~(\ref{eq:IBPs_sp}) one sets instead $v = k$ one finds the relation
\be
(d-\alpha-\beta) I(\alpha,\beta) + \alpha l^2 I(\alpha+1,\beta) +
\beta p^2  I(\alpha,\beta+1) = 0 \, ,
\ee
which, taken together with Eq.~(\ref{eq:relI}), immediately implies 
\be
I(\alpha,\beta) = 0 \, ,
\ee
and therefore in particular, the integral in Eq.~(\ref{eq:IsresM}) vanishes.

\subsection{Collinear Integrals with the Analytic Regulator \label{ap:IcAR}}

In this appendix we describe the calculation of the collinear region integrals 
in Eqs.~(\ref{eq:collregar}). We start by considering the integral $I_{\cc}$. It is useful to employ the following Feynman parameterization
\be \label{eq:FeynPar2}
\frac{1}{abc^{1+\alpha}} = \int_0^\infty dx_1  \int_0^\infty dx_2 
\frac{(2+\alpha) (1+\alpha)}{\left(c +a x_1 +b x_2 \right)^{3+\alpha}} \, ,
\ee
with $a = k^2-m^2$, $b=2 k \cdot l_+$, and $c= (k+p)^2$.
The integral can be rewritten as
\be \label{eq:istep1}
I_{\cc} = i \pi^{-d/2} \mu^{4-d} \left(-\nu^2\right)^\alpha \int_0^\infty dx_1  \int_0^\infty dx_2 \int d^d k \,\frac{(2+\alpha)(1+\alpha)}{\chi^{3+\alpha}(k)} \, ,
\ee
where
\be
\chi(k) = (1+x_1)\left(k^2 +2 k \cdot Q -M^2  \right) \, ,
\ee
with
\be
Q^\mu = \frac{p^\mu + l_+^\mu x_2}{1+x_1} \, , \qquad M^2 = \frac{m^2 x_1}{1+x_1} \, .
\ee
(We remind the reader that here we want to set $p^2 = l^2 = 0$.) The integration over the virtual momentum can be carried out by employing Eq.~(\ref{eq:MF}) to obtain
\be
I_{\cc} = \mu^{2 \ep}   \nu^{2 \alpha} \frac{\Gamma\left( 1+\alpha+\ep \right)}{\Gamma\left(1+\alpha\right)} \int_0^\infty dx_1 \frac{1}{(1+x_1)^{3+\alpha}}\int_0^\infty dx_2 \frac{1}{\left(Q^2+M^2 \right)^{1+\alpha+\ep}}  \,,
\ee
with
\be
Q^2+M^2 = \frac{1}{(1+x_1)^2} \left[ 2 p_-\cdot l_+ x_2 + m^2 x_1 (1+x_1) \right] \, .
\ee
The integration over $x_2$ is of the form
\be
\int_0^\infty dx \frac{1}{\left(F_1 x + F_2 \right)^{1+\alpha+\ep}} = 
\frac{1}{(\alpha +\ep)} \frac{1}{F_1 F_2^{\alpha+\ep}} \, ,
\ee
and therefore
\bea
I_{\cc} &=& \frac{1}{2 p_- \cdot l_+} \left(\frac{\mu^2}{m^2} \right)^\ep 
\left(\frac{\nu^2}{m^2} \right)^\alpha \frac{\Gamma\left(\alpha+\ep \right)}{\Gamma\left(1+\alpha \right)} \int_0^\infty dx_1 \frac{1}{x_1^{\alpha+\ep} (1+x_1)^{1-\ep}} \, \nn \\
&=& \frac{\Gamma(1+\ep)}{Q^2} \left(\frac{\mu^2}{m^2} \right)^\ep 
\left(\frac{\nu^2}{m^2} \right)^\alpha \frac{\Gamma\left(\alpha+\ep \right)\Gamma(\alpha) \Gamma\left(1-\alpha-\ep\right)}{\Gamma\left(1+\alpha \right) \Gamma\left(1+\ep \right)\Gamma\left(1-\ep \right)}  \, .
\eea
Finally, by expanding the $\Gamma$ functions first for $\alpha \to 0$ and then
for $\ep \to 0$ one obtains
\be
\frac{\Gamma\left(\alpha+\ep \right)\Gamma(\alpha) \Gamma\left(1-\alpha-\ep\right)}{\Gamma\left(1+\alpha \right) \Gamma\left(1+\ep \right)\Gamma\left(1-\ep \right)} = -\frac{1}{\ep^2}
+\frac{1}{\alpha \ep} +\frac{\pi^2}{3} + {\mathcal O}\left(\alpha,\ep\right) \,,
\ee
which leads to the expression in Eq.~(\ref{eq:collregarRES}).
We stress the fact that the order in which the expansions for small $\alpha$ and small $\ep$ is taken is important, and one first needs to expand for $\alpha \to 0$ at fixed $\ep$.

We now turn to the calculation of the integral $I_{\cb}$. In this case we apply the Feynman parameterization in Eq.~(\ref{eq:FeynPar2}) with $c = 2 p_- \cdot k$, $a = k^2-m^2$, and $b= (k+l)^2$. The integral can be written exactly in the same form as in Eq.~(\ref{eq:istep1}), except for the fact that in this case the function $\chi$ becomes
\be
\chi(k) = (x_1 +x_2) \left(k^2+2 k\cdot Q -M^2 \right) \, ,
\ee
with
\be
Q^\mu = \frac{p_-^\mu + l^\mu x_2}{(x_1+x_2)} \,,  \qquad M^2 = \frac{m^2 x_1}{(x_1 +x_2)} \, .
\ee
After integration over the virtual momentum by employing Eq.~(\ref{eq:MF}), one finds
\be
I_{\cb} =  \mu^{2 \ep} \nu^{2 \alpha} \frac{\Gamma\left( 1+\alpha+\ep \right)}{\Gamma\left(1+\alpha\right)} \int_0^\infty dx_1\int_0^\infty dx_2 \frac{1}{\left(x_1 +x_2\right)^{3+\alpha}\left( Q^2+M^2\right)^{1+\alpha+\ep}} \, , 
\ee
with 
\be
Q^2 + M^2 = \frac{1}{(x_1 +x_2)^2}\left[\left(2 p_- \cdot l_+ + m^2 x_1 \right) x_2  + m^2 x_1^2\right] \, .
\ee
At this stage it is convenient to change variables by setting $x_1 \to x_2 x_1$; the integral becomes
\be
I_{\cb} = \mu^{2 \ep} \nu^{2 \alpha} \frac{\Gamma\left( 1+\alpha+\ep \right)}{\Gamma\left(1+\alpha\right)} \int_0^\infty dx_1\int_0^\infty dx_2 \frac{x_2^{-1+\ep}}{\left(1 +x_1\right)^{1-\alpha-2 \ep} D(x_1,x_2)^{1+\alpha+\ep}} \, , 
\ee 
where
\be
D(x_1,x_2) = m^2 x_1 (1+x_1) x_2 + 2 p_-\cdot l_+ \, .
\ee
The integration over the parameter $x_2$ is of the form
\be
\int_0^\infty d x \frac{1}{x^{1-\ep} \left( F_1 x +F_2\right)^{1+\alpha+\ep}} = 
\frac{\Gamma(1+\alpha) \Gamma(\ep)}{\Gamma(1+\alpha+\ep)} \frac{1}{F_1^\ep F_2^{1+\alpha}} \, .
\ee
Therefore one finds
\bea
I_{\cb} &=& \frac{1}{2 p_- \cdot l_+}\left(\frac{\mu^2}{m^2} \right)^\ep \left(\frac{\nu^2}{2 p_- \cdot l_+} \right)^\alpha \Gamma(\ep) \int_0^\infty
dx_1 \frac{1}{x_1^\ep (1+x_1)^{1-\al-\ep}} \,  \nonumber \\
&=& \frac{\Gamma(1+\ep)}{Q^2}\left(\frac{\mu^2}{m^2} \right)^\ep \left(\frac{\nu^2}{Q^2} \right)^\alpha \frac{\Gamma(\ep)\Gamma(-\alpha)\Gamma(1-\ep) }{\Gamma(1+\ep) \Gamma(1-\alpha-\ep)} \, .
\eea
Finally, in order to obtain the result in Eq.~(\ref{eq:collregarRES}) it is sufficient to expand  the combination of $\Gamma$ functions in the equation above for $\alpha \to 0$ and then for $\ep \to 0$. Doing so, one obtains 
\be
\frac{\Gamma(\ep)\Gamma(-\alpha)\Gamma(1-\ep) }{\Gamma(1+\ep) \Gamma(1-\alpha-\ep)} = -\frac{1}{\alpha\, \ep} + \frac{\pi^2}{6} + {\mathcal O}(\alpha,\ep) \,.
\ee
%

\subsection{Soft Function in Position Space\label{ap:Sfps}}
In Section \ref{sec:soft} we computed the Drell-Yan soft function in momentum space, using the Feynman rules for the Wilson lines derived in Appendix \ref{ap:Wilson} below. An alternative is to compute the Drell-Yan soft function $\hat{W}_{\textrm{DY}}(x,\mu)$ directly in position space. Indeed, the one and two-loop computations of this function were first carried out in position space \cite{Korchemsky:1993uz,Belitsky:1998tc}. The bare soft function $\hat{W}_{\textrm{DY}}(x)$ can be expressed as a closed Wilson loop\footnote{This can be easily seen using the definition of the soft Wilson line
$$
S_{n}(x)\;=\;\mathbf{P}\exp\Big[i g_s \int^{0}_{-\infty}ds\,n\cdot A_{s}(x+s n)\Big]  \, ,
$$
where $\mathbf{P}$ indicates the path ordering of the color indices.
}
formed by the product of the soft Wilson lines in the two currents, as shown in \cite{Korchemsky:1993uz}:
\bea
\hat{W}_{\textrm{DY}}(x) & = & \frac{1}{N_{c}}\mathrm{tr}\langle 0 |\,\overline{T}\big[S^{\dagger}_{n}(x)S_{\bar{n}}(x)\big]\;T\big[S^{\dagger}_{\bar{n}}(0)S_{n}(0)\big]|0\rangle  \,\nn \\
&=& \langle 0| \mathcal{P}\exp\Big(i g_s \int_{C_{\mathrm{DY}}}dy_{\mu}\,A^{\mu}(y)\Big)|0\rangle   \, ,
\label{eq:wldy}
\eea
where the trace is over color indices, $T$, $\overline{T}$ are the time and anti-time ordering operators.
The operator $\mathcal{P}$, as defined in \cite{Korchemsky:1993uz}, takes care of the path ordering of the color indices, as well as the time and anti-time ordering of the fields. 
\begin{figure}
\begin{center}
\includegraphics[width=12cm]{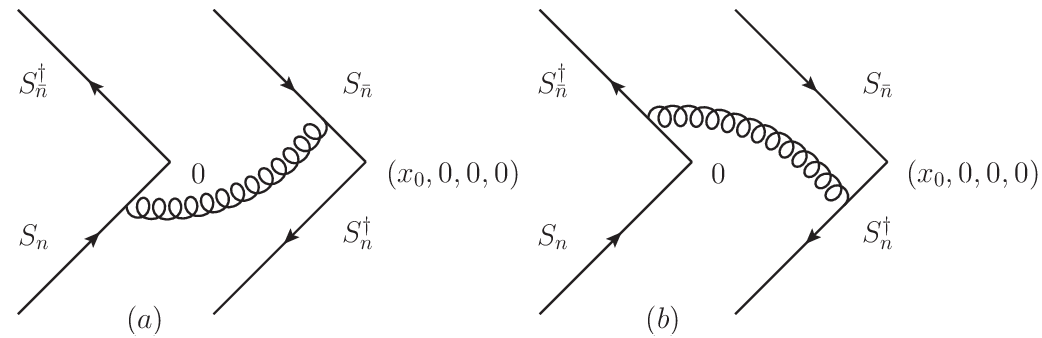}
\label{fig:WDY1}
\caption{Non-zero one-loop diagrams contributing to $\hat{W}_{\textrm{DY}}(x,\mu)$ at order $\alpha_{s}$.}
\end{center}
\end{figure}
We then expand the $\mathcal{P}$ ordered exponential in Eq.~(\ref{eq:wldy}) and write the bare soft functions as
\bea
\hat{W}_{\textrm{DY}}(x) & = & 1+ \frac{1}{2}(i g_s)^{2}C_{F}\int_{C_{\mathrm{DY}}}dx^{\mu}_{1}\int_{C_{\mathrm{DY}}}dx^{\nu}_{2}\,D_{\mu \nu}(x_{1}-x_{2})    \, ,
\label{eq:W1}
\eea
where $D_{\mu \nu}(x_{1}-x_{2})$ is the gluon propagator. This propagator can be a normal loop propagator or a cut propagator depending if the gluon attachment to the path resides on the same side of the cut or not, because the object we compute corresponds to an amplitude squared. A path-integral formalism to compute squared amplitudes is the Keldysh formalism \cite{Schwinger:1960qe,Keldysh:1964ud} which is reviewed in \cite{Belitsky:1998tc,Becher:2007ty}. In a diagrammatic description, the Wilson loop $\hat{W}_{\textrm{DY}}(y,\mu)$ at order $\alpha_{s}$ corresponds to the non-zero one-loop diagrams of the type shown in Fig.~\ref{fig:WDY1} where the curly lines represent the cut gluon propagators. Contributions involving two times the same Wilson line vanish in Feynman gauge because $n^2=\bar{n}^2=0$ and loop corrections are scaleless.
We parametrize the non-vanishing parts of the path from the first diagram of Fig.~\ref{fig:WDY1} as
\begin{align}
x^{\mu}_1(t_1) &= t_1 n^\mu\, , \quad \quad \quad t_1\in[-\infty,0] \, ,\nonumber\\
x^{\nu}_2(t_2) &= x^\nu + t_2 \bar{n}^\nu\, ,\,\,\,\, t_2\in[-\infty,0] \, .
\end{align}
A similar parametrization can be found for the second diagram in Fig.~\ref{fig:WDY1}.
We notice that the two diagrams in Fig.~\ref{fig:WDY1} give the same contribution,
hence we explicitly compute only the first one and we multiply the result by two.
Using the expression for the cut gluon propagator in position space\footnote{
In position space, the cut propagator $D^{\mu\nu}(x)=-g^{\mu\nu}D(x)$ is defined  as
\be\nn
D(x)\;=\; \int\frac{d^{d}k}{(2\pi)^{d}}\, e^{-ik \cdot x}2\pi\theta(k_{0})\delta(k^{2})\;=\; \frac{\Gamma(d/2-1)}{4\pi^{d/2}}\,\left[-(x_{+}-i0)(x_{-}-i0)\right]^{1-d/2}\, ,
\ee
where $x_+\equiv n\cdot x$ and $x_-\equiv \bar{n} \cdot x$.
},
we evaluate the sum of the diagrams in Fig.~\ref{fig:WDY1}:
\be
\hat{W}_{\textrm{DY}}(x)  =  1+ 4 g_s^{2} C_{F} \int^{0}_{-\infty}dt_{1}\int^{0}_{-\infty}dt_{2} \frac{\Gamma(1-\ep)}{4 \pi^{2-\ep}}\left[-(n t_{1} - x - \bar{n}t_{2})^2\right]^{\ep-1} \, .
\ee
By making the change of variable $t_1\to- t_1$ we get
\be
\hat{W}_{\textrm{DY}}(x)  =  1+ 4 g_s^{2} C_{F} \int^{\infty}_{0}dt_{1}\int^{0}_{-\infty}dt_{2} \frac{\Gamma(1-\ep)}{4 \pi^{2-\ep}}\left[-(x + n t_{1} + \bar{n}t_{2})^2\right]^{\ep-1}\, .
\ee
Setting $x=(x_{0},0,0,0)$ we obtain the factorization of the two integrals
\be
\hat{W}_{\textrm{DY}}(x) = 1 + 4 g_s^{2} \;C_{F} \frac{\Gamma(1-\ep)}{4 \pi^{2-\ep}}\int^{\infty}_{0}dt_{1}(x_{0}+2t_{1})^{\ep-1}\,\int^{0}_{-\infty}dt_{2}\, (-(x_{0}+2t_{2}))^{\ep-1}\, .
\ee
By replacing $t_{1}\rightarrow x_{0}t_{1}/2$, $t_{2}\rightarrow x_{0}t_{2}/2$ we find
\bea
\hat{W}_{\textrm{DY}}(x) & = &1 - g_s^{2} \; C_{F} \frac{\Gamma(1-\ep)}{4 \pi^{2-\ep}}(-x^2_0)^{\ep}\int^{\infty}_{0}dt_{1}\,(1+t_{1})^{\ep-1}\,\nn\\
& & \times \int^{0}_{-\infty}dt_{2}\, (1+t_{2})^{\ep-1}\, \nn \\
&=& 1 + C_{F} \frac{\alpha_{s}^0}{\pi}\frac{\Gamma(1-\ep)}{\ep^{2}}\left(-x^{2}_{0}\pi\right)^{\ep}\, .
\label{eq:sfpos}
\eea
As a last step, we express the bare coupling $\alpha_s^0$ in terms of the $\ms$ renormalized coupling constant $\alpha_s(\mu)$ via the usual relation $Z_\alpha\, \alpha_s(\mu) \,\mu^{2\epsilon} = e^{-\ep \gamma_E}(4\pi)^\ep \alpha_s^0$, where $Z_\alpha  = 1+\mathcal{O}(\alpha_s)$. We find
\be
\hat{W}_{\textrm{DY}}(x,\mu) = 1 + C_{F} \frac{\alpha_{s}}{\pi}\frac{\Gamma(1-\ep)}{\ep^{2}}e^{-\ep\gamma_{E}}\left(-\frac{1}{4}\mu^{2}x^{2}_{0} e^{2 \gamma_{E}}\right)^{\ep}\, .
\label{eq:sfposMS}
\ee

\setcounter{figure}{0}
\section{Inverse Derivative Operator \label{sec:definitioninvder}}

The relation in Eq.~(\ref{eq:invder}) must be defined with an infinitesimal positive imaginary part in the operator, and should read 
%
\be
\frac{i}{i n \cdot \partial + i 0^+} \, \phi(x) = \int_{-\infty}^0\! ds\, \phi(x + s n) \, .
\label{eq:invder2}
\ee

In order to check the $+i 0^+$ prescription in the equation above, one can start by rewriting the fields in the coordinate space as the Fourier transform of the field in momentum space
\be
\phi(x) = \int \frac{d^4 k}{(2 \pi)^4} e^{-i k\cdot x} \tilde{\phi}(k) \, .
\ee
The l.h.s. of Eq.~(\ref{eq:invder2}) then becomes 
\be
\frac{i}{i n \cdot \partial + i 0^+} \, \phi(x) = \int \frac{d^4 k}{(2 \pi)^4} \frac{i}{n \cdot k +i 0^+} e^{-i k\cdot x} \tilde{\phi}(k) \,.
\ee
In turn one can write
\be
\frac{1}{n \cdot k +i 0^+} = \bm{P} \left(\frac{1}{n \cdot k}\right) - i \pi \delta(n\cdot k) \, ,
\ee
where $\bm{P}$ indicates Cauchy's principal value and where the relation above can be checked by integrating 
both sides over the $n \cdot k$ complex plane. Finally one finds
\be
\frac{i}{i n \cdot \partial + i 0^+} \, \phi(x) = \int \frac{d^4 k}{(2 \pi)^4} \left[ \bm{P} \left( \frac{i}{n \cdot k} \right) + \pi \delta(n\cdot k)\right]e^{-i k\cdot x} \tilde{\phi}(k) \,. \label{eq:p1}
\ee
 
Similarly, the r.~h.~s. of Eq.~(\ref{eq:invder2}) can be written as 
\be
\int_{-\infty}^0\! ds\, \phi(x + s n)  = \int \frac{d^4 k}{(2 \pi)^4} \left[ \int_{-\infty}^0 ds e^{-i k\cdot n s}\right] e ^{-i k\cdot x} \tilde{\phi}(k) \,.
\ee
The integral over $s$ is the Fourier transform of the Heaviside step function
\be
\int_{-\infty}^{\infty} ds\, e^{-i k\cdot n s} \theta(-s) = \bm{P}\left(  \frac{i}{k \cdot n} \right)
 + \pi \delta \left( k \cdot n \right) \, ,
\ee
so that the r.~h.~s. of Eq.~(\ref{eq:invder2})  becomes
\be
\int_{-\infty}^0\! ds\, \phi(x + s n)  = \int \frac{d^4 k}{(2 \pi)^4} \left[ \bm{P} \left( \frac{i}{n \cdot k} \right) + \pi \delta(n\cdot k)\right]e^{-i k\cdot x} \tilde{\phi}(k) \,. \label{eq:p2}
\ee
Comparing Eqs.~(\ref{eq:p1},\ref{eq:p2}) one proves Eq.~(\ref{eq:invder2}).

By following the same procedure one can also prove that
\be
\frac{-i}{i n \cdot \partial -i 0^+}\phi(x) = \int_0^\infty ds \, \phi(x + s n)\,.
\ee
In reality, the choice of the sign of the infinitesimal imaginary part associated to the inverse operator $1/(i n\cdot \partial)$ is relevant only when $n\cdot p $ is small. However, since $n \cdot p$ is assumed to be the large momentum component, the choice of the infinitesimal part must be physically irrelevant.

\setcounter{figure}{0}
\section{Wilson Lines and Gauge Transformations \label{app:Wil}}

In this appendix, we derive a few fundamental properties of Wilson lines. We start by considering a generic Wilson line connecting two space-time points $y$ and $z$, for an abelian theory such as QED. In the abelian case, no path ordering is needed, and we will indicate a Wilson line as
\be
\left[z,y \right]_A \equiv  \exp\left[ -i e \int_P dx^\mu A_\mu (x)\right] \, ,
\ee 
where $P$ is a path which connects $y$ with $z$, and where $e =-g$ is the  
abelian coupling constant. In most cases, we drop the subscript indicating the gauge field; however, in the following discussion we need to carry out gauge transformations, and  it is therefore convenient to indicate explicitly which  gauge field appears in the Wilson line.
 The Wilson line can be rewritten as
\be
\left[z,y \right]_A  = \exp\left[ -i e \int_{s_y}^{s_z} ds \frac{dx^\mu}{ds} A_\mu \bigl(x(s)\bigr)\right] \, ;
\ee
$s$ is a variable parameterizing the path and $s_y, s_z$ are such that
\be
y \equiv x(s_y) \, , \qquad z \equiv x(s_z) \, .
\ee
The Wilson lines employed in the rest of these lectures involve paths which are straight segments, so that
\be
x(s) = x_0 + s \bar{n}\, , \quad \mbox{and}\quad
 \frac{dx^\mu}{ds} = \bar{n}^\mu \,.
\ee
Moreover we typically choose $s_y=0$ and rewrite $s_z \to s$ and $x_0 \to 0$.
However, in this appendix we will consider the more general case of Wilson lines along arbitrary paths.

Under a gauge transformation $V(x)=e^{i\alpha(x)}$, the field $A$ transforms as
\be
A_\mu(x) \rightarrow A'_\mu(x) = A_\mu(x) +\frac{1}{g} \partial_\mu \alpha(x) \, ,
\ee
and the Wilson line changes to
\bea \label{eq:Sgauge}
\left[z,y \right]_A  &\rightarrow& \left[z,y \right]_{A'}  \,  \nn \\
&=& \exp\left[-i e \int_{s_y}^{s_z} ds \frac{dx^\mu}{ds} A_\mu \bigl(x(s)\bigr) 
+i \int_{s_y}^{s_z} ds \frac{dx^\mu}{ds} \partial_\mu \alpha\left( x(s)\right)\right] \,  \nn \\
&=& \exp\left[-i e \int_{s_y}^{s_z} ds \frac{dx^\mu}{ds} A_\mu \bigl(x(s)\bigr) 
+i \int_{s_y}^{s_z} ds \frac{d}{ds}  \alpha\left( x(s)\right)\right] \, \nn \\
&=& \exp\left[-i e \int_{s_y}^{s_z} ds \frac{dx^\mu}{ds} A_\mu \bigl(x(s)\bigr) 
+i  \alpha\left( z\right) - i  \alpha\left( y\right)\right] \,  \nn \\
&=& V(z) \left[z,y \right]_A  V^\dagger (y) \, .
\eea
From the last line above it is easy to see that if $y=z$ (closed path) the Wilson line is gauge invariant. Such closed Wilson lines are called Wilson loops and play an important role, for example when formulating gauge theory on the lattice.

Next, we will prove that the covariant derivative of the Wilson line along the path is zero. To this end, we consider an intermediate point $x^\mu\equiv x^\mu(s)$ and compute  
\bea \label{eq:Dzero}
\frac{d x^\mu}{d s} D_\mu \left[x,y \right]_A &=&
\frac{d x^\mu}{d s} \left( \partial_\mu + i e A_\mu(x) \right) \left[x,y \right]_A  \,  \nn \\
&=&i e  \frac{d x^\mu}{d s} \left[\left(- \frac{d}{d x^\mu} \int_{s_y}^s dt \frac{d x^\nu}{dt} A_\nu (x) \right) + 
A_\mu(x) \right] \left[x,y \right]_A   \, \nn \\
&=&i e  \left[\left(- \frac{d}{d s} \int_{s_y}^s dt \frac{d x^\nu}{dt} A_\nu (x)\right) + \frac{d x^\mu}{d s} 
A_\mu(x) \right] \left[x,y \right]_A  \, \nn \\
&=&i e  \left[- \frac{d x^\nu}{ds} A_\nu (x(s)) + \frac{d x^\mu}{d s} 
A_\mu(x) \right] \left[x,y \right]_A  \, \nn \\
&=& 0\, .
\eea

The properties shown in Eqs.~(\ref{eq:Sgauge},\ref{eq:Dzero}) are valid also in the non-abelian case as we will show below.
 For the Wilson lines, the only difference to the abelian case is that the exponent is matrix-valued and we therefore need to specify an ordering prescription. The proper prescription is to define 
\be \label{eq:Wilnonab}
\left[z,y \right]_A  = \mathbf{P} \exp \left[i g \int_{s_y}^{s_z} ds \frac{d x^\mu}{d s} A_\mu^b \left( x(s)\right) t^b \right] \, ,
\ee
where $\mathbf{P}$ indicates the path ordering of the matrix-valued integrands in such a way that an integrand evaluated at a given value of $s$ appears to the right of integrands evaluated at larger values of the parameter $s$, while it appears to the left of integrands evaluated at smaller values of the parameter $s$.
In the conjugate Wilson line $\left[z,y \right]^\dagger_A$ the symbols $\mathbf{P}$ indicates
the opposite ordering prescription with respect to the one just described.
In the following, in order to keep the notation compact, we introduce a symbol for the 
argument of the integrand in Eq.~(\ref{eq:Wilnonab}):
\be
\bm{F}(s) \equiv \frac{d x^\mu}{d s} A_\mu^b \left( x(s)\right) t^b \,.
\ee
We use boldface fonts for $\bm{F}$ to indicate that these objects are matrices.
By employing the usual series representation of the exponential
\be
e^x = \sum_{n=0}^\infty \frac{x^n}{n!} \, ,
\ee
one can  rewrite the Wilson line as 
\be \label{eq:Wnawn}
\left[z,y \right]_A=\sum_{n=0}^\infty \frac{(ig)^n}{n!} \int_{s_y}^{s_z}
\!ds_1 \int_{s_y}^{s_z}
\!ds_2 \cdots \int_{s_y}^{s_z}
\!ds_n \mathbf{P}\left\{\bm{F}(s_1) \bm{F}(s_2) \cdots \bm{F}(s_n) \right\} \, . 
\ee
The path ordering prescribes that the non-commuting functions $F$ should be ordered considering the decreasing order of the arguments.
Therefore, 
if $s_1 > s_2 > \cdots > s_n$, the product of $F$'s in the integrand should be 
$F(s_1) F(s_2) \cdots F(s_n)$. The integration region in Eq.~(\ref{eq:Wnawn})
is a $n$-dimensional hypercube. It is possible to subdivide the integration region in $n!$ subregions, which correspond to the $n!$ possible ordering of the elements in the set $\left\{s_1,s_2,\cdots,s_n \right\}$. The $n!$ integration regions are simplexes, as it is easy to see by considering the simple case in which $n=2$,$s_y=0$, and $s_z =1$; in this case
\bea
\int_0^1 ds_1 \int_0^1 ds_2 \mathbf{P} \left\{ \bm{F}(s_1) \bm{F}(s_2)\right\} &=& 
\int_0^1 ds_1 \int_0^{s_1} ds_2 \bm{F}(s_1) \bm{F}(s_2) + 
\int_0^1 ds_2 \int_0^{s_2} ds_1 \bm{F}(s_2) \bm{F}(s_1) \,  \nn \\
&=& 2 \int_0^1 ds_1 \int_0^{s_1} ds_2 \bm{F}(s_1) \bm{F}(s_2) \,.
\eea
This procedure can be generalized to the $n$-dimensional case; each of the integration regions gives the same contribution so that one can eliminate the path ordering and multiply by $n!$ each term\footnote{
Following the same procedure, but taking into account the opposite path ordering prescription, the conjugate Wilson line can be written as
\be
 \left[z,y \right]^\dagger_A=\sum_{n=0}^\infty (-ig)^n \int_{s_y}^{s_z}
\!ds_1 \int_{s_1}^{s_z}
\!ds_2 \cdots \int_{s_{n-1}}^{s_z}
\!ds_n \bm{F}(s_1) \bm{F}(s_2) \cdots \bm{F}(s_n)  \, . 
\ee
 }
 in Eq.~(\ref{eq:Wnawn}):
\be
\left[z,y \right]_A=\sum_{n=0}^\infty (ig)^n \int_{s_y}^{s_z}
\!ds_1 \int_{s_y}^{s_1}
\!ds_2 \cdots \int_{s_y}^{s_{n-1}}
\!ds_n \bm{F}(s_1) \bm{F}(s_2) \cdots \bm{F}(s_n)  \, . 
\ee

We then redefine $s_y \equiv s_0$  and $s_z \equiv s$ and we calculate the derivative of the Wilson line with respect to $s$
\be \label{eq:intermediate}
\frac{d}{ds} \left[x(s),x(s_0) \right]_A  = \frac{d}{ds} \left(\bm{1} +ig
\int_{s_0}^s\! \!ds_1 \bm{F}(s_1) + (ig)^2 \int_{s_0}^s \!\!ds_1 \int_{s_0}^{s_1} \!\!ds_2
\bm{F}(s_1) \bm{F}(s_2) + \cdots 
\right)\, .
\ee
It is easy to take the derivative in each term in the r.h.s.\ of the equation above by observing that for a generic function $g(s)$
\be
\frac{d}{ds} \int_{s_0}^s \!\!dt g(t) = g(s) \, .
\ee
Eq.~(\ref{eq:intermediate}) becomes
\bea
\frac{d}{ds} \left[x(s),x(s_0) \right]_A&=& (ig) \bm{F}(s) +(ig)^2 \bm{F}(s) \int_{s_0}^s
\!\!ds_2 \bm{F}(s_2) \nn \\
& &+ (ig)^3 \bm{F}(s) \int_{s_0}^s
\!\!ds_2 \bm{F}(s_2)  \int_{s_0}^{s_2}
\!\!ds_3 \bm{F}(s_3) +\cdots\, \nn \\
&=& (ig) \bm{F}(s)  \left[x(s),x(s_0) \right]_A \, \nn \\
&=& (ig) \frac{d x^\mu}{d s} A_\mu^b \left( x(s)\right) t^b \left[x(s),x(s_0) \right]_A  \, .
\eea
It is now trivial to see that
\be \label{eq:difeq}
\frac{d x^\mu}{ds} \left(\frac{\partial}{\partial x^\mu} - ig A_\mu^b \left( x(s)\right) t^b \right) \left[x(s),x(s_0) \right]_A   =
\frac{d x^\mu}{ds} D_\mu \left[x(s),x(s_0) \right]_A  = 0 \,,
\ee
and therefore the covariant derivative of the Wilson line along the path is zero also in the non-abelian case. 
Note that this first-order differential equation determines the Wilson line up to an initial condition. The proper initial condition is simply that the Wilson line of zero length is the identity matrix $[x(s_y),y]_A =[y,y]_A=\bm{1}$. The Wilson lines along the path $P$ from $y$ to $z$ is the unique solution of the differential equation in Eq.~(\ref{eq:difeq}) which satisfies the initial condition $[y,y]_A =\bm{1}$.

Finally, we are ready to prove that also in the non-abelian case the Wilson line transforms according to Eq.~(\ref{eq:Sgauge}) under gauge transformations.
%
%
%
Let us define the quantity
\be \label{eq:Wilgana}
[x,y]_{A'}  = V(x) [x,y]_A V^\dagger (y) \, ,
\ee
and prove that it satisfies the differential equation~(\ref{eq:difeq}) when the covariant derivative depends on the field $A'$, which is the gauge transformation of the field $A$. In fact
\bea
\frac{d x^\mu}{ds}  D_\mu (A')[x,y]_{A'} &=&\frac{d x^\mu}{ds}  D_\mu (A') V(x) [x,y]_A V^\dagger (y)\,  \nn \\  
&=&\frac{d x^\mu}{ds} V(x) D_\mu (A)  V^\dagger(x) V(x) [x,y]_A V^\dagger (y) \,  \nn \\
&=&V(x)\frac{d x^\mu}{ds}  D_\mu (A)  [x,y]_{A} V^\dagger (y) =0 \, ,
\eea
where the last equality follows from the fact that   $[x,y]_A$ is the solution of Eq.~(\ref{eq:difeq}).
Our proof is completed by checking that $[x,y]_{A'}$ also satisfies the correct initial condition
\be
[y,y]_{A'} = V(y) \underbrace{[y,y]_A}_{=1} V^\dagger (y) = 1 \,.
\ee
Therefore the non-abelian Wilson lines transform according to Eq.~(\ref{eq:Wilgana}) under gauge transformations.

\setcounter{figure}{0}
\section{Momentum-Space Feynman Rules for Soft Wilson Lines\label{ap:Wilson}}

In this appendix we derive the Feynman rules for soft Wilson lines employed in the calculation of Feynman diagrams in the effective theory.  
These rules were introduced in Eq.~(\ref{eq:wilsonDY}) for the calculation of the Drell-Yan soft matrix element
at order $\alpha_s$. A straightforward approach to extract the Feynman rules consists in  expanding the soft Wilson line. For example, in the case
of an incoming quark (or an outgoing anti-quark), one needs to expand
\be\label{eq:wilsongen}
S_n(x) = 
\mathbf{P} \exp\left[ ig_s \int_{-\infty}^0 \!\!ds \, n \cdot A_s^{a}(x + s n) \, t^a\right] \, 
\ee
to the desired order in the coupling constant. At order $g_s$, only one gluon is emitted from the collinear direction $n$,
and by employing the Fourier representation of the gluon field one finds the well known eikonal vertex approximation:
\begin{align}
\label{eq:onegluon}
S_n(x) & = 1 + i g_s \int_{-\infty}^{0} \!\! ds \, n\cdot A_s^{a}(x + s n) t^a \, + \mathcal{O}(g^2_s) \, \nn\\
& = 1 + i g_s \int_{-\infty}^{0} \!\! ds \int \frac{d^4 k}{(2 \pi)^4} \, e^{-i k\cdot (x+s n)} n\cdot \tilde{A}_s^{a}(k)\,t^a \, + \mathcal{O}(g^2_s) \, \nn\\
& = 1 + \int \frac{d^4 k}{(2 \pi)^4} \, e^{-i k\cdot x} \underbrace{\left(-g_s   \frac{n^{\mu}}{n \cdot k} \,t^a\right)}_{\text{Eikonal Feynman rule}} \tilde{A}_{s\, \mu}^{a}(k) \, + \mathcal{O}(g^2_s)\, .
\end{align}
The same expression for the eikonal vertex can also be found by taking the soft gluon momentum limit of the single gluon emission digram in QCD. 

\begin{figure}
\begin{center}
\begin{tabular}{cl}
\includegraphics[width=5cm]{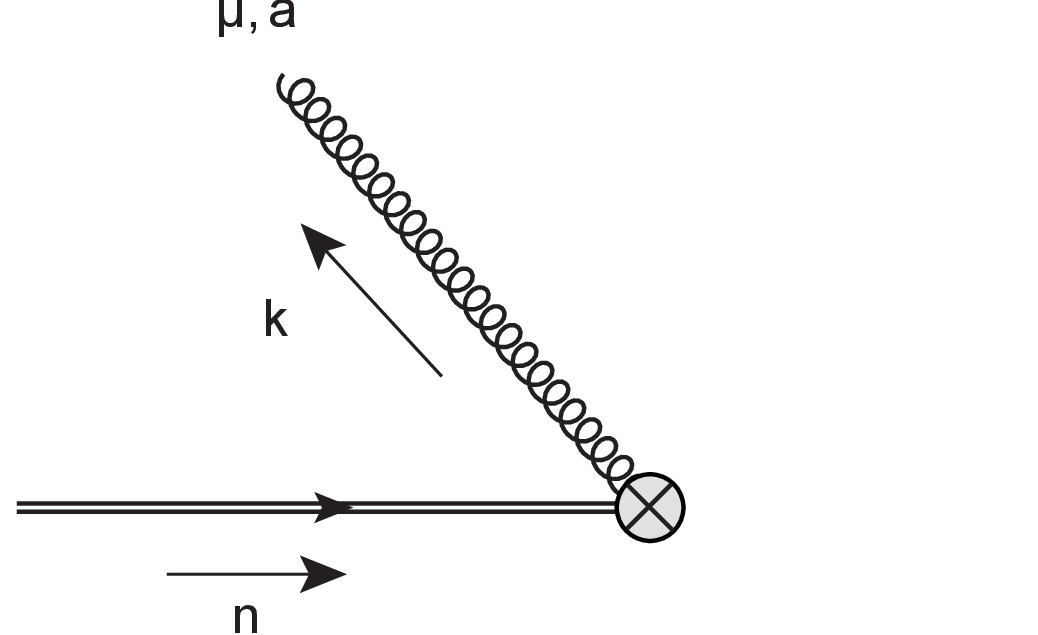} & \begin{tabular}{l}
$\rightarrow\; g_s${\Large $\frac{n^{\mu}}{n \cdot  k}$}$\,\mathbf{T}^a$\\
\\
\\
\\
\\
\end{tabular}\\
\includegraphics[width=5cm]{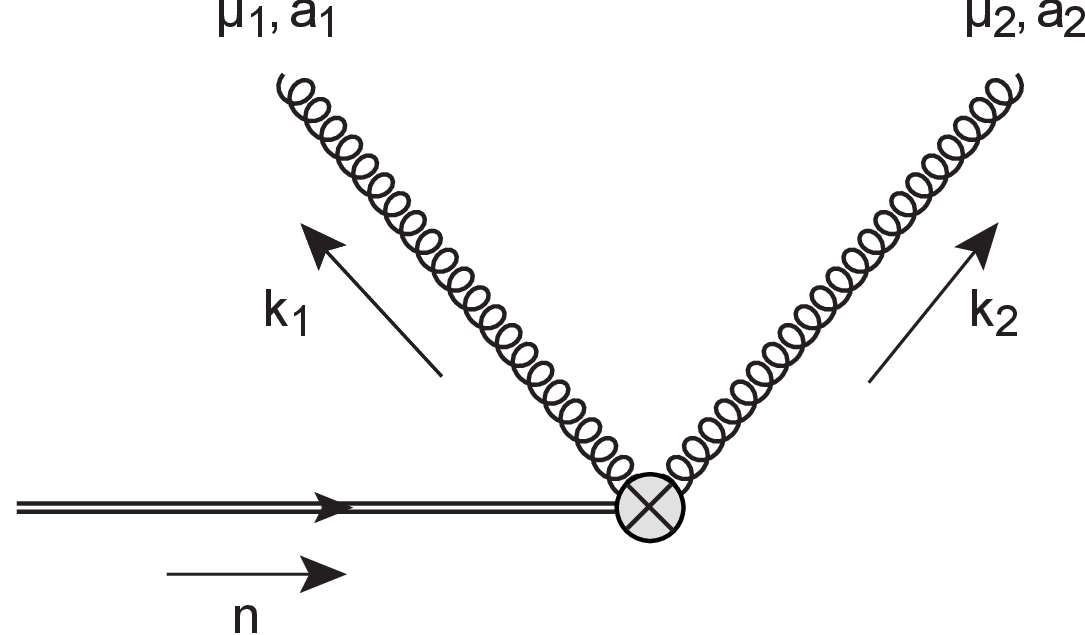} & \begin{tabular}{l}
$\rightarrow\; g_s^2\, n^{\mu_1} n^{\mu_2}\,$ 
{\Large $\left[\frac{\mathbf{T}^{a_1} \mathbf{T}^{a_2}}{n\cdot k_2 \, n\cdot(k_1+k_2)} + \frac{\mathbf{T}^{a_2} \mathbf{T}^{a_1}}{n\cdot k_1 \, n\cdot(k_1+k_2)}\right]$}\\
\\
\\
\\
\\
\end{tabular}
\end{tabular}
\end{center}
\vspace*{-1cm}
\caption{\label{fig:gg} Feynman rule for the emission of one and two gluons from a soft Wilson line.}
\end{figure}

In general, at a given order $g^m_s$ one has to consider the emission of $m$ gluons from a collinear direction. For example, by considering the term of  order $g^2_s$ in the expansion of the Wilson line and by inserting the Fourier representation of the fields one obtains
\begin{align}\label{eq:twogluons}
&-\frac{g^2_s}{2}\left[\int_{-\infty}^{0}\!\!dt \int_{t}^{0}\!\!ds\,  n\cdot A(x+s n) \, n\cdot A(x+t n)\,+\, \left(\begin{array}{c}
s\leftrightarrow t
\end{array}\right)\right]  \nn\\
=&-\frac{g^2_s}{2}\left[\int_{-\infty}^{0}\!\!dt \int_{t}^{0}\!\!ds\,\int_{k_1} \int_{k_2} e^{-i k_1 \cdot (x+s n)}  e^{-i k_2 \cdot (x+t n)} n\cdot \tilde{A}(k_1) \, n\cdot \tilde{A}(k_2)\,+\, \left(\begin{array}{c}
s\leftrightarrow t\\
k_1\leftrightarrow k_2
\end{array}\right)\right]  \, \nn\\
=&\frac{g^2_s}{2}\left[\int_{k_1} \int_{k_2} e^{-i (k_1+k_2) \cdot x}  \left(\frac{n^{\mu_1} n^{\mu_2}}{n\cdot k_2 \, n\cdot(k_1+k_2)}\right) \tilde{A}^{\mu_1}(k_1) \, \tilde{A}^{\mu_2}(k_2)\,+\, \left(\begin{array}{c}
k_1\leftrightarrow k_2\\
\mu_1\leftrightarrow \mu_2
\end{array}\right)\right]\, ,
\end{align}
where we employed the notation $A\equiv A^a\, t^a$ and introduced the symbol
\begin{align}
\int_k \equiv \int \frac{d^4 k}{(2 \pi)^4}\, .
\end{align}
In order to extract the Feynman rule from the last line of Eq.~(\ref{eq:twogluons}) one still needs to sum over the two possible permutations of the gluon momenta and indices. Replacing the $t$ matrices by the color space generators $\mathbf{T}$, one finds the general Feynman rule shown in the second line of Fig.~\ref{fig:gg}. The first line of Fig.~\ref{fig:gg} corresponds to the emission of a single soft gluon.
A similar procedure can be applied to the expansion of collinear Wilson lines.

\begin{figure}
\begin{center}
\begin{tabular}{cl}
\includegraphics[width=5cm]{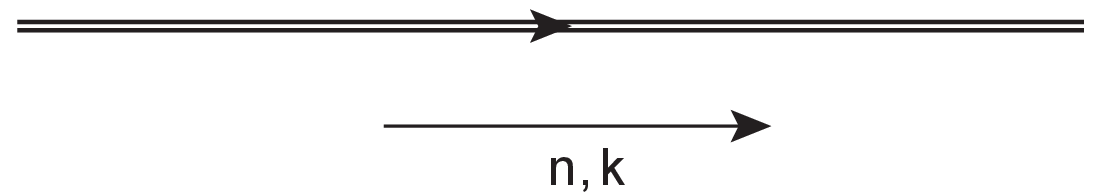} & \begin{tabular}{l}
$\rightarrow\;$ {\Large $\frac{i}{n \cdot k}$}
\\
\\
\\
\end{tabular}\\
\includegraphics[width=5cm]{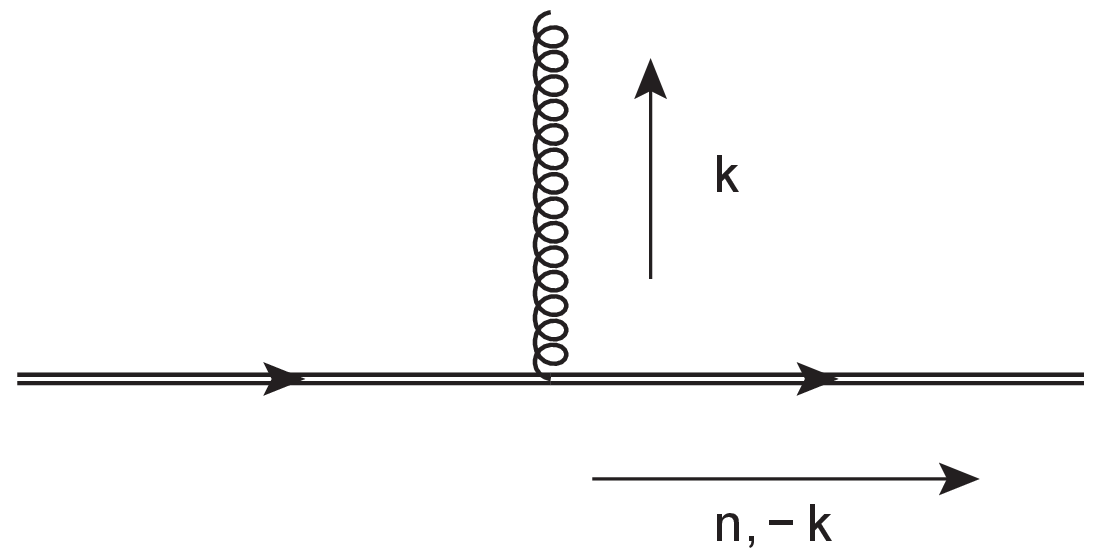}  & \begin{tabular}{l}
$\rightarrow\; i g_s n^\mu \,t^a$\\
\\
\\
\end{tabular}
\end{tabular}
\end{center}
\vspace*{-1cm}
\caption{\label{fig:complexfey} Feynman rules derived from the Lagrangian for the field $\phi$.}
\end{figure}

Formally the Feynman rules arising from the expansion of a Wilson line can be extracted by introducing in the Lagrangian a complex, colored scalar field $\phi_n$ for each collinear direction 
\be
\label{eq:onedimfermions}
\Delta \mathcal{L} = \phi^*_n \left( i n\cdot D\right) \phi_n + \phi^*_{\bar{n}} \left( i \bar{n}\cdot D \right) \phi_{\bar{n}} + j \phi^*_{\bar{n}} \phi_n\, ,
\ee
where the last term in the r.h.s. is a scalar current operator.
From the Lagrangian in Eq.~(\ref{eq:onedimfermions}) one can easily derive the Feynman rules for the scalar field propagators and the interaction vertices of the scalar fields with the gluon field. These rules are shown in Fig.~\ref{fig:complexfey}.

\begin{figure}
\begin{tabular}{c}
\hspace*{1cm} \includegraphics[width=25cm]{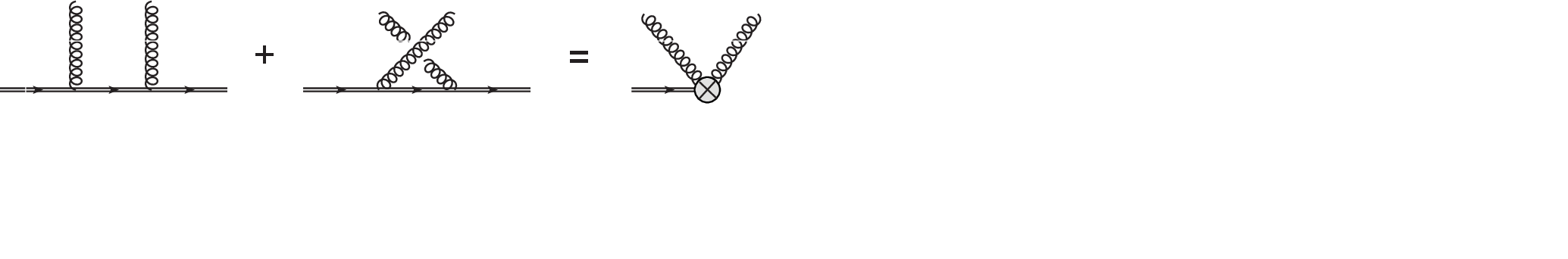}
\end{tabular}
\vspace*{-2cm}
\caption{\label{fig:efflagvswilson} The diagrams on the l.h.s. are constructed by using the Feynman rules for the $\phi$-fields; the r.h.s. coincides with the Feynman rule obtained by expanding the Wilson line.}
\end{figure}

Once one identifies the collinear direction of the Wilson line with the complex scalar field at the Feynman diagram level, it is possible to verify that the Feynman rules obtained by directly expanding the Wilson line coincide with those extracted from the Lagrangian in Eq.~(\ref{eq:onedimfermions}). In fact, by combining the propagator for the field $\phi$ with the $\phi$-gluon vertex one reproduces the rule derived from the last line of Eq.~(\ref{eq:onegluon}). The case in which two gluons are emitted from the scalar field is schematically shown in Fig.~\ref{fig:efflagvswilson}.

In order to prove in a formal way the link between these two apparently independent ways of deriving the Feynman rules, we perform a decoupling transformation of the fields in the Lagrangian
\bea
\phi_n &=& S_n \phi^{(0)}_n \, , \nn \\
\phi_{\bar{n}} &=& S_{\bar{n}} \phi^{(0)}_{\bar{n}}\, ,
\eea
and we obtain from Eq.~(\ref{eq:onedimfermions})
\bea
\Delta \mathcal{L} &=& \phi^{(0)*}_n S^\dagger_n \left( i n\cdot D\right) S_n \phi^{(0)}_n + \phi^{(0)*}_{\bar{n}} S^\dagger_{\bar{n}}\left( i \bar{n}\cdot D \right) S_{\bar{n}} \phi^{(0)}_{\bar{n}} + j \phi^{(0)*}_{\bar{n}} S^\dagger_{\bar{n}} S_n \phi^{(0)}_n \, \nn \\
&=& \phi^{(0)*}_n \left( i n\cdot \partial\right) \phi^{(0)}_n + \phi^{(0)*}_{\bar{n}} \left( i \bar{n}\cdot \partial \right) \phi^{(0)}_{\bar{n}} + j \phi^{(0)*}_{\bar{n}} S^\dagger_{\bar{n}} S_n \phi^{(0)}_n \, ,
\eea
where in the second equality we used the relation $S^\dagger_n \left(i n\cdot D\right) S_n=i n \cdot \partial$.
For example by inserting the scalar current operator in an asymptotic final state one obtains
\bea
\langle \phi_{\bar{n}}| \phi^*_{\bar{n}} \phi_n | \phi_n \rangle &=& \langle \phi_{\bar{n}}| \phi^{(0)*}_{\bar{n}} S^\dagger_{\bar{n}} S_n \phi^{(0)}_n | \phi_n \rangle \, \nn \\
&=& \langle0|S^\dagger_{\bar{n}} S_n |0\rangle\, .
\eea
therefore the two methods are proven to be equivalent.

\setcounter{figure}{0}
 \section{Decoupling Transformation and the Gluon Kinetic Term \label{app:decFmunu}}

In this appendix we want to show that the decoupling of the soft gluons takes place also in the kinetic term involving collinear gluons, which is the last term in Eq.~(\ref{eq:LQCDSCET}). This is necessary because the collinear gluon field strength depends also  on the soft gluon field, as it can be seen from the second of Eqs.~(\ref{eq:fieldstr}) and the last of Eqs.~(\ref{eq:listcd}).
The last term in Eq.~(\ref{eq:LQCDSCET}) can be rewritten as $-1/2 \mbox{Tr} [F_{\mu \nu}^c F^{\mu \nu, c}]$, where
the superscript $c$ indicates that we are dealing with collinear gluons field strengths defined as   
\begin{align}
F_{\mu \nu}^c = F_{\mu \nu}^{c,a}\, t^a & \equiv \frac{1}{ig} \left[i D_\mu, i D_\nu \right] \, . \label{eq:adefF}
\end{align}
In Eq.~(\ref{eq:adefF}) $a$ is an index in the adjoint representation and  the covariant derivative $D_\mu$ is defined as in Eq.~(\ref{eq:DforF}). One then wants to know how the three terms in the r.h.s. of Eq.~(\ref{eq:DforF}) transform under the decoupling transformation $A_c (x) \to S_n (x_{-}) A^{(0)}_c(x)S^\dagger_n (x_-)$.  Since $S_n S_n^\dagger =1$ and 
\begin{align}
\partial_\alpha S_n(x_-) = \frac{\bar{n}_\alpha}{2} n \cdot \partial_- S_n(x_-) \, ,
\end{align}
does not have a transverse component and it vanishes when contracted with $\bar{n}^\alpha$, one can see that under a decoupling transformation
\begin{align}
i \bar{n} \cdot D_c &\to S_n (x_-) i \bar{n} \cdot D^{(0)}_c S^\dagger_n (x_-) \, , \nonumber \\
i  D_{c , \perp} &\to S_n (x_-) i D^{(0)}_{c, \perp} S^\dagger_n (x_-) \, ,
\end{align}
where the superscript $(0)$ indicates that one should replace $A_c \to A_c^{(0)}$ in the second of  Eqs.~(\ref{eq:listcd}).
By using the same procedure adopted in Eq.~(\ref{eq:decforD}) one finds that
\begin{align}
i n \cdot D &\rightarrow i n \cdot D' S_n(x_-) S_n^\dagger(x_-) =  S_n(x_-) i n \cdot D^{(0)}_c S_n^\dagger(x_-) \, .
\end{align}
Therefore, after the decoupling transformation, the covariant derivative depends only on the fields $A_c^{(0)}$ and does not longer depend on $A_s$. Finally, by employing Eq.~(\ref{eq:adefF}) one sees that under a decoupling transformation
\begin{align}
F_{\mu \nu}^c &\rightarrow S_n(x_-) F_{\mu \nu}^{ (0), c} S_n^\dagger(x_-) \, ,
\end{align}
and, using the cyclic property of the trace,
\begin{align}
\mbox{Tr} \left[F_{\mu \nu}^c  F^{\mu \nu, c}\right]& \rightarrow\mbox{Tr} \left[F_{\mu \nu}^{(0) , c}  F^{(0),c,\mu \nu}\right] \, . \label{eq:aux2}
\end{align}
As expected, the r.h.s. of Eq.~(\ref{eq:aux2})  does not depend on the soft gluon field.

\setcounter{figure}{0}
\section{Transverse PDFs at NLO \label{app:TPDFsNLO}}

In this Appendix we briefly sketch the calculation of the  Transverse PDFs (TPDFs)  at NLO by relating it to the evaluation of the 
diagrams in the first line of Fig.~\ref{fig:graphsIfun}. We start from the definition of the $x_T$-dependent PDFs in terms of 
of operators, which was introduced in Eq.~(\ref{Bdef}). As a first step, we replace the hadronic state $N$ by a quark $q$ carrying a collinear momentum $p^\mu \propto n^\mu$. Furthermore, we introduce the identity operator as a sum over a generic intermediate collinear partonic state that we indicate with $X$, so that we find
\begin{align}
{\mathcal B}_{q/q} \left(z, x_T^2, \mu \right) &= \frac{1}{2 \pi} \int \! dt \, e^{- i z t \bar{n} \cdot p}
\frac{\nbsl_{ij}}{2} \sum\hspace{-0.5cm}\int\limits_X\;  \langle q(p)| \bar{\chi}_i(t \bar{n} + x_\perp) |X\rangle \langle X| \chi_j(0)|q(p)\rangle \, ,
\label{eq:Boppar}
\end{align}
where $i,j$ in the Eq.~(\ref{eq:Boppar}) are Dirac indices.
Since the partonic PDFs are trivial, $f_{i/j}(z) = \delta(1-z)\, \delta_{ij}$, we immediately obtain the coefficient functions in Eq.~(\ref{eq:OPE}) from the partonic TPDFs: ${\cal I}_{i\leftarrow j}(z,x_T^2, \mu)={\mathcal B}_{i/j}(z,x_T^2, \mu)$.

One can then use the momentum operator $P$ to rewrite
\be
\bar{\chi}_i(t \bar{n} + x_\perp)= e^{i P\cdot (t \bar{n} + x_\perp)} \bar{\chi}_i(0) e^{-i P\cdot (t \bar{n} + x_\perp)} \, ,
\ee
so that Eq.~(\ref{eq:Boppar}) can be rewritten as
\begin{align}
{\mathcal B}_{q/q} \left(z, x_T^2, \mu \right) &= \frac{1}{2 \pi} \int \! dt \, e^{- i z t \bar{n} \cdot p}
\sum\hspace{-0.5cm}\int\limits_X\; e^{i (p-k) \cdot (t \bar{n} + x_\perp)}
\frac{\nbsl_{ij}}{2}   \langle q(p)| \bar{\chi}_i(0) |X\rangle \langle X| \chi_j(0)|q(p)\rangle \, ,
\label{eq:Boppar2}
\end{align}
where $k$ is the momentum of the intermediate state $X$. By observing that $p \cdot x_\perp =0$ and by integrating over 
$t$ one finds
\begin{align}
{\mathcal B}_{q/q} \left(z, x_T^2, \mu \right) &= 
 \sum\hspace{-0.5cm}\int\limits_X\; e^{-i k_\perp \cdot x_\perp} 
 \delta\left(\bar{n}\cdot (k - (1-z) p) \right)
 \frac{\nbsl_{ij}}{2} \langle q(p)| \bar{\chi}_i(0) |X\rangle \langle X| \chi_j(0)|q(p)\rangle \, .
\label{eq:Boppar3}
\end{align}
The two matrix elements in Eq.~(\ref{eq:Boppar3}) form a sort of``squared'' amplitude which can be interpreted diagrammatically in perturbation theory. By expanding ${\mathcal B}$ in powers of $\alpha_s$ according to
\be
{\mathcal B}_{q/q} \left(z, x_T^2, \mu \right) = \sum_{n = 0}^{\infty} \left( \frac{\alpha_s}{4 \pi}\right)^n  {\mathcal B}^{(n)}_{q/q} \left(z, x_T^2, \mu \right) \, ,
\ee
one can evaluate the TPDFs up to the desired perturbative order. A calculation of the quark TPDF was carried out in \cite{Gehrmann:2012ze}. At leading order, the intermediate state is the vacuum, therefore $k=0$, so that
\be
{\mathcal B}^{(0)}_{q/q} \left(z, x_T^2,\mu \right) = \frac{1}{2 N_c} \text{Tr} [\bm{1}] \frac{1}{\bar{n} \cdot p} \delta \left(1-z \right) \mbox{Tr} \left[ \psl \frac{\nbsl}{2}\right] = \delta\left(1-z \right) \, . \label{eq:B0}
\ee 
In Eq.~(\ref{eq:B0}), the factor $1/2/N_c$ arises from the average over the spin and color of the quark and the trace over the identity in the fundamental representation of $\text{SU}(3)$ gives a further factor of $N_c$.

At NLO, only one gluon of momentum $k$ contributes to the intermediate state $X$,
and the integral over $X$ is the integral over the on-shell four momentum of the gluon. The relevant diagrams are the ones shown in Fig.~\ref{fig:graphsIfun}. In the figure, the Wilson lines in $\chi$ and $\bar{\chi}$ are represented by the symbol $\otimes$. 
Before UV renormalization, the contribution of each of the graphs contributing to the bare NLO  TPDFs ${\mathcal B}_{q/q}$ can be obtained
from the relation
\be
 {\mathcal B}^{(i)}_{q/q} \left(z, x_T^2 \right) \equiv \int \, \frac{d^d k}{(2 \pi)^d} \delta^+(k^2) (2 \pi) \delta\left(\bar{n} \cdot (k - (1-z) p) \right) \left(\frac{\nu}{n \cdot k} \right)^\alpha e^{i k_T \cdot x_T} {\mathcal M}^{(i)} \, ,
\label{eq:mathBi}
\ee
where the factor raised to the power $\alpha$ is the analytic regulator as introduced in \cite{Becher:2011dz}, and the integrand ${\mathcal M}^{(i)}$ depends on the diagram $i$ which we are considering. For convenience, in the following we will make use of the notation
$q_+ \equiv n \cdot q$, $q_- \equiv \bar{n} \cdot q$ ($q_{\pm}$ are scalars, not to be confused with the vectors $q_{\pm}^\mu$ employed in other parts of this work). 

For the first diagram in Fig.~(\ref{fig:graphsIfun}) (diagram $a$), in which the gluon is attached to the quark line on both sides of the cut, the integrand ${\mathcal M}$
is 
\begin{align}
{\mathcal M}^{(a)} &= -\frac{g_s^2}{2 N_c} C_F N_c \text{Tr} \left[ \left(\psl - \ksl \right) \gamma^\mu \psl \gamma_\mu \left(\psl - \ksl \right) \frac{\nbsl}{2}\right]  \frac{1}{\left(p -k\right)^4} \,  \nonumber \\
&=2 g_s^2  C_F  (1-\ep) \frac{k_-}{ p_- k_ +} \, ,  \label{eq:Ma}
\end{align}
where regular QCD vertices and propagators could be  employed, since the matrix element involves only collinear fields.
 By inserting ${\mathcal M}^{(a)}$ in Eq.~(\ref{eq:mathBi}) one finds
\begin{align}
 {\mathcal B}^{(a)}_{q/q} \left(z, x_T^2 \right) &=\frac{2 g_s^2 C_F}{(2 \pi)^{3-2 \ep}}
\left(1- \ep \right) \int d^d k  \, \delta^{+} \left(k_+ k_- - k_T^2 \right)   \nonumber \\ & 
 \,\,\, \,\,\times \delta\left(k_- - (1-z) p_- \right) 
e^{i k_T \cdot x_T}
\left(\frac{\nu}{k_+}\right)^\alpha \frac{k_-}{p_- k_+} \, .
\end{align}
One can rewrite the integration measure in terms of $k_+, k_-$ and $k_T$ according to
\begin{align}
\int d^d k \, \delta^{+}\left(k_+ k_- - k_T^2 \right) &= 
\frac{1}{2} \int_0^\infty \! d k_+ \, \int_0^{\infty}\! d k_- \, \int d^{d-2}  k_T
\delta\left(k_+ k_- - k_T^2 \right) \, ,
\end{align}
and then integrate over $k_-$ and $k_+$ by using the two delta functions to find
\begin{align}
 {\mathcal B}^{(a)}_{q/q} \left(z, x_T^2 \right) &=
\frac{g_s^2 C_F}{(2 \pi)^{3-2 \ep}} \left(1 -\ep\right) \left( 1-z \right)^{1+\alpha} \left(\nu p_-\right)^\alpha
\int d^{2-2 \ep}  k_T \left( k_T^2\right)^{-1-\alpha} e^{i k_T \cdot x_T} \, .
\end{align}
The integral over the transverse momentum can be calculated as follows
\begin{align}
\int d^{2-2 \ep}  k_T \left( k_T^2\right)^{\omega} e^{i k_T \cdot x_T}  &= \Omega_{1 - 2 \ep}
\int_0^\infty \! d u \, u^{1-2 \ep + 2 \omega} \int_{-1}^{1} \!  dr \left( 1- r^2\right)^{-\frac{1}{2} - \ep}
e^{i u  |\vec{x}_T| r} \,  \nonumber \\
& = \pi^{1-\ep} \, \frac{\Gamma\left(1 - \ep + \omega\right)}{\Gamma\left( - \omega\right)}
\left(\frac{x_T^2}{4} \right)^{-1+\ep -\omega} \, , \label{eq:kTmas}
\end{align}
where in the first line of Eq.~(\ref{eq:kTmas}) we set $u \equiv |\vec{k}_T|$, and $r$ represents the cosine of the angle between the vectors $k_T$ and $x_T$. By employing Eq.~(\ref{eq:kTmas}) one finds
\begin{align}\label{eq:barecoeff}
{\mathcal B}^{(a)}_{q/q} \left(z, x_T^2 \right) &=
\frac{ 4 \pi^{2 -\ep} \alpha_s^0  C_F}{(2 \pi)^{3- 2 \ep}} \left(1 -\ep\right) \left( 1-z \right)^{1+\alpha} \left(\nu p_-\right)^\alpha
\frac{\Gamma\left(- \ep - \alpha\right)}{\Gamma\left( 1+ \alpha\right)}
\left(\frac{x_T^2}{4} \right)^{\ep + \alpha} \, .
\end{align}
Next we replace the bare coupling constant $\alpha_{s}^{0} = g_s^2/(4\pi)$ by the coupling
$\alpha_s\equiv \alpha_s(\mu)$  in the $\ms$ scheme using the relation $Z_\alpha\, \alpha_s \,\mu^{2\ep} = e^{-\ep \gamma_E}(4\pi)^\ep \alpha_s^0$, with $Z_\alpha  = 1+\mathcal{O}(\alpha_s)$, 
and make use of the identity
\begin{align}
\left(\frac{x_T^2}{4} \right)^{\ep + \alpha}  \mu^{2 \ep}  e^{\ep \gamma_E} &= \mu^{-2 \alpha}
e^{(\ep+\alpha) L_\perp - (\ep +2 \alpha) \gamma_E} \, ,
\end{align}
where we introduced the symbol $L_\perp = \ln( x_T^2 \mu^2/ (4 e^{-2 \gamma_E}))$. We then find
\begin{align}
{\mathcal B}^{(1,a)}_{q/q} \left(z, x_T^2, \mu \right) &= C_F e^{(\ep+\alpha) L_\perp - (\ep +2 \alpha) \gamma_E} 
\frac{\Gamma\left(- \ep - \alpha\right)}{\Gamma\left( 1+ \alpha\right)}
\left( \frac{\nu p_-}{\mu^2}\right)^\alpha (1-z)^{-1+\alpha} \left[2 (1-z)^2 (1-\ep) \right] \, ,
\end{align}
where ${\mathcal B}^{(1,a)}_{q/q} \left(z, x_T^2, \mu \right)$ denotes the coefficient of the renormalized coupling $\alpha_s(\mu)$, which is scale dependent (in contrast to the result for the bare diagram). The result above matches the notation employed in \cite{Gehrmann:2012ze}.  We notice that the calculation of diagram $a$ did not require the introduction of an analytic regulator; consequently, the $\alpha \to 0$ limit is finite, as long as $\ep$ kept different from zero. One can then extract the corresponding bare coefficient ${\mathcal I}_{q \leftarrow q}^{(a)}$ which is found to be
\be
{\mathcal I}_{q \leftarrow q}^{(a)} (z, x_T^2, \mu^2) = -\frac{C_F \alpha_s}{ 2 \pi} (1-z) \left(\frac{1}{\ep} + L_\perp -1 \right) \, .
\ee

The second and third diagram in Fig.~\ref{fig:graphsIfun}  (diagrams $b$ and $c$) give identical results. The integrand for diagram $b$ is
\begin{align}
{\mathcal M}^{(b)} &= \frac{g_s^2 }{2 N_c} C_F N_c \text{Tr} \left[ \left(\psl - \ksl \right) \nbsl \psl   \frac{\nbsl}{2}\right]  \frac{1}{\left(p -k\right)^2 \bar{n} \cdot k} =2 g_s^2 C_F \frac{p_- -k_-}{ k_- k_ +} \, ,  \label{eq:Mb}
\end{align}
where one can use the first Feynman rule in Fig.~\ref{fig:gg} in order to describe the gluon connecting to the Wilson line, 
provided that one replaces $n \to \bar{n}$, since in the diagrams we want to calculate we deal with collinear Wilson lines rather then with soft Wilson lines as in Fig.~\ref{fig:gg}.
The integral to be evaluated is then 
\begin{align}
 {\mathcal B}^{(b)}_{q/q} \left(z, x_T^2 \right) &=\frac{2 g_s^2  C_F}{(2 \pi)^{3-2 \ep}}
 \int d^d k  \, \delta^{+} \left(k_+ k_- - k_T^2 \right) 
\times  \nonumber \\ & 
\times 
 \delta\left(k_- - (1-z) p_- \right) 
e^{i k_T \cdot x_T}
\left(\frac{\nu}{k_+}\right)^\alpha \frac{p_- - k_-}{k_- k_+} \, .
\end{align}
By following the same procedure employed in order to evaluate diagram $a$ one finds
\begin{align}
{\mathcal B}^{(1,b)}_{q/q} \left(z, x_T^2, \mu \right) &= C_F e^{(\ep+\alpha) L_\perp - (\ep +2 \alpha) \gamma_E} 
\frac{\Gamma\left(- \ep - \alpha\right)}{\Gamma\left( 1+ \alpha\right)}
\left( \frac{\nu p_-}{\mu^2}\right)^\alpha (1-z)^{-1+\alpha} \left[2 z\right] \, . \label{eq:B1b}
\end{align}
The result in Eq.~(\ref{eq:B1b}) matches what was found in \cite{Gehrmann:2012ze}. The singularity in $\alpha$
arises from the expansion the factor
\be
(1-z)^{-1+\alpha} = \frac{1}{\alpha} \delta(1-z) + \left[ \frac{1}{1-z}\right]_+ + {\mathcal O}(\alpha) \, .
\ee

The fourth diagram in Fig.~(\ref{fig:graphsIfun}) evaluates to zero because the two Wilson lines give rise to  a factor $\bar{n}^2 = 0$.

The TPDFs for the antiquark carrying momentum $l$ can be calculated following the same steps outlined above for the quark case, provided that one exchanges $n \leftrightarrow \bar{n}$ everywhere in the integrands, except for the analytic regulator factor $(\nu/k_+)^\alpha$, which is not changed. At NLO, one finds that the integral corresponding to the first diagram in 
Fig.~(\ref{fig:graphsIfun}) gives
\be
\bar{{\mathcal B}}^{(1,a)}_{\bar{q}/\bar{q}} \left(z, x_T^2, \mu \right) = C_F e^{\ep L_\perp - \ep  \gamma_E} 
\Gamma\left(- \ep \right)
\left( \frac{\nu}{l_+}\right)^\alpha (1-z)^{-1-\alpha} \left[2 (1-z)^2 (1-\ep) \right] \, ,
\ee
while by evaluating the second (and third) diagram one finds
\be
\bar{{\mathcal B}}^{(1,b)}_{\bar{q}/\bar{q}} \left(z, x_T^2, \mu \right) = C_F e^{\ep L_\perp - \ep  \gamma_E} 
\Gamma\left(- \ep \right)
\left( \frac{\nu}{l_+}\right)^\alpha (1-z)^{-1-\alpha} \left[2 z \right] \, .
\ee
The diagrams with the gluon connected to the Wilson line on both sides of the cut vanishes also in the antiquark case.

By calculating the product of the quark and the antiquark TPDFs and then by expanding first in $\alpha$ and then in $\ep$,
one sees that the poles in $\alpha$ in the NLO terms cancel out
\begin{align}
\left[{\mathcal B}_{q/q}\left(z_1,x_T^2\right)  \bar{{\mathcal B}}_{\bar{q}/\bar{q}} \left(z_2, x_T^2 \right) \right]_{q^2}
& = \left[ {\cal I}_{q\leftarrow q}
\left(z_1,x_T^2\right) \bar{{\cal I}}_{\bar{q}\leftarrow \bar{q}}\left(z_2,x_T^2\right) \right]_{q^2} \nonumber\\
&= \delta(1-z_1) \delta(1-z_2) \Biggl[1 + \frac{C_F \alpha_s}{4 \pi} \Biggl( \frac{4}{\ep^2}  -\frac{4}{\ep} \ln\left( \frac{q^2}{\mu^2}\right)   \nonumber  \\ 
&\,\,\, - 4  L_\perp \ln\left( \frac{q^2}{\mu^2}\right) - 2 L_\perp^2 - \frac{\pi^2}{3}   \Biggr) \Biggr]  -\frac{C_F \alpha_s}{4 \pi} \Biggl\{2 \delta(1-z_1)  \nonumber \\
&\,\,\, \times \Biggl[\left(\frac{1}{\ep} + L_\perp\right) \frac{1+z_2^2}{\left[ 1-z_2\right]_+} - (1-z_2)  \Biggr] + 
\left(z_1 \leftrightarrow z_2 \right)\Biggr\}  \nonumber \\
&\,\,\,   + {\mathcal O}\left( \alpha_s^2\right) \, . \label{eq:BBbare}
\end{align}
The subscript $q^2$ in the l.h.s. of  Eq.~(\ref{eq:BBbare}) indicates the hidden $q^2$ dependence induced by the collinear anomaly. After $\ms$ renormalization, this result matches Eq.~(38) in\footnote{In \cite{Becher:2010tm} the result involves the plus distribution $[(1+z^2)/(1-z)]_+$. In oder to match Eq.~(\ref{eq:BBbare}), which is written in terms of the plus distribution
$[1/(1-z)]_+$, one needs to make use of the relation
\be
\left[f(z)\frac{\ln^n (1-z)}{1-z} \right]_+ = f(z)\left[\frac{\ln^n (1-z)}{1-z}  \right]_++\delta(1-z) \int_0^1 dx\left( f(1) - f(x) \right) \frac{\ln^n (1-x)}{1-x} \, .
\ee
} \cite{Becher:2010tm}.

\setcounter{figure}{0}
 \section{Color Space Formalism \label{app:colorspace}}

Writing out the gauge-group indices of $n$-particle amplitudes explicitly can be tedious and is inefficient, because many relations hold independent of the explicit representation of the partons. The color-space formalism provides a convenient language to discuss general properties of amplitudes. 
To unify the treatment of color,
 we introduce for each QCD process with $n$ external legs an orthonormal basis of  vectors indicated by $|a_1,a_2, \cdots ,a_n \rangle $ where the indices in the string $\{a\}$ refer to the colors of the external particles. The indices can be in the fundamental representation (quarks) or adjoint representation (gluons).
The amplitude for the process with fixed color indices $\{a\}$ for the external particles, which we indicate with ${\mathcal M}_{\{a\}} $, is related to $|{\mathcal M} \rangle$, an abstract vector representing the process in color space, by the relation
\be
{\mathcal M}_{\{a\}} \equiv \langle a_1, a_2, \cdots a_n | {\mathcal M} \rangle 
= \langle \{a \} | {\mathcal M} \rangle \, .
\ee

In order to describe the color algebra associated with the emission of a soft gluon from one of the external particles, for example the one carrying color index $a_i$, one introduces the color generators $\mathbf{T}^c_i$. These matrices in color space act on the color index of the $i$-th parton in the basis vectors as follows
\be
\mathbf{T}^c_i | \cdots, a_i, \cdots \rangle = (\mathbf{T}^c_i)_{b_i a_i} 
 | \cdots, b_i, \cdots \rangle \, .
\ee
If the $i$-th parton is a final-state quark or an initial-state antiquark, 
the matrices $\mathbf{T}$ are defined by $(\mathbf{T}^c)_{b a}  = t^c_{b a}$, where the $t$ matrices are the usual $\text{SU}(N)$ generators. For a final-state antiquark or an initial state quark instead one defines $(\mathbf{T}^c)_{b a}  = -t^c_{a b}$
 (we remind the reader that $t_{a b} =  t_{b a}^{{\small *}}$), while  for gluons
$(\mathbf{T}^c)_{b a} = i f^{a b c}$.
We also employ the notation $\mathbf{T}_i \cdot \mathbf{T}_j \equiv \mathbf{T}^c_i \mathbf{T}^c_j$. Therefore $\mathbf{T}_i^2$ denotes the Casimir operator representing the $i$-th parton, with eigenvalues $C_F = (N_c^2-1)/2N_c$ for quarks and $C_A = N_c$ for gluons. This  can be seen easily seen  by using the relations
\be
t^a  t^b  = \frac{1}{2} \left[ \frac{1}{N_c} \delta^{a b} \mathbf{I}  
+\left(d^{abc} + i f^{abc} \right) t^c \right] \,  ,
\ee 
with $a =b$ in the quark case and
\be
f^{abc} f^{abd} = N_c \delta^{cd} \, ,
\ee
for the gluon case. If one considers color singlet amplitudes, color conservation implies that
\be \label{eq:colorcons}
\sum_{i=1}^n \mathbf{T}^a_i |{\mathcal M}\rangle = 0\, .
\ee

%

This property can be shown in a similar way both for on-shell amplitudes $|{\mathcal M}\rangle$ and for Wilson coefficients $|{\mathcal C}\rangle$, because of the relation in Eq.~(\ref{eq:ampvswilson}). In particular, we focus on the Wilson coefficient case and, in order to simplify the notation, we only keep the color indices $a_i$ and we drop the Dirac and Lorentz indices from the expressions.
To prove the relation in Eq.~(\ref{eq:colorcons}) we apply a generic gauge transformation of the fields
\be
(\phi_i)_{a_i}\,\rightarrow\,  \text{exp}[i \alpha^c(x) \mathbf{T}^c_i] (\phi_i)_{a_i}(x)
\ee
to  the Hamiltonian in Eq.~(\ref{eq:Heff}). By expanding the exponential we obtain:
\bea\label{eq:gtrasfo}
\mathcal{H}_n^{\mathrm{eff}}\,&\rightarrow&\, \int dt_1\ldots dt_n \mathcal{C}_{a_1\ldots a_n}(t_1,\ldots,t_n,\mu)\nn\\
&& \times \left(\mathbf{1}+i \alpha(x)^c \left((\mathbf{T}^c_1)_{a_1\, b_1}\delta_{a_2\, b_2}\ldots \delta_{a_n\,b_n} +\ldots + \delta_{a_1\, b_1}\ldots \delta_{a_{n-1}\,b_{n-1}}(\mathbf{T}^c_n)_{a_n\, b_n}\right)\right) \nn \\
&& \times (\phi_1)_{b_1}(x+t_1 \bar{n}_1)\ldots (\phi_n)_{b_n}(x+t_n \bar{n}_n) \nn \\
 &=& \int dt_1 \cdots dt_n \langle O_n(\{\underline{t}\},\mu) |\left(\mathbf{1}+i \alpha^c(x) \sum_i \mathbf{T}^c_i\right) |{{\mathcal C}}(\{\underline{t}\},\mu) \rangle \, ,
\eea
where in the last line we used the color-space notation.
Due to the gauge invariance of the Hamiltonian the following equation holds
\bea
\int dt_1 \cdots dt_n \langle O_n(\{\underline{t}\},\mu) |\left(\mathbf{1}+i \alpha^c(x) \sum_i \mathbf{T}^c_i\right) |{{\mathcal C}}(\{\underline{t}\},\mu) \rangle \nn \\ =
\int dt_1 \cdots dt_n \langle O_n(\{\underline{t}\},\mu) |{{\mathcal C}}(\{\underline{t}\},\mu) \rangle \, . 
\eea
Since this equality is valid for arbitrary configurations of the fields $\phi_i$, it follows that
\be
\sum_i \mathbf{T}^c_i  \,|{{\mathcal C}}(\{\underline{t}\},\mu) \rangle\,=\, 0\, .
\ee

A useful relation which follows from color conservation Eq.~(\ref{eq:colorcons}) is
\be
\sum_{i \neq j} \mathbf{T}_i \cdot \mathbf{T}_j  = - \sum_i  \mathbf{T}^2_i = 
- \sum_i C_i \, ,
\ee
where $C_i = C_F$ for quarks and $C_A$ for gluons. In deriving it, we implicitly assume that the operator acts on an amplitude.

When considering color singlet amplitudes, it is possible to decompose the vector representing the amplitude as follows
\be
|{\mathcal M}\rangle  = \sum_{I} {\mathcal M}_I \sum_{\{a\}} \left(c_I \right)_{\{a\}} | \{a\} \rangle \equiv \sum_{I} {\mathcal M}_I |c_I\rangle \,,
\ee
where the factors ${\mathcal M}_I$ are combinations of Dirac matrices, external vectors, spinors, and polarization vectors. The coefficients $\left(c_I\right)_{\{a\}}$ 
are the sets of independent color structures which can appear in the amplitude.
They satisfy the relation
\be 
\sum_{\{a\}} \left[(c_J)_{\{a\}}\right]^* (c_I)_{\{a\}} = 0 \quad \mbox{if} \quad I \neq J \, .
\ee
With this definition the vectors $|c_I\rangle $ form an orthogonal but not orthonormal basis. Consequently, to project out the coefficients ${\mathcal M}_I$ one must use 
\be
{\mathcal M}_I = \frac{1}{\langle c_I | c_I \rangle} \langle c_I | {\mathcal M}\rangle \, .
\ee  

When dealing with IR poles of QCD amplitudes it is often necessary to calculate object of the form 
\be
\langle {\mathcal M} | \mathbf{T}_i \cdot \mathbf{T}_j | {\mathcal M} \rangle = 
{\mathcal M}^*_{a_1, \cdots, a_{i-1}, b_i,  a_{i+1}, \cdots, a_{j-1}, b_j, a_{j+1}, \cdots, a_n }\left( \mathbf{T}^a_i \right)_{b_i a_i} \left( \mathbf{T}^a_j \right)_{b_j a_j} {\mathcal M}_{a_1, \cdots, a_n} \, .
\ee
Of course it is convenient to know how the products $\mathbf{T}_i \cdot \mathbf{T}_j$ act on the basis of vector $|c_I \rangle $:
\be
\mathbf{T}_i \cdot \mathbf{T}_j | c_I \rangle  = \left[\mathbf{T}_i \cdot \mathbf{T}_j \right]_{IJ} | c_J \rangle \, ,
\ee 
where on the r.~h.~s. one has the matrix elements of indices $IJ$ 
\be
\left[\mathbf{T}_i \cdot \mathbf{T}_j \right]_{IJ}  = 
\frac{1}{\langle c_I | c_J \rangle } \langle c_I|\mathbf{T}_i \cdot \mathbf{T}_j | c_J\rangle \, .
\ee
While the generators act in the abstract color space, in the l.h.s. of the equation above the symbol $\left[\mathbf{T}_i \cdot \mathbf{T}_j \right]$ is just employed to indicate a matrix acting on the space of color-singlet structures.

\setcounter{figure}{0}
\section{Anomalous Dimensions \label{app:AnDim}}

For the convenience of the reader, we collect here the explicit expressions of
the factors appearing in Eqs.~(\ref{eq:conj},\ref{eq:Gampr},\ref{eq:Gbexp}).
The QCD beta function up to three loops is given by
\begin{align}
  \beta_0 &= \frac{11}{3} C_A - \frac{4}{3} T_F n_f \, ,
  \nonumber\\
  \beta_1 &= \frac{34}{3} C_A^2 - \frac{20}{3} C_A T_F n_f - 4 C_F T_F n_f \, ,
  \nonumber\\
  \beta_2 &= \frac{2857}{54} C_A^3 + \left( 2 C_F^2 - \frac{205}{9} C_F C_A -
    \frac{1415}{27} C_A^2 \right) T_F n_f + \left( \frac{44}{9} C_F + \frac{158}{27} C_A
  \right) T_F^2 n_f^2 \, , 
\end{align}
where $T_F = 1/2$ and $n_f$ is the number of active quark flavors.

The quantities $\gamma_{i}$ (where $i \in \{ \cusp, q, g\}$) have the following 
expansion in the strong coupling constant
\be \label{eq:expsmallg}
\gamma_i(\alpha_s) = \gamma^i_0 \frac{\alpha_s}{4 \pi} + \gamma^i_1 \left(\frac{\alpha_s}{4 \pi} \right)^2
 + \gamma^i_2 \left(\frac{\alpha_s}{4 \pi} \right)^3  + \cdots \, ,
\ee
The coefficients for the cusp anomalous dimension are \cite{Moch:2004pa}
\begin{eqnarray}
   \gamma_0^{\cusp} &=& 4 \,, \nonumber\\
   \gamma_1^{\cusp} &=& \left( \frac{268}{9} 
    - \frac{4\pi^2}{3} \right) C_A - \frac{80}{9}\,T_F n_f \,,
    \nonumber\\
   \gamma_2^{\cusp} &=& C_A^2 \left( \frac{490}{3} 
    - \frac{536\pi^2}{27}
    + \frac{44\pi^4}{45} + \frac{88}{3}\,\zeta_3 \right) 
    + C_A T_F n_f  \left( - \frac{1672}{27} + \frac{160\pi^2}{27}
    - \frac{224}{3}\,\zeta_3 \right) \nonumber\\
   &&\mbox{}+ C_F T_F n_f \left( - \frac{220}{3} + 64\zeta_3 \right) 
    - \frac{64}{27}\,T_F^2 n_f^2 \,.
\end{eqnarray}

The anomalous dimension $\gamma_q=\gamma_{\bar q}$ can be determined from the three-loop expression for the divergent part of the on-shell quark form factor in QCD \cite{Moch:2005id}. The result was extracted in \cite{Becher:2006mr}. In the notation of this paper $2\gamma_q=\gamma_V$. One obtains
\begin{eqnarray}
   \gamma_0^q &=& -3 C_F \,, \nonumber\\
   \gamma_1^q &=& C_F^2 \left( -\frac{3}{2} + 2\pi^2
    - 24\zeta_3 \right)
    + C_F C_A \left( - \frac{961}{54} - \frac{11\pi^2}{6} 
    + 26\zeta_3 \right)
    + C_F T_F n_f \left( \frac{130}{27} + \frac{2\pi^2}{3} \right) ,
    \nonumber\\
   \gamma_2^q &=& C_F^3 \left( -\frac{29}{2} - 3\pi^2
    - \frac{8\pi^4}{5}
    - 68\zeta_3 + \frac{16\pi^2}{3}\,\zeta_3 + 240\zeta_5 \right) 
    \nonumber\\
   &&\mbox{}+ C_F^2 C_A \left( - \frac{151}{4} + \frac{205\pi^2}{9}
    + \frac{247\pi^4}{135} - \frac{844}{3}\,\zeta_3
    - \frac{8\pi^2}{3}\,\zeta_3 - 120\zeta_5 \right) \nonumber\\
   &&\mbox{}+ C_F C_A^2 \left( - \frac{139345}{2916} - \frac{7163\pi^2}{486}
    - \frac{83\pi^4}{90} + \frac{3526}{9}\,\zeta_3
    - \frac{44\pi^2}{9}\,\zeta_3 - 136\zeta_5 \right) \nonumber\\
   &&\mbox{}+ C_F^2 T_F n_f \left( \frac{2953}{27} - \frac{26\pi^2}{9} 
    - \frac{28\pi^4}{27} + \frac{512}{9}\,\zeta_3 \right) 
    \nonumber\\
   &&\mbox{}+ C_F C_A T_F n_f \left( - \frac{17318}{729}
    + \frac{2594\pi^2}{243} + \frac{22\pi^4}{45} 
    - \frac{1928}{27}\,\zeta_3 \right) \nonumber\\
   &&\mbox{}+ C_F T_F^2 n_f^2 \left( \frac{9668}{729} 
    - \frac{40\pi^2}{27} - \frac{32}{27}\,\zeta_3 \right) .
\end{eqnarray}
Similarly, the expression for the gluon anomalous dimension can be extracted from the divergent part of the gluon form factor obtained in \cite{Moch:2005id}. In terms of the anomalous dimensions given in \cite{Becher:2007ty}, we have $2\gamma_g(\alpha_s)=\gamma_t(\alpha_s)+\gamma_S(\alpha_s)+\beta(\alpha_s)/\alpha_s$. One finds
\begin{eqnarray}
   \gamma_0^g &=& - \beta_0 
    = - \frac{11}{3}\,C_A + \frac{4}{3}\,T_F n_f \,, \nonumber\\
   \gamma_1^g &=& C_A^2 \left( -\frac{692}{27} + \frac{11\pi^2}{18}
    + 2\zeta_3 \right) 
    + C_A T_F n_f \left( \frac{256}{27} - \frac{2\pi^2}{9} \right)
    + 4 C_F T_F n_f \,, \nonumber\\
   \gamma_2^g &=& C_A^3 \left( - \frac{97186}{729} 
    + \frac{6109\pi^2}{486} - \frac{319\pi^4}{270} 
    + \frac{122}{3}\,\zeta_3 - \frac{20\pi^2}{9}\,\zeta_3 
    - 16\zeta_5 \right) \nonumber\\
   &&\mbox{}+ C_A^2 T_F n_f \left( \frac{30715}{729}
    - \frac{1198\pi^2}{243} + \frac{82\pi^4}{135} 
    + \frac{712}{27}\,\zeta_3 \right) \nonumber\\
   &&\mbox{}+ C_A C_F T_F n_f \left( \frac{2434}{27} 
    - \frac{2\pi^2}{3} - \frac{8\pi^4}{45} 
    - \frac{304}{9}\,\zeta_3 \right) 
    - 2 C_F^2 T_F n_f \nonumber\\
   &&\mbox{}+ C_A T_F^2 n_f^2 \left( - \frac{538}{729}
    + \frac{40\pi^2}{81} - \frac{224}{27}\,\zeta_3 \right) 
    - \frac{44}{9}\,C_F T_F^2 n_f^2 \,.
\end{eqnarray}
These results for $\gamma_q$ and $\gamma_g$ are valid in conventional dimensional regularization, where polarization vectors and spinors of all particles are treated as $d$-dimensional objects (so that gluons have $(2-2\ep)$ helicity states). 

The anomalous  dimension for massive quarks appearing in Eq.~(\ref{eq:G2parmass}) has an expansion of the same form of Eq.~(\ref{eq:expsmallg}) where the first two coefficients are \cite{Becher:2009kw}.
\bea
\gamma^Q_0 &=& - 2 C_F \, ,
  \nonumber
  \\
  \gamma^Q_1 &=& C_F C_A \left( -\frac{98}{9} + \frac{2\pi^2}{3} - 4 \zeta_3 \right) +
  \frac{40}{9} C_F T_F n_f \, .
\eea
The cusp anomalous dimension for massive partons, which depends on hyperbolic angles $\beta_{IJ}$ and appears in Eq.~(\ref{eq:G2parmass}) also has an expansion of the form shown in Eq.~(\ref{eq:expsmallg}), where the first two coefficients are \cite{Korchemsky:1987wg, Korchemsky:1991zp, Kidonakis:2009ev, Becher:2009kw} 
\bea
\gamma^{\text{cusp}}_0(\beta) &=& \gamma^{\text{cusp}}_0 \beta \coth \beta
  \, ,
  \nonumber
  \\
  \gamma^{\text{cusp}}_1(\beta) &=& \gamma^{\text{cusp}}_1 \beta \coth \beta
  + 8 C_A \Bigg\{ \frac{\pi^2}{6} + \zeta_3 + \beta^2 \nonumber
  \\
  &\quad& + \coth^2\beta \bigg[ \Li_3(e^{-2\beta}) + \beta
  \Li_2(e^{-2\beta}) - \zeta_3 + \frac{\pi^2}{6} \beta + \frac{\beta^3}{3}
  \bigg]
  \nonumber
  \\
  &\quad& + \coth\beta \bigg[ \Li_2(e^{-2\beta}) - 2\beta
  \ln(1-e^{-2\beta}) - \frac{\pi^2}{6} (1+\beta) - \beta^2 -
  \frac{\beta^3}{3} \bigg] \Bigg\} \, .
\eea
The anomalous dimension describing the evolution of the quark  PDFs near
$x = 1$, employed  in Eq.~(\ref{eq:splitting}), is \cite{Becher:2009th}
\bea
\gamma^{f_q}_0 &=& 3 C_F \, , \nn \\
\gamma^{f_q}_1 &=& C_F^2 \left(\! \frac{3}{2} \!-\! 2 \pi^2 \!+\!24 \zeta(3)\!\!\right) \!+\! C_F C_A
\left(\!\frac{17}{6} \!+\! \frac{22 \pi^2}{9} \!-\! 12 \zeta_3\! \right) \!-\! C_F T_F n_f \left(\!\frac{2}{3} 
\!+\! \frac{8 \pi^2}{9} \!\right) \, ,
\eea
where the coefficients above refer to an expansion of the type in Eq.~(\ref{eq:expsmallg}).
Similarly, for the gluon case, one finds
\bea
\gamma^{f_g}_0 &=& \beta_0 \, , \nn \\
\gamma^{f_g}_1 &=& C_A^2 \left(\! \frac{32}{3} \!+\! 12 \zeta_3\!\!\right) -
\frac{16}{3} C_A T_F n_f -  4 C_F T_F n_f  \, .
\eea
%

%% file: scet_sv.bbl
\providecommand{\href}[2]{#2}\begingroup\raggedright\begin{thebibliography}{100}

\bibitem{Bauer:2000ew}
C.~W. Bauer, S.~Fleming, and M.~E. Luke, {\it {Summing Sudakov logarithms in $B
  \to X_s \gamma$ in effective field theory}},  {\em Phys.Rev.} {\bf D63}
  (2000) 014006, [\href{http://arxiv.org/abs/hep-ph/0005275}{{\tt
  hep-ph/0005275}}].

\bibitem{Bauer:2000yr}
C.~W. Bauer, S.~Fleming, D.~Pirjol, and I.~W. Stewart, {\it {An effective field
  theory for collinear and soft gluons: Heavy to light decays}},  {\em Phys.
  Rev.} {\bf D63} (2001) 114020,
  [\href{http://arxiv.org/abs/hep-ph/0011336}{{\tt hep-ph/0011336}}].

\bibitem{Bauer:2001ct}
C.~W. Bauer and I.~W. Stewart, {\it {Invariant operators in collinear effective
  theory}},  {\em Phys.Lett.} {\bf B516} (2001) 134--142,
  [\href{http://arxiv.org/abs/hep-ph/0107001}{{\tt hep-ph/0107001}}].

\bibitem{Bauer:2001yt}
C.~W. Bauer, D.~Pirjol, and I.~W. Stewart, {\it {Soft-Collinear Factorization
  in Effective Field Theory}},  {\em Phys. Rev.} {\bf D65} (2002) 054022,
  [\href{http://arxiv.org/abs/hep-ph/0109045}{{\tt hep-ph/0109045}}].

\bibitem{Beneke:2002ph}
M.~Beneke, A.~P. Chapovsky, M.~Diehl, and T.~Feldmann, {\it {Soft-collinear
  effective theory and heavy-to-light currents beyond leading power}},  {\em
  Nucl. Phys.} {\bf B643} (2002) 431--476,
  [\href{http://arxiv.org/abs/hep-ph/0206152}{{\tt hep-ph/0206152}}].

\bibitem{Beneke:2002ni}
M.~Beneke and T.~Feldmann, {\it {Multipole-expanded soft-collinear effective
  theory with non-abelian gauge symmetry}},  {\em Phys. Lett.} {\bf B553}
  (2003) 267--276, [\href{http://arxiv.org/abs/hep-ph/0211358}{{\tt
  hep-ph/0211358}}].

\bibitem{Hill:2002vw}
R.~J. Hill and M.~Neubert, {\it {Spectator interactions in soft collinear
  effective theory}},  {\em Nucl.Phys.} {\bf B657} (2003) 229--256,
  [\href{http://arxiv.org/abs/hep-ph/0211018}{{\tt hep-ph/0211018}}].

\bibitem{Collins:1989gx}
J.~C. Collins, D.~E. Soper, and G.~F. Sterman, {\it {Factorization of Hard
  Processes in QCD}},  {\em Adv.Ser.Direct.High Energy Phys.} {\bf 5} (1988)
  1--91, [\href{http://arxiv.org/abs/hep-ph/0409313}{{\tt hep-ph/0409313}}].

\bibitem{Collins:2011zzd}
J.~Collins, {\em {Foundations of perturbative QCD}}.
\newblock Cambridge University Press, Cambridge U.K., 2011.

\bibitem{Becher:2007ty}
T.~Becher, M.~Neubert, and G.~Xu, {\it {Dynamical Threshold Enhancement and
  Resummation in Drell- Yan Production}},  {\em JHEP} {\bf 07} (2008) 030,
  [\href{http://arxiv.org/abs/0710.0680}{{\tt arXiv:0710.0680}}].

\bibitem{Becher:2010tm}
T.~Becher and M.~Neubert, {\it {Drell-Yan production at small $q_T$, transverse
  parton distributions and the collinear anomaly}},  {\em Eur. Phys. J.} {\bf
  C71} (2011) 1665, [\href{http://arxiv.org/abs/1007.4005}{{\tt
  arXiv:1007.4005}}].

\bibitem{Chiu:2011qc}
J.-y. Chiu, A.~Jain, D.~Neill, and I.~Z. Rothstein, {\it {The Rapidity
  Renormalization Group}},  {\em Phys.Rev.Lett.} {\bf 108} (2012) 151601,
  [\href{http://arxiv.org/abs/1104.0881}{{\tt arXiv:1104.0881}}].

\bibitem{Becher:2011dz}
T.~Becher and G.~Bell, {\it {Analytic Regularization in Soft-Collinear
  Effective Theory}},  {\em Phys.Lett.} {\bf B713} (2012) 41--46,
  [\href{http://arxiv.org/abs/1112.3907}{{\tt arXiv:1112.3907}}].

\bibitem{Chiu:2012ir}
J.-Y. Chiu, A.~Jain, D.~Neill, and I.~Z. Rothstein, {\it {A Formalism for the
  Systematic Treatment of Rapidity Logarithms in Quantum Field Theory}},  {\em
  JHEP} {\bf 1205} (2012) 084, [\href{http://arxiv.org/abs/1202.0814}{{\tt
  arXiv:1202.0814}}].

\bibitem{Beneke:1997zp}
M.~Beneke and V.~A. Smirnov, {\it {Asymptotic expansion of Feynman integrals
  near threshold}},  {\em Nucl. Phys.} {\bf B522} (1998) 321--344,
  [\href{http://arxiv.org/abs/hep-ph/9711391}{{\tt hep-ph/9711391}}].

\bibitem{Smirnov:2002pj}
V.~A. Smirnov, {\it {Applied asymptotic expansions in momenta and masses}},
  {\em Springer Tracts Mod. Phys.} {\bf 177} (2002) 1--262.

\bibitem{Manohar:2006nz}
A.~V. Manohar and I.~W. Stewart, {\it {The Zero-Bin and Mode Factorization in
  Quantum Field Theory}},  {\em Phys.Rev.} {\bf D76} (2007) 074002,
  [\href{http://arxiv.org/abs/hep-ph/0605001}{{\tt hep-ph/0605001}}].

\bibitem{Smirnov:1999bza}
V.~A. Smirnov, {\it {Problems of the strategy of regions}},  {\em Phys. Lett.}
  {\bf B465} (1999) 226--234, [\href{http://arxiv.org/abs/hep-ph/9907471}{{\tt
  hep-ph/9907471}}].

\bibitem{Jantzen:2011nz}
B.~Jantzen, {\it {Foundation and generalization of the expansion by regions}},
  {\em JHEP} {\bf 1112} (2011) 076, [\href{http://arxiv.org/abs/1111.2589}{{\tt
  arXiv:1111.2589}}].

\bibitem{Cheung:2009sg}
W.~M.-Y. Cheung, M.~Luke, and S.~Zuberi, {\it {Phase Space and Jet Definitions
  in SCET}},  {\em Phys.Rev.} {\bf D80} (2009) 114021,
  [\href{http://arxiv.org/abs/0910.2479}{{\tt arXiv:0910.2479}}].

\bibitem{Hornig:2009kv}
A.~Hornig, C.~Lee, and G.~Ovanesyan, {\it {Infrared Safety in Factorized Hard
  Scattering Cross-Sections}},  {\em Phys.Lett.} {\bf B677} (2009) 272--277,
  [\href{http://arxiv.org/abs/0901.1897}{{\tt arXiv:0901.1897}}].

\bibitem{Hornig:2009vb}
A.~Hornig, C.~Lee, and G.~Ovanesyan, {\it {Effective Predictions of Event
  Shapes: Factorized, Resummed, and Gapped Angularity Distributions}},  {\em
  JHEP} {\bf 0905} (2009) 122, [\href{http://arxiv.org/abs/0901.3780}{{\tt
  arXiv:0901.3780}}].

\bibitem{Chiu:2007yn}
J.-y. Chiu, F.~Golf, R.~Kelley, and A.~V. Manohar, {\it {Electroweak Sudakov
  Corrections using Effective Field Theory}},  {\em Phys. Rev. Lett.} {\bf 100}
  (2008) 021802, [\href{http://arxiv.org/abs/0709.2377}{{\tt
  arXiv:0709.2377}}].

\bibitem{Chiu:2007dg}
J.-y. Chiu, F.~Golf, R.~Kelley, and A.~V. Manohar, {\it {Electroweak
  Corrections in High Energy Processes using Effective Field Theory}},  {\em
  Phys. Rev.} {\bf D77} (2008) 053004,
  [\href{http://arxiv.org/abs/0712.0396}{{\tt arXiv:0712.0396}}].

\bibitem{Becher:2011pf}
T.~Becher, G.~Bell, and M.~Neubert, {\it {Factorization and Resummation for Jet
  Broadening}},  {\em Phys. Lett.} {\bf B704} (2011) 276--283,
  [\href{http://arxiv.org/abs/1104.4108}{{\tt arXiv:1104.4108}}].

\bibitem{Smirnov:1997gx}
V.~A. Smirnov, {\it {Asymptotic expansions of two-loop Feynman diagrams in the
  Sudakov limit}},  {\em Phys. Lett.} {\bf B404} (1997) 101--107,
  [\href{http://arxiv.org/abs/hep-ph/9703357}{{\tt hep-ph/9703357}}].

\bibitem{Beneke_lectures}
M.~Beneke, {\it {Helmholtz International Summer School on Heavy Quark
  Physics}},  {\em Dubna} (2005).

\bibitem{Bohm:2001yx}
M.~Bohm, A.~Denner, and H.~Joos, {\em {Gauge theories of the strong and
  electroweak interaction}}.
\newblock Teubner, Stuttgart, 2001.

\bibitem{Luke:1992cs}
M.~E. Luke and A.~V. Manohar, {\it {Reparametrization invariance constraints on
  heavy particle effective field theories}},  {\em Phys.Lett.} {\bf B286}
  (1992) 348--354, [\href{http://arxiv.org/abs/hep-ph/9205228}{{\tt
  hep-ph/9205228}}].

\bibitem{Manohar:2002fd}
A.~V. Manohar, T.~Mehen, D.~Pirjol, and I.~W. Stewart, {\it {Reparameterization
  invariance for collinear operators}},  {\em Phys.Lett.} {\bf B539} (2002)
  59--66, [\href{http://arxiv.org/abs/hep-ph/0204229}{{\tt hep-ph/0204229}}].

\bibitem{Heinonen:2012km}
J.~Heinonen, R.~J. Hill, and M.~P. Solon, {\it {Lorentz invariance in heavy
  particle effective theories}},  {\em Phys.Rev.} {\bf D86} (2012) 094020,
  [\href{http://arxiv.org/abs/1208.0601}{{\tt arXiv:1208.0601}}].

\bibitem{Bauer:2003mga}
C.~W. Bauer, D.~Pirjol, and I.~W. Stewart, {\it {On Power suppressed operators
  and gauge invariance in SCET}},  {\em Phys.Rev.} {\bf D68} (2003) 034021,
  [\href{http://arxiv.org/abs/hep-ph/0303156}{{\tt hep-ph/0303156}}].

\bibitem{Becher:2005fg}
T.~Becher, R.~J. Hill, and M.~Neubert, {\it {Factorization in $B \to V \gamma$
  decays}},  {\em Phys.Rev.} {\bf D72} (2005) 094017,
  [\href{http://arxiv.org/abs/hep-ph/0503263}{{\tt hep-ph/0503263}}].

\bibitem{Baikov:2009bg}
P.~A. Baikov, K.~G. Chetyrkin, A.~V. Smirnov, V.~A. Smirnov, and
  M.~Steinhauser, {\it {Quark and gluon form factors to three loops}},  {\em
  Phys. Rev. Lett.} {\bf 102} (2009) 212002,
  [\href{http://arxiv.org/abs/0902.3519}{{\tt arXiv:0902.3519}}].

\bibitem{Gehrmann:2010tu}
T.~Gehrmann, E.~W.~N. Glover, T.~Huber, N.~Ikizlerli, and C.~Studerus, {\it
  {The quark and gluon form factors to three loops in QCD through to
  $O(\epsilon^2)$}},  {\em JHEP} {\bf 11} (2010) 102,
  [\href{http://arxiv.org/abs/1010.4478}{{\tt arXiv:1010.4478}}].

\bibitem{Polyakov:1980ca}
A.~M. Polyakov, {\it {Gauge Fields as Rings of Glue}},  {\em Nucl. Phys.} {\bf
  B164} (1980) 171--188.

\bibitem{Brandt:1981kf}
R.~A. Brandt, F.~Neri, and M.-a. Sato, {\it {Renormalization of Loop Functions
  for All Loops}},  {\em Phys. Rev.} {\bf D24} (1981) 879.

\bibitem{Korchemskaya:1992je}
I.~A. Korchemskaya and G.~P. Korchemsky, {\it {On lightlike Wilson loops}},
  {\em Phys. Lett.} {\bf B287} (1992) 169--175.

\bibitem{Sterman:1986aj}
G.~F. Sterman, {\it {Summation of Large Corrections to Short Distance Hadronic
  Cross-Sections}},  {\em Nucl. Phys.} {\bf B281} (1987) 310.

\bibitem{Catani:1989ne}
S.~Catani and L.~Trentadue, {\it {Resummation of the QCD Perturbative Series
  for Hard Processes}},  {\em Nucl. Phys.} {\bf B327} (1989) 323.

\bibitem{Becher:2006mr}
T.~Becher, M.~Neubert, and B.~D. Pecjak, {\it {Factorization and momentum-space
  resummation in deep- inelastic scattering}},  {\em JHEP} {\bf 01} (2007) 076,
  [\href{http://arxiv.org/abs/hep-ph/0607228}{{\tt hep-ph/0607228}}].

\bibitem{Ahrens:2008qu}
V.~Ahrens, T.~Becher, M.~Neubert, and L.~L. Yang, {\it {Origin of the Large
  Perturbative Corrections to Higgs Production at Hadron Colliders}},  {\em
  Phys.Rev.} {\bf D79} (2009) 033013,
  [\href{http://arxiv.org/abs/0808.3008}{{\tt arXiv:0808.3008}}].

\bibitem{Ahrens:2008nc}
V.~Ahrens, T.~Becher, M.~Neubert, and L.~L. Yang, {\it {Renormalization-Group
  Improved Prediction for Higgs Production at Hadron Colliders}},  {\em
  Eur.Phys.J.} {\bf C62} (2009) 333--353,
  [\href{http://arxiv.org/abs/0809.4283}{{\tt arXiv:0809.4283}}].

\bibitem{Ahrens:2010rs}
V.~Ahrens, T.~Becher, M.~Neubert, and L.~L. Yang, {\it {Updated Predictions for
  Higgs Production at the Tevatron and the LHC}},  {\em Phys.Lett.} {\bf B698}
  (2011) 271--274, [\href{http://arxiv.org/abs/1008.3162}{{\tt
  arXiv:1008.3162}}].

\bibitem{Becher:2012qa}
T.~Becher and M.~Neubert, {\it {Factorization and NNLL Resummation for Higgs
  Production with a Jet Veto}},  {\em JHEP} {\bf 1207} (2012) 108,
  [\href{http://arxiv.org/abs/1205.3806}{{\tt arXiv:1205.3806}}].

\bibitem{Becher:2009th}
T.~Becher and M.~D. Schwartz, {\it {Direct photon production with effective
  field theory}},  {\em JHEP} {\bf 02} (2010) 040,
  [\href{http://arxiv.org/abs/0911.0681}{{\tt arXiv:0911.0681}}].

\bibitem{Ahrens:2009uz}
V.~Ahrens, A.~Ferroglia, M.~Neubert, B.~D. Pecjak, and L.~L. Yang, {\it
  {Threshold expansion at order $\alpha_s^4$ for the t-tbar invariant mass
  distribution at hadron colliders}},  {\em Phys. Lett.} {\bf B687} (2010)
  331--337, [\href{http://arxiv.org/abs/0912.3375}{{\tt arXiv:0912.3375}}].

\bibitem{Ahrens:2010zv}
V.~Ahrens, A.~Ferroglia, M.~Neubert, B.~D. Pecjak, and L.~L. Yang, {\it
  {Renormalization-Group Improved Predictions for Top-Quark Pair Production at
  Hadron Colliders}},  {\em JHEP} {\bf 09} (2010) 097,
  [\href{http://arxiv.org/abs/1003.5827}{{\tt arXiv:1003.5827}}].

\bibitem{Ahrens:2011mw}
V.~Ahrens, A.~Ferroglia, M.~Neubert, B.~D. Pecjak, and L.-L. Yang, {\it
  {RG-improved single-particle inclusive cross sections and forward-backward
  asymmetry in $t\bar t$ production at hadron colliders}},  {\em JHEP} {\bf 09}
  (2011) 070, [\href{http://arxiv.org/abs/1103.0550}{{\tt arXiv:1103.0550}}].

\bibitem{Becher:2011xn}
T.~Becher, M.~Neubert, and D.~Wilhelm, {\it {Electroweak Gauge-Boson Production
  at Small $q_T$: Infrared Safety from the Collinear Anomaly}},  {\em JHEP}
  {\bf 1202} (2012) 124, [\href{http://arxiv.org/abs/1109.6027}{{\tt
  arXiv:1109.6027}}].

\bibitem{Becher:2012xr}
T.~Becher, C.~Lorentzen, and M.~D. Schwartz, {\it {Precision Direct Photon and
  $W$-Boson Spectra at High $p_T$ and Comparison to LHC Data}},  {\em
  Phys.Rev.} {\bf D86} (2012) 054026,
  [\href{http://arxiv.org/abs/1206.6115}{{\tt arXiv:1206.6115}}].

\bibitem{Broggio:2011bd}
A.~Broggio, M.~Neubert, and L.~Vernazza, {\it {Soft-gluon resummation for
  slepton-pair production at hadron colliders}},  {\em JHEP} {\bf 1205} (2012)
  151, [\href{http://arxiv.org/abs/1111.6624}{{\tt arXiv:1111.6624}}].

\bibitem{Broggio:2013uba}
A.~Broggio, A.~Ferroglia, M.~Neubert, L.~Vernazza, and L.~L. Yang, {\it
  {Approximate NNLO Predictions for the Stop-Pair Production Cross Section at
  the LHC}},  {\em JHEP} {\bf 1307} (2013) 042,
  [\href{http://arxiv.org/abs/1304.2411}{{\tt arXiv:1304.2411}}].

\bibitem{Broggio:2013cia}
A.~Broggio, A.~Ferroglia, M.~Neubert, L.~Vernazza, and L.~L. Yang, {\it {NNLL
  Momentum-Space Resummation for Stop-Pair Production at the LHC}},  {\em JHEP}
  {\bf 1403} (2014) 066, [\href{http://arxiv.org/abs/1312.4540}{{\tt
  arXiv:1312.4540}}].

\bibitem{Becher:2006nr}
T.~Becher and M.~Neubert, {\it {Threshold resummation in momentum space from
  effective field theory}},  {\em Phys.Rev.Lett.} {\bf 97} (2006) 082001,
  [\href{http://arxiv.org/abs/hep-ph/0605050}{{\tt hep-ph/0605050}}].

\bibitem{Bonvini:2013td}
M.~Bonvini, S.~Forte, M.~Ghezzi, and G.~Ridolfi, {\it {The scale of soft
  resummation in SCET vs perturbative QCD}},  {\em Nucl.Phys.Proc.Suppl.} {\bf
  241-242} (2013) 121--126, [\href{http://arxiv.org/abs/1301.4502}{{\tt
  arXiv:1301.4502}}].

\bibitem{Almeida:2014uva}
L.~G. Almeida, S.~D. Ellis, C.~Lee, G.~Sterman, I.~Sung, et~al., {\it
  {Comparing and counting logs in direct and effective methods of QCD
  resummation}},  {\em JHEP} {\bf 1404} (2014) 174,
  [\href{http://arxiv.org/abs/1401.4460}{{\tt arXiv:1401.4460}}].

\bibitem{Sterman:2013nya}
G.~Sterman and M.~Zeng, {\it {Quantifying Comparisons of Threshold
  Resummations}},  {\em JHEP} {\bf 1405} (2014) 132,
  [\href{http://arxiv.org/abs/1312.5397}{{\tt arXiv:1312.5397}}].

\bibitem{Bonvini:2014qga}
M.~Bonvini, S.~Forte, G.~Ridolfi, and L.~Rottoli, {\it {Resummation
  prescriptions and ambiguities in SCET vs. direct QCD: Higgs production as a
  case study}},  {\em JHEP} {\bf 1501} (2015) 046,
  [\href{http://arxiv.org/abs/1409.0864}{{\tt arXiv:1409.0864}}].

\bibitem{Soper:1996sn}
D.~E. Soper, {\it {Parton distribution functions}},  {\em Nucl. Phys. Proc.
  Suppl.} {\bf 53} (1997) 69--80,
  [\href{http://arxiv.org/abs/hep-lat/9609018}{{\tt hep-lat/9609018}}].

\bibitem{Belitsky:1998tc}
A.~V. Belitsky, {\it {Two-loop renormalization of Wilson loop for Drell-Yan
  production}},  {\em Phys. Lett.} {\bf B442} (1998) 307--314,
  [\href{http://arxiv.org/abs/hep-ph/9808389}{{\tt hep-ph/9808389}}].

\bibitem{DDT}
Y.~L. Dokshitzer, D.~I. Dyakonov, and S.~I. Troyan, {\it {Hard processes in
  quantum chromodynamics}},  {\em Phys. Rep.} {\bf 58} (1980) 269.

\bibitem{Parisi:1979se}
G.~Parisi and R.~Petronzio, {\it {Small Transverse Momentum Distributions in
  Hard Processes}},  {\em Nucl.Phys.} {\bf B154} (1979) 427.

\bibitem{Curci:1979bg}
G.~Curci, M.~Greco, and Y.~Srivastava, {\it {{QCD} Jets From Coherent States}},
   {\em Nucl.Phys.} {\bf B159} (1979) 451.

\bibitem{Collins:1984kg}
J.~C. Collins, D.~E. Soper, and G.~F. Sterman, {\it {Transverse Momentum
  Distribution in Drell-Yan Pair and W and Z Boson Production}},  {\em
  Nucl.Phys.} {\bf B250} (1985) 199.

\bibitem{Collins:1981uk}
J.~C. Collins and D.~E. Soper, {\it {Back-To-Back Jets in QCD}},  {\em
  Nucl.Phys.} {\bf B193} (1981) 381.

\bibitem{Collins:1981uw}
J.~C. Collins and D.~E. Soper, {\it {Parton Distribution and Decay Functions}},
   {\em Nucl.Phys.} {\bf B194} (1982) 445.

\bibitem{Stewart:2009yx}
I.~W. Stewart, F.~J. Tackmann, and W.~J. Waalewijn, {\it {Factorization at the
  LHC: From PDFs to Initial State Jets}},  {\em Phys.Rev.} {\bf D81} (2010)
  094035, [\href{http://arxiv.org/abs/0910.0467}{{\tt arXiv:0910.0467}}].

\bibitem{Becher:2013xia}
T.~Becher, M.~Neubert, and L.~Rothen, {\it {Factorization and
  $N^{3}LL_{p}$+NNLO predictions for the Higgs cross section with a jet veto}},
   {\em JHEP} {\bf 1310} (2013) 125,
  [\href{http://arxiv.org/abs/1307.0025}{{\tt arXiv:1307.0025}}].

\bibitem{Becher:2013iya}
T.~Becher and G.~Bell, {\it {Enhanced non-perturbative effects through the
  collinear anomaly}},  {\em Phys.Rev.Lett.} {\bf 112} (2014) 182002,
  [\href{http://arxiv.org/abs/1312.5327}{{\tt arXiv:1312.5327}}].

\bibitem{CMS:2014sma}
{\bf CMS} Collaboration, {\it {Measurement of Z production as a function of
  $p_T$, $Y$}},  {\em CMS-PAS-SMP-13-013} (2014).

\bibitem{Aad:2014xaa}
{\bf ATLAS Collaboration} Collaboration, G.~Aad et~al., {\it {Measurement of
  the $Z/\gamma^*$ boson transverse momentum distribution in $pp$ collisions at
  $\sqrt{s}$ = 7 TeV with the ATLAS detector}},  {\em JHEP} {\bf 1409} (2014)
  145, [\href{http://arxiv.org/abs/1406.3660}{{\tt arXiv:1406.3660}}].

\bibitem{Becher:2009cu}
T.~Becher and M.~Neubert, {\it {Infrared singularities of scattering amplitudes
  in perturbative QCD}},  {\em Phys. Rev. Lett.} {\bf 102} (2009) 162001,
  [\href{http://arxiv.org/abs/0901.0722}{{\tt arXiv:0901.0722}}].

\bibitem{Becher:2009qa}
T.~Becher and M.~Neubert, {\it {On the Structure of Infrared Singularities of
  Gauge-Theory Amplitudes}},  {\em JHEP} {\bf 06} (2009) 081,
  [\href{http://arxiv.org/abs/0903.1126}{{\tt arXiv:0903.1126}}].

\bibitem{Ahrens:2012he}
V.~Ahrens, M.~Neubert, and L.~Vernazza, {\it {Structure of Infrared
  Singularities of Gauge-Theory Amplitudes at Three and Four Loops}},  {\em
  JHEP} {\bf 1209} (2012) 138, [\href{http://arxiv.org/abs/1208.4847}{{\tt
  arXiv:1208.4847}}].

\bibitem{Gardi:2009qi}
E.~Gardi and L.~Magnea, {\it {Factorization constraints for soft anomalous
  dimensions in QCD scattering amplitudes}},  {\em JHEP} {\bf 03} (2009) 079,
  [\href{http://arxiv.org/abs/0901.1091}{{\tt arXiv:0901.1091}}].

\bibitem{Dixon:2009ur}
L.~J. Dixon, E.~Gardi, and L.~Magnea, {\it {On soft singularities at three
  loops and beyond}},  {\em JHEP} {\bf 02} (2010) 081,
  [\href{http://arxiv.org/abs/0910.3653}{{\tt arXiv:0910.3653}}].

\bibitem{DelDuca:2011ae}
V.~Del~Duca, C.~Duhr, E.~Gardi, L.~Magnea, and C.~D. White, {\it {The Infrared
  structure of gauge theory amplitudes in the high-energy limit}},  {\em JHEP}
  {\bf 1112} (2011) 021, [\href{http://arxiv.org/abs/1109.3581}{{\tt
  arXiv:1109.3581}}].

\bibitem{DelDuca:2012qg}
V.~Del~Duca, C.~Duhr, E.~Gardi, L.~Magnea, and C.~D. White, {\it {Infrared
  Singularities and the High-Energy Limit}},  {\em PoS} {\bf RADCOR2011} (2011)
  038, [\href{http://arxiv.org/abs/1201.2841}{{\tt arXiv:1201.2841}}].

\bibitem{Becher:2009kw}
T.~Becher and M.~Neubert, {\it {Infrared singularities of QCD amplitudes with
  massive partons}},  {\em Phys. Rev.} {\bf D79} (2009) 125004,
  [\href{http://arxiv.org/abs/0904.1021}{{\tt arXiv:0904.1021}}].
  [Erratum-ibid.D80:109901,2009].

\bibitem{Bassetto:1984ik}
A.~Bassetto, M.~Ciafaloni, and G.~Marchesini, {\it {Jet Structure and Infrared
  Sensitive Quantities in Perturbative QCD}},  {\em Phys. Rept.} {\bf 100}
  (1983) 201--272.

\bibitem{Catani:1996jh}
S.~Catani and M.~H. Seymour, {\it {The Dipole Formalism for the Calculation of
  QCD Jet Cross Sections at Next-to-Leading Order}},  {\em Phys. Lett.} {\bf
  B378} (1996) 287--301, [\href{http://arxiv.org/abs/hep-ph/9602277}{{\tt
  hep-ph/9602277}}].

\bibitem{Sen:1982bt}
A.~Sen, {\it {Asymptotic Behavior of the Wide Angle On-Shell Quark Scattering
  Amplitudes in Nonabelian Gauge Theories}},  {\em Phys. Rev.} {\bf D28} (1983)
  860.

\bibitem{Kidonakis:1998nf}
N.~Kidonakis, G.~Oderda, and G.~F. Sterman, {\it {Evolution of color exchange
  in QCD hard scattering}},  {\em Nucl. Phys.} {\bf B531} (1998) 365--402,
  [\href{http://arxiv.org/abs/hep-ph/9803241}{{\tt hep-ph/9803241}}].

\bibitem{Aybat:2006wq}
S.~M. Aybat, L.~J. Dixon, and G.~F. Sterman, {\it {The Two-loop anomalous
  dimension matrix for soft gluon exchange}},  {\em Phys.Rev.Lett.} {\bf 97}
  (2006) 072001, [\href{http://arxiv.org/abs/hep-ph/0606254}{{\tt
  hep-ph/0606254}}].

\bibitem{Aybat:2006mz}
S.~M. Aybat, L.~J. Dixon, and G.~F. Sterman, {\it {The Two-loop soft anomalous
  dimension matrix and resummation at next-to-next-to leading pole}},  {\em
  Phys.Rev.} {\bf D74} (2006) 074004,
  [\href{http://arxiv.org/abs/hep-ph/0607309}{{\tt hep-ph/0607309}}].

\bibitem{Moch:2005tm}
S.~Moch, J.~A.~M. Vermaseren, and A.~Vogt, {\it {Three-loop results for quark
  and gluon form factors}},  {\em Phys. Lett.} {\bf B625} (2005) 245--252,
  [\href{http://arxiv.org/abs/hep-ph/0508055}{{\tt hep-ph/0508055}}].

\bibitem{Garland:2001tf}
L.~W. Garland, T.~Gehrmann, E.~W.~N. Glover, A.~Koukoutsakis, and E.~Remiddi,
  {\it {The Two-Loop QCD Matrix Element for $e^+e^- \to 3$ Jets}},  {\em Nucl.
  Phys.} {\bf B627} (2002) 107--188,
  [\href{http://arxiv.org/abs/hep-ph/0112081}{{\tt hep-ph/0112081}}].

\bibitem{Garland:2002ak}
L.~W. Garland, T.~Gehrmann, E.~W.~N. Glover, A.~Koukoutsakis, and E.~Remiddi,
  {\it {Two-Loop QCD Helicity Amplitudes for $e^+e^- \to 3$~Jets}},  {\em Nucl.
  Phys.} {\bf B642} (2002) 227--262,
  [\href{http://arxiv.org/abs/hep-ph/0206067}{{\tt hep-ph/0206067}}].

\bibitem{Anastasiou:2000kg}
C.~Anastasiou, E.~N. Glover, C.~Oleari, and M.~Tejeda-Yeomans, {\it {Two-loop
  QCD corrections to the scattering of massless distinct quarks}},  {\em
  Nucl.Phys.} {\bf B601} (2001) 318--340,
  [\href{http://arxiv.org/abs/hep-ph/0010212}{{\tt hep-ph/0010212}}].

\bibitem{Anastasiou:2000ue}
C.~Anastasiou, E.~N. Glover, C.~Oleari, and M.~Tejeda-Yeomans, {\it {Two loop
  QCD corrections to massless identical quark scattering}},  {\em Nucl.Phys.}
  {\bf B601} (2001) 341--360, [\href{http://arxiv.org/abs/hep-ph/0011094}{{\tt
  hep-ph/0011094}}].

\bibitem{Anastasiou:2001sv}
C.~Anastasiou, E.~N. Glover, C.~Oleari, and M.~Tejeda-Yeomans, {\it {Two loop
  QCD corrections to massless quark gluon scattering}},  {\em Nucl.Phys.} {\bf
  B605} (2001) 486--516, [\href{http://arxiv.org/abs/hep-ph/0101304}{{\tt
  hep-ph/0101304}}].

\bibitem{Bern:2002tk}
Z.~Bern, A.~De~Freitas, and L.~J. Dixon, {\it {Two loop helicity amplitudes for
  gluon-gluon scattering in QCD and supersymmetric Yang-Mills theory}},  {\em
  JHEP} {\bf 0203} (2002) 018, [\href{http://arxiv.org/abs/hep-ph/0201161}{{\tt
  hep-ph/0201161}}].

\bibitem{Bern:2003ck}
Z.~Bern, A.~De~Freitas, and L.~J. Dixon, {\it {Two loop helicity amplitudes for
  quark gluon scattering in QCD and gluino gluon scattering in supersymmetric
  Yang-Mills theory}},  {\em JHEP} {\bf 0306} (2003) 028,
  [\href{http://arxiv.org/abs/hep-ph/0304168}{{\tt hep-ph/0304168}}].

\bibitem{Bern:2005iz}
Z.~Bern, L.~J. Dixon, and V.~A. Smirnov, {\it {Iteration of planar amplitudes
  in maximally supersymmetric Yang-Mills theory at three loops and beyond}},
  {\em Phys. Rev.} {\bf D72} (2005) 085001,
  [\href{http://arxiv.org/abs/hep-th/0505205}{{\tt hep-th/0505205}}].

\bibitem{Caron-Huot:2013fea}
S.~Caron-Huot, {\it {When does the gluon reggeize?}},
  \href{http://arxiv.org/abs/1309.6521}{{\tt arXiv:1309.6521}}.

\bibitem{Gatheral:1983cz}
J.~G.~M. Gatheral, {\it {Exponentiation of the Eikonal Cross-sections in
  Nonabelian Gauge Theories}},  {\em Phys. Lett.} {\bf B133} (1983) 90.

\bibitem{Frenkel:1984pz}
J.~Frenkel and J.~C. Taylor, {\it {Nonabelian Eikonal Exponentiation}},  {\em
  Nucl. Phys.} {\bf B246} (1984) 231.

\bibitem{Mitov:2010rp}
A.~Mitov, G.~Sterman, and I.~Sung, {\it {Diagrammatic Exponentiation for
  Products of Wilson Lines}},  {\em Phys.Rev.} {\bf D82} (2010) 096010,
  [\href{http://arxiv.org/abs/1008.0099}{{\tt arXiv:1008.0099}}].

\bibitem{Gardi:2010rn}
E.~Gardi, E.~Laenen, G.~Stavenga, and C.~D. White, {\it {Webs in multiparton
  scattering using the replica trick}},  {\em JHEP} {\bf 1011} (2010) 155,
  [\href{http://arxiv.org/abs/1008.0098}{{\tt arXiv:1008.0098}}].

\bibitem{Gardi:2011wa}
E.~Gardi and C.~D. White, {\it {General properties of multiparton webs: Proofs
  from combinatorics}},  {\em JHEP} {\bf 1103} (2011) 079,
  [\href{http://arxiv.org/abs/1102.0756}{{\tt arXiv:1102.0756}}].

\bibitem{Gardi:2011yz}
E.~Gardi, J.~M. Smillie, and C.~D. White, {\it {On the renormalization of
  multiparton webs}},  {\em JHEP} {\bf 1109} (2011) 114,
  [\href{http://arxiv.org/abs/1108.1357}{{\tt arXiv:1108.1357}}].

\bibitem{Berends:1987me}
F.~A. Berends and W.~T. Giele, {\it {Recursive Calculations for Processes with
  n Gluons}},  {\em Nucl. Phys.} {\bf B306} (1988) 759.

\bibitem{Mangano:1990by}
M.~L. Mangano and S.~J. Parke, {\it {Multi-Parton Amplitudes in Gauge
  Theories}},  {\em Phys. Rept.} {\bf 200} (1991) 301--367,
  [\href{http://arxiv.org/abs/hep-th/0509223}{{\tt hep-th/0509223}}].

\bibitem{Bern:1995ix}
Z.~Bern and G.~Chalmers, {\it {Factorization in one loop gauge theory}},  {\em
  Nucl. Phys.} {\bf B447} (1995) 465--518,
  [\href{http://arxiv.org/abs/hep-ph/9503236}{{\tt hep-ph/9503236}}].

\bibitem{Kosower:1999xi}
D.~A. Kosower, {\it {All-order collinear behavior in gauge theories}},  {\em
  Nucl. Phys.} {\bf B552} (1999) 319--336,
  [\href{http://arxiv.org/abs/hep-ph/9901201}{{\tt hep-ph/9901201}}].

\bibitem{Catani:2003vu}
S.~Catani, D.~de~Florian, and G.~Rodrigo, {\it {The triple collinear limit of
  one-loop QCD amplitudes}},  {\em Phys. Lett.} {\bf B586} (2004) 323--331,
  [\href{http://arxiv.org/abs/hep-ph/0312067}{{\tt hep-ph/0312067}}].

\bibitem{Bret:2011xm}
V.~Del~Duca, C.~Duhr, E.~Gardi, L.~Magnea, and C.~D. White, {\it {An infrared
  approach to Reggeization}},  {\em Phys.Rev.} {\bf D85} (2012) 071104,
  [\href{http://arxiv.org/abs/1108.5947}{{\tt arXiv:1108.5947}}].

\bibitem{Balitsky:1979ap}
I.~Balitsky, L.~Lipatov, and V.~S. Fadin, {\it {Regge processes in non-abelian
  gauge theories. (In russian)}},  {\em Leningrad 1979, Proceedings, Physics Of
  Elementary Particles} (1979) 109--149.

\bibitem{Fadin:2006bj}
V.~Fadin, R.~Fiore, M.~Kozlov, and A.~Reznichenko, {\it {Proof of the
  multi-Regge form of QCD amplitudes with gluon exchanges in the NLA}},  {\em
  Phys.Lett.} {\bf B639} (2006) 74--81,
  [\href{http://arxiv.org/abs/hep-ph/0602006}{{\tt hep-ph/0602006}}].

\bibitem{Bogdan:2006af}
A.~Bogdan and V.~Fadin, {\it {A Proof of the reggeized form of amplitudes with
  quark exchanges}},  {\em Nucl.Phys.} {\bf B740} (2006) 36--57,
  [\href{http://arxiv.org/abs/hep-ph/0601117}{{\tt hep-ph/0601117}}].

\bibitem{DelDuca:2001gu}
V.~Del~Duca and E.~N. Glover, {\it {The High-energy limit of QCD at two
  loops}},  {\em JHEP} {\bf 0110} (2001) 035,
  [\href{http://arxiv.org/abs/hep-ph/0109028}{{\tt hep-ph/0109028}}].

\bibitem{Bern:2008pv}
Z.~Bern, J.~J.~M. Carrasco, L.~J. Dixon, H.~Johansson, and R.~Roiban, {\it
  {Manifest Ultraviolet Behavior for the Three-Loop Four- Point Amplitude of
  $N=8$ Supergravity}},  {\em Phys. Rev.} {\bf D78} (2008) 105019,
  [\href{http://arxiv.org/abs/0808.4112}{{\tt arXiv:0808.4112}}].

\bibitem{Moch:2004pa}
S.~Moch, J.~A.~M. Vermaseren, and A.~Vogt, {\it {The three-loop splitting
  functions in QCD: The non-singlet case}},  {\em Nucl. Phys.} {\bf B688}
  (2004) 101--134, [\href{http://arxiv.org/abs/hep-ph/0403192}{{\tt
  hep-ph/0403192}}].

\bibitem{Armoni:2006ux}
A.~Armoni, {\it {Anomalous dimensions from a spinning $D5-$brane}},  {\em JHEP}
  {\bf 11} (2006) 009, [\href{http://arxiv.org/abs/hep-th/0608026}{{\tt
  hep-th/0608026}}].

\bibitem{Alday:2007hr}
L.~F. Alday and J.~M. Maldacena, {\it {Gluon scattering amplitudes at strong
  coupling}},  {\em JHEP} {\bf 06} (2007) 064,
  [\href{http://arxiv.org/abs/0705.0303}{{\tt arXiv:0705.0303}}].

\bibitem{Alday:2007mf}
L.~F. Alday and J.~M. Maldacena, {\it {Comments on operators with large spin}},
   {\em JHEP} {\bf 11} (2007) 019, [\href{http://arxiv.org/abs/0708.0672}{{\tt
  arXiv:0708.0672}}].

\bibitem{Grozin:2014hna}
A.~Grozin, J.~M. Henn, G.~P. Korchemsky, and P.~Marquard, {\it {The three-loop
  cusp anomalous dimension in QCD}},
  \href{http://arxiv.org/abs/1409.0023}{{\tt arXiv:1409.0023}}.

\bibitem{vanRitbergen:1998pn}
T.~van Ritbergen, A.~Schellekens, and J.~Vermaseren, {\it {Group theory factors
  for Feynman diagrams}},  {\em Int.J.Mod.Phys.} {\bf A14} (1999) 41--96,
  [\href{http://arxiv.org/abs/hep-ph/9802376}{{\tt hep-ph/9802376}}].

\bibitem{Badger:2004uk}
S.~Badger and E.~N. Glover, {\it {Two loop splitting functions in QCD}},  {\em
  JHEP} {\bf 0407} (2004) 040, [\href{http://arxiv.org/abs/hep-ph/0405236}{{\tt
  hep-ph/0405236}}].

\bibitem{Feige:2014wja}
I.~Feige and M.~D. Schwartz, {\it {Hard-Soft-Collinear Factorization to All
  Orders}},  {\em Phys.Rev.} {\bf D90} (2014), no.~10 105020,
  [\href{http://arxiv.org/abs/1403.6472}{{\tt arXiv:1403.6472}}].

\bibitem{Yennie:1961ad}
D.~R. Yennie, S.~C. Frautschi, and H.~Suura, {\it {The infrared divergence
  phenomena and high-energy processes}},  {\em Ann. Phys.} {\bf 13} (1961)
  379--452.

\bibitem{Weinberg:1965nx}
S.~Weinberg, {\it {Infrared photons and gravitons}},  {\em Phys. Rev.} {\bf
  140} (1965) B516--B524.

\bibitem{Neubert:1993mb}
M.~Neubert, {\it {Heavy quark symmetry}},  {\em Phys. Rept.} {\bf 245} (1994)
  259--396, [\href{http://arxiv.org/abs/hep-ph/9306320}{{\tt hep-ph/9306320}}].

\bibitem{Mitov:2009sv}
A.~Mitov, G.~F. Sterman, and I.~Sung, {\it {The Massive Soft Anomalous
  Dimension Matrix at Two Loops}},  {\em Phys. Rev.} {\bf D79} (2009) 094015,
  [\href{http://arxiv.org/abs/0903.3241}{{\tt arXiv:0903.3241}}].

\bibitem{Ferroglia:2009ep}
A.~Ferroglia, M.~Neubert, B.~D. Pecjak, and L.~L. Yang, {\it {Two-loop
  divergences of scattering amplitudes with massive partons}},  {\em Phys. Rev.
  Lett.} {\bf 103} (2009) 201601, [\href{http://arxiv.org/abs/0907.4791}{{\tt
  arXiv:0907.4791}}].

\bibitem{Ferroglia:2009ii}
A.~Ferroglia, M.~Neubert, B.~D. Pecjak, and L.~L. Yang, {\it {Two-loop
  divergences of massive scattering amplitudes in non-abelian gauge theories}},
   {\em JHEP} {\bf 11} (2009) 062, [\href{http://arxiv.org/abs/0908.3676}{{\tt
  arXiv:0908.3676}}].

\bibitem{Mitov:2006xs}
A.~Mitov and S.~Moch, {\it {The singular behavior of massive QCD amplitudes}},
  {\em JHEP} {\bf 05} (2007) 001,
  [\href{http://arxiv.org/abs/hep-ph/0612149}{{\tt hep-ph/0612149}}].

\bibitem{Becher:2007cu}
T.~Becher and K.~Melnikov, {\it {Two-loop QED corrections to Bhabha
  scattering}},  {\em JHEP} {\bf 06} (2007) 084,
  [\href{http://arxiv.org/abs/0704.3582}{{\tt arXiv:0704.3582}}].

\bibitem{Manohar:2000dt}
A.~V. Manohar and M.~B. Wise, {\it {Heavy quark physics}},  {\em
  Camb.Monogr.Part.Phys.Nucl.Phys.Cosmol.} {\bf 10} (2000) 1--191.

\bibitem{Bigi:1993ex}
I.~I. Bigi, M.~A. Shifman, N.~Uraltsev, and A.~Vainshtein, {\it {On the motion
  of heavy quarks inside hadrons: Universal distributions and inclusive
  decays}},  {\em Int.J.Mod.Phys.} {\bf A9} (1994) 2467--2504,
  [\href{http://arxiv.org/abs/hep-ph/9312359}{{\tt hep-ph/9312359}}].

\bibitem{Neubert:1993um}
M.~Neubert, {\it {Analysis of the photon spectrum in inclusive $\bar{B}\to X_s
  \gamma$ decays}},  {\em Phys.Rev.} {\bf D49} (1994) 4623--4633,
  [\href{http://arxiv.org/abs/hep-ph/9312311}{{\tt hep-ph/9312311}}].

\bibitem{Korchemsky:1994jb}
G.~P. Korchemsky and G.~F. Sterman, {\it {Infrared factorization in inclusive B
  meson decays}},  {\em Phys.Lett.} {\bf B340} (1994) 96--108,
  [\href{http://arxiv.org/abs/hep-ph/9407344}{{\tt hep-ph/9407344}}].

\bibitem{Bauer:2003pi}
C.~W. Bauer and A.~V. Manohar, {\it {Shape function effects in $B \to X_s
  \gamma$ and $B \to X_u \ell \bar{\nu}$ decays}},  {\em Phys.Rev.} {\bf D70}
  (2004) 034024, [\href{http://arxiv.org/abs/hep-ph/0312109}{{\tt
  hep-ph/0312109}}].

\bibitem{Beneke:2004in}
M.~Beneke, F.~Campanario, T.~Mannel, and B.~Pecjak, {\it {Power corrections to
  $\bar{B} \to X_u \ell \bar{\nu}\, (X_s \gamma)$ decay spectra in the
  'shape-function' region}},  {\em JHEP} {\bf 0506} (2005) 071,
  [\href{http://arxiv.org/abs/hep-ph/0411395}{{\tt hep-ph/0411395}}].

\bibitem{Bosch:2004th}
S.~Bosch, B.~Lange, M.~Neubert, and G.~Paz, {\it {Factorization and shape
  function effects in inclusive $B$ meson decays}},  {\em Nucl.Phys.} {\bf
  B699} (2004) 335--386, [\href{http://arxiv.org/abs/hep-ph/0402094}{{\tt
  hep-ph/0402094}}].

\bibitem{Lee:2004ja}
K.~S. Lee and I.~W. Stewart, {\it {Factorization for power corrections to $B
  \to X_s \gamma$ and $B \to X_u \ell \bar{nu}$}},  {\em Nucl.Phys.} {\bf B721}
  (2005) 325--406, [\href{http://arxiv.org/abs/hep-ph/0409045}{{\tt
  hep-ph/0409045}}].

\bibitem{Lee:2006wn}
S.~J. Lee, M.~Neubert, and G.~Paz, {\it {Enhanced Non-local Power Corrections
  to the $\bar{B} \to X_s \gamma$ Decay Rate}},  {\em Phys.Rev.} {\bf D75}
  (2007) 114005, [\href{http://arxiv.org/abs/hep-ph/0609224}{{\tt
  hep-ph/0609224}}].

\bibitem{Benzke:2010js}
M.~Benzke, S.~J. Lee, M.~Neubert, and G.~Paz, {\it {Factorization at Subleading
  Power and Irreducible Uncertainties in $\bar B\to X_s\gamma$ Decay}},  {\em
  JHEP} {\bf 1008} (2010) 099, [\href{http://arxiv.org/abs/1003.5012}{{\tt
  arXiv:1003.5012}}].

\bibitem{Bonciani:2008wf}
R.~Bonciani and A.~Ferroglia, {\it {Two-Loop QCD Corrections to the
  Heavy-to-Light Quark Decay}},  {\em JHEP} {\bf 0811} (2008) 065,
  [\href{http://arxiv.org/abs/0809.4687}{{\tt arXiv:0809.4687}}].

\bibitem{Asatrian:2008uk}
H.~Asatrian, C.~Greub, and B.~Pecjak, {\it {NNLO corrections to $\bar{B} \to
  X_u \ell \bar{\nu}$ in the shape-function region}},  {\em Phys.Rev.} {\bf
  D78} (2008) 114028, [\href{http://arxiv.org/abs/0810.0987}{{\tt
  arXiv:0810.0987}}].

\bibitem{Beneke:2008ei}
M.~Beneke, T.~Huber, and X.-Q. Li, {\it {Two-loop QCD correction to
  differential semi-leptonic $b \to u$ decays in the shape-function region}},
  {\em Nucl.Phys.} {\bf B811} (2009) 77--97,
  [\href{http://arxiv.org/abs/0810.1230}{{\tt arXiv:0810.1230}}].

\bibitem{Bell:2008ws}
G.~Bell, {\it {NNLO corrections to inclusive semileptonic $B$ decays in the
  shape-function region}},  {\em Nucl.Phys.} {\bf B812} (2009) 264--289,
  [\href{http://arxiv.org/abs/0810.5695}{{\tt arXiv:0810.5695}}].

\bibitem{Greub:2009sv}
C.~Greub, M.~Neubert, and B.~Pecjak, {\it {NNLO corrections to $\bar{B} \to X_u
  l \bar{\nu} \ell$ and the determination of $|V_{ub}|$}},  {\em Eur.Phys.J.}
  {\bf C65} (2010) 501--515, [\href{http://arxiv.org/abs/0909.1609}{{\tt
  arXiv:0909.1609}}].

\bibitem{Becher:2006qw}
T.~Becher and M.~Neubert, {\it {Toward a NNLO calculation of the $\bar{B} \to
  X_s \gamma$ decay rate with a cut on photon energy. II. Two-loop result for
  the jet function}},  {\em Phys.Lett.} {\bf B637} (2006) 251--259,
  [\href{http://arxiv.org/abs/hep-ph/0603140}{{\tt hep-ph/0603140}}].

\bibitem{Becher:2005pd}
T.~Becher and M.~Neubert, {\it {Toward a NNLO calculation of the $\bar{B} \to
  X_s \gamma$ decay rate with a cut on photon energy: I. Two-loop result for
  the soft function}},  {\em Phys.Lett.} {\bf B633} (2006) 739--747,
  [\href{http://arxiv.org/abs/hep-ph/0512208}{{\tt hep-ph/0512208}}].

\bibitem{Neubert:2004dd}
M.~Neubert, {\it {Renormalization-group improved calculation of the $B \to X_s
  \gamma$ branching ratio}},  {\em Eur.Phys.J.} {\bf C40} (2005) 165--186,
  [\href{http://arxiv.org/abs/hep-ph/0408179}{{\tt hep-ph/0408179}}].

\bibitem{Becher:2006pu}
T.~Becher and M.~Neubert, {\it {Analysis of ${\rm Br}(\bar{B}\to X_s \gamma)$
  at NNLO with a cut on photon energy}},  {\em Phys.Rev.Lett.} {\bf 98} (2007)
  022003, [\href{http://arxiv.org/abs/hep-ph/0610067}{{\tt hep-ph/0610067}}].

\bibitem{Lange:2005yw}
B.~O. Lange, M.~Neubert, and G.~Paz, {\it {Theory of charmless inclusive B
  decays and the extraction of $V_{ub}$}},  {\em Phys.Rev.} {\bf D72} (2005)
  073006, [\href{http://arxiv.org/abs/hep-ph/0504071}{{\tt hep-ph/0504071}}].

\bibitem{Lee:2005pk}
K.~S. Lee and I.~W. Stewart, {\it {Shape-function effects and split matching in
  $B \to X_s \ell^+ \ell^-$}},  {\em Phys.Rev.} {\bf D74} (2006) 014005,
  [\href{http://arxiv.org/abs/hep-ph/0511334}{{\tt hep-ph/0511334}}].

\bibitem{Lee:2005pwa}
K.~S. Lee, Z.~Ligeti, I.~W. Stewart, and F.~J. Tackmann, {\it {Universality and
  $m_X$ cut effects in $B\to X_s \ell^+ \ell^-$}},  {\em Phys.Rev.} {\bf D74}
  (2006) 011501, [\href{http://arxiv.org/abs/hep-ph/0512191}{{\tt
  hep-ph/0512191}}].

\bibitem{Lee:2008xc}
K.~S. Lee and F.~J. Tackmann, {\it {Nonperturbative $m_X$ cut effects in $B\to
  X_s \ell^+ \ell^-$ observables}},  {\em Phys.Rev.} {\bf D79} (2009) 114021,
  [\href{http://arxiv.org/abs/0812.0001}{{\tt arXiv:0812.0001}}].

\bibitem{Bell:2010mg}
G.~Bell, M.~Beneke, T.~Huber, and X.-Q. Li, {\it {Heavy-to-light currents at
  NNLO in SCET and semi-inclusive $\bar{B} \to X_s \ell^+ \ell^-$ decay}},
  {\em Nucl.Phys.} {\bf B843} (2011) 143--176,
  [\href{http://arxiv.org/abs/1007.3758}{{\tt arXiv:1007.3758}}].

\bibitem{Beneke:1999br}
M.~Beneke, G.~Buchalla, M.~Neubert, and C.~T. Sachrajda, {\it {QCD
  factorization for $B\to \pi \pi$ decays: Strong phases and CP violation in
  the heavy quark limit}},  {\em Phys.Rev.Lett.} {\bf 83} (1999) 1914--1917,
  [\href{http://arxiv.org/abs/hep-ph/9905312}{{\tt hep-ph/9905312}}].

\bibitem{Efremov:1979qk}
A.~Efremov and A.~Radyushkin, {\it {Factorization and Asymptotical Behavior of
  Pion Form-Factor in QCD}},  {\em Phys.Lett.} {\bf B94} (1980) 245--250.

\bibitem{Lepage:1980fj}
G.~P. Lepage and S.~J. Brodsky, {\it {Exclusive Processes in Perturbative
  Quantum Chromodynamics}},  {\em Phys.Rev.} {\bf D22} (1980) 2157.

\bibitem{Beneke:2001ev}
M.~Beneke, G.~Buchalla, M.~Neubert, and C.~T. Sachrajda, {\it {QCD
  factorization in $B \to \pi K$, $\pi \pi$ decays and extraction of
  Wolfenstein parameters}},  {\em Nucl.Phys.} {\bf B606} (2001) 245--321,
  [\href{http://arxiv.org/abs/hep-ph/0104110}{{\tt hep-ph/0104110}}].

\bibitem{Beneke:2003zv}
M.~Beneke and M.~Neubert, {\it {QCD factorization for $B \to PP$ and $B\to PV$
  decays}},  {\em Nucl.Phys.} {\bf B675} (2003) 333--415,
  [\href{http://arxiv.org/abs/hep-ph/0308039}{{\tt hep-ph/0308039}}].

\bibitem{Beneke:2000ry}
M.~Beneke, G.~Buchalla, M.~Neubert, and C.~T. Sachrajda, {\it {QCD
  factorization for exclusive, nonleptonic B meson decays: General arguments
  and the case of heavy light final states}},  {\em Nucl.Phys.} {\bf B591}
  (2000) 313--418, [\href{http://arxiv.org/abs/hep-ph/0006124}{{\tt
  hep-ph/0006124}}].

\bibitem{Bauer:2001cu}
C.~W. Bauer, D.~Pirjol, and I.~W. Stewart, {\it {A Proof of factorization for
  $B \to D \pi$}},  {\em Phys.Rev.Lett.} {\bf 87} (2001) 201806,
  [\href{http://arxiv.org/abs/hep-ph/0107002}{{\tt hep-ph/0107002}}].

\bibitem{Mantry:2003uz}
S.~Mantry, D.~Pirjol, and I.~W. Stewart, {\it {Strong phases and factorization
  for color suppressed decays}},  {\em Phys.Rev.} {\bf D68} (2003) 114009,
  [\href{http://arxiv.org/abs/hep-ph/0306254}{{\tt hep-ph/0306254}}].

\bibitem{Chay:2003zp}
J.-g. Chay and C.~Kim, {\it {Factorization of B decays into two light mesons in
  soft collinear effective theory}},  {\em Phys.Rev.} {\bf D68} (2003) 071502,
  [\href{http://arxiv.org/abs/hep-ph/0301055}{{\tt hep-ph/0301055}}].

\bibitem{Chay:2003ju}
J.~Chay and C.~Kim, {\it {Nonleptonic B decays into two light mesons in soft
  collinear effective theory}},  {\em Nucl.Phys.} {\bf B680} (2004) 302--338,
  [\href{http://arxiv.org/abs/hep-ph/0301262}{{\tt hep-ph/0301262}}].

\bibitem{Bauer:2004tj}
C.~W. Bauer, D.~Pirjol, I.~Z. Rothstein, and I.~W. Stewart, {\it {$B \to M_1
  M_2$: Factorization, charming penguins, strong phases, and polarization}},
  {\em Phys.Rev.} {\bf D70} (2004) 054015,
  [\href{http://arxiv.org/abs/hep-ph/0401188}{{\tt hep-ph/0401188}}].

\bibitem{Bauer:2005kd}
C.~W. Bauer, I.~Z. Rothstein, and I.~W. Stewart, {\it {SCET analysis of $B \to
  K \pi$, $B\to K \bar{K}$, and $B \to \pi \pi$ decays}},  {\em Phys.Rev.} {\bf
  D74} (2006) 034010, [\href{http://arxiv.org/abs/hep-ph/0510241}{{\tt
  hep-ph/0510241}}].

\bibitem{Williamson:2006hb}
A.~R. Williamson and J.~Zupan, {\it {Two body B decays with isosinglet final
  states in SCET}},  {\em Phys.Rev.} {\bf D74} (2006) 014003,
  [\href{http://arxiv.org/abs/hep-ph/0601214}{{\tt hep-ph/0601214}}].

\bibitem{Beneke:2005vv}
M.~Beneke and S.~Jager, {\it {Spectator scattering at NLO in non-leptonic b
  decays: Tree amplitudes}},  {\em Nucl.Phys.} {\bf B751} (2006) 160--185,
  [\href{http://arxiv.org/abs/hep-ph/0512351}{{\tt hep-ph/0512351}}].

\bibitem{Beneke:2006mk}
M.~Beneke and S.~Jager, {\it {Spectator scattering at NLO in non-leptonic B
  decays: Leading penguin amplitudes}},  {\em Nucl.Phys.} {\bf B768} (2007)
  51--84, [\href{http://arxiv.org/abs/hep-ph/0610322}{{\tt hep-ph/0610322}}].

\bibitem{Kivel:2006xc}
N.~Kivel, {\it {Radiative corrections to hard spectator scattering in $B \to
  \pi \pi$ decays}},  {\em JHEP} {\bf 0705} (2007) 019,
  [\href{http://arxiv.org/abs/hep-ph/0608291}{{\tt hep-ph/0608291}}].

\bibitem{Bell:2007tv}
G.~Bell, {\it {NNLO vertex corrections in charmless hadronic B decays:
  Imaginary part}},  {\em Nucl.Phys.} {\bf B795} (2008) 1--26,
  [\href{http://arxiv.org/abs/0705.3127}{{\tt arXiv:0705.3127}}].

\bibitem{Pilipp:2007mg}
V.~Pilipp, {\it {Hard spectator interactions in $B \to \pi \pi$ at order
  $\alpha_s^2$}},  {\em Nucl.Phys.} {\bf B794} (2008) 154--188,
  [\href{http://arxiv.org/abs/0709.3214}{{\tt arXiv:0709.3214}}].

\bibitem{Bell:2009nk}
G.~Bell, {\it {NNLO vertex corrections in charmless hadronic B decays: Real
  part}},  {\em Nucl.Phys.} {\bf B822} (2009) 172--200,
  [\href{http://arxiv.org/abs/0902.1915}{{\tt arXiv:0902.1915}}].

\bibitem{Beneke:2009ek}
M.~Beneke, T.~Huber, and X.-Q. Li, {\it {NNLO vertex corrections to
  non-leptonic B decays: Tree amplitudes}},  {\em Nucl.Phys.} {\bf B832} (2010)
  109--151, [\href{http://arxiv.org/abs/0911.3655}{{\tt arXiv:0911.3655}}].

\bibitem{Ali:2001ez}
A.~Ali and A.~Parkhomenko, {\it {Branching ratios for $B \to K^* gamma and B
  \to \rho \gamma$ decays in next-to-leading order in the large energy
  effective theory}},  {\em Eur.Phys.J.} {\bf C23} (2002) 89--112,
  [\href{http://arxiv.org/abs/hep-ph/0105302}{{\tt hep-ph/0105302}}].

\bibitem{Beneke:2001at}
M.~Beneke, T.~Feldmann, and D.~Seidel, {\it {Systematic approach to exclusive
  $B\to V l^+ l^-,\, V \gamma$ decays}},  {\em Nucl.Phys.} {\bf B612} (2001)
  25--58, [\href{http://arxiv.org/abs/hep-ph/0106067}{{\tt hep-ph/0106067}}].

\bibitem{Bosch:2001gv}
S.~W. Bosch and G.~Buchalla, {\it {The Radiative decays $B\to V \gamma$ at
  next-to-leading order in QCD}},  {\em Nucl.Phys.} {\bf B621} (2002) 459--478,
  [\href{http://arxiv.org/abs/hep-ph/0106081}{{\tt hep-ph/0106081}}].

\bibitem{Ali:2006ew}
A.~Ali, G.~Kramer, and G.-h. Zhu, {\it {$B^+ \to K^+l^+l^-$ decay in
  soft-collinear effective theory}},  {\em Eur.Phys.J.} {\bf C47} (2006)
  625--641, [\href{http://arxiv.org/abs/hep-ph/0601034}{{\tt hep-ph/0601034}}].

\bibitem{Ali:2007sj}
A.~Ali, B.~D. Pecjak, and C.~Greub, {\it {$B \to V \gamma$ Decays at NNLO in
  SCET}},  {\em Eur.Phys.J.} {\bf C55} (2008) 577--595,
  [\href{http://arxiv.org/abs/0709.4422}{{\tt arXiv:0709.4422}}].

\bibitem{Bauer:2002aj}
C.~W. Bauer, D.~Pirjol, and I.~W. Stewart, {\it {Factorization and endpoint
  singularities in heavy to light decays}},  {\em Phys.Rev.} {\bf D67} (2003)
  071502, [\href{http://arxiv.org/abs/hep-ph/0211069}{{\tt hep-ph/0211069}}].

\bibitem{Beneke:2003pa}
M.~Beneke and T.~Feldmann, {\it {Factorization of heavy to light form-factors
  in soft collinear effective theory}},  {\em Nucl.Phys.} {\bf B685} (2004)
  249--296, [\href{http://arxiv.org/abs/hep-ph/0311335}{{\tt hep-ph/0311335}}].

\bibitem{Lange:2003pk}
B.~O. Lange and M.~Neubert, {\it {Factorization and the soft overlap
  contribution to heavy to light form-factors}},  {\em Nucl.Phys.} {\bf B690}
  (2004) 249--278, [\href{http://arxiv.org/abs/hep-ph/0311345}{{\tt
  hep-ph/0311345}}].

\bibitem{Becher:2003qh}
T.~Becher, R.~J. Hill, and M.~Neubert, {\it {Soft collinear messengers: A New
  mode in soft collinear effective theory}},  {\em Phys.Rev.} {\bf D69} (2004)
  054017, [\href{http://arxiv.org/abs/hep-ph/0308122}{{\tt hep-ph/0308122}}].

\bibitem{Beneke:2008pi}
M.~Beneke and L.~Vernazza, {\it {$B \to \chi_{cJ} K$ decays revisited}},  {\em
  Nucl.Phys.} {\bf B811} (2009) 155--181,
  [\href{http://arxiv.org/abs/0810.3575}{{\tt arXiv:0810.3575}}].

\bibitem{Arnesen:2006vb}
C.~M. Arnesen, Z.~Ligeti, I.~Z. Rothstein, and I.~W. Stewart, {\it {Power
  Corrections in Charmless Nonleptonic B-Decays: Annihilation is Factorizable
  and Real}},  {\em Phys.Rev.} {\bf D77} (2008) 054006,
  [\href{http://arxiv.org/abs/hep-ph/0607001}{{\tt hep-ph/0607001}}].

\bibitem{Dawson:2013lya}
S.~Dawson, I.~M. Lewis, and M.~Zeng, {\it {Threshold resummed and approximate
  next-to-next-to-leading order results for $W^+W^-$ pair production at the
  LHC}},  {\em Phys.Rev.} {\bf D88} (2013), no.~5 054028,
  [\href{http://arxiv.org/abs/1307.3249}{{\tt arXiv:1307.3249}}].

\bibitem{Wang:2014mqt}
Y.~Wang, C.~S. Li, Z.~L. Liu, and D.~Y. Shao, {\it {Threshold Resummation for
  $WZ$ and $ZZ$ Pair Production at the LHC}},  {\em Phys.Rev.} {\bf D90} (2014)
  034008, [\href{http://arxiv.org/abs/1406.1417}{{\tt arXiv:1406.1417}}].

\bibitem{Li:2014ria}
Y.~Li and X.~Liu, {\it {High precision predictions for exclusive $VH$
  production at the LHC}},  {\em JHEP} {\bf 1406} (2014) 028,
  [\href{http://arxiv.org/abs/1401.2149}{{\tt arXiv:1401.2149}}].

\bibitem{Becher:2011fc}
T.~Becher, C.~Lorentzen, and M.~D. Schwartz, {\it {Resummation for $W$ and $Z$
  production at large $p_T$}},  {\em Phys.Rev.Lett.} {\bf 108} (2012) 012001,
  [\href{http://arxiv.org/abs/1106.4310}{{\tt arXiv:1106.4310}}].

\bibitem{Becher:2013vva}
T.~Becher, G.~Bell, C.~Lorentzen, and S.~Marti, {\it {Transverse-momentum
  spectra of electroweak bosons near threshold at NNLO}},  {\em JHEP} {\bf
  1402} (2014) 004, [\href{http://arxiv.org/abs/1309.3245}{{\tt
  arXiv:1309.3245}}].

\bibitem{Becher:2014tsa}
T.~Becher, G.~Bell, C.~Lorentzen, and S.~Marti, {\it {The transverse-momentum
  spectrum of Higgs bosons near threshold at NNLO}},  {\em JHEP} {\bf 1411}
  (2014) 026, [\href{http://arxiv.org/abs/1407.4111}{{\tt arXiv:1407.4111}}].

\bibitem{Becher:2010pd}
T.~Becher and G.~Bell, {\it {The gluon jet function at two-loop order}},  {\em
  Phys.Lett.} {\bf B695} (2011) 252--258,
  [\href{http://arxiv.org/abs/1008.1936}{{\tt arXiv:1008.1936}}].

\bibitem{Becher:2012za}
T.~Becher, G.~Bell, and S.~Marti, {\it {NNLO soft function for electroweak
  boson production at large transverse momentum}},  {\em JHEP} {\bf 1204}
  (2012) 034, [\href{http://arxiv.org/abs/1201.5572}{{\tt arXiv:1201.5572}}].

\bibitem{Beneke:2009ye}
M.~Beneke, M.~Czakon, P.~Falgari, A.~Mitov, and C.~Schwinn, {\it {Threshold
  expansion of the $gg (q\bar{q}) \to Q\bar{Q}+ X$ cross section at ${\cal
  O}(\alpha^4_s)$}},  {\em Phys.Lett.} {\bf B690} (2010) 483--490,
  [\href{http://arxiv.org/abs/0911.5166}{{\tt arXiv:0911.5166}}].

\bibitem{Beneke:2010da}
M.~Beneke, P.~Falgari, and C.~Schwinn, {\it {Threshold resummation for pair
  production of coloured heavy (s)particles at hadron colliders}},  {\em
  Nucl.Phys.} {\bf B842} (2011) 414--474,
  [\href{http://arxiv.org/abs/1007.5414}{{\tt arXiv:1007.5414}}].

\bibitem{Beneke:2011mq}
M.~Beneke, P.~Falgari, S.~Klein, and C.~Schwinn, {\it {Hadronic top-quark pair
  production with NNLL threshold resummation}},  {\em Nucl.Phys.} {\bf B855}
  (2012) 695--741, [\href{http://arxiv.org/abs/1109.1536}{{\tt
  arXiv:1109.1536}}].

\bibitem{Broggio:2014yca}
A.~Broggio, A.~S. Papanastasiou, and A.~Signer, {\it {Renormalization-group
  improved fully differential cross sections for top pair production}},  {\em
  JHEP} {\bf 1410} (2014) 98, [\href{http://arxiv.org/abs/1407.2532}{{\tt
  arXiv:1407.2532}}].

\bibitem{Ferroglia:2012ku}
A.~Ferroglia, B.~D. Pecjak, and L.~L. Yang, {\it {Soft-gluon resummation for
  boosted top-quark production at hadron colliders}},  {\em Phys.Rev.} {\bf
  D86} (2012) 034010, [\href{http://arxiv.org/abs/1205.3662}{{\tt
  arXiv:1205.3662}}].

\bibitem{Ferroglia:2012uy}
A.~Ferroglia, B.~D. Pecjak, L.~L. Yang, B.~D. Pecjak, and L.~L. Yang, {\it {The
  NNLO soft function for the pair invariant mass distribution of boosted top
  quarks}},  {\em JHEP} {\bf 1210} (2012) 180,
  [\href{http://arxiv.org/abs/1207.4798}{{\tt arXiv:1207.4798}}].

\bibitem{Ferroglia:2013zwa}
A.~Ferroglia, B.~D. Pecjak, and L.~L. Yang, {\it {Top-quark pair production at
  high invariant mass: an NNLO soft plus virtual approximation}},  {\em JHEP}
  {\bf 1309} (2013) 032, [\href{http://arxiv.org/abs/1306.1537}{{\tt
  arXiv:1306.1537}}].

\bibitem{Ferroglia:2013awa}
A.~Ferroglia, S.~Marzani, B.~D. Pecjak, and L.~L. Yang, {\it {Boosted top
  production: factorization and resummation for single-particle inclusive
  distributions}},  {\em JHEP} {\bf 1401} (2014) 028,
  [\href{http://arxiv.org/abs/1310.3836}{{\tt arXiv:1310.3836}}].

\bibitem{Falgari:2012sq}
P.~Falgari, C.~Schwinn, and C.~Wever, {\it {Finite-width effects on threshold
  corrections to squark and gluino production}},  {\em JHEP} {\bf 1301} (2013)
  085, [\href{http://arxiv.org/abs/1211.3408}{{\tt arXiv:1211.3408}}].

\bibitem{Beneke:2013opa}
M.~Beneke, P.~Falgari, J.~Piclum, C.~Schwinn, and C.~Wever, {\it {Higher-order
  soft and Coulomb corrections to squark and gluino production at the LHC}},
  {\em PoS} {\bf RADCOR2013} (2013) 051,
  [\href{http://arxiv.org/abs/1312.0837}{{\tt arXiv:1312.0837}}].

\bibitem{Bauer:2010vu}
C.~W. Bauer, N.~D. Dunn, and A.~Hornig, {\it {Factorization of Boosted Multijet
  Processes for Threshold Resummation}},  {\em Phys.Rev.} {\bf D82} (2010)
  054012, [\href{http://arxiv.org/abs/1002.1307}{{\tt arXiv:1002.1307}}].

\bibitem{Bauer:2010jv}
C.~W. Bauer, N.~D. Dunn, and A.~Hornig, {\it {On the effectiveness of threshold
  resummation away from hadronic endpoint}},
  \href{http://arxiv.org/abs/1010.0243}{{\tt arXiv:1010.0243}}.

\bibitem{Liu:2012zg}
X.~Liu, S.~Mantry, and F.~Petriello, {\it {Gauge-Boson Production with Multiple
  Jets Near Threshold}},  {\em Phys.Rev.} {\bf D86} (2012) 074004,
  [\href{http://arxiv.org/abs/1205.4465}{{\tt arXiv:1205.4465}}].

\bibitem{Kelley:2010fn}
R.~Kelley and M.~D. Schwartz, {\it {1-loop matching and NNLL resummation for
  all partonic 2 to 2 processes in QCD}},  {\em Phys.Rev.} {\bf D83} (2011)
  045022, [\href{http://arxiv.org/abs/1008.2759}{{\tt arXiv:1008.2759}}].

\bibitem{Kelley:2010qs}
R.~Kelley and M.~D. Schwartz, {\it {Threshold hadronic event shapes with
  effective field theory}},  {\em Phys.Rev.} {\bf D83} (2011) 033001,
  [\href{http://arxiv.org/abs/1008.4355}{{\tt arXiv:1008.4355}}].

\bibitem{Kelley:2011tj}
R.~Kelley, M.~D. Schwartz, and H.~X. Zhu, {\it {Resummation of jet mass with
  and without a jet veto}},  \href{http://arxiv.org/abs/1102.0561}{{\tt
  arXiv:1102.0561}}.

\bibitem{Kelley:2011aa}
R.~Kelley, M.~D. Schwartz, R.~M. Schabinger, and H.~X. Zhu, {\it {Jet Mass with
  a Jet Veto at Two Loops and the Universality of Non-Global Structure}},  {\em
  Phys.Rev.} {\bf D86} (2012) 054017,
  [\href{http://arxiv.org/abs/1112.3343}{{\tt arXiv:1112.3343}}].

\bibitem{Chien:2012ur}
Y.-T. Chien, R.~Kelley, M.~D. Schwartz, and H.~X. Zhu, {\it {Resummation of Jet
  Mass at Hadron Colliders}},  {\em Phys.Rev.} {\bf D87} (2013) 014010,
  [\href{http://arxiv.org/abs/1208.0010}{{\tt arXiv:1208.0010}}].

\bibitem{Broggio:2014hoa}
A.~Broggio, A.~Ferroglia, B.~D. Pecjak, and Z.~Zhang, {\it {NNLO hard functions
  in massless QCD}},  {\em JHEP} {\bf 1412} (2014) 005,
  [\href{http://arxiv.org/abs/1409.5294}{{\tt arXiv:1409.5294}}].

\bibitem{GarciaEchevarria:2011rb}
M.~G. Echevarria, A.~Idilbi, and I.~Scimemi, {\it {Factorization Theorem For
  Drell-Yan At Low $q_T$ And Transverse Momentum Distributions
  On-The-Light-Cone}},  {\em JHEP} {\bf 1207} (2012) 002,
  [\href{http://arxiv.org/abs/1111.4996}{{\tt arXiv:1111.4996}}].

\bibitem{Becher:2012fx}
T.~Becher, M.~Neubert, and D.~Wilhelm, {\it {Massive Boson Production at Small
  $q_T$ in Soft-Collinear Effective Theory}},  {\em Nucl.Phys.Proc.Suppl.} {\bf
  234} (2013) 85--88, [\href{http://arxiv.org/abs/1209.4630}{{\tt
  arXiv:1209.4630}}].

\bibitem{Becher:2012yn}
T.~Becher, M.~Neubert, and D.~Wilhelm, {\it {Higgs-Boson Production at Small
  Transverse Momentum}},  {\em JHEP} {\bf 1305} (2013) 110,
  [\href{http://arxiv.org/abs/1212.2621}{{\tt arXiv:1212.2621}}].

\bibitem{Wang:2013qua}
Y.~Wang, C.~S. Li, Z.~L. Liu, D.~Y. Shao, and H.~T. Li, {\it
  {Transverse-Momentum Resummation for Gauge Boson Pair Production at the
  Hadron Collider}},  {\em Phys.Rev.} {\bf D88} (2013) 114017,
  [\href{http://arxiv.org/abs/1307.7520}{{\tt arXiv:1307.7520}}].

\bibitem{Zhu:2012ts}
H.~X. Zhu, C.~S. Li, H.~T. Li, D.~Y. Shao, and L.~L. Yang, {\it
  {Transverse-momentum resummation for top-quark pairs at hadron colliders}},
  {\em Phys.Rev.Lett.} {\bf 110} (2013) 082001,
  [\href{http://arxiv.org/abs/1208.5774}{{\tt arXiv:1208.5774}}].

\bibitem{Li:2013mia}
H.~T. Li, C.~S. Li, D.~Y. Shao, L.~L. Yang, and H.~X. Zhu, {\it {Top quark pair
  production at small transverse momentum in hadronic collisions}},  {\em
  Phys.Rev.} {\bf D88} (2013) 074004,
  [\href{http://arxiv.org/abs/1307.2464}{{\tt arXiv:1307.2464}}].

\bibitem{Catani:2014qha}
S.~Catani, M.~Grazzini, and A.~Torre, {\it {Transverse-momentum resummation for
  heavy-quark hadroproduction}},  {\em Nucl.Phys.} {\bf B890} (2015) 518--538,
  [\href{http://arxiv.org/abs/1408.4564}{{\tt arXiv:1408.4564}}].

\bibitem{Gehrmann:2012ze}
T.~Gehrmann, T.~Lubbert, and L.~L. Yang, {\it {Transverse parton distribution
  functions at next-to-next-to-leading order: the quark-to-quark case}},  {\em
  Phys.Rev.Lett.} {\bf 109} (2012) 242003,
  [\href{http://arxiv.org/abs/1209.0682}{{\tt arXiv:1209.0682}}].

\bibitem{Gehrmann:2014yya}
T.~Gehrmann, T.~Luebbert, and L.~L. Yang, {\it {Calculation of the transverse
  parton distribution functions at next-to-next-to-leading order}},  {\em JHEP}
  {\bf 1406} (2014) 155, [\href{http://arxiv.org/abs/1403.6451}{{\tt
  arXiv:1403.6451}}].

\bibitem{Gaunt:2014xga}
J.~R. Gaunt, M.~Stahlhofen, and F.~J. Tackmann, {\it {The Quark Beam Function
  at Two Loops}},  {\em JHEP} {\bf 1404} (2014) 113,
  [\href{http://arxiv.org/abs/1401.5478}{{\tt arXiv:1401.5478}}].

\bibitem{Gaunt:2014cfa}
J.~Gaunt, M.~Stahlhofen, and F.~J. Tackmann, {\it {The Gluon Beam Function at
  Two Loops}},  {\em JHEP} {\bf 1408} (2014) 020,
  [\href{http://arxiv.org/abs/1405.1044}{{\tt arXiv:1405.1044}}].

\bibitem{Jain:2011iu}
A.~Jain, M.~Procura, and W.~J. Waalewijn, {\it {Fully-Unintegrated Parton
  Distribution and Fragmentation Functions at Perturbative $k_T$}},  {\em JHEP}
  {\bf 1204} (2012) 132, [\href{http://arxiv.org/abs/1110.0839}{{\tt
  arXiv:1110.0839}}].

\bibitem{Gaunt:2014xxa}
J.~R. Gaunt and M.~Stahlhofen, {\it {The Fully-Differential Quark Beam Function
  at NNLO}},  {\em JHEP} {\bf 1412} (2014) 146,
  [\href{http://arxiv.org/abs/1409.8281}{{\tt arXiv:1409.8281}}].

\bibitem{D'Alesio:2014vja}
U.~D'Alesio, M.~G. Echevarria, S.~Melis, and I.~Scimemi, {\it {Non-perturbative
  QCD effects in $q_{T}$ spectra of Drell-Yan and Z-boson production}},  {\em
  JHEP} {\bf 1411} (2014) 098, [\href{http://arxiv.org/abs/1407.3311}{{\tt
  arXiv:1407.3311}}].

\bibitem{Heister:2003aj}
{\bf ALEPH} Collaboration, A.~Heister et~al., {\it {Studies of QCD at $e^+e^-$
  centre-of-mass energies between 91-GeV and 209-GeV}},  {\em Eur.Phys.J.} {\bf
  C35} (2004) 457--486.

\bibitem{Fleming:2007qr}
S.~Fleming, A.~H. Hoang, S.~Mantry, and I.~W. Stewart, {\it {Jets from massive
  unstable particles: Top-mass determination}},  {\em Phys.Rev.} {\bf D77}
  (2008) 074010, [\href{http://arxiv.org/abs/hep-ph/0703207}{{\tt
  hep-ph/0703207}}].

\bibitem{Schwartz:2007ib}
M.~D. Schwartz, {\it {Resummation and NLO matching of event shapes with
  effective field theory}},  {\em Phys.Rev.} {\bf D77} (2008) 014026,
  [\href{http://arxiv.org/abs/0709.2709}{{\tt arXiv:0709.2709}}].

\bibitem{Bauer:2008dt}
C.~W. Bauer, S.~P. Fleming, C.~Lee, and G.~F. Sterman, {\it {Factorization of
  $e^+ e^-$ Event Shape Distributions with Hadronic Final States in Soft
  Collinear Effective Theory}},  {\em Phys.Rev.} {\bf D78} (2008) 034027,
  [\href{http://arxiv.org/abs/0801.4569}{{\tt arXiv:0801.4569}}].

\bibitem{Becher:2008cf}
T.~Becher and M.~D. Schwartz, {\it {A Precise determination of $\alpha_s$ from
  LEP thrust data using effective field theory}},  {\em JHEP} {\bf 0807} (2008)
  034, [\href{http://arxiv.org/abs/0803.0342}{{\tt arXiv:0803.0342}}].

\bibitem{GehrmannDeRidder:2007hr}
A.~Gehrmann-De~Ridder, T.~Gehrmann, E.~Glover, and G.~Heinrich, {\it {NNLO
  corrections to event shapes in $e^+ e^-$ annihilation}},  {\em JHEP} {\bf
  0712} (2007) 094, [\href{http://arxiv.org/abs/0711.4711}{{\tt
  arXiv:0711.4711}}].

\bibitem{Weinzierl:2009ms}
S.~Weinzierl, {\it {Event shapes and jet rates in electron-positron
  annihilation at NNLO}},  {\em JHEP} {\bf 0906} (2009) 041,
  [\href{http://arxiv.org/abs/0904.1077}{{\tt arXiv:0904.1077}}].

\bibitem{Abbate:2010xh}
R.~Abbate, M.~Fickinger, A.~H. Hoang, V.~Mateu, and I.~W. Stewart, {\it {Thrust
  at N$^3$LL with Power Corrections and a Precision Global Fit for
  $\alpha_s(m_Z)$}},  {\em Phys.Rev.} {\bf D83} (2011) 074021,
  [\href{http://arxiv.org/abs/1006.3080}{{\tt arXiv:1006.3080}}].

\bibitem{Agashe:2014kda}
{\bf Particle Data Group} Collaboration, K.~Olive et~al., {\it {Review of
  Particle Physics}},  {\em Chin.Phys.} {\bf C38} (2014) 090001.

\bibitem{Salam:2001bd}
G.~Salam and D.~Wicke, {\it {Hadron masses and power corrections to event
  shapes}},  {\em JHEP} {\bf 0105} (2001) 061,
  [\href{http://arxiv.org/abs/hep-ph/0102343}{{\tt hep-ph/0102343}}].

\bibitem{Mateu:2012nk}
V.~Mateu, I.~W. Stewart, and J.~Thaler, {\it {Power Corrections to Event Shapes
  with Mass-Dependent Operators}},  {\em Phys.Rev.} {\bf D87} (2013) 014025,
  [\href{http://arxiv.org/abs/1209.3781}{{\tt arXiv:1209.3781}}].

\bibitem{Gritschacher:2013pha}
S.~Gritschacher, A.~H. Hoang, I.~Jemos, and P.~Pietrulewicz, {\it {Secondary
  Heavy Quark Production in Jets through Mass Modes}},  {\em Phys.Rev.} {\bf
  D88} (2013) 034021, [\href{http://arxiv.org/abs/1302.4743}{{\tt
  arXiv:1302.4743}}].

\bibitem{Pietrulewicz:2014qza}
P.~Pietrulewicz, S.~Gritschacher, A.~H. Hoang, I.~Jemos, and V.~Mateu, {\it
  {Variable Flavor Number Scheme for Final State Jets in Thrust}},  {\em
  Phys.Rev.} {\bf D90} (2014), no.~11 114001,
  [\href{http://arxiv.org/abs/1405.4860}{{\tt arXiv:1405.4860}}].

\bibitem{Chien:2010kc}
Y.-T. Chien and M.~D. Schwartz, {\it {Resummation of heavy jet mass and
  comparison to LEP data}},  {\em JHEP} {\bf 1008} (2010) 058,
  [\href{http://arxiv.org/abs/1005.1644}{{\tt arXiv:1005.1644}}].

\bibitem{Becher:2012qc}
T.~Becher and G.~Bell, {\it {NNLL Resummation for Jet Broadening}},  {\em JHEP}
  {\bf 1211} (2012) 126, [\href{http://arxiv.org/abs/1210.0580}{{\tt
  arXiv:1210.0580}}].

\bibitem{Lee:2006nr}
C.~Lee and G.~F. Sterman, {\it {Momentum Flow Correlations from Event Shapes:
  Factorized Soft Gluons and Soft-Collinear Effective Theory}},  {\em
  Phys.Rev.} {\bf D75} (2007) 014022,
  [\href{http://arxiv.org/abs/hep-ph/0611061}{{\tt hep-ph/0611061}}].

\bibitem{Dokshitzer:1998kz}
Y.~L. Dokshitzer, A.~Lucenti, G.~Marchesini, and G.~Salam, {\it {On the QCD
  analysis of jet broadening}},  {\em JHEP} {\bf 9801} (1998) 011,
  [\href{http://arxiv.org/abs/hep-ph/9801324}{{\tt hep-ph/9801324}}].

\bibitem{Larkoski:2014uqa}
A.~J. Larkoski, D.~Neill, and J.~Thaler, {\it {Jet Shapes with the Broadening
  Axis}},  {\em JHEP} {\bf 1404} (2014) 017,
  [\href{http://arxiv.org/abs/1401.2158}{{\tt arXiv:1401.2158}}].

\bibitem{Stewart:2010tn}
I.~W. Stewart, F.~J. Tackmann, and W.~J. Waalewijn, {\it {N-Jettiness: An
  Inclusive Event Shape to Veto Jets}},  {\em Phys.Rev.Lett.} {\bf 105} (2010)
  092002, [\href{http://arxiv.org/abs/1004.2489}{{\tt arXiv:1004.2489}}].

\bibitem{Berger:2010xi}
C.~F. Berger, C.~Marcantonini, I.~W. Stewart, F.~J. Tackmann, and W.~J.
  Waalewijn, {\it {Higgs Production with a Central Jet Veto at NNLL$+$NNLO}},
  {\em JHEP} {\bf 1104} (2011) 092, [\href{http://arxiv.org/abs/1012.4480}{{\tt
  arXiv:1012.4480}}].

\bibitem{Jouttenus:2013hs}
T.~T. Jouttenus, I.~W. Stewart, F.~J. Tackmann, and W.~J. Waalewijn, {\it {Jet
  mass spectra in Higgs boson plus one jet at next-to-next-to-leading
  logarithmic order}},  {\em Phys.Rev.} {\bf D88} (2013), no.~5 054031,
  [\href{http://arxiv.org/abs/1302.0846}{{\tt arXiv:1302.0846}}].

\bibitem{Kang:2012zr}
Z.-B. Kang, S.~Mantry, and J.-W. Qiu, {\it {N-Jettiness as a Probe of Nuclear
  Dynamics}},  {\em Phys.Rev.} {\bf D86} (2012) 114011,
  [\href{http://arxiv.org/abs/1204.5469}{{\tt arXiv:1204.5469}}].

\bibitem{Kang:2013nha}
D.~Kang, C.~Lee, and I.~W. Stewart, {\it {Using 1-Jettiness to Measure 2 Jets
  in DIS 3 Ways}},  {\em Phys.Rev.} {\bf D88} (2013) 054004,
  [\href{http://arxiv.org/abs/1303.6952}{{\tt arXiv:1303.6952}}].

\bibitem{Kang:2013lga}
Z.-B. Kang, X.~Liu, and S.~Mantry, {\it {The 1-Jettiness DIS event shape: NNLL
  + NLO results}},  {\em Phys.Rev.} {\bf D90} (2014) 014041,
  [\href{http://arxiv.org/abs/1312.0301}{{\tt arXiv:1312.0301}}].

\bibitem{Kang:2014qba}
D.~Kang, C.~Lee, and I.~W. Stewart, {\it {Analytic calculation of 1-jettiness
  in DIS at $ \mathcal{O}\left({\alpha}_s\right) $}},  {\em JHEP} {\bf 1411}
  (2014) 132, [\href{http://arxiv.org/abs/1407.6706}{{\tt arXiv:1407.6706}}].

\bibitem{Cacciari:2008gp}
M.~Cacciari, G.~P. Salam, and G.~Soyez, {\it {The Anti-$k_t$ jet clustering
  algorithm}},  {\em JHEP} {\bf 0804} (2008) 063,
  [\href{http://arxiv.org/abs/0802.1189}{{\tt arXiv:0802.1189}}].

\bibitem{Salam:2007xv}
G.~P. Salam and G.~Soyez, {\it {A Practical Seedless Infrared-Safe Cone jet
  algorithm}},  {\em JHEP} {\bf 0705} (2007) 086,
  [\href{http://arxiv.org/abs/0704.0292}{{\tt arXiv:0704.0292}}].

\bibitem{Salam:2009jx}
G.~P. Salam, {\it {Towards Jetography}},  {\em Eur.Phys.J.} {\bf C67} (2010)
  637--686, [\href{http://arxiv.org/abs/0906.1833}{{\tt arXiv:0906.1833}}].

\bibitem{Bauer:2008jx}
C.~W. Bauer, A.~Hornig, and F.~J. Tackmann, {\it {Factorization for generic jet
  production}},  {\em Phys.Rev.} {\bf D79} (2009) 114013,
  [\href{http://arxiv.org/abs/0808.2191}{{\tt arXiv:0808.2191}}].

\bibitem{Dasgupta:2001sh}
M.~Dasgupta and G.~Salam, {\it {Resummation of nonglobal QCD observables}},
  {\em Phys.Lett.} {\bf B512} (2001) 323--330,
  [\href{http://arxiv.org/abs/hep-ph/0104277}{{\tt hep-ph/0104277}}].

\bibitem{Banfi:2002hw}
A.~Banfi, G.~Marchesini, and G.~Smye, {\it {Away from jet energy flow}},  {\em
  JHEP} {\bf 0208} (2002) 006, [\href{http://arxiv.org/abs/hep-ph/0206076}{{\tt
  hep-ph/0206076}}].

\bibitem{Kelley:2011ng}
R.~Kelley, M.~D. Schwartz, R.~M. Schabinger, and H.~X. Zhu, {\it {The two-loop
  hemisphere soft function}},  {\em Phys.Rev.} {\bf D84} (2011) 045022,
  [\href{http://arxiv.org/abs/1105.3676}{{\tt arXiv:1105.3676}}].

\bibitem{Hornig:2011iu}
A.~Hornig, C.~Lee, I.~W. Stewart, J.~R. Walsh, and S.~Zuberi, {\it {Non-global
  Structure of the ${\mathcal O}({\alpha}_s^2)$ Dijet Soft Function}},  {\em
  JHEP} {\bf 1108} (2011) 054, [\href{http://arxiv.org/abs/1105.4628}{{\tt
  arXiv:1105.4628}}].

\bibitem{Schwartz:2014wha}
M.~D. Schwartz and H.~X. Zhu, {\it {Nonglobal logarithms at three loops, four
  loops, five loops, and beyond}},  {\em Phys.Rev.} {\bf D90} (2014), no.~6
  065004, [\href{http://arxiv.org/abs/1403.4949}{{\tt arXiv:1403.4949}}].

\bibitem{Trott:2006bk}
M.~Trott, {\it {Jets in Effective Theory: Summing Phase Space Logs}},  {\em
  Phys.Rev.} {\bf D75} (2007) 054011,
  [\href{http://arxiv.org/abs/hep-ph/0608300}{{\tt hep-ph/0608300}}].

\bibitem{Kelley:2012kj}
R.~Kelley, J.~R. Walsh, and S.~Zuberi, {\it {Abelian Non-Global Logarithms from
  Soft Gluon Clustering}},  {\em JHEP} {\bf 1209} (2012) 117,
  [\href{http://arxiv.org/abs/1202.2361}{{\tt arXiv:1202.2361}}].

\bibitem{Kelley:2012zs}
R.~Kelley, J.~R. Walsh, and S.~Zuberi, {\it {Disentangling Clustering Effects
  in Jet Algorithms}},  \href{http://arxiv.org/abs/1203.2923}{{\tt
  arXiv:1203.2923}}.

\bibitem{Ellis:2009wj}
S.~D. Ellis, A.~Hornig, C.~Lee, C.~K. Vermilion, and J.~R. Walsh, {\it
  {Consistent Factorization of Jet Observables in Exclusive Multijet
  Cross-Sections}},  {\em Phys.Lett.} {\bf B689} (2010) 82--89,
  [\href{http://arxiv.org/abs/0912.0262}{{\tt arXiv:0912.0262}}].

\bibitem{Ellis:2010rwa}
S.~D. Ellis, C.~K. Vermilion, J.~R. Walsh, A.~Hornig, and C.~Lee, {\it {Jet
  Shapes and Jet Algorithms in SCET}},  {\em JHEP} {\bf 1011} (2010) 101,
  [\href{http://arxiv.org/abs/1001.0014}{{\tt arXiv:1001.0014}}].

\bibitem{Banfi:2010pa}
A.~Banfi, M.~Dasgupta, K.~Khelifa-Kerfa, and S.~Marzani, {\it {Non-global
  logarithms and jet algorithms in high-$p_T$ jet shapes}},  {\em JHEP} {\bf
  1008} (2010) 064, [\href{http://arxiv.org/abs/1004.3483}{{\tt
  arXiv:1004.3483}}].

\bibitem{vonManteuffel:2013vja}
A.~von Manteuffel, R.~M. Schabinger, and H.~X. Zhu, {\it {The Complete Two-Loop
  Integrated Jet Thrust Distribution In Soft-Collinear Effective Theory}},
  {\em JHEP} {\bf 1403} (2014) 139, [\href{http://arxiv.org/abs/1309.3560}{{\tt
  arXiv:1309.3560}}].

\bibitem{Thaler:2010tr}
J.~Thaler and K.~Van~Tilburg, {\it {Identifying Boosted Objects with
  $N$-subjettiness}},  {\em JHEP} {\bf 1103} (2011) 015,
  [\href{http://arxiv.org/abs/1011.2268}{{\tt arXiv:1011.2268}}].

\bibitem{Banfi:2012yh}
A.~Banfi, G.~P. Salam, and G.~Zanderighi, {\it {NLL+NNLO predictions for
  jet-veto efficiencies in Higgs-boson and Drell-Yan production}},  {\em JHEP}
  {\bf 1206} (2012) 159, [\href{http://arxiv.org/abs/1203.5773}{{\tt
  arXiv:1203.5773}}].

\bibitem{Banfi:2012jm}
A.~Banfi, P.~F. Monni, G.~P. Salam, and G.~Zanderighi, {\it {Higgs and Z-boson
  production with a jet veto}},  {\em Phys.Rev.Lett.} {\bf 109} (2012) 202001,
  [\href{http://arxiv.org/abs/1206.4998}{{\tt arXiv:1206.4998}}].

\bibitem{Alioli:2013hba}
S.~Alioli and J.~R. Walsh, {\it {Jet Veto Clustering Logarithms Beyond Leading
  Order}},  {\em JHEP} {\bf 1403} (2014) 119,
  [\href{http://arxiv.org/abs/1311.5234}{{\tt arXiv:1311.5234}}].

\bibitem{Tackmann:2012bt}
F.~J. Tackmann, J.~R. Walsh, and S.~Zuberi, {\it {Resummation Properties of Jet
  Vetoes at the LHC}},  {\em Phys.Rev.} {\bf D86} (2012) 053011,
  [\href{http://arxiv.org/abs/1206.4312}{{\tt arXiv:1206.4312}}].

\bibitem{Stewart:2013faa}
I.~W. Stewart, F.~J. Tackmann, J.~R. Walsh, and S.~Zuberi, {\it {Jet $p_T$
  Resummation in Higgs Production at $NNLL'+NNLO$}},  {\em Phys.Rev.} {\bf D89}
  (2014) 054001, [\href{http://arxiv.org/abs/1307.1808}{{\tt
  arXiv:1307.1808}}].

\bibitem{Liu:2013hba}
X.~Liu and F.~Petriello, {\it {Reducing theoretical uncertainties for exclusive
  Higgs-boson plus one-jet production at the LHC}},  {\em Phys.Rev.} {\bf D87}
  (2013), no.~9 094027, [\href{http://arxiv.org/abs/1303.4405}{{\tt
  arXiv:1303.4405}}].

\bibitem{Liu:2012sz}
X.~Liu and F.~Petriello, {\it {Resummation of jet-veto logarithms in hadronic
  processes containing jets}},  {\em Phys.Rev.} {\bf D87} (2013) 014018,
  [\href{http://arxiv.org/abs/1210.1906}{{\tt arXiv:1210.1906}}].

\bibitem{Boughezal:2013oha}
R.~Boughezal, X.~Liu, F.~Petriello, F.~J. Tackmann, and J.~R. Walsh, {\it
  {Combining Resummed Higgs Predictions Across Jet Bins}},  {\em Phys.Rev.}
  {\bf D89} (2014) 074044, [\href{http://arxiv.org/abs/1312.4535}{{\tt
  arXiv:1312.4535}}].

\bibitem{Moult:2014pja}
I.~Moult and I.~W. Stewart, {\it {Jet Vetoes interfering with $H \to WW$}},
  {\em JHEP} {\bf 1409} (2014) 129, [\href{http://arxiv.org/abs/1405.5534}{{\tt
  arXiv:1405.5534}}].

\bibitem{Shao:2013uba}
D.~Y. Shao, C.~S. Li, and H.~T. Li, {\it {Resummation Prediction on Higgs and
  Vector Boson Associated Production with a Jet Veto at the LHC}},  {\em JHEP}
  {\bf 1402} (2014) 117, [\href{http://arxiv.org/abs/1309.5015}{{\tt
  arXiv:1309.5015}}].

\bibitem{Jaiswal:2014yba}
P.~Jaiswal and T.~Okui, {\it {Explanation of the $WW$ excess at the LHC by
  jet-veto resummation}},  {\em Phys.Rev.} {\bf D90} (2014), no.~7 073009,
  [\href{http://arxiv.org/abs/1407.4537}{{\tt arXiv:1407.4537}}].

\bibitem{Banfi:2004yd}
A.~Banfi, G.~P. Salam, and G.~Zanderighi, {\it {Principles of general
  final-state resummation and automated implementation}},  {\em JHEP} {\bf
  0503} (2005) 073, [\href{http://arxiv.org/abs/hep-ph/0407286}{{\tt
  hep-ph/0407286}}].

\bibitem{Bauer:2006mk}
C.~W. Bauer and M.~D. Schwartz, {\it {Event Generation from Effective Field
  Theory}},  {\em Phys.Rev.} {\bf D76} (2007) 074004,
  [\href{http://arxiv.org/abs/hep-ph/0607296}{{\tt hep-ph/0607296}}].

\bibitem{Bauer:2011uc}
C.~W. Bauer, F.~J. Tackmann, J.~R. Walsh, and S.~Zuberi, {\it {Factorization
  and Resummation for Dijet Invariant Mass Spectra}},  {\em Phys.Rev.} {\bf
  D85} (2012) 074006, [\href{http://arxiv.org/abs/1106.6047}{{\tt
  arXiv:1106.6047}}].

\bibitem{Bauer:2008qh}
C.~W. Bauer, F.~J. Tackmann, and J.~Thaler, {\it {GenEvA. I. A New framework
  for event generation}},  {\em JHEP} {\bf 0812} (2008) 010,
  [\href{http://arxiv.org/abs/0801.4026}{{\tt arXiv:0801.4026}}].

\bibitem{Bauer:2008qj}
C.~W. Bauer, F.~J. Tackmann, and J.~Thaler, {\it {GenEvA. II. A Phase space
  generator from a reweighted parton shower}},  {\em JHEP} {\bf 0812} (2008)
  011, [\href{http://arxiv.org/abs/0801.4028}{{\tt arXiv:0801.4028}}].

\bibitem{Alioli:2012fc}
S.~Alioli, C.~W. Bauer, C.~J. Berggren, A.~Hornig, F.~J. Tackmann, et~al., {\it
  {Combining Higher-Order Resummation with Multiple NLO Calculations and Parton
  Showers in GENEVA}},  {\em JHEP} {\bf 1309} (2013) 120,
  [\href{http://arxiv.org/abs/1211.7049}{{\tt arXiv:1211.7049}}].

\bibitem{Alioli:2013hqa}
S.~Alioli, C.~W. Bauer, C.~Berggren, F.~J. Tackmann, J.~R. Walsh, et~al., {\it
  {Matching Fully Differential NNLO Calculations and Parton Showers}},  {\em
  JHEP} {\bf 1406} (2014) 089, [\href{http://arxiv.org/abs/1311.0286}{{\tt
  arXiv:1311.0286}}].

\bibitem{Bauer:2002ie}
C.~W. Bauer, A.~V. Manohar, and M.~B. Wise, {\it {Enhanced nonperturbative
  effects in jet distributions}},  {\em Phys.Rev.Lett.} {\bf 91} (2003) 122001,
  [\href{http://arxiv.org/abs/hep-ph/0212255}{{\tt hep-ph/0212255}}].

\bibitem{Bauer:2003di}
C.~W. Bauer, C.~Lee, A.~V. Manohar, and M.~B. Wise, {\it {Enhanced
  nonperturbative effects in $Z$ decays to hadrons}},  {\em Phys.Rev.} {\bf
  D70} (2004) 034014, [\href{http://arxiv.org/abs/hep-ph/0309278}{{\tt
  hep-ph/0309278}}].

\bibitem{Neubert:2007je}
M.~Neubert, {\it {Factorization analysis for the fragmentation functions of
  hadrons containing a heavy quark}},
  \href{http://arxiv.org/abs/0706.2136}{{\tt arXiv:0706.2136}}.

\bibitem{Procura:2009vm}
M.~Procura and I.~W. Stewart, {\it {Quark Fragmentation within an Identified
  Jet}},  {\em Phys.Rev.} {\bf D81} (2010) 074009,
  [\href{http://arxiv.org/abs/0911.4980}{{\tt arXiv:0911.4980}}].

\bibitem{Jain:2011xz}
A.~Jain, M.~Procura, and W.~J. Waalewijn, {\it {Parton Fragmentation within an
  Identified Jet at NNLL}},  {\em JHEP} {\bf 1105} (2011) 035,
  [\href{http://arxiv.org/abs/1101.4953}{{\tt arXiv:1101.4953}}].

\bibitem{Procura:2011aq}
M.~Procura and W.~J. Waalewijn, {\it {Fragmentation in Jets: Cone and Threshold
  Effects}},  {\em Phys.Rev.} {\bf D85} (2012) 114041,
  [\href{http://arxiv.org/abs/1111.6605}{{\tt arXiv:1111.6605}}].

\bibitem{Jain:2012uq}
A.~Jain, M.~Procura, B.~Shotwell, and W.~J. Waalewijn, {\it {Fragmentation with
  a Cut on Thrust: Predictions for B-factories}},  {\em Phys.Rev.} {\bf D87}
  (2013), no.~7 074013, [\href{http://arxiv.org/abs/1207.4788}{{\tt
  arXiv:1207.4788}}].

\bibitem{Bauer:2013bza}
C.~W. Bauer and E.~Mereghetti, {\it {Heavy Quark Fragmenting Jet Functions}},
  {\em JHEP} {\bf 1404} (2014) 051, [\href{http://arxiv.org/abs/1312.5605}{{\tt
  arXiv:1312.5605}}].

\bibitem{Waalewijn:2012sv}
W.~J. Waalewijn, {\it {Calculating the Charge of a Jet}},  {\em Phys.Rev.} {\bf
  D86} (2012) 094030, [\href{http://arxiv.org/abs/1209.3019}{{\tt
  arXiv:1209.3019}}].

\bibitem{Chang:2013rca}
H.-M. Chang, M.~Procura, J.~Thaler, and W.~J. Waalewijn, {\it {Calculating
  Track-Based Observables for the LHC}},  {\em Phys.Rev.Lett.} {\bf 111} (2013)
  102002, [\href{http://arxiv.org/abs/1303.6637}{{\tt arXiv:1303.6637}}].

\bibitem{Chang:2013iba}
H.-M. Chang, M.~Procura, J.~Thaler, and W.~J. Waalewijn, {\it {Calculating
  Track Thrust with Track Functions}},  {\em Phys.Rev.} {\bf D88} (2013)
  034030, [\href{http://arxiv.org/abs/1306.6630}{{\tt arXiv:1306.6630}}].

\bibitem{Chiu:2008vv}
J.-y. Chiu, R.~Kelley, and A.~V. Manohar, {\it {Electroweak Corrections using
  Effective Field Theory: Applications to the LHC}},  {\em Phys.Rev.} {\bf D78}
  (2008) 073006, [\href{http://arxiv.org/abs/0806.1240}{{\tt
  arXiv:0806.1240}}].

\bibitem{Chiu:2009mg}
J.-y. Chiu, A.~Fuhrer, R.~Kelley, and A.~V. Manohar, {\it {Factorization
  Structure of Gauge Theory Amplitudes and Application to Hard Scattering
  Processes at the LHC}},  {\em Phys.Rev.} {\bf D80} (2009) 094013,
  [\href{http://arxiv.org/abs/0909.0012}{{\tt arXiv:0909.0012}}].

\bibitem{Chiu:2009ft}
J.-y. Chiu, A.~Fuhrer, R.~Kelley, and A.~V. Manohar, {\it {Soft and Collinear
  Functions for the Standard Model}},  {\em Phys.Rev.} {\bf D81} (2010) 014023,
  [\href{http://arxiv.org/abs/0909.0947}{{\tt arXiv:0909.0947}}].

\bibitem{Becher:2013zua}
T.~Becher and X.~Garcia~i Tormo, {\it {Electroweak Sudakov effects in W, Z and
  $\gamma$ production at large transverse momentum}},  {\em Phys.Rev.} {\bf
  D88} (2013), no.~1 013009, [\href{http://arxiv.org/abs/1305.4202}{{\tt
  arXiv:1305.4202}}].

\bibitem{Fuhrer:2010vi}
A.~Fuhrer, A.~V. Manohar, and W.~J. Waalewijn, {\it {Electroweak radiative
  Corrections to Higgs Production via Vector Boson Fusion using Soft-Collinear
  Effective Theory}},  {\em Phys.Rev.} {\bf D84} (2011) 013007,
  [\href{http://arxiv.org/abs/1011.1505}{{\tt arXiv:1011.1505}}].

\bibitem{Siringo:2012mi}
F.~Siringo and G.~Buccheri, {\it {Electroweak Radiative Corrections to Higgs
  Production via Vector Boson Fusion using SCET: Numerical Results}},  {\em
  Phys.Rev.} {\bf D86} (2012) 053013,
  [\href{http://arxiv.org/abs/1207.1906}{{\tt arXiv:1207.1906}}].

\bibitem{Manohar:2012rs}
A.~V. Manohar and M.~Trott, {\it {Electroweak Sudakov Corrections and the Top
  Quark Forward-Backward Asymmetry}},  {\em Phys.Lett.} {\bf B711} (2012)
  313--316, [\href{http://arxiv.org/abs/1201.3926}{{\tt arXiv:1201.3926}}].

\bibitem{Ciafaloni:2000df}
M.~Ciafaloni, P.~Ciafaloni, and D.~Comelli, {\it {Bloch-Nordsieck violating
  electroweak corrections to inclusive TeV scale hard processes}},  {\em
  Phys.Rev.Lett.} {\bf 84} (2000) 4810--4813,
  [\href{http://arxiv.org/abs/hep-ph/0001142}{{\tt hep-ph/0001142}}].

\bibitem{Bell:2010gi}
G.~Bell, J.~Kuhn, and J.~Rittinger, {\it {Electroweak Sudakov Logarithms and
  Real Gauge-Boson Radiation in the TeV Region}},  {\em Eur.Phys.J.} {\bf C70}
  (2010) 659--671, [\href{http://arxiv.org/abs/1004.4117}{{\tt
  arXiv:1004.4117}}].

\bibitem{Manohar:2014vxa}
A.~Manohar, B.~Shotwell, C.~Bauer, and S.~Turczyk, {\it {Non-cancellation of
  electroweak logarithms in high-energy scattering}},  {\em Phys.Lett.} {\bf
  B740} (2015) 179--187, [\href{http://arxiv.org/abs/1409.1918}{{\tt
  arXiv:1409.1918}}].

\bibitem{Collins:1981tt}
J.~C. Collins, D.~E. Soper, and G.~F. Sterman, {\it {Does the Drell-Yan cross
  section factorize?}},  {\em Phys.Lett.} {\bf B109} (1982) 388.

\bibitem{Bodwin:1984hc}
G.~T. Bodwin, {\it {Factorization of the Drell-Yan Cross-Section in
  Perturbation Theory}},  {\em Phys.Rev.} {\bf D31} (1985) 2616.

\bibitem{Collins:1985ue}
J.~C. Collins, D.~E. Soper, and G.~F. Sterman, {\it {Factorization for Short
  Distance Hadron - Hadron Scattering}},  {\em Nucl.Phys.} {\bf B261} (1985)
  104.

\bibitem{Bauer:2010cc}
C.~W. Bauer, B.~O. Lange, and G.~Ovanesyan, {\it {On Glauber modes in
  Soft-Collinear Effective Theory}},  {\em JHEP} {\bf 1107} (2011) 077,
  [\href{http://arxiv.org/abs/1010.1027}{{\tt arXiv:1010.1027}}].

\bibitem{Donoghue:2014mpa}
J.~F. Donoghue, B.~K. El-Menoufi, and G.~Ovanesyan, {\it {Regge behavior in
  effective field theory}},  {\em Phys.Rev.} {\bf D90} (2014), no.~9 096009,
  [\href{http://arxiv.org/abs/1405.1731}{{\tt arXiv:1405.1731}}].

\bibitem{Fleming:2014rea}
S.~Fleming, {\it {The role of Glauber exchange in soft collinear effective
  theory and the Balitsky–Fadin–Kuraev–Lipatov Equation}},  {\em
  Phys.Lett.} {\bf B735} (2014) 266--271,
  [\href{http://arxiv.org/abs/1404.5672}{{\tt arXiv:1404.5672}}].

\bibitem{Fadin:1975cb}
V.~S. Fadin, E.~Kuraev, and L.~Lipatov, {\it {On the Pomeranchuk Singularity in
  Asymptotically Free Theories}},  {\em Phys.Lett.} {\bf B60} (1975) 50--52.

\bibitem{Balitsky:1978ic}
I.~Balitsky and L.~Lipatov, {\it {The Pomeranchuk Singularity in Quantum
  Chromodynamics}},  {\em Sov.J.Nucl.Phys.} {\bf 28} (1978) 822--829.

\bibitem{Gaunt:2014ska}
J.~R. Gaunt, {\it {Glauber Gluons and Multiple Parton Interactions}},  {\em
  JHEP} {\bf 1407} (2014) 110, [\href{http://arxiv.org/abs/1405.2080}{{\tt
  arXiv:1405.2080}}].

\bibitem{Idilbi:2008vm}
A.~Idilbi and A.~Majumder, {\it {Extending Soft-Collinear-Effective-Theory to
  describe hard jets in dense QCD media}},  {\em Phys.Rev.} {\bf D80} (2009)
  054022, [\href{http://arxiv.org/abs/0808.1087}{{\tt arXiv:0808.1087}}].

\bibitem{D'Eramo:2010ak}
F.~D'Eramo, H.~Liu, and K.~Rajagopal, {\it {Transverse Momentum Broadening and
  the Jet Quenching Parameter, Redux}},  {\em Phys.Rev.} {\bf D84} (2011)
  065015, [\href{http://arxiv.org/abs/1006.1367}{{\tt arXiv:1006.1367}}].

\bibitem{Ovanesyan:2011xy}
G.~Ovanesyan and I.~Vitev, {\it {An effective theory for jet propagation in
  dense QCD matter: jet broadening and medium-induced bremsstrahlung}},  {\em
  JHEP} {\bf 1106} (2011) 080, [\href{http://arxiv.org/abs/1103.1074}{{\tt
  arXiv:1103.1074}}].

\bibitem{Benzke:2012sz}
M.~Benzke, N.~Brambilla, M.~A. Escobedo, and A.~Vairo, {\it {Gauge invariant
  definition of the jet quenching parameter}},  {\em JHEP} {\bf 1302} (2013)
  129, [\href{http://arxiv.org/abs/1208.4253}{{\tt arXiv:1208.4253}}].

\bibitem{Manohar:2012jr}
A.~V. Manohar and W.~J. Waalewijn, {\it {A QCD Analysis of Double Parton
  Scattering: Color Correlations, Interference Effects and Evolution}},  {\em
  Phys.Rev.} {\bf D85} (2012) 114009,
  [\href{http://arxiv.org/abs/1202.3794}{{\tt arXiv:1202.3794}}].

\bibitem{Manohar:2012pe}
A.~V. Manohar and W.~J. Waalewijn, {\it {What is Double Parton Scattering?}},
  {\em Phys.Lett.} {\bf B713} (2012) 196--201,
  [\href{http://arxiv.org/abs/1202.5034}{{\tt arXiv:1202.5034}}].

\bibitem{Chang:2012nw}
H.-M. Chang, A.~V. Manohar, and W.~J. Waalewijn, {\it {Double Parton
  Correlations in the Bag Model}},  {\em Phys.Rev.} {\bf D87} (2013), no.~3
  034009, [\href{http://arxiv.org/abs/1211.3132}{{\tt arXiv:1211.3132}}].

\bibitem{Beneke:2012xa}
M.~Beneke and G.~Kirilin, {\it {Soft-collinear gravity}},  {\em JHEP} {\bf
  1209} (2012) 066, [\href{http://arxiv.org/abs/1207.4926}{{\tt
  arXiv:1207.4926}}].

\bibitem{Akhoury:2011kq}
R.~Akhoury, R.~Saotome, and G.~Sterman, {\it {Collinear and Soft Divergences in
  Perturbative Quantum Gravity}},  {\em Phys.Rev.} {\bf D84} (2011) 104040,
  [\href{http://arxiv.org/abs/1109.0270}{{\tt arXiv:1109.0270}}].

\bibitem{Schwartz:2013pla}
M.~D. Schwartz, {\em {Quantum Field Theory and the Standard Model}}.
\newblock Cambridge University Press, Cambridge U.K., 2013.

\bibitem{Baumgart:2014vma}
M.~Baumgart, I.~Z. Rothstein, and V.~Vaidya, {\it {On the Annihilation Rate of
  WIMPs}},  \href{http://arxiv.org/abs/1409.4415}{{\tt arXiv:1409.4415}}.

\bibitem{Bauer:2014ula}
M.~Bauer, T.~Cohen, R.~J. Hill, and M.~P. Solon, {\it {Soft Collinear Effective
  Theory for Heavy WIMP Annihilation}},  {\em JHEP} {\bf 1501} (2015) 099,
  [\href{http://arxiv.org/abs/1409.7392}{{\tt arXiv:1409.7392}}].

\bibitem{Ovanesyan:2014fwa}
G.~Ovanesyan, T.~R. Slatyer, and I.~W. Stewart, {\it {Heavy Dark Matter
  Annihilation from Effective Field Theory}},
  \href{http://arxiv.org/abs/1409.8294}{{\tt arXiv:1409.8294}}.

\bibitem{Georgi:2013lza}
H.~Georgi, G.~Kestin, and A.~Sajjad, {\it {The Photon Propagator in Light-Shell
  Gauge}},  \href{http://arxiv.org/abs/1312.1741}{{\tt arXiv:1312.1741}}.

\bibitem{Georgi:2014sxa}
H.~Georgi, G.~Kestin, and A.~Sajjad, {\it {Towards an Effective Field Theory on
  the Light-Shell}},  \href{http://arxiv.org/abs/1401.7667}{{\tt
  arXiv:1401.7667}}.

\bibitem{Korchemsky:1993uz}
G.~Korchemsky and G.~Marchesini, {\it {Resummation of large infrared
  corrections using Wilson loops}},  {\em Phys.Lett.} {\bf B313} (1993)
  433--440.

\bibitem{Schwinger:1960qe}
J.~S. Schwinger, {\it {Brownian motion of a quantum oscillator}},  {\em
  J.Math.Phys.} {\bf 2} (1961) 407--432.

\bibitem{Keldysh:1964ud}
L.~Keldysh, {\it {Diagram technique for nonequilibrium processes}},  {\em
  Zh.Eksp.Teor.Fiz.} {\bf 47} (1964) 1515--1527.

\bibitem{Moch:2005id}
S.~Moch, J.~A.~M. Vermaseren, and A.~Vogt, {\it {The quark form factor at
  higher orders}},  {\em JHEP} {\bf 08} (2005) 049,
  [\href{http://arxiv.org/abs/hep-ph/0507039}{{\tt hep-ph/0507039}}].

\bibitem{Korchemsky:1987wg}
G.~Korchemsky and A.~Radyushkin, {\it {Renormalization of the Wilson Loops
  Beyond the Leading Order}},  {\em Nucl.Phys.} {\bf B283} (1987) 342--364.

\bibitem{Korchemsky:1991zp}
G.~Korchemsky and A.~Radyushkin, {\it {Infrared factorization, Wilson lines and
  the heavy quark limit}},  {\em Phys.Lett.} {\bf B279} (1992) 359--366,
  [\href{http://arxiv.org/abs/hep-ph/9203222}{{\tt hep-ph/9203222}}].

\bibitem{Kidonakis:2009ev}
N.~Kidonakis, {\it {Two-loop soft anomalous dimensions and NNLL resummation for
  heavy quark production}},  {\em Phys.Rev.Lett.} {\bf 102} (2009) 232003,
  [\href{http://arxiv.org/abs/0903.2561}{{\tt arXiv:0903.2561}}].

\end{thebibliography}\endgroup
